# Surface-code Superconducting Quantum Processors: From Calibration to Logical Performance

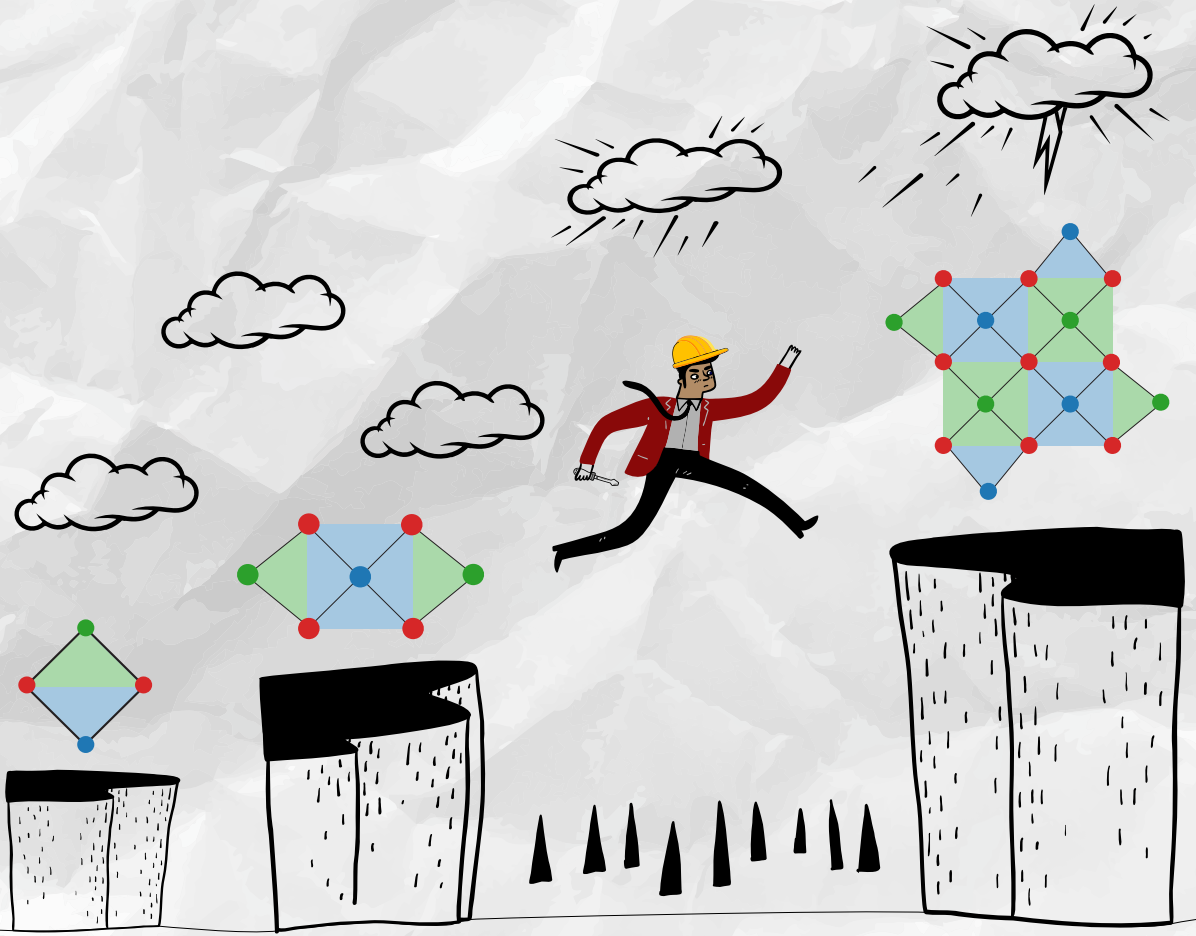

**Hany Ali**

# Surface-code Superconducting Quantum Processors: From Calibration to Logical Performance

# SURFACE-CODE SUPERCONDUCTING QUANTUM PROCESSORS: FROM CALIBRATION TO LOGICAL PERFORMANCE

**Proefschrift**

ter verkrijging van de graad van doctor
aan de Technische Universiteit Delft,
op gezag van de Rector Magnificus prof. dr. ir. T.H.J.J. van der Hagen,
voorzitter van het College voor Promoties,
in het openbaar te verdedigen op Maandag 14 April 2025 om 12:30 uur

door

**Hany ALI**

Ingenieur in Technische Natuurkunde,
Technische Universiteit Delft, Nederland,
geboren te Ismailia, Egypt.

Dit proefschrift is goedgekeurd door de promotoren

  Prof. dr. L. DiCarlo
  Prof. dr. B. M. Terhal

Samenstelling promotiecommissie:

| | |
|---|---|
| Rector Magnificus, | voorzitter |
| Prof. dr. L. DiCarlo | Technische Universiteit Delft, promotor |
| Prof. dr. B. M. Terhal | Technische Universiteit Delft, promotor |
| | |
| *Onafhankelijke leden:* | |
| Prof. dr. J. Bylander | Chalmers University, Sweden |
| Prof. dr. M. P. M. Möttönen | Aalto University, Finland |
| | VTT Technical Research Centre, Finland |
| Dr. A. Potočnik | IMEC, Belgium |
| Prof. dr. ir. L. M. K. Vandersypen | Technische Universiteit Delft |
| Prof. dr. S. Groeblacher, | Technische Universiteit Delft, Reserve member |

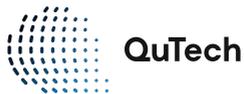

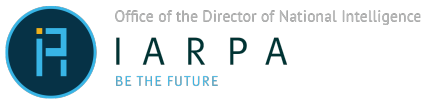

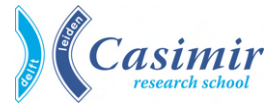

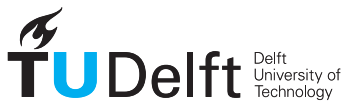

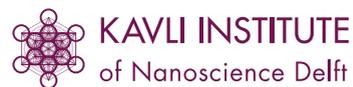

Casimir PhD Series, Delft-Leiden

Cover image credit: Abdallah, Eugene, and Figen.

An electronic version of this dissertation is available at http://repository.tudelft.nl/.

# DEDICATION

I would like to dedicate this work to the soul of my mother, whose love, and memory continue to inspire me every day.

Your son,
Hany Ali

## SUMMARY

Quantum computers hold great potential to solve complex physics and chemistry problems that are beyond the capabilities of classical computers. Unlike classical systems that use bits represented as deterministic 0s and 1s, quantum computers operate with quantum bits or qubits, which can exist in a superposition of states. In addition to superposition, quantum computers harness the power of entanglement and quantum interference that enable quantum computers to encode and process information in ways that classical systems fundamentally cannot. These properties allow quantum computers to explore vast solution spaces in parallel, solving certain computational problems far more efficiently than classical systems. This has numerous potential applications ranging from the factoring of large numbers, optimizing complex systems, computational chemistry, machine learning, cryptography, and artificial intelligence.

Yet, current quantum processors are fragile, noisy and fairly limited in both quantity and quality with tens of qubits and physical error rates of around $10^{-}$. To realize practical quantum applications, however, error rates need to be below $10^{-}$ across millions of qubits. To bridge this gap and fully harness the potential of quantum computers, quantum error correction (QEC) is essential. QEC codes are designed to protect quantum information by redundantly encoding it onto multiple physical qubits. This encoding allows for the detection and correction of local errors affecting individual qubits, e.g., through stabilizer measurements. Importantly, if the physical error rates are below a specific threshold, QEC codes can exponentially suppress logical error rates by increasing the number of physical qubits involved. This is essential for achieving fault-tolerant computations, which are key to unlocking the full potential of quantum computers.

The work presented in this thesis focuses on the implementation and optimization of small-scale QEC experiments using the surface code and flux-tunable superconducting qubits (Transmons). It addresses several key challenges: enhancing two-qubit gate fidelity in Surface-4 (Chapter 2), implementing an error-detection code with Surface-7 (Chapter 3), automating the calibration and benchmarking of the building blocks in Surface-17 (Chapter 4), reducing leakage into higher excited states with leakage reduction units (Chapter 5), assessing and enhancing the performance of logical qubits (Chapters 7 and 8).

Chapter 2 focuses on improving two-qubit gates, a major source of errors in many quantum algorithms. This chapter introduces a diabatic two-qubit scheme known as Sudden Net-Zero (SNZ) for implementing controlled-phase gates in a Surface-4. The SNZ approach achieves outstanding performance, demonstrating two-qubit gate fidelity of $99.93\%$ and minimal leakage of $0.1\%$. A key advantage of the SNZ approach is tuneup simplicity due to the regular structure of gate landscapes as function of the SNZ pulse parameters.

Scaling to a distance-2 surface code (Surface-7) in Chapter 3, we successfully implemented a quantum error detection code. Additionally, we demonstrate a comprehensive suite of logical operations, including initialization, measurement, and arbitrary single-qubit gates. Notably, the performance of fault-tolerant variants outperforms that of non-fault-tolerant variants.

In order to realize a logical-qubit in the surface-code architecture, we extend our quantum device to Surface-17, a distance-3 surface code. Chapter 4 presents automatic calibration strategies for maintaining high performance of single-qubit gates, two-qubit gates, and readout on a 17-transmon device using a homebuilt framework known as Graph-Based Tuneup (GBT). A significant experimental challenge that is addressed in this chapter is mitigating the impact of two-level systems, which can severely compromise the operational fidelities of one- and two-qubit gates as well as readout.

Leakage poses a significant threat to QEC codes due to its non-discretizable nature and the potential spread of correlated errors over time. Chapter 5 details the implementation and extension of a leakage reduction scheme that effectively reduces leakage by up to $99\%$, while minimizing its impact on the qubit subspace. This approach has proven crucial for quickly stabilizing the detection error probability in repeated stabilizers experiment at lower error rates.

We assemble the QEC cycle from the individual building blocks calibrated as detailed in Chapter 4. Constructing the QEC cycle merely from isolated calibrations can result in suboptimal performance. To address this, Chapter 6 presents highly parallelized and orthogonal calibration methods that are designed to optimally calibrate stabilizer measurements as parallel block units. This approach significantly enhances performance by effectively absorbing coherent phase errors caused by residual ZZ couplings and flux crosstalk. Additionally, the chapter presents the benchmarking the performance of repeated weight-2 and weight-4 stabilizer measurements. This serves as a valuable testbed for exploring optimal decoding strategies and for understanding the relationship between logical and physical qubit performance.

Chapter 7 explores error decoding with soft information from analog readout signals. This method stands in contrast to traditional binary decoding and is shown to reduce the logical error rate. The chapter presents experimental results that highlight the advantages of using soft information into the decoding process of a distance-3 bit-flip code.

Finally, Chapter 8 summarizes the findings from each chapter, reflecting on the key challenges and issues that must be addressed to advance the development of practical quantum computers. This chapter offers personal interpretations of the work presented, identifies open questions, and outlines future directions for research.

Kwantumcomputers hebben de potentie om complexe problemen in de natuurkunde en schei-kunde op te lossen die buiten de mogelijkheden van klassieke computers vallen. In tegen-stelling tot klassieke computers die bits gebruiken die deterministisch worden voorgesteld als 0'en en 1'en, werken kwantumcomputers met kwantum bits of qubits, deze kunnen bestaan in een superpositie van toestanden. Naast superpositie, maakt een kwantum computer gebruik van de verstrengeling en kwantuminterferentie die het mogelijk maken om informatie te code-ren en te verwerken op manieren die fundamenteel onmogelijk zijn voor klassieke computers. Deze eigenschappen stellen kwantumcomputers in staat om enorme oplossingsruimtes pa-rallel te verkennen, bepaalde computationele problemen veel efficiënter op te lossen dan klassieke computers. Dit heeft talloze potentiële toepassingen, variërend van het factorise-ren van grote getallen, het optimaliseren van complexe systemen, computationele chemie, machine learning, cryptografie en kunstmatige intelligentie.

Toch zijn huidige kwantumprocessors beperkt in zowel kwantiteit als kwaliteit met tientallen qubits en fysieke foutpercentages van ongeveer $10^-$. Om praktische kwantumtoepassingen te realiseren, moeten deze foutpercentages echter onder $10^-$ liggen voor miljoenen van deze qubits. Om deze kloof te overbruggen en de volledige potentie van kwantumcompu-ters te benutten, is QEC essentieel. QEC-codes zijn ontworpen om kwantuminformatie te beschermen door deze meervoudig te coderen over meerdere qubits. Deze codering maakt het mogelijk om lokale fouten die individuele qubits beïnvloeden te detecteren en te corrige-ren, bijvoorbeeld door stabilisatiemetingen. Wanneer de fysieke foutpercentages onder een specifieke niveau worden gehouden, dan kunnen QEC-codes de logische foutpercentages exponentieel onderdrukken door het aantal betrokken fysieke qubits te verhogen. Dit is es-sentieel voor het bereiken van fouttolerante berekeningen. Dit is cruciaal om het volledige potentieel van kwantumcomputers te ontsluiten.

Het werk dat in deze scriptie wordt gepresenteerd, richt zich op de implementatie en optimalisatie van kleinschalige QEC-experimenten met de surface-code en fluxafstembaar supergeleidende qubits (Transmons). We bespreken de volgende belangrijke punten: het verbeteren van de betrouwbaarheid van twee-qubit operaties in Surface-4 (Chapter 2), het implementeren van een foutdetectiecode met Surface-7 (Chapter 3), het automatiseren van de kalibratie en benchmarking van de bouwstenen in Surface-17 (Chapter 4), het verminderen van lekkage naar hoger energie toestanden met lekkagereductie-eenheden (Chapter 5), en het beoordelen en verbeteren van de prestaties van logische qubits (Hoofdstukken 7 en 8).

Chapter 2 richt zich op het verbeteren van twee-qubit operaties, een belangrijke bron van fouten in veel kwantumalgoritmes. Dit hoofdstuk introduceert een diabatisch twee-qubitschema, bekend als Sudden Net-Zero (SNZ), voor het implementeren van gecontroleerde-fase ope-raties in een Surface-4. De SNZ-aanpak bereikt uitstekende prestaties, met een twee-qubit poortbetrouwbaarheid van $99\ 93\%$ en minimale lekkage van $0\ 1\%$. Een belangrijk voordeel van de SNZ-aanpak is de eenvoud in het kalibreren vanwege een hoge hoeveelheid structuur in de operatie eigenschappen als functie van de SNZ pulse parameters.

Door op te schalen naar een afstand-twee surface code (Surface-7) in Chapter 3, implementeren we met success kwantumfoutdetectie. Daarnaast demonstreren we een uitgebreide reeks logische operaties, waaronder initialisatie, meting en willekeurige enkele-qubit operaties. Opmerkelijk is dat de prestaties van fouttolerante varianten beter zijn dan die van niet-fouttolerante varianten.

Om een logische qubit in de surface code-architectuur te realiseren, moeten we uiteindelijk ons kwantumapparaat uitbreiden naar Surface-17, een afstand-drie surface code. Chapter 4 presenteert automatische kalibratiestrategieën voor het behouden van hoge resultaten van enkele-qubit operaties, twee-qubit operaties en uitlezing op een 17-transmon apparaat met een zelfgebouwd systeem dat bekend staat als Graph-Based Tuneup (GBT). Een belangrijke experimentele uitdaging die GBT adresseert, is het verminderen van twee-staten systeem invloeden, die de operationele betrouwbaarheid van een- en twee-qubit operaties ernstig kunnen aantasten, evenals uitlezing.

Lekkage vormt een aanzienlijke bedreiging voor QEC-codes vanwege de niet-discretiseerbare aard en de potentiële verspreiding van gecorreleerde fouten over tijd. Chapter 5 beschrijft de implementatie en uitbreiding van een lekkagereductieschema dat de lekkage effectief reduceert tot $99\%$, terwijl de impact op de qubit-subruimte wordt geminimaliseerd. Deze aanpak is cruciaal gebleken voor het snel stabiliseren van de foutdetectiewaarschijnlijkheid bij lagere foutpercentages.

We bouwen de QEC-cyclus uit de individueel gekalibreerde bouwstenen zoals gedetailleerd beschreven in Chapter 4. Het opbouwen van de QEC-cyclus van uit geïsoleerde kalibraties kan resulteren in suboptimale resultaten. Om dit aan te pakken, presenteert Chapter 6 geparalleliseerde en orthogonale kalibratiemethoden die zijn ontworpen om stabilisatormetingen als parallelle blokeenheden optimaal te kalibreren. Deze aanpak verbetert de prestaties aanzienlijk door coherentefasefouten die worden veroorzaakt door residuale ZZ-koppelingen en fluxcrosstalk effectief te absorberen. Daarnaast benchmark ik de resultaten van herhaalde gewicht-2 en gewicht-4 stabilisatormetingen. Deze analyse dient als een waardevol testbed voor het verkennen van optimale decoderingsstrategieën en voor het begrijpen van de relatie tussen logische en fysieke qubitprestaties.

Chapter 7 verkent foutdecodering met zachte informatie van analoge uitleessignalen. Deze methode staat in contrast met traditionele binaire decodering en is aangetoond dat het de logische foutenratio vermindert. Het hoofdstuk presenteert experimentele resultaten die de voordelen van het gebruik van zachte informatie in het decoderingsproces van een afstand-drie bit-flip code benadrukken.

Ten slotte vat Chapter 8 de bevindingen van elk hoofdstuk samen, reflecteert op de belangrijkste uitdagingen en kwesties die moeten worden aangepakt om de ontwikkeling van praktische kwantumcomputers te bevorderen. Dit hoofdstuk biedt persoonlijke interpretaties van het gepresenteerde werk, identificeert open vragen en schetst toekomstige richtingen voor onderzoek.

# 1

## 1.1  The power of a quantum computer

I would like to start this introduction with a question: what is a quantum computer? A quantum computer is a physical system engineered to leverage the principles of quantum mechanics to process and manipulate information. Unlike classical computers, which operate with bits represented as deterministic 0s and 1s, quantum computers use qubits, which can exist in a superposition of states. This means a qubit can simultaneously represent a combination of 0 and 1 and can be visualized using a Bloch sphere for a single qubit [1]. Superposition allows quantum systems to encode and process information in ways that classical systems fundamentally cannot.

In addition to superposition, quantum computers harness the power of entanglement—a phenomenon where qubits become correlated in such a way that the state of one qubit cannot be described independently of the state of another, even if they are separated by large distances. Together with quantum interference, these properties enable quantum computers to explore vast solution spaces in parallel, solving certain computational problems far more efficiently than classical systems.

This brings us to the question: why will quantum computing be useful? The field of quantum computing emerged in the late 20th century, inspired by the groundbreaking success of quantum mechanics in explaining the physical world. Physicists in the 1980s and 1990s, including Richard Feynman and David Deutsch, recognized that classical computers face fundamental limitations when addressing certain class of problems. While classical algorithms have achieved extraordinary success, they have significant challenges in solving problems that scale exponentially with size. Examples include factoring large integers, simulating quantum systems, and optimizing complex problems. In these cases, the computational resources required by classical systems—memory, processing power, and time—grow prohibitively large, rendering many problems effectively unsolvable.

Quantum computers, by contrast, promise exponential or quadratic speedups for solving specific problems [2]. One of the first major breakthroughs in quantum computing was Shor's algorithm, developed by Peter Shor in 1994 [3]. This algorithm efficiently solves the integer factorization problem, which is computationally infeasible for classical systems at large scale.

**1**

The best classical algorithm for this task has a heuristic runtime of $\exp(\ ((\log\ )\ (\log\log\ )\ ))$, requiring exponential resources as grows larger. In comparison, Shor's algorithm solves the problem in polynomial time, specifically $((\log\ )\ )$ [4].

The implications of Shor's algorithm extend far beyond mathematics. Factoring large numbers is the foundation of widely used cryptographic protocols, such as RSA encryption [5], which secures modern digital communication and commerce. A sufficiently large quantum computer running Shor's algorithm could break RSA encryption, exposing secure communications to significant risks. Although current quantum hardware is not yet capable of performing such large-scale computations, as will be explained in the next section, the potential impact on global security has driven intense research into quantum computing and post-quantum cryptography.

An early application of quantum computing is quantum simulation—proposed by Richard Feynman in the 1980s [6]. Quantum simulation typically involves calculating the dynamical properties of a system by evolving its state under a given Hamiltonian , as described by the Schrödinger equation, where the evolved state is $\ =\ ^{-}\quad$ for a given time . The exponential complexity of describing general quantum states makes such tasks intractable for classical computers, as no efficient algorithms exist for most cases. Quantum computers, by contrast, can naturally mimic the dynamics of quantum systems that obey the Schrödinger equation, defined by continuous variables, termed "analog quantum simulation". They can also simulate physical quantum systems by executing tailored algorithms using a discrete set of quantum operations, termed "digital quantum simulation". This ability to precisely simulate chemical reactions, molecular properties, and quantum phenomena opens up new frontiers in drug discovery, material science, and energy solutions—areas that remain active fields of research in quantum computing.

Another significant application of quantum computing is the search over unsorted databases, which achieves a quadratic speedup through Grover's algorithm [7]. This algorithm efficiently identifies a solution such that $(\ )=1$ with $(\ ^{\overline{\ }})$ evaluations of , compared to the $(\ )$ evaluations required by classical exhaustive search. Beyond unstructured search, Grover's algorithm has potential applications across a wide range of search tasks and complex optimization problems [8].

To sum up, there exist numerous quantum algorithms demonstrating the potential benefits of quantum computation [8–10]. These algorithms can be applied across various fields, including computational chemistry, machine learning, cryptography, and artificial intelligence.

## 1.2   Building a quantum computer at a scale

Large-scale quantum computing is essential to unlock the full potential of quantum algorithms. For example, factoring a 2000-bit prime number using Shor's algorithm would approximately require a billion physical qubit operations operating over a full day, assuming reasonable estimates of gate time and underlying physical performance [11]. This is far beyond the capabilities of today's quantum processors.

Current quantum computers are categorized as noisy intermediate-scale quantum (NISQ) devices [12]. These devices feature tens to a few hundred noisy qubits with high error rates, $10^-$ to $10^-$ per quantum operation. Over the past few years, NISQ devices have demonstrated significant progress, including the notable realization of quantum advantage [13] and active exploration of a wide range of potential applications, from optimization problems to quantum chemistry [14]. However, these systems remain constrained by noise and errors that limit their ability to outperform classical computers in most practical applications.

I devote this section to the following question: how to experimentally realize a scalable quantum computer? A key guideline for answering this question comes from David DiVincenzo, who outlined five essential criteria for realizing a universal quantum computer [15]. The first requirement is a physical platform that provides a well-defined quantum two-level system. This typically represents the ground, $0$, and excited, $1$, states of a qubit. These states must be sufficiently isolated from environmental noise to prevent energy decay, i.e. losing stored information, and their energy splitting must remain stable to preserve coherence. If a qubit has additional energy levels, we need to keep those higher levels mostly unoccupied during computation. Another key aspect is ensuring that qubits can couple to external fields and to each other for controlled operations. However, this requires a careful balance between making qubits accessible for control and protecting them from noise for preserving coherence.

The second requirement emphasizes the ability to initialize the qubit in a well-defined state, often the ground state, prior to computations. Initialization is part for any quantum algorithm and can sometimes be time-intensive, particularly as coherence times improve. For applications requiring mid-circuit resets, active initialization using projective measurement or external driving becomes essential [16–20].

The third requirement is long coherence and faster gate times. Coherence time refers to the duration a qubit retains its quantum state before decoherence sets in, disrupting the information it encodes. Coherence is characterized by two key timescales: $T$ is the time that a qubit takes to decay from the excited state $1$ to the ground state $0$, driven by unwanted coupling with the surrounding environment. $T$ is the time that a generic superposition state $(0 + 1)$ remains coherent before becoming a mixed state, typically caused by fluctuations in the qubit's energy levels or frequency. Short coherence times directly limit the number of computational steps a quantum computer can perform. This is one of the main bottlenecks in quantum computation, as decoherence errors are almost inevitable during computation.

To mitigate decoherence errors, Peter Shor proposed a quantum error correction (QEC) scheme in 1995 [21]. This employs redundancy to encode quantum information across multiple physical qubits, forming a protected state known as a logical qubit. This protected state can tolerate certain types of errors without compromising the integrity of the computation [4, 11, 22–24]. Assuming errors are small, local, and stable, QEC exponentially sup-

**1**

presses logical qubit errors [25–27]. This makes fault-tolerant quantum computation (FTQC) achievable, unlocking the full potential of quantum systems. Further details on QEC will be discussed in Chapter 6.

While long coherence times are essential, they must be complemented by fast gate operations to optimize the clock speed of a quantum computer. Ideally, the ratio of coherence time to gate time should be high, typically $10$    $10$ . This ensures that a quantum computer can perform a significant number of gate operations within the coherence time. Without faster gates, long coherence times alone are more suited to applications like quantum memory rather than computation. However, faster gates come with a trade-off: they require stronger coupling between qubits or between qubits and external control fields. This increased coupling introduces additional channels for decoherence, creating stringent requirements for the physical implementation of quantum computers.

The fourth requirement is a universal set of quantum gates, comprising single-qubit and two-qubit gates. Together, these gates can approximate any unitary transformation for implementing a certain quantum algorithm, enabling a quantum computer to perform any theoretically possible computation with a specific level of accuracy. Achieving high-fidelity gates is challenging and involves addressing bottlenecks ranging from precise gate calibration to mitigating crosstalk for improving simultaneous operations. Further discussion and techniques can be found in Section 1.4 and Chapter 4.

Finally, the system must provide a reliable readout mechanism to accurately measure qubit states after computation. Measurement causes the qubit state to collapse from a superposition into one of its basis states, $0$  or $1$ . A reliable readout mechanism must ensure that this collapse reflects the true probability distribution of the qubit state at the time of measurement. Readout performance is typically quantified by readout fidelity, which measures the probability of correctly identifying the prepared state. However, fidelity alone does not fully capture readout performance, as measurement can also disturb the qubit state. To address this, readout should ideally be quantum non-demolition (QND), meaning the measurement does not alter or disturb the qubit state being measured. QND readout is particularly crucial for applications requiring mid-circuit measurements, such as QEC schemes. Further details on readout mechanism, calibration, and benchmarking are provided in Chapters 4 and 7.

A wide variety of physical platforms have been proposed to meet these requirements, each with distinct advantages and challenges. Early quantum computing demonstrations used nuclear magnetic resonance (NMR) on complex molecules [28, 29], followed by neutral atoms [30], trapped ions [31], nitrogen-vacancy centers in diamond [32], photonic systems [33], and spin qubits [34]. Another significant avenue for quantum hardware is the family of superconducting qubits [35–38]. In this thesis, the focus is on one specific member of that family: the transmon qubit [36]. Transmons offer reduced sensitivity to charge noise, straightforward scalability through established fabrication techniques, and can achieve large electrical dipole moments due to their relatively large size [36, 39]. This larger dipole moment enhances the interaction between the qubit and external fields, enabling stronger coupling and thus faster gate operations. These features make transmon qubits one of the leading platforms for large-scale quantum computing. In Section 1.4, I discuss how superconducting transmons operate within the circuit quantum electrodynamics framework and how they can be engineered to satisfy DiVincenzo's criteria.

## 1.3 Full-stack superconducting-based quantum computers

Building a full-stack quantum computer is a deeply interdisciplinary effort, requiring expertise from diverse fields to design, fabricate, characterize and optimize both the hardware and software layers needed to realize a scalable, reliable, and high-performance quantum computer. Each layer of the stack provides critical functionality, ranging from device layout to high-level quantum applications. An overview of these layers, which together constitute a full-stack superconducting quantum computer, is presented in Figure 1.1. This figure, inspired by [40], also highlights the challenges associated with each layer. Furthermore, I also choose to mention the Delft quantum startups that are actively addressing the engineering and innovation challenges at each layer of this stack. These include Qblox, Orange Quantum Systems, and QuantWare, all co-founded by graduates from the DiCarlo lab, while Delft Circuits was founded by Sal Jua Bosman, a former graduate of the Steele lab.

The foundational layer of a superconducting quantum computer is the quantum processor, which houses the quantum hardware and serves as the core for qubit measurement and control. This layer involves two main goals: design and fabrication. The design phase focuses on creating the qubit layout and architecture tailored to a target Hamiltonian, ensuring scalability, long coherence times, and high-fidelity operations. The fabrication phase translates these designs into physical devices using micro- and nano-fabrication techniques. QuantWare specializes in designing and fabricating scalable and high-performance quantum processors (QPs). They envision to scale QPs using vertical interconnect (VIO) technology to route all input/output (I/O) connections to a single quantum plane [41]. This allows quantum processors of any size to be constructed by duplicating unit cells and adjusting at the boundaries as needed. However, realizing this ambitious vision introduces numerous engineering challenges, some of which are discussed below.

Unfortunately, superconducting qubit devices require cooling to millikelvin temperatures for characterization, as they cannot be measured at room temperature. The need to wait and measure at cryogenic temperatures introduces time-intensive feedback cycles for design iterations and fabrication optimization. Design inaccuracies and fabrication variations further exacerbate these issues by causing targeting errors in qubit and resonator frequencies, which can lead to significant crosstalk errors (discussed in Chapter 4). Such errors are a substantial obstacle to scaling quantum processors.

**1**

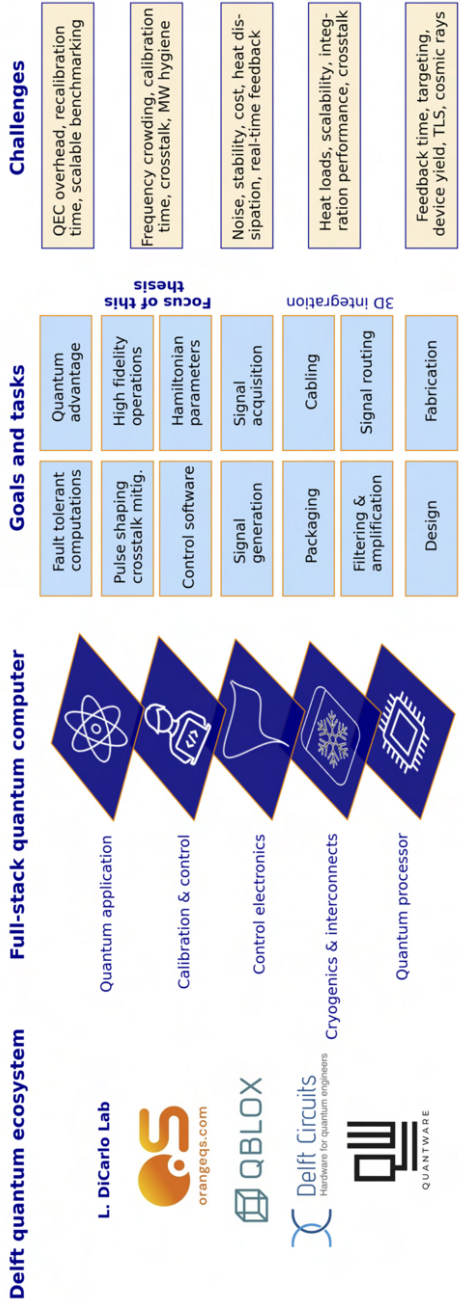

Figure 1.1: **A full-stack quantum computer** consists of multiple layers that must work together to implement a superconducting quantum computer. Various startups within the Delft quantum ecosystem contribute to different layers of this stack. The figure also highlights the goals, tasks, and challenges associated with each layer. A detailed discussion of these layers is provided in the main text. This thesis focuses on calibration and optimization of quantum operations for the implementation of small-scale QEC experiments.

Another critical issue is device yield, which is already a big issue for devices with just a few tens of qubits, as discussed in Section 4.1.2. Superconducting device yield can be affected by frequency crowding, material defects, wafer-scale variations, and other fabrication inconsistencies [42]. In particular, material defects and unintended dielectrics near the qubit or at surface interfaces can behave like two-level system (TLS) [43]. These TLSs can strongly couple to the qubit, reducing coherence times and posing significant challenges for coherent qubit control [44]. Additionally, the device is not shielded from high-energy particles, such as cosmic rays, which can generate quasiparticles that tunnel through the Josephson junctions, induce TLS defects, and reduce coherence globally, leading to correlated errors [45, 46]. Addressing each of these challenges requires independent studies and innovative approaches, as outlined in relevant reviews and references [47–49].

The second layer of the stack focuses on cryogenics and interconnects [Figure 1.1], both of which are essential for the operation of superconducting qubits. The cryogenic setup, typically a dilution refrigerator, ensures that the quantum device operates at millikelvin temperatures to maintain coherence and suppress thermal noise. Interconnects provide signal routing between room-temperature control electronics and the quantum chip at cryogenic temperatures. The cryogenic system incorporates multi-layered shielding and packaging to thermally isolate the device and shield it from environmental noise. Signal routing involves a combination of coaxial cables, attenuators, and filters distributed across various temperature stages, which are necessary for reducing thermal loads and suppressing noise before it reaches the quantum chip. This substantial attenuation weakens the signals, requiring amplification at different cryogenic stages—especially at the output lines—to improve the signal-to-noise ratio.

Delft Circuits, a startup in the Delft quantum ecosystem, has developed flexible stripline technology that integrates filtering and attenuation components directly into cryogenic cabling [50]. This technology provides miniaturized and scalable interconnect solutions. A comparative study with semi-rigid coaxial cables at microwave frequencies showed that flexible striplines do not degrade device coherence [51].

As the qubit count increases, realizing scalable interconnects becomes a significant challenge. Each additional qubit requires dedicated control and readout lines, which introduces substantial passive heat loads. This issue is further exacerbated by active heat loads generated by control signals passing through cables and attenuators at different stages. Together, these loads place stringent requirements on the cryogenic system's cooling capacity and thermal management. Krinner *et al.* nicely reported a thermally optimized, robust cabling scheme and cryogenic setup capable of operating 50 qubits at 14 mK, demonstrating a practical approach for 100-qubit scale devices [52]. However, scaling this approach to systems with hundreds or thousands of qubits is not straightforward, as heat loads and dissipation become increasingly significant.

Another critical challenge in scaling interconnects is maintaining signal integrity while avoiding performance degradation. High-density interconnects increase the risk of crosstalk, where unintended interactions between signal lines lead to noise, reduced qubit fidelity, and errors during operations. This challenge is especially pronounced for integrated components, where compactness comes often at the price of isolation and thermal management. Developing scalable interconnect solutions that balance thermal management, low noise, and high microwave isolation is essential for enabling large-scale superconducting quantum computers.

The third layer of the stack is classical control electronics, responsible for generating, processing, and acquiring signals for qubit control and readout. Control electronics are generally composed of two main modules. First, signal control units generate microwave signals for arbitrary single-qubit gates, provide DC offsets for static biasing, e.g. for flux-tunable transmons, and produce baseband flux pulses for two-qubit gates. Microwave signals are typically synthesized through IQ mixing with local oscillators (LOs), while baseband signals are produced using arbitrary waveform generators (AWGs) [53]. Second, signal acquisition units handle qubit readout by generating baseband signals for readout pulse envelopes, performing IQ modulation and demodulation, and integrating signals to extract qubit state information.

As the quantum device scales, control electronics face major challenges. Signal integrity, noise and signal stability are critical for minimizing phase noise and ensuring high-fidelity operations. Heat dissipation from room-temperature electronics, including DACs, AWGs, and amplifiers, becomes a growing concern as the superconducting device scales. Additionally, real-time feedback and error correction demand ultra-low-latency processing (on the order of hundreds of nanoseconds) to support fault-tolerant architectures.

Qblox, a Delft-based quantum startup, is developing modular, scalable control electronics that integrate microwave signal synthesis, RF modules, DC biasing, and acquisition in a single cluster [54]. Their FPGA-based solutions improve phase coherence and synchronization across multiple modules, offering practical solutions for scaling superconducting qubit control. As quantum processors continue to advance, innovations in control electronics will be essential for addressing these scalability and performance challenges.

The next layer of the stack focuses on calibrating and controlling quantum operations to enable the execution of specific quantum circuits, which is the main topic of this thesis. This involves tuning and optimizing the building blocks of quantum algorithms—such as single-qubit gates, two-qubit gates, and quantum measurement—to achieve high-fidelity operations. Calibration relies on software that sweeps system parameters, acquires data, and analyzes key performance metrics to control quantum hardware. It requires prior characterization of fundamental Hamiltonian parameters, including qubit transition frequencies, resonator frequencies, anharmonicities, coherence times, and coupling strengths. Calibration procedures often incorporate pulse shaping techniques to minimize leakage into higher energy states and reduce crosstalk between qubits or readout resonators.

As the qubit number scales, calibration and control present increasing challenges. For example, frequency crowding, where qubits or resonator frequencies overlap, can induce significant crosstalk errors [55]. Moreover, calibration times grow with system size, necessitating efficient, automated routines to reduce downtime and manual intervention. Maintaining microwave hygiene and mitigating crosstalk are also crucial for preserving high performance during simultaneous operations. Further details on these challenges in a 17-qubit device are discussed in Chapter 4.

The Delft quantum ecosystem is actively tackling these challenges. For example, Orange Quantum Systems (OQS) develops scalable calibration protocols that automate tuning processes for large quantum processors. These automated systems not only save time but also improve the consistency and reliability of calibration. Such advancements, combined with modular control software, are essential for realizing fault-tolerant quantum computation and scaling quantum hardware to hundreds or thousands of qubits.

The final layer of the stack is quantum applications, as discussed in Section 1.1. Current quantum computers remain far from executing these applications, which require fault-tolerant quantum computation and large-scale QEC to realize quantum advantage—i.e., exponential speedups. At the DiCarlo Lab, we develop and optimize small-scale QEC experiments using surface code architectures and flux-tunable transmons. These efforts provide proof-of-concept demonstrations, assess error sensitivity, identify limiting error mechanisms, and guide the development of large-scale quantum computers. I think Delft has a unique quantum ecosystem that together can build a full-stack quantum computer. The OpenSuperQPlus initiative is a perfect example of this effort, aimed at building a full-stack 100-qubit system [56].

## 1.4 Introduction to circuit quantum electrodynamics

Circuit quantum electrodynamics (cQED) studies the interaction of quantized electromagnetic fields in the microwave frequency domain and superconducting circuits [39]. Inspired by advancements in cavity QED, cQED has emerged as a foundational framework for studying light-matter interactions for quantum information and quantum computing [57]. Perhaps, the most simple superconducting circuit is a quantum LC resonator, composed of an inductor with an inductance $L$, and a capacitor with a capacitance $C$ [Figure 1.2. a]. The resonator angular frequency is given by $\omega = 1/\sqrt{LC}$, and its characteristics impedance is $Z = \sqrt{L/C}$. The total energy of this circuit is described by the following Hamiltonian:

$$\hat{H} = 4E_C \hat{n}^2 + \frac{1}{2}E_L \hat{\phi}^2 \tag{1.1}$$

where $E_C = e^2/(2C)$ is the charging energy for adding one electron with charge $e$ on the capacitor islands, and $E_L = (\Phi_0/2\pi)^2/L$ is the inductive energy with $\Phi_0$ being the superconducting magnetic flux quantum. The operators $\hat{n} = \hat{Q}/2e$ (reduced charge) and $\hat{\phi} = 2\pi\hat{\Phi}/\Phi_0$ (reduced flux) are conjugate observables, satisfying the commutation relation $[\hat{\phi}, \hat{n}] = i$. This Hamiltonian is formally equivalent to that of a particle in a one-dimensional quadratic potential, characteristic of a quantum harmonic oscillator (QHO). It is often conve-

**1**

nient to rewrite Equation (1.1) in a more compact form using creation $^{\dagger}$, and annihilation operators:

$$= \quad ^{\dagger} \quad + \frac{1}{2} \tag{1.2}$$

The eigenstates of Equation (1.2) are plotted as a function of the superconducting phase, [Figure 1.2. b]. The energy spectrum exhibits an evenly spaced ladder of eigenvalues, with a frequency spacing given by $= \overline{8} = 1$. This equidistant spacing implies that a QHO cannot function as a qubit, as it lacks a well-defined computational subspace ($0$ and $1$), thereby failing to satisfy the first DiVincenzo's criteria. However, an LC resonator can be used as a quantum bus and to perform quantum measurements via dispersive readout, as discussed below.

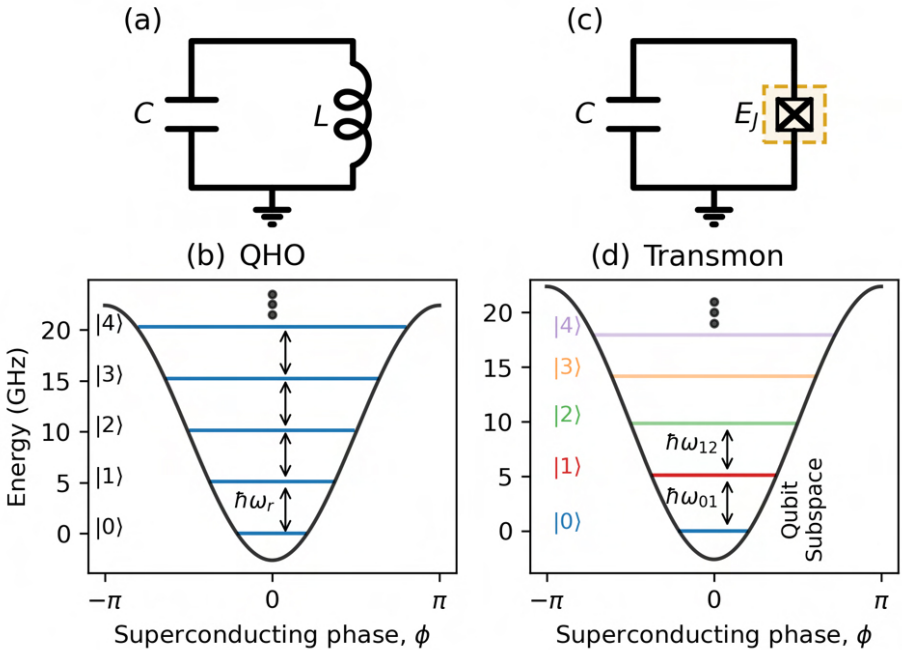

Figure 1.2: **Energy levels of QHO and transmon** as a function of the superconducting phase, . (a) Circuit diagram of a QHO consisting of an inductor and a capacitor. (b) Energy potential and eigenstates of the QHO, showing evenly spaced energy levels ( ) characteristic of a QHO. (c) Circuit diagram of a transmon qubit, composed of a Josephson Junction shunted with a capacitance. (d) Energy potential and eigenstates of the transmon qubit, illustrating an anharmonic energy spectrum with distinct energy spacings. The anharmonicity ( = ) allows a well-defined computational subspace formed by the two lowest energy states ($0$ and $1$). This figure is adapted from reference [58].

A transmon qubit replaces the linear inductor in the QHO with a Josephson junction (JJ) [Figure 1.2. c]. A JJ is superconductor-insulator-superconductor junction that allows dissipationless current to tunnel through [59]. This introduces non-linearity into the potential energy leading to the Hamiltonian:

$$\hat{H} = 4E_C(\hat{n} - n_g)^2 - E_J\cos(\hat{\phi}) \tag{1.3}$$

where $E_C$ is the charging energy as defined earlier, $n_g$ is the offset charge, and $E_J = \frac{I_c\Phi_0}{2\pi}$ is the Josephson energy with $I_c$ as the critical current of the junction. The cosine term in the Hamiltonian results in an anharmonic (non-linear) energy spectrum, breaking the equidistant level spacing of the QHO energy spectrum [Figure 1.2. d]. This anharmonicity enables a well-defined computational subspace using the two lowest energy states, ensuring that transition frequencies between levels are distinct: $\omega_{12} = \omega_{01} + \alpha$, where $\alpha = -\left(\frac{8E_C}{\hbar}\right)$.

The transmon qubit operates in a regime where $E_J/E_C \gg 1$, significantly reducing charge dispersion—the dependence of energy eigenvalues on the charge offset $n_g$. Charge dispersion decreases exponentially with increasing $E_J/E_C$ while the anharmonicity reduces according to a weak power law. This provides a reasonable $\alpha$ (typically a few hundred MHz) for addressing individual transitions during qubit control [58, 60]. In this regime, $n_g$ can often be ignored during standard qubit operations, as its effect is minimal. However, $n_g$ plays a critical role in scenarios involving leakage outside the computational subspace or when measurement induces transitions [61, 62].

### 1.4.1  Dispersive readout

When an LC resonator is capacitively coupled to a transmon qubit, their interaction is described by the Jaynes-Cummings Hamiltonian [39]. The total Hamiltonian of the system includes the individual Hamiltonian of the two circuit elements and their interaction term, under the assumption that the transmon qubit can be approximated as a two-level system and within the rotating-wave approximation (RWA). The Jaynes-Cummings Hamiltonian is given by:

$$\hat{H} = \hbar\omega_r\left(\hat{a}^\dagger\hat{a} + \frac{1}{2}\right) + \frac{\hbar\omega_q}{2}\hat{\sigma}_z + \hbar g\left(\hat{a}\hat{\sigma}_+ + \hat{a}^\dagger\hat{\sigma}_-\right) \tag{1.4}$$

where the first term describes the Hamiltonian of the LC resonator (Equation (1.2)), the second term represents the transmon Hamiltonian (Equation (1.3)) truncated to its first two levels, and the third term describes the interaction between the resonator and the qubit. Here, $\hat{\sigma}_z = |e\rangle\langle e| - |g\rangle\langle g|$ is the Pauli-z operator, $\hat{\sigma}_+ = |e\rangle\langle g|$, and $\hat{\sigma}_- = |g\rangle\langle e|$ are the raising and lowering operators, respectively, that add or remove an excitation from the qubit. The parameter $g$ represents the coupling strength between the resonator and the transmon qubit.

**1**

In the dispersive limit, where the detuning between the qubit and resonator frequencies $\Delta = \omega_q - \omega_r$ is much larger than the coupling strength ($\Delta \gg g$), the interaction between the two systems occur only via virtual photons, enabling non-destructive measurement of the qubit state [63]. In this limit, the dispersive approximation can be applied, and the Hamiltonian of the system can be rewritten as:

$$H = \hbar(\omega_r + \chi\sigma_z)a^\dagger a + \frac{1}{2}\hbar\tilde{\omega}_q\sigma_z \tag{1.5}$$

where $\chi = g^2/\Delta$ is the qubit-state-dependent dispersive shift of the resonator frequency, and $\tilde{\omega}_q = \omega_q + g^2/\Delta$ represents the slight renormalization of the qubit frequency due to zero-point fluctuations of photons in the resonator (also known as the Lamb-shifted frequency of the qubit). Higher energy levels of the transmon modify this dispersive shift. For example, considering the second excited state of the transmon changes the dispersive shift to take this form:

$$\chi = \chi_{01} + \frac{\chi_{12}}{2} = -\frac{g^2}{\Delta}\frac{1}{1 + \Delta/\alpha} \tag{1.6}$$

The dispersive approximation remains valid as long as the photon number in the resonator stays below the critical photon number:

$$n_{crit} = \frac{\Delta^2}{4g^2} \tag{1.7}$$

When the photon number approaches $n_{crit}$, the approximation breaks down due to non-linearities in the transmon-resonator interaction, causing measurement-induced transitions (as discussed in Chapter 5).

### 1.4.2   Single-qubit gates

Single-qubit gates are implemented by driving the qubit transition frequency, $\omega_q$ using precisely controlled microwave pulses [58]. By truncating the energy spectrum to the lowest two levels, the system Hamiltonian can be effectively represented using Pauli matrices $\sigma_x$, $\sigma_y$, and $\sigma_z$. In the RWA, the system Hamiltonian is expressed as:

$$H = \frac{\hbar\omega_q}{2}\sigma_z + \hbar\Omega(t) \tag{1.8}$$

where $\Omega$ is the Rabi frequency, which defines the strength of the applied microwave drive, and $\Omega(t)$ is the time-dependent drive applied to the qubit. After moving into the rotating frame at the qubit frequency, it becomes evident that this Hamiltonian describes coherent rotations of the qubit state on the Bloch sphere [58]. Arbitrary rotations around  and  can be achieved by changing the amplitude and phase of the applied microwave pulse. When approximating the transmon qubit as a two-level system, an  gate of a $180°$ rotation excites the qubit from the ground state $0$ to the excited state $1$ , as illustrated on the Bloch sphere [Figure 1.3. a].

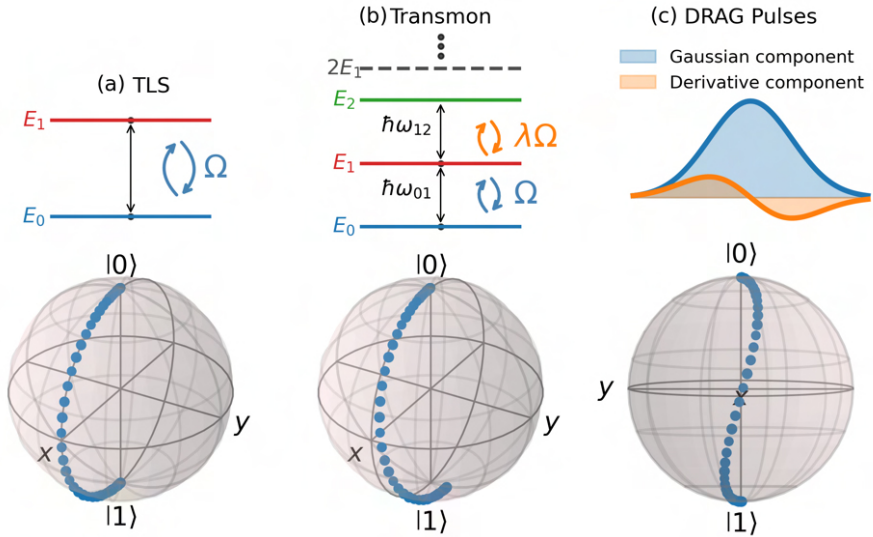

Figure 1.3: **Single-qubit gates and DRAG pulses.** (a) The transmon qubit approximated as a two-level system, showing coherent qubit rotations on the Bloch sphere using microwave drives at the transition frequency . An  gate excites the qubit from the ground state $0$ to the excited state $1$ . (b) Due to the weak anharmonicity of the transmon, pulse imperfections can cause leakage to higher states, such as $2$ , and induce phase errors due to the residual drive at the  transition. (c) The Derivative Removal by Adiabatic Gate (DRAG) pulse envelopes, combining a Gaussian component with a derivative term, effectively minimize these errors by correcting the qubit trajectory on the Bloch sphere and suppressing leakage into non-computational states.

However, the transmon is not a true two-level system. Its weakly anharmonic nature can lead to leakage into higher energy states, such as $2$ , due to pulse imperfections. The weak anharmonicity of the transmon, typically a few hundred MHz, has a non-zero probability of spectral overlap between the applied microwave pulse and the leakage transition , resulting in excitation outside the computational subspace. This undesired residual drive to the $2$ state, characterized by a coupling term  [Figure 1.3. b], also causes a repulsion between the $1$ and $2$ states. Consequently, the drive at  is distorted, leading to phase errors on the Bloch sphere.

To address these issues, pulse-shaping techniques such as the Derivative Removal by Adiabatic Gate (DRAG) are employed [64, 65]. The DRAG approach modifies the control pulses by adding a derivative component to the Gaussian drive pulse [Figure 1.3. c]. This changes the qubit trajectory during an     gate around the Bloch sphere, effectively minimizing leakage and phase errors. More details about single-qubit gate calibration with DRAG pulses can be found in Chapter 4.

### 1.4.3  Flux-tunable transmon and two-qubit gates

Flux-tunable transmons extend the standard transmon design by incorporating a superconducting quantum interference device (SQUID) instead of a single JJ. This SQUID consists of two JJs in a loop, allowing the effective Josephson energy to be controlled by an external magnetic flux threading through the loop [Figure 1.4. a]. The Hamiltonian for this system, when the JJs are identical, is given by:

$$= 4 \quad ( \qquad ) \quad 2 \quad \cos \quad \underbrace{\frac{\text{ext}}{}}_{\text{eff}} \quad \cos \qquad (1.9)$$

where   $_{\text{ext}}$ is the externally applied magnetic flux, and    is the average phase difference across the the two junctions. By varying   $_{\text{ext}}$, the effective Josephson energy         becomes tunable, thereby tuning the energy transition frequencies. An example of this tunability shows the dependencies of the transition energies         and       on   $_{\text{ext}}$ [ Figure 1.4. b]. It is important to note that the Josephson energies         and       might not be identical, due either to unintentional fabrication variations or intentionally by targeting asymmetric junctions. The latter has a profound effect on the qubit frequency sensitivity to the applied flux [58].

Although this tunability provides an extra degree of freedom essential for achieving fast quantum gates, it comes with the downside of additional loss channel due to flux noise, which leads to qubit dephasing. To minimize the impact of flux noise, it is preferable to operate the qubit at the flux symmetry point—often referred to as the sweetspot—where the sensitivity to flux noise is minimized to second order [Figure 1.4. b]. However, if a TLS defect strongly couples to the qubit at this sweetspot, coherence can be significantly reduced. To mitigate this, the qubit may be intentionally detuned from the sweetspot to retrieve good coherence, as discussed in Chapter 4.

For achieving high-fidelity two-qubit gates, our group utilizes dynamic baseband flux pulses to implement controlled-Z gates, harnessing the transverse coupling       [Figure 1.4. d] between a computational state $11$ and a non-computational state such as $02$ [66, 67]. Compared to other approaches, baseband flux pulses achieve the fastest controlled-   (a special case of CPHASE), operating near or at the speed limit       =       [68]. This is evident from the energy levels in the one- and two-excitation manifold, where the CPHASE avoided crossing comes first as a function of external magnetic flux [Figure 1.4. c]. Further details on the gate Hamiltonian, our flux pulse strategy, and the gate tune-up procedures are discussed in Chapter 2.

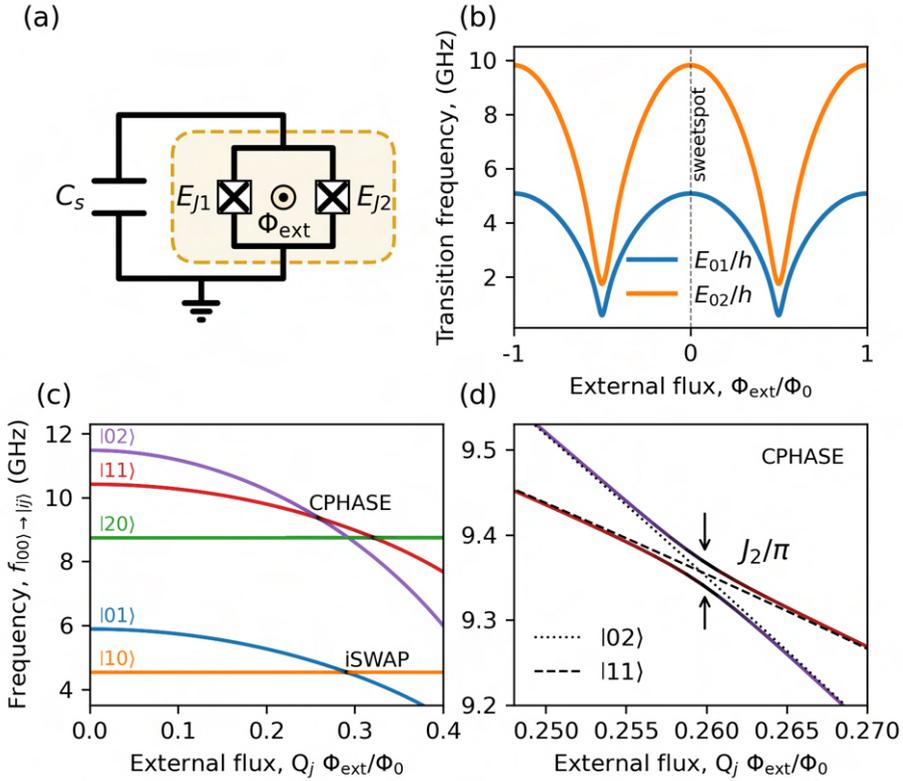

Figure 1.4: **Flux-tunable transmons and flux-based two-qubit gates** (a) Circuit diagram of a flux-tunable transmon featuring a SQUID loop shunted with a capacitor and controlled by an external flux, $\Phi_{ext}$. (b) Transition frequencies, $E_{01}$ and $E_{02}$, as a function of the external flux, showing the sweetspot where sensitivity to flux noise is minimized. (c) Transition frequencies from $00$ to levels $ij$ in the one- and two-excitation manifold for a two-qubit pair as a function of magnetic flux through the SQUID loop of the high-frequency qubit, normalized to the flux quantum $\Phi_0$. Two avoided crossings are highlighted, corresponding to operational points for controlled-phase (CPHASE) and iSWAP interactions. (d) Zoom-in to the $11$ - $02$ avoided crossing. The minimum frequency splitting corresponds to $1$ $J_2$ = $\pi$. The dashed and dotted lines indicate the bare levels of the states $11$ and $02$, respectively.

## 1.5 Thesis overview

This thesis focuses on the calibration and optimization of small-scale QEC experiments using flux-tunable transmons embedded into the surface code layout. The work addresses essential

**1**

challenges, ranging from the calibration of high-fidelity quantum operations to enhancing logical qubit performance in a distance-3 surface code.

Chapter 1 introduces the fundamental concepts of quantum computation, cQED framework, and the building blocks of a small-scale quantum processor based on superconducting transmons. Quantum computers hold the promise of solving complex problems in physics and chemistry that are beyond the capabilities of current classical computers. However, significant challenges exist in building a full-stack quantum computer. In particular, the fragile nature of qubits leads to the loss of quantum information after only a few cycles. To improve the integrity of quantum computation and unlock the full potential of quantum computers, this chapter discusses the need for a QEC scheme, which provides a promising solution for fault-tolerant quantum computations.

Chapter 2 focuses on improving two-qubit gates, a major source of errors in many quantum algorithms. The chapter presents the Sudden Net-Zero (SNZ) variant of the Net Zero scheme, enabling the realization of high-fidelity controlled-Z gates. A key advantage of the SNZ approach is its simplicity, as it facilitates easy tuning due to the regular structure of conditional phase and leakage as functions of flux pulse control parameters. The SNZ scheme is compatible with QEC protocols and can be generalized to arbitrary conditional-phase gates, making it an good choice for both fault-tolerant quantum computing and near-term quantum applications.

Chapter 3 demonstrates the implementation of a quantum error detection code using a distance-2 surface code, referred to as Surface-7. This chapter highlights the realization of logical operations, including initialization, measurement, and a universal set of single-qubit gates. It also shows that fault-tolerant operations outperform non-fault-tolerant variants, with detailed characterization using process tomography of logical gates.

Chapter 4 addresses the challenges of automating the calibration and benchmarking of superconducting quantum computers. The chapter introduces an automatic framework for calibrating single-qubit gates, two-qubit gates, and readout on a 17-transmon device. This framework allows for a largely hands-off approach to calibration, while also exploring the manual interventions required when parasitic interactions with TLS defects occur. A comparative analysis of individual and simultaneous device performance is included, highlighting the impact of crosstalk on device performance and emphasizing the need for robust calibration methods as quantum systems scale.

Chapter 5 tackles the issue of leakage in multi-level systems, such as transmons. Leakage is particularly problematic for QEC as it falls outside the stabilizer formalism. The chapter implements and extends the leakage reduction unit scheme, which effectively reduces leakage to the second- and third-excited transmon states with minimal impact on the qubit subspace. As an immediate application of the leakage reduction unit in a QEC setup, the chapter demonstrates how they can suppress leakage buildup and reduce error detection rates during repeated stabilizer measurements.

Chapter 6 introduces calibration strategies for optimizing the QEC cycle in a distance-3 surface code. The chapter describes the realization of a bit-flip distance-3 surface code, known as Surface-13. This serves as a testbed for exploring decoding strategies and understanding the relationship between logical and physical qubit performance. Additionally, the chapter investigates the stabilization of logical states using X and Z stabilizers and evaluates measured defect rates over multiple rounds, offering valuable insights into optimizing QEC codes for larger systems.

Chapter 7 explores improvements in error decoding by incorporating soft information from analog readout signals. This approach, which contrasts with traditional binary decoding, reduces the logical error rate. The chapter presents experimental results demonstrating the advantages of soft information in the decoding process, offering further insights into optimizing the performance of surface codes and highlighting its potential for broader applicability across different quantum platforms.

Finally, in Chapter 8, the thesis concludes with a summary of the findings from each chapter, reflecting on the key challenges and issues that must be addressed to develop practical quantum computers. The chapter provides personal interpretations on the work presented and identifies open questions and future directions for research.

# HIGH-FIDELITY CONTROLLED-Z GATE WITH MAXIMAL INTERMEDIATE LEAKAGE OPERATING AT THE SPEED LIMIT IN A SUPERCONDUCTING QUANTUM PROCESSOR

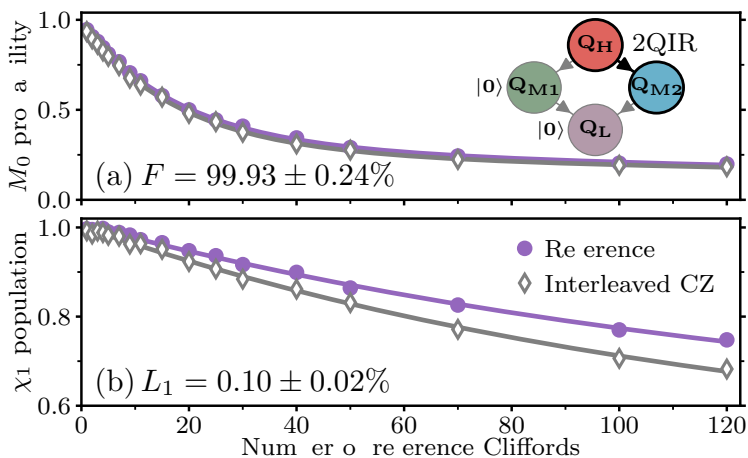

Simple tuneup of fast two-qubit gates is essential for the scaling of quantum processors. We introduce the sudden variant (SNZ) of the Net Zero scheme realizing controlled- (CZ) gates by flux control of transmon frequency. SNZ CZ gates realized in a multi-transmon processor operate at the speed limit of transverse coupling between computational and non-computational states by maximizing intermediate leakage. Beyond speed, the key advantage of SNZ is tuneup simplicity, owing to the regular structure of conditional phase and leakage as a function of two control parameters. SNZ is compatible with scalable schemes for quantum error correction and adaptable to generalized conditional-phase gates useful in intermediate-scale applications.

**2**

## 2.1 Historical context

Superconducting quantum processors have recently reached important milestones [70], notably the demonstration of quantum supremacy on a 53-transmon processor [13]. On the path to quantum error correction (QEC) and fault tolerance [71], recent experiments have used repetitive parity measurements to stabilize two-qubit entanglement [72, 73] and to perform surface-code quantum error detection in a 7-transmon processor [74]. These developments have relied on two-qubit controlled-phase (CPhase) gates realized by dynamical flux control of transmon frequency, harnessing the transverse coupling $J_2$ between a computational state $|11\rangle$ and a non-computational state such as $|02\rangle$ [66, 67]. Compared to other implementations, e.g., cross-resonance using microwave-frequency pulses [75] and parametric radio-frequency pulsing [76], baseband flux pulses achieve the fastest controlled-$Z$ (CZ) gates (a special case of CPhase), operating near the speed limit $t_p = \sqrt{2}/J_2$ [68].

Over the last decade, baseband flux pulsing for two-qubit gating has evolved in an effort to increase gate fidelity and to reduce leakage and residual $ZZ$ coupling. In particular, leakage became a main focus for its negative impact on QEC, adding complexity to error-decoder design [77] and requiring hardware and operational overhead to seep [78–82]. To reduce leakage from linear-dynamical distortion in flux-control lines and limited time resolution in AWGs, unipolar square pulses [67, 83] have been superseded by softened counterparts [84, 85] based on fast-adiabatic theory [86]. In parallel, coupling strengths have reduced to $J_2/2\pi \approx 10-20\,\mathrm{MHz}$ to mitigate residual $ZZ$ coupling, which affects single-qubit gates and idling at bias points, and produces crosstalk from spectator qubits [87]. Many groups are actively developing tunable coupling schemes to suppress residual coupling without incurring slowdown [88–92].

A main limitation to the fidelity of flux-based CPhase gates is dephasing from flux noise, as one qubit is displaced $0.5-1\,\mathrm{GHz}$ below its flux-symmetry point (i.e., sweetspot [93]) to reach the $|11\rangle$ - $|02\rangle$ resonance. To address this limitation, Ref. [94] introduced a bipolar variant [termed Net Zero (NZ)] of the fast-adiabatic scheme, which provides a built-in echo reducing the impact of low-frequency flux noise. The double use of the transverse interaction also reduces leakage by destructive interference, as understood by analogy to a Mach-Zehnder interferometer (MZI). Finally, the zero-average characteristic avoids the buildup of long-timescale distortions in the flux-control lines, significantly improving gate repeatability. NZ pulsing has been successfully used in several recent experiments [72, 74, 95], elevating the state of the art for CZ gate fidelity to $99.72 \pm 0.35\%$ [95]. However, NZ suffers from complicated tuneup, owing to the complex dependence of conditional phase and leakage on fast-adiabatic pulse parameters. This limits the use of NZ for two-qubit gating as processors grow in qubit count.

In this chapter, we introduce the sudden variant (SNZ) of the NZ scheme implementing CZ, which offers two advantages while preserving the built-in echo, destructive leakage interference, and repeatability characteristic of conventional NZ (CNZ). First, SNZ operates at the speed limit of transverse coupling by maximizing intermediate leakage to the non-computational state. The second and main advantage is greatly simplified tuneup: the landscapes of conditional phase and leakage as a function of two pulse parameters have regular structure and interrelation, easily understood by exact analogy to the MZI. We realize SNZ CZ gates among four pairs of nearest neighbors in a seven-transmon processor and characterize their performance using two-qubit interleaved randomized benchmarking (2QIRB) with modifications to quantify leakage [94, 96, 97]. The highest performance achieved from one 2QIRB characterization has $99.93 \pm 0.24\%$ fidelity and $0.10 \pm 0.02\%$ leakage. SNZ CZ gates are fully compatible with scalable approaches to QEC [98]. The generalization of SNZ to arbitrary CPhase gates is straightforward and useful for optimization [99], quantum simulation [100], and NISQ applications [12].

## 2.2 Concept

A flux pulse to the $11$-$02$ interaction implements the unitary

$$
U = \begin{pmatrix}
1 & 0 & 0 & 0 & 0 \\
0 & e^{i\phi_{01}} & 0 & 0 & 0 \\
0 & 0 & e^{i\phi_{10}} & 0 & 0 \\
0 & 0 & 0 & \sqrt{1-4} \, e^{i\phi_{11}} & \sqrt{4} \, e^{i\phi_{02,11}} \\
0 & 0 & 0 & \sqrt{4} \, e^{i\phi_{11,02}} & \sqrt{1-4} \, e^{i\phi_{02}}
\end{pmatrix}
$$

in the $\{00, 01, 10, 11, 02\}$ subspace, neglecting decoherence and residual interaction between far off-resonant levels. Here, $\phi_{01}$ and $\phi_{10}$ are the single-qubit phases, $\phi = \phi_{11} + \phi_{01} + \phi_{10}$, where $\phi$ is the conditional phase, and is the leakage, The ideal CZ gate simultaneously achieves $\phi = \phi = 0 \,(\mathrm{mod}\, 2\pi)$, $\phi = \pi \,(\mathrm{mod}\, 2\pi)$ (phase condition PC), and $= 0$ (leakage condition LC), with arbitrary .

The SNZ CZ gate is realized with two square half pulses with equal and opposite amplitude and duration $2$ each. To understand its action, consider first the ideal scenario with perfectly square half pulses (infinite bandwidth), infinite time resolution, $=$, and $= 1$ (corresponding to $11$ and $02$ on resonance). The unitary action of each complete half pulse (rising edge, steady level, and falling edge combined) implements one of two beamsplitters in the MZI analogy: $\mathrm{BS1}$ fully transmits $11$ to $02$ (producing maximal intermediate leakage), and $\mathrm{BS2}$ fully transmits $02$ to $11$, yielding an ideal CZ gate. SNZ adds an idling period between the half pulses to perfect the analogy to the MZI, allowing accrual of relative phase between $02$ and $11$ in between the beamsplitters.

The key advantage of SNZ over CNZ is the straightforward procedure to simultaneously meet PC and LC. To appreciate this, consider the landscapes of and as a function of and [Fig. 1(c, d)] in this ideal scenario. The landscapes have a clear structure and link to each other. The landscape shows a vertical leakage valley at $= 1$ arising from perfect transmission at each beamsplitter (LC1), and also two vertical valleys arising from perfect reflection (LC2). Leakage interference gives rise to additional diagonal valleys (LC3). Crucially,

**2**

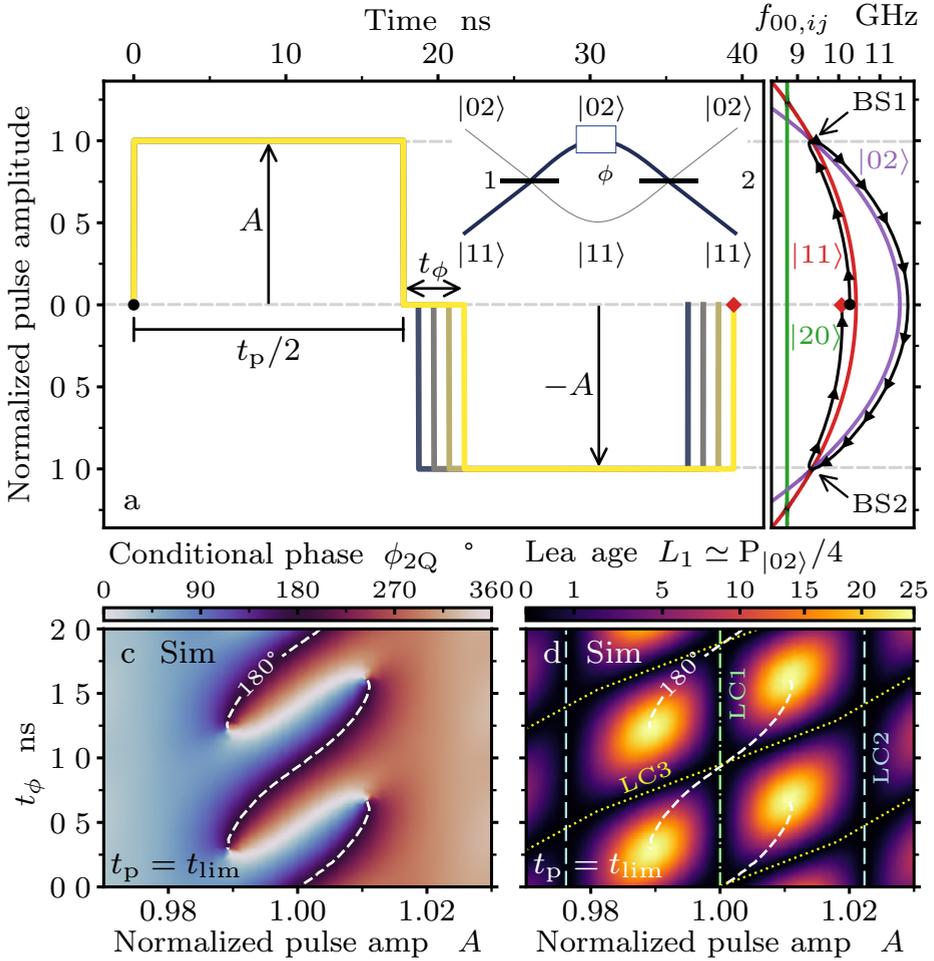

Figure 2.1: Numerical simulation of an ideal SNZ pulse (infinite bandwidth and time resolution) using parameters for pair Q  -Q    (see Table 1). (a) Schematic of the ideal SNZ flux pulse, with      =      and variable       and    . The amplitude       is normalized to the   11 - 02 resonance. Inset: MZI analogy for      = 1. (b) Transition frequency from    00   to levels        in the two-excitation manifold as a function of instantaneous pulse amplitude. (c, d) Landscapes of conditional phase       (b) and leakage       (c) as a function of    and    .

juxtaposing the $\varphi = 180°$ contour shows that PC is met periodically, at the crossing of LC1 and LC3 valleys, where $\zeta = 0 \pmod{2\pi}$ ($\zeta$ is the detuning between $\overline{02}$ and $\overline{11}$ at the bias point). This regular leakage landscape therefore provides useful crosshairs for simultaneously achieving PC and LC. We note that $\varphi(\tau)$ changes monotonically along the LC1 valley, allowing for CPhase gates with arbitrary $\varphi$. We leave this generalization for future work.

There are practical reasons to include $\tau$ in experiment: any flux-pulse distortion remaining from the first half pulse (e.g., due to finite pulse decay time) will break the symmetry between BS1 and BS2. Due to the time resolution $\delta t$ of the AWG used for flux control, $\tau$ can only increment in steps of $\delta t$. Typically $\omega/2\pi = 0.5{-}1~\text{GHz}$ and $\delta t \sim 1~\text{ns}$, so the number of intermediate sampling points only provides coarse control. For fine control, we propose to use the amplitude $A_s$ of the first and last sampling points during $\tau$ [101].

## 2.3 Experimental results

We now turn to the experimental realization of SNZ CZ gates between nearest-neighbor pairs among four transmons. High- and low-frequency transmons ($Q_H$ and $Q_L$, respectively) connect to two mid-frequency transmons ($Q_M$ and $Q_{M'}$) using bus resonators dedicated to each pair [connectivity diagram shown in Fig. 4(a) inset]. Each transmon has a flux-control line for two-qubit gating, a microwave-drive line for single-qubit gating, and dedicated readout resonators [72, 102] (see [101] for details). Each transmon is statically flux-biased at its sweetspot to counter residual offsets. Flux pulsing is performed using a Zurich Instruments HDAWG-8 ($\delta t = 1/2.4~\text{ns}$). Following prior work [94, 103], we compensate the bandwidth-limiting effect of attenuation in the flux-control coaxial line (skin effect) and cryogenic reflective and absorptive low-pass filters using real-time digital filters in the AWG. In this way, we produce on-chip flux waveforms with rise time $t_r$ on par with that of the AWG ($0.5~\text{ns}$).

We exemplify the tuneup of SNZ using pair $Q_H$-$Q_M$ (Fig. 2). We first identify $\tau$ for the $\overline{11}$-$\overline{02}$ interaction and amplitude $A$ bringing the two levels on resonance. Both are extracted from the characteristic chevron pattern of $\overline{02}$-population $P_{|2\rangle}$ in $Q_H$ as a function of the amplitude and duration of a unipolar square flux pulse acting on $\overline{11}$ [Fig. 2(a)]. The symmetry axis corresponds to $A = 1$. The difference in consecutive pulse durations achieving $P_{|2\rangle}$ maxima along this axis gives an accurate estimate of $\tau$ unaffected by initial transients. We set $N_p = 2$, where $N_p$ is the number of sampling points achieving the first $P_{|2\rangle}$ maximum. Using the measured positive difference $\tau$ and numerical simulation (data not shown), we estimate $t_r \approx 0.5~\text{ns}$. Next, we use standard conditional-oscillation experiments [94] to measure the landscapes of $\varphi$ and leakage estimate $L_1$ for SNZ pulses over amplitude ranges $[0.9, 1.1]$ and $[0, \pi]$, keeping $N_p = 3$. As expected, the landscape of $\varphi$ [Fig. 2(c)] reveals a vertical valley at $A = 1$ and a diagonal valley. Juxtaposing the $\varphi = 180°$ contour from Fig. 2(b), we observe the matching of PC at the crossing of these valleys, in excellent agreement with a numerical two-qutrit simulation [Fig. 2(d)].

**2**

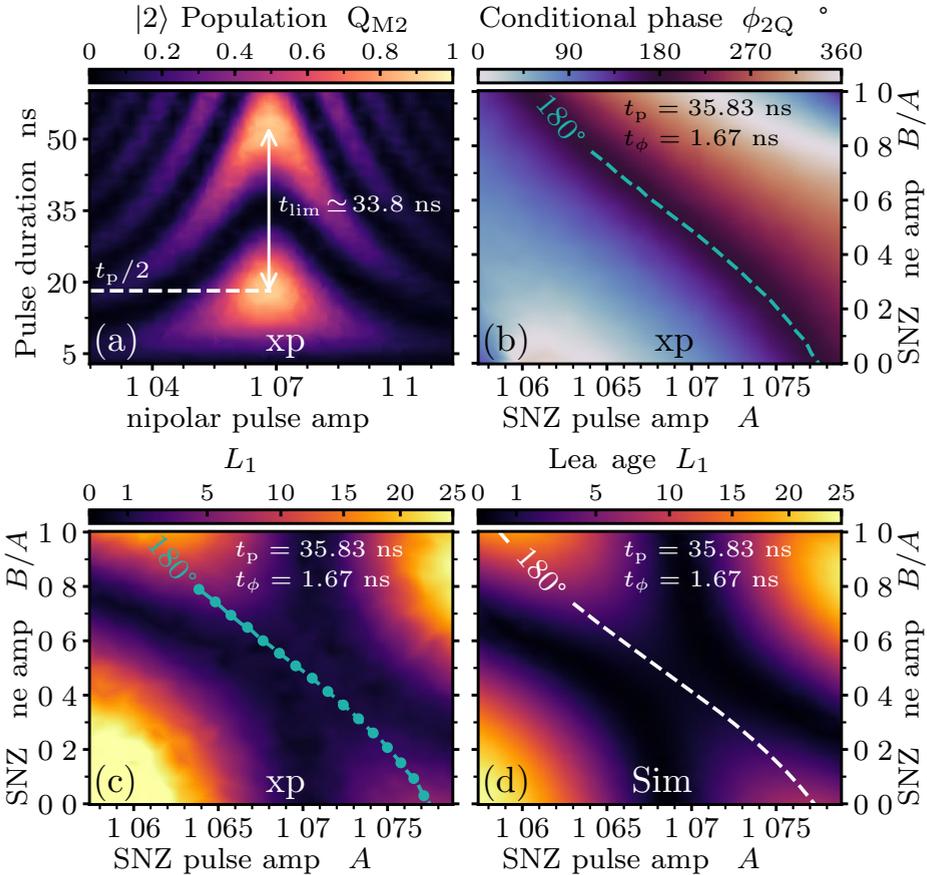

Figure 2.2: Calibration of the SNZ pulse for pair $Q$ -$Q$    and comparison to simulation. (a) $2$ -state population of $Q$    as a function of the amplitude and duration of a unipolar square pulse making $11$  interact with $02$ . (b,c) Landscapes of conditional phase    and leakage estimate    as a function of SNZ pulse amplitudes    and   , with    = $1$ 67 ns. The juxtaposed    = $180°$ contour runs along the opposite diagonal compared to Figs. 1(b,c) because increasing    (which decreases   ) changes    in the opposite direction from   . Data points marked with dots are measured with extra averaging for examination in Fig. 3. (d) Numerical simulation of leakage    landscape and    = $180°$ contour with parameters and flux-pulse distortions from experiment. All landscapes (also in Fig. 3) are sampled using an adaptive algorithm based on [104].

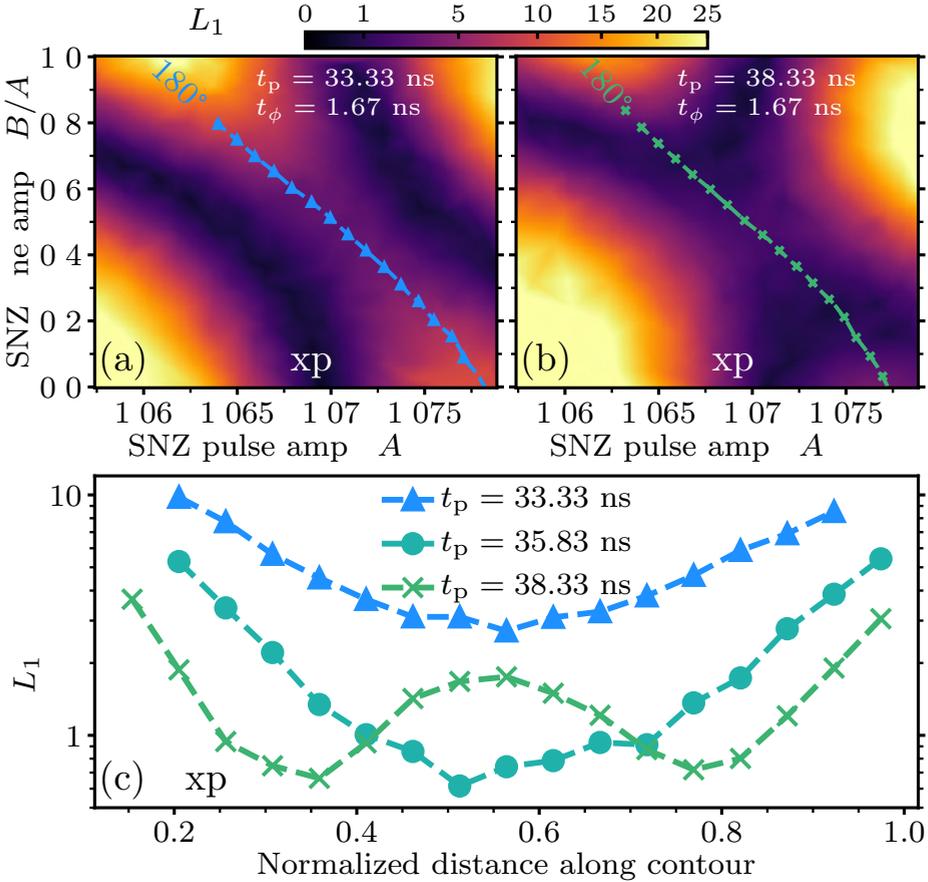

Figure 2.3: (a,b) Landscapes of the leakage estimate    for intentionally short and long SNZ half pulses on Q   . (c) Extracted    along the    = 180° contours from (a), (b), and Fig. 2(c).

Experimentally, due to the discreteness of   , it is unlikely to precisely match    2 to the half-pulse duration that truly maximizes   | ⟩. To understand the consequences, we examine the    and    landscapes for SNZ pulses upon intentionally changing    by  6   (Fig. 3). While the PC contour remains roughly unchanged in both cases, there are significant effects on   . In both cases, we observe that    lifts at the prior crossing of LC1 and LC3 valleys where    = 180°. For too-short pulses [Fig. 3(a)], there remain two valleys of minimal   , but these are now curved and do not cross    = 180°. For too-long pulses [Fig. 3(b)], there are also two curved valleys. Crucially, these cross the    = 180° contour, and it remains possible to achieve PC and minimize leakage at two (   ) settings. Extracting    along the    = 180° contours [Fig. 3(c)] confirms that too-long pulses can achieve the same minimal    as when using the nominal   . The impossibility to achieve minimal leakage at    = 180° for too-short pulses manifests the speed limit set by   . In turn, the demonstrated possibility to do so for too-long pulses (even overshooting by several sampling points) proves the viability of the SNZ pulse in practice.

| Parameter | $Q_{M1}$-$Q_H$ | $Q_{M2}$-$Q_H$ | $Q_L$-$Q_{M1}$ | $Q_L$-$Q_{M2}$ |
|---|---|---|---|---|
| $_{\text{lim}}$ (ns) | 31 0 | 27 6 | 38 4 | 33 8 |
| $_p$ $_\phi$ (ns) | 32 50  2 92 | 29 10  3 75 | 40 83  1 25 | 35 83  1 67 |
| $_{\text{total}}$ (ns) | 45 42 | 42 91 | 52 08 | 47 50 |
| Interaction | $|11\rangle$-$|02\rangle$ | $|11\rangle$-$|02\rangle$ | $|11\rangle$-$|20\rangle$ | $|11\rangle$-$|02\rangle$ |
| Parked qubit | $Q_{M2}$ | $Q_{M1}$ | – | – |
| Avg.  (%) | 98 89 $\pm$ 0 35 | 99 54 $\pm$ 0 27 | 93 72 $\pm$ 2 10 | 97 14 $\pm$ 0 72 |
| Avg.  $_1$ (%) | 0 13 $\pm$ 0 02 | 0 18 $\pm$ 0 04 | 0 78 $\pm$ 0 32 | 0 63 $\pm$ 0 11 |
| Max.  (%) | 99 77 $\pm$ 0 23 | 99 93 $\pm$ 0 24 | 99 15 $\pm$ 1 20 | 98 56 $\pm$ 0 70 |
| Min.  $_1$ (%) | 0 07 $\pm$ 0 04 | 0 10 $\pm$ 0 02 | 0 04 $\pm$ 0 08 | 0 41 $\pm$ 0 10 |

Table 2.1: Summary of SNZ CZ pulse parameters and achieved performance for the four transmon pairs. Single-qubit phase corrections are included in  . Gate fidelities and leakage are obtained from 2QIRB keeping the other two qubits in  0 . Statistics (average and standard deviation) are taken from repeated 2QIRB runs (see [101] for technical details). The maximum  and minimum  quoted are not necessarily from the same run.

With these insights, we proceed to tune the remaining SNZ CZ gates following similar procedures. We use final weak bipolar pulses of total duration  $= 10 \text{ ns}$ to null the single-qubit phases in the frame of microwave drives. Since our codeword-based control electronics has a $20 \text{ ns}$ timing grid, and $40 \text{ ns}$  $=$  $+$  $+$  $60 \text{ ns}$ for all pairs, we allocate $60 \text{ ns}$ to every CZ gate. Some pair-specific details must be noted. Owing to the frequency overlap of $Q$  and $Q$ , implementing CZ between $Q$  and $Q$  ($Q$ ) requires a bipolar parking flux pulse on $Q$  ($Q$ ) during the SNZ pulse on $Q$  [74, 98]. For most pairs, we employ the  11 - 02  interaction, which requires the smallest flux amplitude (reducing the impact of dephasing from flux noise) and does not require crossing any other interaction. However, for $Q$ -$Q$ , we cannot reliably use this interaction as there is a flickering two-level system (TLS) overlapping with the  0 - 1  transition in $Q$  at this amplitude [101]. For this pair, we therefore employ the  11 - 20  interaction. Here, SNZ offers a side benefit: it crosses the $Q$ -TLS,  11 - 02 , and  01 - 10  resonances as suddenly as possible, minimizing population exchange.

Table 1 summarizes the timing parameters and performance attained for the four SNZ CZ gates. The CZ gate fidelity  and leakage  are extracted using a 2QIRB protocol [94, 97]. For each pair, we report the average and standard deviation of both based on at least 10 repetitions of the protocol spanning more than $8 \text{ h}$ [101]. Several observations can be drawn. First, CZ gates involving $Q$  perform better on average than those involving $Q$ . This is likely due to the shorter  and correspondingly longer time $60 \text{ ns}$  spent near the sweetspot. Additionally, the frequency downshifting required of $Q$  to interact with the mid-frequency transmons is roughly half that required of the latter to interact with $Q$ . This reduces the impact of dephasing from flux noise during the pulse. Not surprisingly, performance is worst for $Q$ -$Q$ . Here, the pulse must downshift $Q$ the most to reach the distant 11 - 20 interaction, increasing dephasing from flux noise. Also, there may be residual exchange at the crossed resonances. Overall, there is significant temporal variation in performance as gleaned by repeated 2QIRB characterizations. We believe this reflects the underlying variability of qubit relaxation and dephasing times and flux offsets, which however were not tracked simultaneously. In addition to having the best average performance, pair $Q$ -$Q$ displays the maximum of $99.93 \pm 0.24\%$ (Fig. 4) extracted from a single 2QIRB characterization. To the best of our knowledge, this is the highest CZ fidelity extracted from one 2QIRB characterization in a multi-transmon processor.

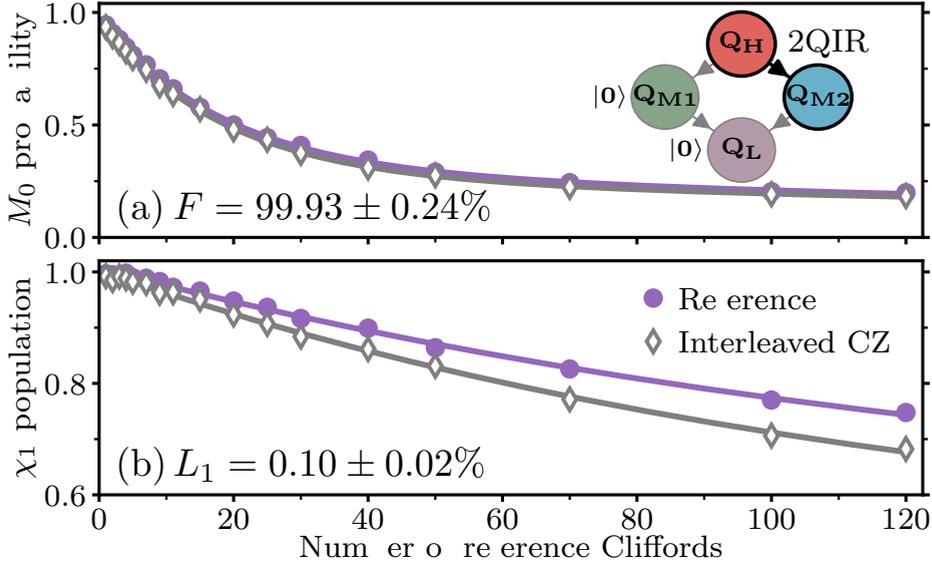

Figure 2.4: Best SNZ CZ gate performance achieved from a single run of 2QIRB. (a) Reference and CZ-interleaved return probability to 00 and (b) population in the computational space as a function of the number of two-qubit Cliffords in the reference curve. Errors bars in and are obtained from the uncertainty of exponential-decay fits.

To understand the dominant sources of infidelity $= 1$ and leakage, we run numerical simulations [94], for both SNZ and CNZ, with experimental input parameters for pair $Q$ -$Q$ . We dissect an error budget versus various models finding similar contributions for both gates (see [101]). Nevertheless, the results suggest that SNZ slightly outperforms CNZ, likely due to a shorter time spent away from the sweetspot during the fixed 60 ns allocated for both variants. This confirms that the temporary full transfer from 11 to 02 does not compromise the gate fidelity.

**2**

### 2.4   Conclusions

In summary, we have proposed and realized high-fidelity CZ gates using the sudden version of the Net Zero bipolar fluxing scheme. SNZ CZ gates operate ever closer to the speed limit of transverse coupling by maximizing intermediate leakage to the non-computational state. Control architectures without a timing grid will benefit most from the speedup of SNZ over CNZ by reducing total gate time and thereby minimizing the impact of decoherence. A demonstrated second key advantage of SNZ over CNZ is ease of tuneup, owing to the simple structure of error landscapes as a function of pulse parameters. Harnessing the tuning simplicity, we already employ SNZ CZ gates in the Starmon-5 processor publicly available via the QuTech Quantum Inspire platform [105]. Moving forward, the compatibility of SNZ with our scalable scheme [98] for surface coding makes SNZ our choice for CZ gates for quantum error correction. Finally, the straightforward extension of SNZ to arbitrary conditional-phase gates will find immediate use in NISQ applications.

### 2.5   Data availability

Interested readers can reproduce our figures by using the processed data of the figures. The processed data can be found at https://github.com/DiCarloLab-Delft/High_Fidelity_ControlledZ_Data/.

### 2.6   Supplemental material

This supplement provides additional information supporting statements and claims made in the main text. Section 3.6.1 summarizes the main differences between conventional NZ (CNZ) pulses and SNZ pulses. Section 2.6.2 provides further details on the device used and measured transmon parameters. Section 2.6.3 presents the characterization of single-qubit gate performance. Section 2.6.4 provides evidence for the two-level system affecting the realization of SNZ CZ gates in pair $Q$ -$Q$    using the $11$ - $02$  interaction. Section 2.6.5 presents a characterization of the residual    coupling between qubits at the bias point. Section 2.6.6 summarizes the technical details of the CZ characterization by repeated 2QIRB. Section 2.6.7 presents the numerical simulation of the error budget for SNZ and CNZ CZ gates on pair $Q$   -$Q$  .

### 2.6.1   Comparison of conventional NZ pulses and SNZ pulses

This section highlights the main differences between CNZ pulses and the SNZ pulses introduced here. For reference, Fig. S1 illustrates the relevant energy-level structure for a pair of coupled transmons (here $Q$   and $Q$  ) as a function of magnetic flux on the higher-frequency transmon (here $Q$    ). CNZ and SNZ CZ gates both exploit the avoided crossing in the two-excitation manifold between the computational state $11$ and a non-computational state. Most often this non-computational state is $02$ as reaching the avoided crossing requires the smallest flux-pulse amplitude and does not require passing through any other avoided crossings. In contrast, reaching the $11$ - $20$ avoided crossing requires passing through the $01$ - $10$ avoided crossing in the one-excitation manifold.

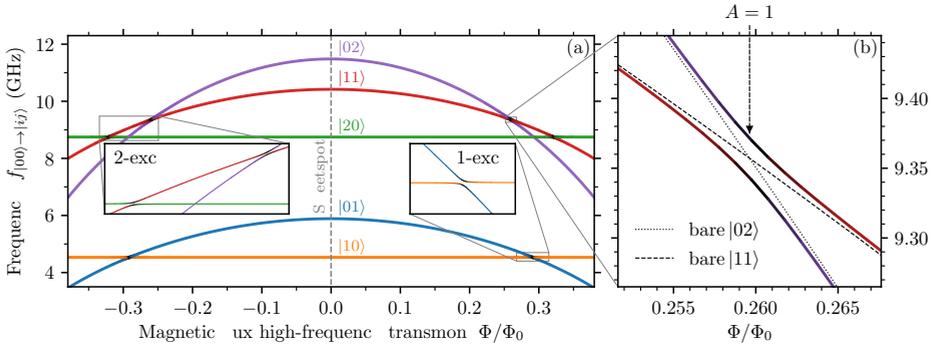

Figure 2.5: (a) Transition frequencies from $00$ to levels $ij$ in the one- and two-excitation manifold for transmon pair $Q$ -$Q$ as a function of magnetic flux through the SQUID loop of $Q$ , normalized to the flux quantum $\Phi_0$. Insets: Zoom-ins to the avoided crossings in the (left) two-excitation and (right) one-excitation manifolds. (b) Zoom-in to the $11$ - $02$ avoided crossing occurring at $A = 1$. The minimum frequency splitting corresponds to $1 =$ .

CNZ implements a CZ gate based on two back-to-back half strong flux pulses [Fig. S2(a)] of duration $2$ each, applied on the higher-frequency transmon. Typically, $11$ $16$. The strong half pulses are formally parametrized as in [86]. For the purposes of illustration, here we can loosely lump this parametrization as affecting the amplitude ( ) and curvature ( $'$) of the strong half pulses. Immediately following the strong pulse, weak bipolar pulses of duration are applied on both the higher- and lower-frequency transmons with amplitudes and , respectively, in order to null the single-qubit phases acquired by each. Typically, $= 10\,\mathrm{ns}$. In CNZ there is no intermediate idling period between the strong half pulses, so the analogy to the MZI is not exact [Fig. S2(c)]. During tuneup, one searches the ( $'$) space to achieve a conditional phase (PC) of by only affecting the unitary action of the two beamsplitters. Because for typical CNZ produces significant leakage at the first strong pulse, achieving minimal leakage relies on meeting LC3 (leakage interference). The structure of the ( $'$) and ( $'$) landscapes and especially their interrelation are not straightforward, so the search for an ( $'$) setting satisfying both PC and LC3 is not easily guided. We point the interested reader to [94] for examples.

The SNZ pulses introduced here [Fig. S2(b)] differ in two key ways. First, the strong half pulses are replaced by square half pulses each with duration $2$ maximizing the transfer from $11$ to $02$ and viceversa. Second, an intermediate idling period is added to accrue relative phase between $02$ and $11$ , perfecting the analogy to the MZI [Fig. S2(d)]. We use the

**2**

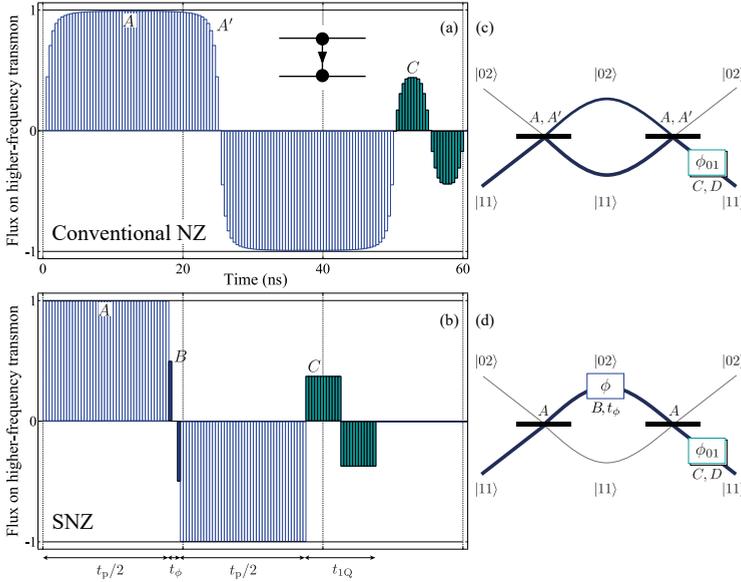

Figure 2.6: Comparison of conventional NZ and SNZ pulses for CZ gates. (a) Conventional NZ CZ pulses consist of two back-to-back strong half pulses of duration $t_p/2$ each, followed by two weak back-to-back half pulses of duration $t_{1Q}/2$ each on the higher-frequency qubit. The amplitude ($A$) and curvature ($A'$) of the strong pulses are jointly tuned to set the conditional phase $\phi$ at minimal leakage $L_1$, while the amplitude $C$ of the weak pulses is used to null the single-qubit phase on the higher-frequency transmon. Weak pulses (amplitude $D$) on the lower-frequency qubit (not shown here) are also used to null its single-qubit phase. (b) In SNZ, the strong pulses are replaced by square pulses with $t_p$ as close as possible to but not shorter. Also, an intermediate idling period $t_\phi$ is added to accrue relative phase $\phi$ between $02$ and $11$. The amplitude $B$ of the first and last sampling points in $t_p$ and the number of intermediate zero-amplitude points provide fine and coarse control of this relative phase, respectively. SNZ CZ gates also use weak bipolar pulses (now square) of total duration $t_{1Q}$ to null single-qubit phases. (c) The MZI analogy for conventional NZ pulses is incomplete. Each strong half pulse implements a beamsplitter (ideally identical) with scattering parameters affected by $A$ and $A'$. However, there is no possibility to independently control the relative phase in the two arms between the beamsplitters. (d) The MZI analogy is exact for SNZ pulse. The scattering at the beamsplitters is controlled by $A$ and the relative phase $\phi$ is controlled finely using $B$ and coarsely using $t_\phi$.

amplitude $B$ of the first and last sampling points in $t_p$ and the number of intermediate zero-amplitude points to achieve fine and coarse control of $\phi$, respectively. As in CNZ, we use weak bipolar pulses on both transmons (also with $t_{1Q} = 10\ \mathrm{ns}$) to null the single-qubit phases. During tuneup, we search the ($B, t_\phi$) space to achieve $\phi = 180°$. As shown in the main text, the SNZ pulse design gives a very simple structure to the ($B, t_\phi$) and $(\quad)$ landscapes. Crucially, the crossing point of LC1 and LC3 leakage valleys matches $= 180°$. This simplicity of tuneup is the key advantage of SNZ over CNZ.

Another advantage of SNZ over CNZ is the reduced total time $=\ +\ +$ required to achieve a CZ gate. However, due to the $20\,\mathrm{ns}$ timing grid of our control electronics and the transverse coupling strengths in our device, this speedup is insufficient to reduce the total time allocated per CZ gate from $60$ to $40\,\mathrm{ns}$. Nonetheless, in SNZ, the fluxed transmon spends more time at its sweetspot, which reduces the dephasing due to flux noise.

Figure S3 illustrates the qualitative difference in the trajectory of level populations in the two-excitation manifold implemented by strong CNZ and SNZ pulses acting on the $|11\rangle$ state. A CNZ pulse [Fig. S3(a)] uses the interaction point fast-adiabatically, keeping most population in $|11\rangle$ after the first strong half pulse. In contrast, a SNZ pulse [Fig. S3(b)] uses the interaction suddenly to transfer most (ideally all) of the population to $|02\rangle$ with the first strong half pulse.

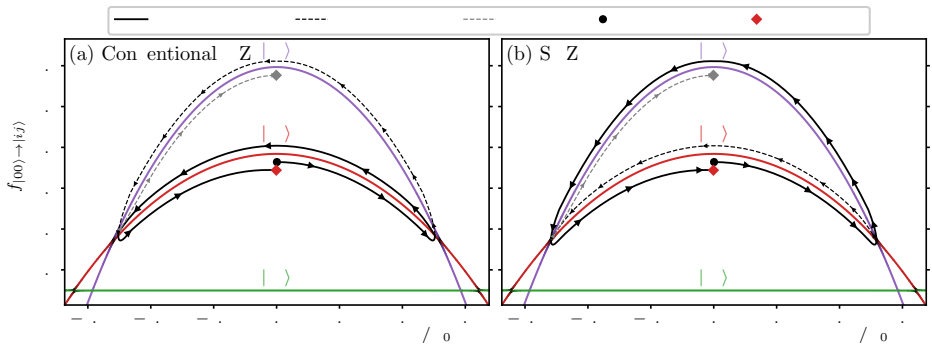

Figure 2.7: Schematic comparison of the trajectory of level populations in the two-excitation manifold for CNZ and SNZ strong pulses acting on $|11\rangle$. Note that in both cases, most of the time is spent close or at the $|11\rangle$ - $|02\rangle$ avoided crossing. (a) Trajectory for a CNZ pulse. (b) Trajectory for an SNZ pulse.

### 2.6.2 Device and transmon parameters

Our experimental study focuses on four transmons in a patch of our 7-qubit processor. An optical image of the device, zoomed in to these four transmons, is shown in Fig. S4. Transmons $\mathrm{Q}$ and $\mathrm{Q}$ both connect to $\mathrm{Q}$ and $\mathrm{Q}$ with a dedicated coupling bus resonator for each connection. Every transmon has a dedicated microwave-drive line for single-qubit gating, a flux-control line used for CZ gating, and a dispersively-coupled readout resonator with dedicated Purcell filter [72, 102] for readout. Readout is performed by frequency multiplexed measurement of a common feedline capacitively connected to all four Purcell filters. Table S1 provides a summary of measured parameters for the four transmons.

**2**

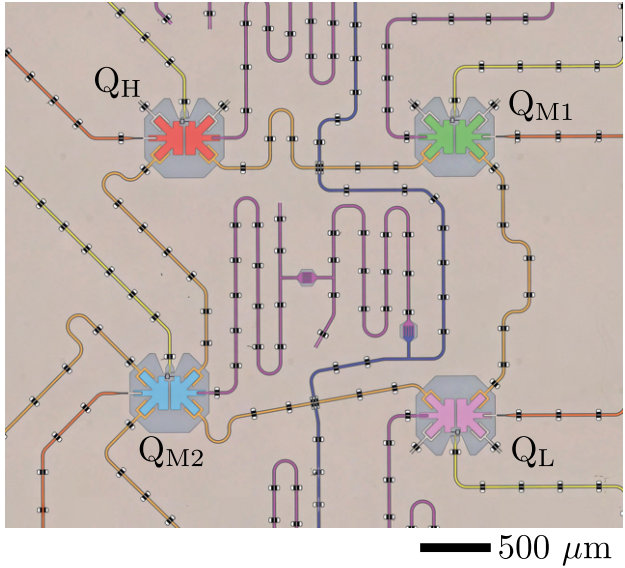

Figure 2.8: Optical image of the device, zoomed in to the four transmons used in this study and with added false color to help identify circuit elements. Transmons $Q_H$ (red) and $Q_L$ (pink) each connect to $Q_{M1}$ (green) and $Q_{M2}$ (cyan) using dedicated coupling bus resonators for each pair (light orange). Each transmon has a flux-control line for two-qubit gating (yellow), a microwave-drive line for single-qubit gating (dark orange), and dispersively-coupled resonator with Purcell filter for readout (purple) [72, 102]. The readout-resonator/Purcell-filter pair for $Q_L$ is visible at the center of this image. The vertically running common feedline (blue) connects to all Purcell filters, enabling simultaneous readout of the four transmons by frequency multiplexing. Air-bridge crossovers enable the routing of all input and output lines to the edges of the chip, where they connect to a printed circuit board through aluminum wirebonds.

### 2.6.3 Single-qubit gate performance

All single-qubit gates are implemented as DRAG-type [64, 65] microwave pulses with total duration $4\sigma = 20$ ns, where $\sigma$ is the Gaussian width of the main-quadrature Gaussian pulse envelope. We perform two sets of experiments to jointly quantify the infidelity $\epsilon$ and leakage $L$ of these gates. First, we perform individual single-qubit randomized benchmarking (1QRB) keeping the other three qubits in $|0\rangle$. Second, we perform simultaneous single-qubit randomized benchmarking (S1QRB) on pairs of qubits, keeping the other two qubits in $|0\rangle$. The results obtained from both types of experiment are reported as diagonal and off-diagonal elements in the matrices presented in Fig. S5.

| | | Q$_1$ | Q$_2$ | Q$_3$ | Q$_4$ |
|---|---|---|---|---|---|
| Qubit transition frequency at sweetspot, $\omega_Q/2\pi$ (GHz) | | 6.4329 | 5.7707 | 5.8864 | 4.5338 |
| Transmon anharmonicity, $\alpha/2\pi$ (MHz) | | 280 | 290 | 285 | 320 |
| Readout frequency, $\omega_r/2\pi$ (GHz) | | 7.4925 | 7.2248 | 7.0584 | 6.9132 |
| Relaxation time, $T_1$ ($\mu$s) | | 37 $\pm$ 1 | 40 $\pm$ 1 | 47 $\pm$ 1 | 66 $\pm$ 1 |
| Ramsey dephasing time, $T_2^*$ ($\mu$s) | | 38 $\pm$ 1 | 49 $\pm$ 1 | 47 $\pm$ 1 | 64 $\pm$ 1 |
| Echo dephasing time, $T_2^e$ ($\mu$s) | | 54 $\pm$ 2 | 68 $\pm$ 1 | 77 $\pm$ 1 | 94 $\pm$ 2 |
| Residual qubit excitation, $p_e$ (%) | | 1.4 | 1.2 | 4.3 | 1.7 |
| Best readout fidelity, $F$ (%) | | 99.1 | 98.5 | 99.4 | 97.8 |

Table 2.2: Summary of frequency, coherence, residual excitation, and readout parameters of the four transmons. The statistics of coherence times for each transmon are obtained from 30 repetitions of standard time-domain measurements [58] taken over $\sim 4$ h. The residual excitation is extracted from double-Gaussian fits of single-shot readout histograms with the qubit nominally prepared in $|0\rangle$. The readout fidelity quoted is the average assignment fidelity [106], extracted from single-shot readout histograms after mitigating residual excitation by post-selection on a pre-measurement.

### 2.6.4  Flickering two-level system

As mentioned in the main text, we were unable to realize the SNZ CZ gate between pair Q$_1$ - Q$_2$ using the $|11\rangle$ - $|02\rangle$ interaction due to the presence of a TLS interacting intermittently with Q$_2$ at the flux amplitude placing $|11\rangle$ and $|02\rangle$ on resonance. Figures S6(a,b) show the negative impact of this TLS when attempting to characterize the $|11\rangle$ - $|02\rangle$ interaction by the standard time-domain chevron measurement. While experience shows that it is probable that such a TLS could be displaced or eliminated by thermal cycling at least above the critical temperature of aluminum, we chose instead to use the more flux distant $|11\rangle$ - $|20\rangle$ interaction to realize the SNZ CZ gate for this pair. For this interaction, a standard, stable chevron pattern is observed [Figs. S6(c,d)].

### 2.6.5  Residual ZZ coupling at bias point

Coupling between nearest-neighbor transmons in our device is realized using dedicated coupling bus resonators. The non-tunability of said couplers leads to residual $ZZ$ coupling between the transmons at the bias point. We quantify the residual $ZZ$ coupling between every pair of qubits as the shift in frequency of one qubit when the state of the other changes from $|0\rangle$ to $|1\rangle$. We extract this frequency shift using a simple time-domain measurement: we perform a standard echo experiment on one qubit (the echo qubit), but add a $\pi$ pulse on the other qubit

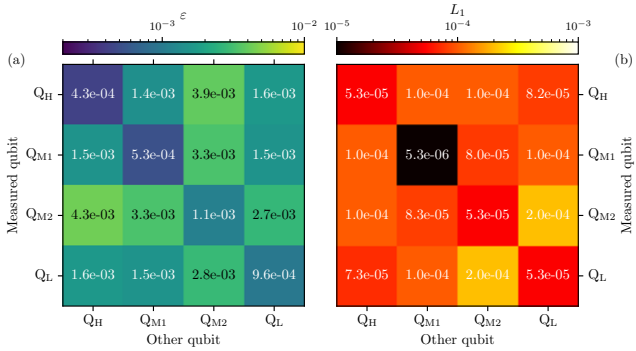

Figure 2.9: Characterization of single-qubit gate infidelity $\varepsilon$ (a) and leakage $L_1$ (b) using randomized benchmarking ($100$ randomization seeds). Diagonal elements are extracted from individual single-qubit randomized benchmarking keeping the other 3 qubits in $0$. Off-diagonal elements are extracted from simultaneous one-qubit randomized benchmarking on pairs of qubits, keeping the other two qubits in $0$.

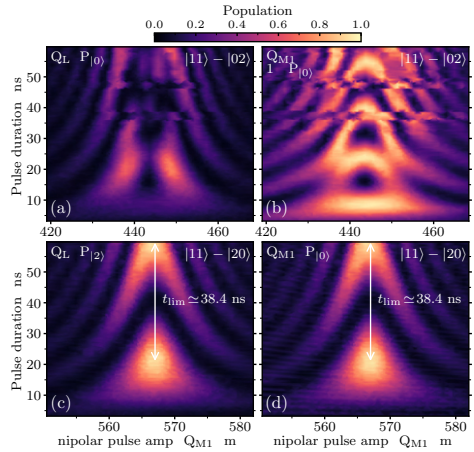

Figure 2.10: Time-domain characterization of the $11$-$02$ and (c,d) $11$-$20$ interactions for pair $Q$ -$Q$. (a,b) Landscapes of (a) ground-state population $|\rangle$ of $Q$ and (b) total excited-state population $1$ $|\rangle$ of $Q$ as a function of the amplitude and duration of a unipolar square pulse near the $11$-$02$ resonance. The absence of the expected chevron pattern in these landscapes reflects a flickering TLS resonant with the qubit transition of $Q$ at this pulse amplitude. Horizontally shifting fringes in (a) and (b) are due to flickering of the TLS on the scale of a few minutes. These observations preclude the use of the $11$-$02$ interaction to realize the CZ gate. In contrast, the landscapes of (c) two-state population $|\rangle$ of $Q$ and (d) $|\rangle$ of $Q$ and as a function of unipolar square pulse parameters near the $11$-$20$ resonance reveal a standard, stable chevron pattern. All landscapes were sampled using an adaptive algorithm based on [104].

(control qubit) halfway through the free-evolution period simultaneous with the refocusing pulse on the echo qubit. The results are presented as a matrix in Fig. S7. We observe that the residual $ZZ$ coupling is highest between $Q_H$ and the mid-frequency qubits $Q_{M1}$ and $Q_{M2}$. This is consistent with the higher (lower) absolute detuning between $Q_H$ ($Q_L$) and the mid-frequency transmons, and the higher (lower) transverse coupling $J = J_2$ for the upper (lower) pairs.

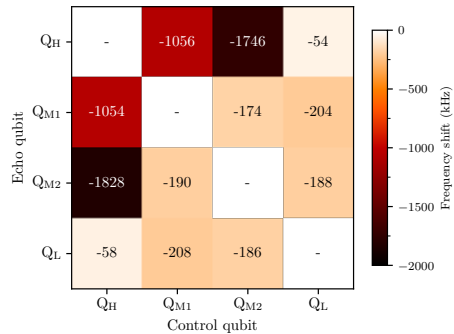

Figure 2.11: Extracted residual $ZZ$ coupling between all pairs of qubits at their bias points. We report the frequency shift in one qubit (named echo qubit) when the computational state of another qubit (named control qubit) is shifted from $|0\rangle$ to $|1\rangle$.

An alternative way to evidence this residual $ZZ$ coupling is to extract the fidelity of idling using 2QIRB and to compare this fidelity to that of CZ. To this end, we perform 2QIRB of idling (for $60\,\text{ns}$) on pairs $Q_H$-$Q_{M2}$ and $Q_L$-$Q_{M2}$. The results, shown in Fig. S8, show striking differences for the two pairs. For $Q_H$-$Q_{M2}$, the pair with strongest residual coupling, the idling fidelity is significantly lower than the CZ fidelity. This is because the residual $ZZ$ coupling is a source of error during idling but is absorbed into the tuneup of SNZ. For $Q_L$-$Q_{M2}$, for which the residual coupling is one order of magnitude lower, this trend is not observed.

### 2.6.6   Technical details on 2QIRB

Table S2 details technical aspects of the characterization of CZ gates by repeated 2QIRB runs.

### 2.6.7   Simulation results for SNZ and conventional NZ CZ gates versus different error models

To identify the dominant sources of infidelity $L = 1$ and leakage for SNZ CZ gates, we perform a two-qutrit numerical simulation for pair $Q_H$-$Q_{M2}$ with incremental addition of measured error sources [Fig. S9], as in our previous work on conventional NZ [94]. The simulation incrementally adds: (A) no noise; (B) relaxation; (C) Markovian dephasing; (D)

**2**

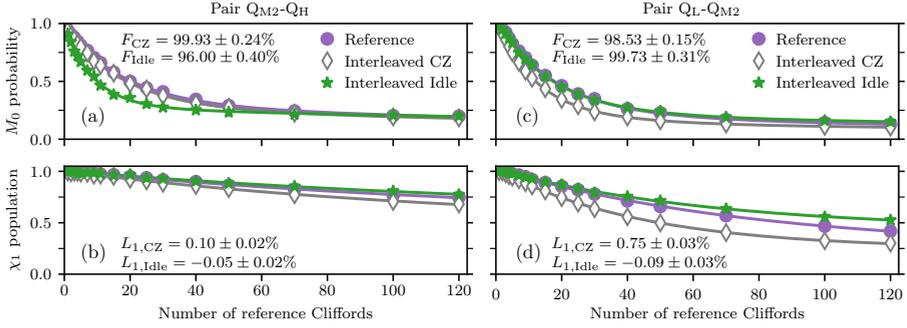

Figure 2.12: Comparison by 2QIRB of the fidelity and leakage of SNZ CZ versus idling (for an equivalent $60\,\mathrm{ns}$) for pairs $Q_{\text{M2}}$-$Q_{\text{H}}$ and $Q_{\text{L}}$-$Q_{\text{M2}}$. SNZ gate parameters are provided in Table 1 of the main text. (a,c) Return probability to $00$ as a function of the number $m$ of two-qubit Clifford operations in the reference curve. For $Q_{\text{M2}}$-$Q_{\text{H}}$, the extracted idling fidelity is significantly lower than the SNZ CZ fidelity. This is due to the high residual $ZZ$ coupling between these two qubits as reported in Fig. S7, which is not refocused during idling but absorbed into the tuneup of the SNZ CZ gate. For $Q_{\text{L}}$-$Q_{\text{M2}}$, idling fidelity exceeds SNZ CZ fidelity as the residual coupling is one order of magnitude weaker. (b,d) Population in the computational subspace as a function of $m$. Leakage as a function of $m$ is weakest when interleaving idling steps, leading to negative $L_{1,\text{Idle}}$. This is due to seepage (during idling) of the leakage produced by the reference two-qubit Cliffords.

| Parameter | $Q_{\text{M2}}$-$Q_{\text{H}}$ | $Q_{\text{L}}$-$Q_{\text{M2}}$ | $Q_{\text{H}}$-$Q_{\text{M1}}$ | $Q_{\text{M1}}$-$Q_{\text{L}}$ |
|---|---|---|---|---|
| Number of 2QIRB runs | 39 | 10 | 88 | 35 |
| Number of randomization seeds | 100 | 300 | 100 | 100 |
| Same randomization seeds | No | No | Yes | No |
| Avg. time per 2QIRB run $(\mathrm{min})$ | 17 | 50 | 9 | 17 |
| Total wall-clock time $(\mathrm{h})$ | 28 8 | 16 9 | 16 7 | 14 8 |

Table 2.3: Technical details of the characterization of CZ gates by repeated 2QIRB. The average time per 2QIRB run is the time required to perform back-to-back measurements of the reference and the CZ-interleaved curves. The total wall-clock time includes the overhead from compilation of RB sequences and other measurements performed in between the CZ 2QIRB runs, e.g., idling 2QIRB (Fig. S8).

dephasing from quasistatic flux noise; and (E) flux-pulse distortion. The experimental inputs for models B, C and D combine measured qubit relaxation time $T_1$ at the bias point, and measured echo and Ramsey dephasing times ($T_2$ and $T_2^*$) as a function of qubit frequency. The input to E consists of a final Cryoscope measurement of the flux step response using all real-time filters. The simulation suggests that the main source of $F$ is Markovian dephasing (as in [94]), while the dominant contribution to $L_1$ is low-frequency flux noise. The latter contrasts with Ref. [94], where simulation identified flux-pulse distortion as the dominant leakage source. We identify two possible reasons for this difference: in the current experiment, the $1$ low-frequency flux noise is $4$ times larger (in units of $\overline{\mathrm{Hz}}$) and the achieved flux step response is noticeably sharper. Finally, we use the simulation to compare performance of SNZ to conventional NZ CZ. For the latter, we fix $= 0$, $= 60\,\mathrm{ns}$ , and use the fast-adiabatic pulse shape and $= 45\,83\,\mathrm{ns}$ optimized by simulation. Overall, the error sources contribute very similarly to the error budget for both cases. The marginally higher overall performance found for SNZ is likely due to the increased time spent at the sweetspot during the gate time.

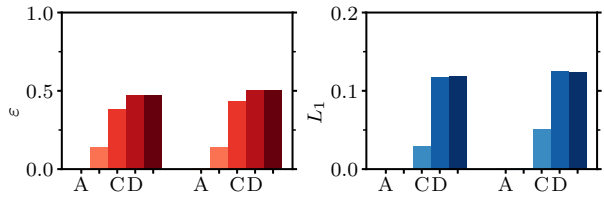

Figure 2.13: Error budgets for infidelity $\varepsilon$ (a) and leakage $L_1$ (b) obtained by a numerical simulation (as in [94]) of the $Q$ -$Q$ SNZ CZ gate with parameters in Fig. 4 and for a conventional NZ gate with optimized parameters (see text for details). The simulation incrementally adds errors using experimental input parameters for this pair: (A) no noise; (B) relaxation; (C) Markovian dephasing; (D) dephasing from quasistatic flux noise; and (E) flux-pulse distortion.

We emphasize that our two-qutrit simulation includes 9 energy levels (from $00$ to $22$ ). Therefore, it also captures leakage to $20$ ( $02$ ) when using the $11$ - $02$ ( $11$ - $20$ ) avoided crossing. For pair $Q$ -$Q$ , for which we use $11$ - $02$ , the simulation gives a final $20$ -population of $0\,005\%$, merely $4\%$ of the total leakage .

Finally, we use this numerical simulation with full error model E to illustrate that the SNZ pulse preserves the resilience of the conventional NZ scheme [94] to low-frequency (quasistatic) flux offsets. Figure S10 shows that the single-qubit phase of the fluxed higher-frequency qubit ($Q$ ) and leakage are second-order sensitive to the offset. The conditional phase shows a very weak first-order dependence at zero offset. Numerical simulations with model D (not shown) show that the negative shift of the local maximum in originates from the finite flux-pulse rise time.

**2**

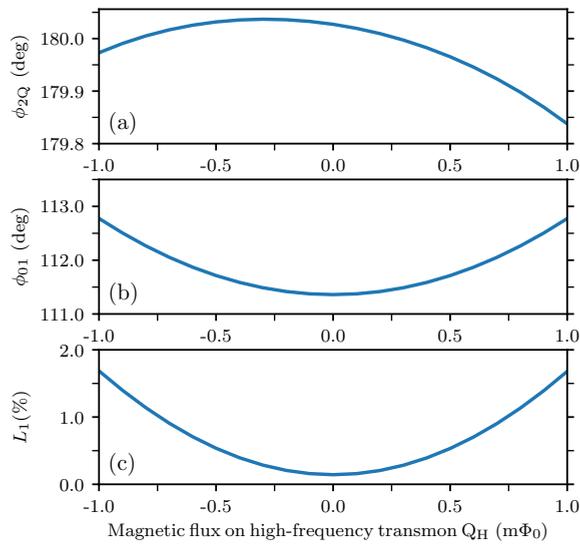

Figure 2.14: Numerical simulation (using model E and pair Q_H-Q_L) of the dependence of conditional phase $\phi_{2Q}$, single-qubit phase $\phi_{01}$, and leakage $L_1$ on quasistatic flux noise.

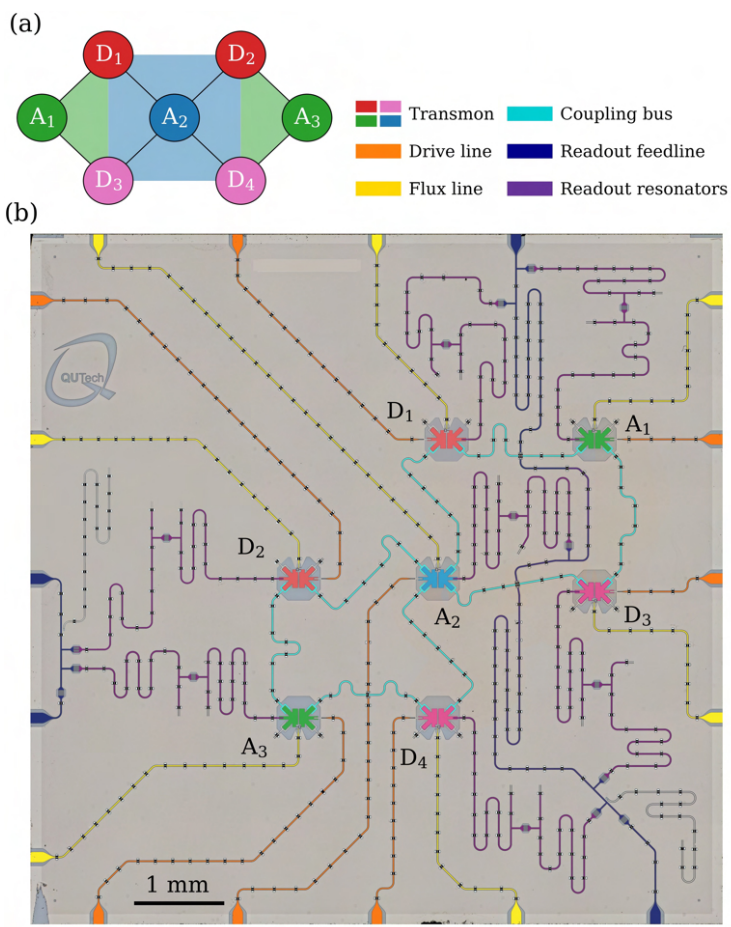

Future fault-tolerant quantum computers will require storing and processing quantum data in logical qubits. We realize a suite of logical operations on a distance-two logical qubit stabilized using repeated error detection cycles. Logical operations include initialization into arbitrary states, measurement in the cardinal bases of the Bloch sphere, and a universal set of single-qubit gates. For each type of operation, we observe higher performance for fault-tolerant variants over non-fault-tolerant variants, and quantify the difference through detailed characterization. In particular, we demonstrate process tomography of logical gates, using the notion of a logical Pauli transfer matrix. This integration of high-fidelity logical operations with a scalable scheme for repeated stabilization is a milestone on the road to quantum error correction with higher-distance superconducting surface codes.

## 3.1  Historical context

Two key capabilities will distinguish an error-corrected quantum computer from present-day noisy intermediate-scale quantum (NISQ) processors [12]. First, it will initialize, transform, and measure quantum information encoded in logical qubits rather than physical qubits. A logical qubit is a highly entangled two-dimensional subspace in the larger Hilbert space of many more physical qubits. Second, it will use repetitive quantum parity checks to discretize, signal, and (with aid of a decoder) correct errors occurring in the constituent physical qubits without destroying the encoded information [24]. Provided the incidence of physical errors is below a code-specific threshold and the quantum circuits for logical operations and stabilization are fault-tolerant, the logical error rate can be exponentially suppressed by increasing the distance (redundancy) of the quantum error correction (QEC) code employed [108]. At present, the exponential suppression for specific physical qubit errors (bit-flip or phase-flip) has been experimentaly demonstrated [25, 85, 109, 110] for repetition codes [111–113].

Leading experimental quantum platforms have taken key steps towards implementing QEC codes protecting logical qubits from general physical qubit errors. In particular, trapped-ion systems have demonstrated logical-level initialization, gates and measurements for single logical qubits in the Calderbank-Shor-Steane [114] and Bacon-Shor [31] codes. Most recently, entangling operations between two logical qubits have been demonstrated in the surface code using lattice surgery [115]. However, except for smaller-scale experiments using two ion species [116], trapped-ion experiments in QEC have so far been limited to a single round of stabilization.

In parallel, taking advantage of highly-non-demolition measurement in circuit quantum electrodynamics [63], superconducting circuits have taken key strides in repetitive stabilization of two-qubit entanglement [72, 73] and logical qubits. Quantum memories based on 3D-cavity logical qubits in cat [117, 118] and Gottesman-Kitaev-Preskill [119] codes have crossed the memory break-even point. Meanwhile, monolithic architectures have focused on logical qubit stabilization in a surface code realized with a 2D lattice of transmon qubits. Currently, the surface code [71] is the most attractive QEC code for solid-state implementation owing to its practical nearest-neighbor-only connectivity requirement and high error threshold. Recent experiments [25, 74] have demonstrated repetitive stabilization by post-selection in a surface code which, owing to its small size, is capable of quantum error detection but not correction. In particular, Ref. [74] has demonstrated the preparation of logical cardinal states and logical measurement in two cardinal bases. Here, we go beyond previous work by demonstrating a complete suite of logical-qubit operations for this small (distance-2) surface code while preserving multi-round stabilization. Our logical operations include initialization anywhere on the logical Bloch sphere (with significant improvement over previously reported fidelities), measurement in all cardinal bases, and a universal set of single-qubit logical gates. For each type of operation, we quantify the increased performance of fault-tolerant variants over non-fault-tolerant ones. We use a logical Pauli transfer matrix to describe a logical gate, analogous to the procedure commonly used to describe gates on physical qubits [120]. Finally, we perform logical state stabilization by means of repeated error detection where we compare the performance of two scalable, fault-tolerant stabilizer measurement schemes compatible with our quantum hardware architecture [98].

The distance-2 surface code (Fig. 3.1a) uses four data qubits ($D$  through $D$ ) to encode one logical qubit, whose two-dimensional codespace is the even-parity (i.e., eigenvalue +1) subspace of the stabilizer set

$$\mathscr{S} = \tag{3.1}$$

This codespace has logical Pauli operators

$$= \quad\quad\quad\quad\quad\quad \text{and} \tag{3.2}$$

$$= \quad\quad \text{and} \tag{3.3}$$

that anti-commute with each other and commute with $\mathscr{S}$, and logical computational basis

$$0 \quad = \frac{1}{2} ( 0000 + 1111 ) \tag{3.4}$$

$$1 \quad = \frac{1}{2} ( 0101 + 1010 ) \tag{3.5}$$

Measuring the stabilizers using three ancilla qubits ($A$ , $A$  and $A$  in Fig. 3.1a) allows detection of all physical errors that change the outcome of one or more stabilizers to $= 1$. This list includes all errors on any one single qubit. However, no error syndrome is unique to a specific physical error. For instance, a phase flip in any one data qubit triggers the same syndrome: $= 1$.

Consequently, this code cannot be used to correct such errors. We thus perform state stabilization by post-selecting runs in which no error is detected by the stabilizer measurements in any cycle. In this error-detection context, an operation is fault-tolerant if any single-fault produces a non-trivial syndrome and can therefore be post-selected out [121] (see Suppl. Material).

## 3.2  Method

### 3.2.1  Device

We use a superconducting circuit-QED processor (Fig. 3.1b) featuring the quantum hardware architecture proposed in Ref. [98]. Seven flux-tunable transmons are arranged in three frequency groups: a high-frequency group for $D$  and $D$ ; a middle-frequency group for $A$ , $A$  and $A$ ; and a low-frequency group for $D$  and $D$ . Similar to the device in Ref. [74], each transmon is transversely coupled to its nearest neighbors using a coupling bus resonator dedicated to each pair. This simplest and minimal connectivity minimizes multi-qubit crosstalk. Also, every transmon has a dedicated flux line for two-qubit gating, and a dispersively coupled readout resonator with Purcell filter enabling frequency-multiplexed readout [72, 102] using two feedlines. In contrast to Ref. [74], every transmon has a dedicated microwave drive line for single-qubit gating, avoiding the need to drive any via a feedline and thus reducing driving crosstalk.

All transmons are flux biased to their maximal frequency (i.e., flux sweetspot [93]), where measured qubit relaxation ($T_1$) and dephasing ($T_2$) times lie in the range 27—102 $\mu$s and 55—117 $\mu$s, respectively. Detailed information on the implementation and performance of single- and two-qubit gates for this same device can be found in Ref. [69]. Device characteristics are also summarized in Table 5.1.

The device was fabricated on a high-resistivity intrinsic Si<100> wafer that was first des-cummed using UV-ozone cleaner and stripped of native oxides using buffered oxide etch solution (BOE 7:1). The wafer was subjected to vapor of hexamethyldisilazane (HMDS) at $150^\circ$C and sputtered with $200\,\mathrm{nm}$ of niobium titanium nitride (NbTiN). Post dicing into smaller dies, a layer of hydrogen silsesquioxane (HSQ) was spun and baked at $300^\circ$C to serve as an inorganic sacrificial mask for wet etching of NbTiN. This layer was removed post base-patterning steps. The quantum plane was defined using electron-beam (e-beam) lithography of a high-contrast, positive-tone resist spun on top of the NbTiN-HSQ stack. Post development, the exposed region was first dry etched using SF6/O2 mixture and then wet etched to remove any residual metal. Dolan-bridge-style Al/AlOx/Al Josephson junctions were then fabricated using standard double-angle e-beam evaporation. Airbridges and crossovers were added using a two-step process. The first step involved patterning galvanic contact using e-beam resist ($\sim 6\,\mu$m thick) subjected to reflow. In the second step, the airbridges and crossovers were patterned with e-beam evaporated Al ($450\,\mathrm{nm}$ thick). Finally, the device underwent dicing, resist lift-off and Al wirebonding to a printed circuit board.

### 3.2.2 State tomography

To perform state tomography on the prepared logical states, we measure the $4^L - 1$ expectation values of data-qubit Pauli observables, $P_i = \bigotimes_j \sigma_j$ (except $\bigotimes_j I$). Interleaved with this measurement we also characterize the measurement POVM used to correct for readout errors in $P_i$. These are then used to construct the density matrix

$$\rho = \frac{\sum_i^{4^L-1} c_i P_i}{2^L} \tag{3.6}$$

with $c_0 = 1$, corresponding to $P_0 = \bigotimes_j I$. Due to statistical uncertainty in the measurement, the constructed state, $\rho$, might lack the physicality characteristic of a density matrix, that is, $\mathrm{Tr}(\rho) = 1$ and $\rho \geq 0$. Specifically, $\rho$ might not satisfy the latter constraint, while the former is automatically satisfied by $= 1$. To enforce these constraints, we use a maximum-likelihood method [120] to find the physical density matrix, , that is closest to the measured state, where closeness is defined in terms of best matching the measurement results. We thus minimize the cost function $^4 - \mathrm{Tr}( \ )$ , subject to $\mathrm{Tr}( \ ) = 1$ and $0$. We find the optimal using the convex-optimization package *cvxpy* via *cvx-fit* in Qiskit [122]. The fidelity to a target pure state, , is then computed as

$$= \tag{3.7}$$

One can further project onto the codespace to obtain a logical state using

$$= \frac{1}{2} \ \frac{\mathrm{Tr}( \quad )}{\mathrm{Tr}( \quad )} \tag{3.8}$$

where is the projector onto the codespace. Here, we can compute the logical fidelity using Eq. 3.7.

### 3.2.3  Process tomography in the codespace

A general single-qubit gate can be described [120] by a Pauli transfer matrix (PTM) $\mathcal{R}$ that maps an input state described by $=$ , with $= 1$, to an output state $'$:

$$' = \ \mathcal{R} \tag{3.9}$$

To construct $\mathcal{R}$ in the codespace, we use an overcomplete set of input states, $\ 0 \quad 1 \ $, $+ \quad + \quad$ , and their corresponding output states and perform linear inversion. The input and output logical states are characterized using state tomography of the data qubits to find the four-qubit state , which is then projected to the codespace using:

$$= \frac{\mathrm{Tr}( \quad )}{\mathrm{Tr}( \quad )} \tag{3.10}$$

We find that all the measured logical states already satisfy the constraints of a physical density matrix. This is likely to happen as one-qubit states that are not very pure usually lie within the Bloch sphere even within the uncertainty in the measurement. The constructed LPTM, however, might not satisfy the constraints of a physical quantum channel, that is, trace preservation and complete positivity (TPCP). These are better expressed by switching from the PTM representation to the Choi representation. The Choi state $^\mathcal{R}$ can be computed as

$$^\mathcal{R} = \frac{1}{4} \ \mathcal{R} \tag{3.11}$$

where the first tensor-product factor corresponds to an auxiliary subsystem. The TPCP constraints are $\mathrm{Tr}( \ ^\mathcal{R} ) = 1$, $^\mathcal{R} \quad 0$ and $\mathrm{Tr} \ ( \ ^\mathcal{R} ) = 1 \ 2$, where $\mathrm{Tr}$ is the partial trace over the auxiliary subsystem. In other words, $^\mathcal{R}$ is a density matrix satisfying an extra constraint.

We then find the optimal $\mathcal{R}$ using the same convex-optimization methods as for state tomography and adding this extra constraint [120, 123]. We compute the corresponding LPTM via

$$(\mathcal{R}\quad) = \mathrm{Tr}(\quad\mathcal{R}\quad\quad) \tag{3.12}$$

and the average logical gate fidelity using

$$= \frac{\mathrm{Tr}(\mathcal{R}^{\dagger}\quad\mathcal{R}\quad) + 2}{6} \tag{3.13}$$

where $\mathcal{R}$ is the LPTM of the ideal target gate.

### 3.2.4  Extraction of error-detection rate

The fraction of post-selected data in the repetitive error detection experiment (Fig. 3.4b) decays exponentially with the number of cycles . This is consistent with a constant error-detection rate per cycle . We extract this rate by fitting the function

$$(\ ) = \ (1\quad) \tag{3.14}$$

## 3.3  Results

### 3.3.1  Stabilizer measurements

Achieving high performance in a code hinges on performing projective quantum parity (stabilizer) measurements with high assignment fidelity, meaning one can accurately discriminate parity, and low additional backaction such that the state of the qubits after the measurement is properly projected onto the parity subspace. We implement each of the stabilizers in $\mathscr{S}$ using a standard indirect-measurement scheme [124, 125] with a dedicated ancilla. We benchmark the accuracy of each parity measurement by preparing the data-qubits in a computational state and measuring the probability of ancilla outcome $= 1$. As a fidelity metric, we calculate the average probability to correctly assign the parity ,

and , finding $94\ 2\%$, $86\ 1\%$ and $97\ 2\%$, respectively (see Suppl. Material Fig. 3.6).

### 3.3.2  Logical state initialization using stabilizer measurements

A practical means to quantify the backaction of stabilizer measurements is using them to initialize logical states. As proposed in Ref. [74], we can prepare arbitrary logical states by first initializing the data-qubit register in the product state

$$= \quad 0 + \quad 1\ 0 \quad\quad 0 + \quad 1\ 0 \tag{3.15}$$

**3**

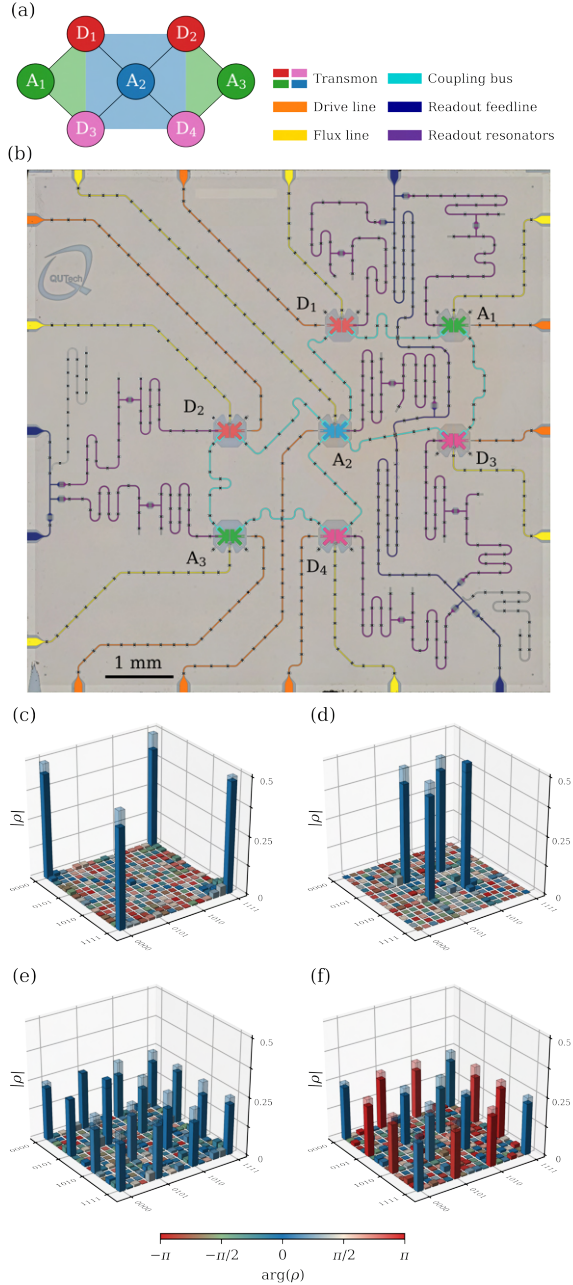

Figure 3.1: **Surface-7 quantum processor and initialization of logical cardinal states.** (a) Distance-two surface code. (b) Optical image of the quantum hardware with added false-color to emphasize different circuit elements. (c-f) Estimated physical density matrices, , after targeting the preparation of the logical cardinal states $0$   (c), $1$   (d), $+$   (e) and (f). Each state is measured after preparing the data qubits in $0000$ , $1010$ , $++++$ and $++$     , respectively. The ideal target state density matrix is shown in the shaded wireframe.

using single-qubit rotations $\theta$ on $D_a$ and $\phi$ on $D_c$ acting on $|0000\rangle$ ($\alpha = \cos\theta$ and $\beta = \sin\phi$). A follow-up round of stabilizer measurements ideally projects the four-qubit state onto the logical state

$$|\psi_L\rangle = \alpha|0_L\rangle + \beta|1_L\rangle \qquad \overline{\qquad\qquad+\qquad} \qquad (3.16)$$

with probability

$$P = \frac{1}{2}\left(\quad + \quad\right) \qquad (3.17)$$

We use this procedure to target initialization of the logical cardinal states $|0_L\rangle$, $|1_L\rangle$, $|+_L\rangle =$ $(|0_L\rangle + |1_L\rangle)/\sqrt{2}$, and $|-_L\rangle = (|0_L\rangle - |1_L\rangle)/\sqrt{2}$. For the first two states, the procedure is fault-tolerant according to the definition above. We characterize the produced states using full four-qubit state tomography including readout calibration and maximum-likelihood estimation (MLE) (Fig. 3.1c-f). The fidelity $F$ to the ideal four-qubit target states is $90.0 \pm 0.3\%$, $92.9 \pm 0.2\%$, $77.3 \pm 0.5\%$, and $77.1 \pm 0.5\%$, respectively. For each state, we can extract a logical fidelity $F_L$ by further projecting the obtained four-qubit density matrix onto the codespace [74], finding $99.83 \pm 0.08\%$, $99.97 \pm 0.04\%$, $96.82 \pm 0.55\%$, and $95.54 \pm 0.55\%$, respectively (see Methods). This sharp increase from $F$ to $F_L$ demonstrates that the vast majority of errors introduced by the parity check are weight-1 and detectable. A simple modification makes the initialization of $|+_L\rangle$ ($|-_L\rangle$) also fault-tolerant: initialize the data-qubit register in a different product state, namely $|++++\rangle$ ($|++--\rangle$), before performing the stabilizer measurements. With this modification, $F$ increases to $85.4 \pm 0.3\%$ ($84.6 \pm 0.3\%$) and $F_L$ to $99.78 \pm 0.09\%$ ($99.64 \pm 0.17\%$), matching the performance achieved when targetting $|0_L\rangle$ and $|1_L\rangle$.

### 3.3.3 Logical measurement of arbitrary states

A key feature of a code is the ability to measure logical operators. In the surface code, we can measure $Z_L$ ($X_L$) fault-tolerantly, albeit destructively, by simultaneously measuring all data qubits in the $Z$ ($X$) basis to obtain a string of data-qubit outcomes (each $+1$ or $-1$). The value assigned to the logical operator is the computed product of data-qubit outcomes as prescribed by Eq. 3.3 (3.2). Additionally, the outcome string is used to compute a value for the stabilizer(s) $S$ ($Z_aZ_b$ and $Z_cZ_d$), enabling a final step of error detection (Fig. 3.2a). Measurement of $Y_L = iX_LZ_L$ is not fault-tolerant. However, we lower the logical assignment error by also measuring $D_b$ in the $Y$ basis to compute a value for $X_aX_b$ and thereby detect bit-flip errors in $D_a$ and $D_b$.

We demonstrate $X_L$, $Y_L$ and $Z_L$ measurements on logical states prepared on two orthogonal planes of the logical Bloch sphere. Setting $\theta = \pi/2$ and sweeping $\phi$, we ideally prepare logical states on the equator (Fig. 3.2d)

$$|\psi_L\rangle = (|0_L\rangle + e^{i\phi}|1_L\rangle)/\sqrt{2} \qquad (3.18)$$

We measure the produced states in the $X$, $Y$ and $Z$ bases and obtain experimental averages $\langle X_L\rangle$, $\langle Y_L\rangle$ and $\langle Z_L\rangle$. As expected, we observe sinusoidal oscillations in and       and near-zero       . The reduced range of the       oscillation evidences the non-fault-tolerant nature of       measurement. A second manifestation is the higher fraction       of post-selected data for       (Fig. 3.2b). To quantify the logical assignment fidelity       with correction for initialization error, it is tempting to apply the formula

$$\frac{\phantom{xx}}{2} = (2 \qquad 1)(2 \qquad 1) \qquad (3.19)$$

inspired by the standard method to quantify readout fidelity of physical qubits from Rabi oscillations with limited initialization fidelity (described in the Supplementary Material). This method suggests       $= 95\,8\%$ for       and $87\,5\%$ for       . However, this method is not accurate for a logical qubit because not all input states outside the codespace are rejected by the limited set of stabilizer checks computable from the data-qubit outcome string and, moreover, detectable initialization errors can become undetectable when compounded with data-qubit readout errors. An accurate method to extract       based on the measured $16 \quad 16$ data-qubit assignment probability matrix (detailed in the Supplementary Material) gives       $=$ $98\,7\%$ for       and $91\,4\%$ for       .

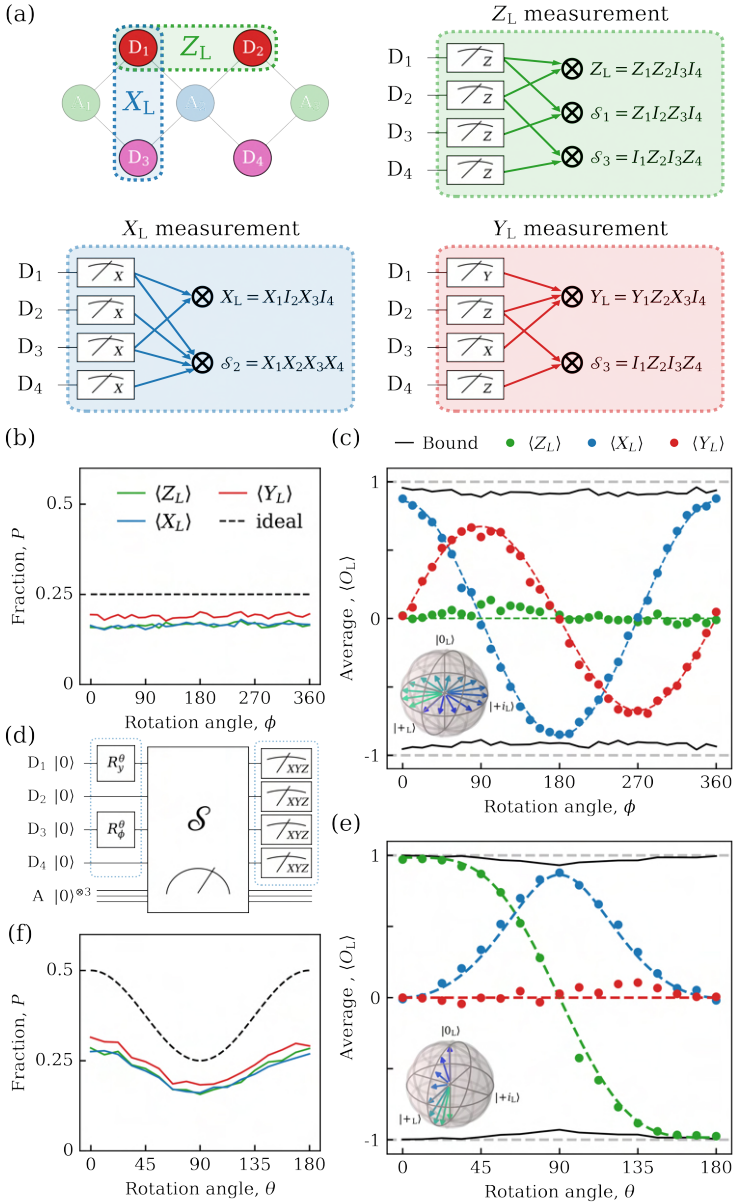

Figure 3.2: **Arbitrary logical-state initialization and measurement in the logical cardinal bases.** (a) Assembly of data-qubit measurements used to evaluate logical operators , and with additional error detection. (d) Initialization of logical states using the procedure described in Eq. 3.15. (c, e) , and logical measurement results as a function of the gate angles (c) and (e). The colored dashed curves show a fit of the analytical prediction based on Eqs. 3.18 and 3.20 to the data and the dark curve denotes a bound based on the measured of each state. (b, f) Total fraction of post-selected data as a function of the input angle for each logical measurement. The dashed curve shows the ideal fraction given by Eq. 3.17.

Setting       $= 0$ and sweeping       , we then prepare logical states on the       -       plane (Fig. 3.2e), ideally

$$= \qquad 0 \quad + \quad 1 \qquad \overline{\phantom{xxx} + \phantom{xxx}} \qquad (3.20)$$

Note that due to the changing overlap of the initial product state with the codespace,       is now a function of       (Eq. 3.17). The approximate extraction method based on the range of       suggests       $= 99\,4\%$, while the accurate method gives $99\,8\%$. Note that while both are fault-tolerant, the       measurement has higher fidelity than the       measurement as the former is only vulnerable to vertical double bit-flip errors while the latter is vulnerable to both horizontal and diagonal double phase-flip errors.

### 3.3.4 Logical gates

Finally, we demonstrate a suite of gates enabling universal logical-qubit control (Fig. 3.3). Full control of the logical qubit requires a gate set comprising Clifford and non-Clifford logical gates. Some Clifford gates, like       $_L$ and       $_L$ (where       $_L = {}^{-}\quad _L$ ), can be implemented transversally and therefore fault-tolerantly (Fig. 3.3d). We perform arbitrary rotations (generally non-fault-tolerant) about the       axis using the standard gate-by-measurement circuit [126] shown in Fig. 3.3a. In our case, the ancilla is physical ($A$ ), while the qubit transformed is our logical qubit. The rotation angle       is set by the initial ancilla state       $= ( 0 \quad +$ $1 ) \quad \overline{2}$. Since we cannot do binary-controlled       rotations, we simply post-select runs in which the measurement outcome is       $= +1$. However, we note that these gates can be performed deterministicaly using repeat-until-success [127]. Choosing       $= \quad 4$ implements the non-Clifford       $= \quad _L$ gate. A similar circuit (Fig. 3.3b) can be used to perform arbitrary rotations around the       axis. We compile both circuits using our hardware-native gateset (Figs. 3.3c,d). To assess logical-gate performance, we perform logical process tomography using the procedure illustrated in Fig. 3.3e for ___. First, we initialize into each of the six logical cardinal states __0__  __1__  __+__     __+__           . We characterize each actual input state by four-qubit state tomography and project to the codespace to obtain a logical density matrix. Next, we similarly characterize each output state produced by the logical gate and a second round of stabilizer measurements to detect errors occurred in the gate (full data in Fig. 3.7). Using this over-complete set of input-output logical-state pairs, combined with MLE (see Methods), we extract a logical Pauli transfer matrix (LPTM). The resulting LPTMs for the non-fault-tolerant ____ and ___$_\text{L}$ gates as well as the fault-tolerant ___$_\text{L}$ and ____ are shown in Fig. 3.3e. From the LPTMs, we extract average logical gate fidelities ___ (Eq. 3.13) 97.3%, 95.6%, 97.9%, and 98.1%, respectively.

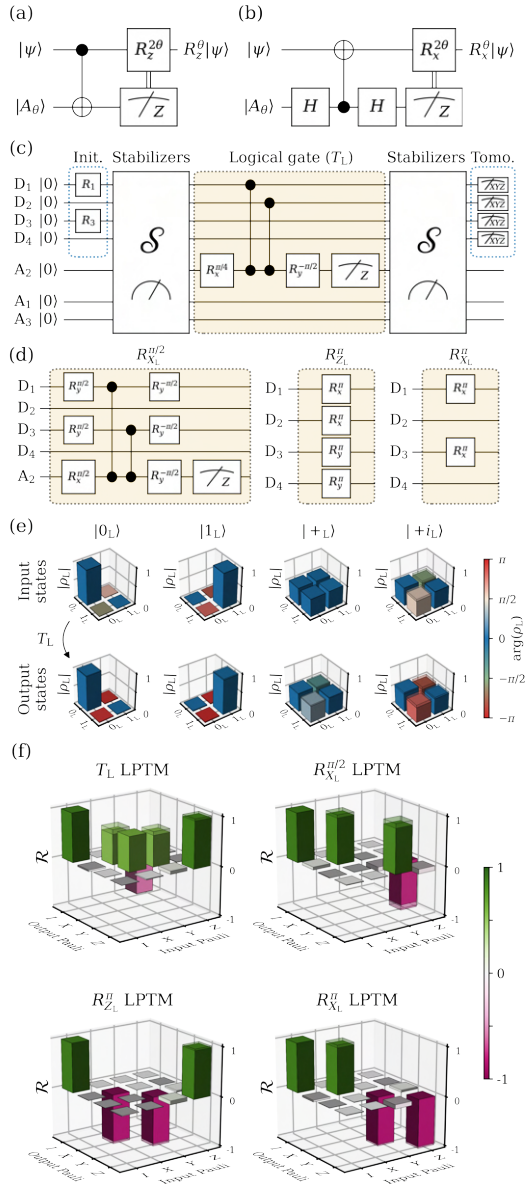

Figure 3.3: **Logical gates and their characterization.** (a, b) General gate-by-measurement schemes realizing arbitrary rotations around the (a) and (b) axis of the Bloch sphere. (c) Process tomography experiment of the gate. Input cardinal logical states are initialized using the method of Fig. 3.2. Output states are measured following a second round of stabilizer measurements. (d) Logical $_L$, $_L$ and $_L$ gates compiled using our hardware-native gateset. (e) Logical state tomography of input and output states of the gate. These logical density matrices are obtained by performing four-qubit tomography of the data qubits and then projecting onto the codespace. (f) Extracted (solid) and ideal (wireframe) logical Pauli transfer matrices.

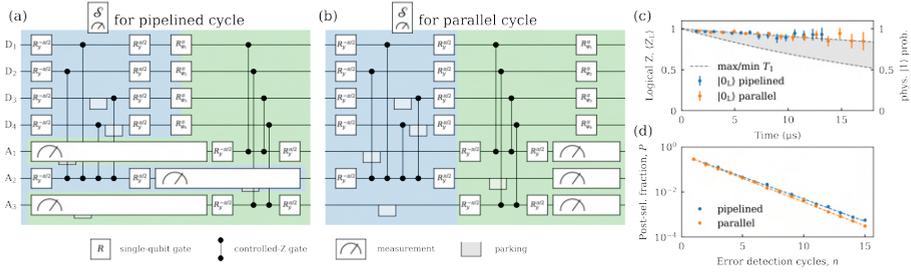

Figure 3.4: **Repetitive error detection using pipelined and parallel stabilizer measurement schemes.** (a, b) Gate sequences used to implement the pipelined (a) and parallel (b) stabilizer measurement schemes. Gate duration is $20\,\text{ns}$ for single-qubit gates, $60\,\text{ns}$ for controlled-Z (CZ) gates and parking [73, 98], and $540\,\text{ns}$ for ancilla readout. The order of CZs in the ___________ stabilizer (blue shaded region) prevents the propagation of ancilla errors into logical qubit errors [121]. The total cycle duration for the pipelined (parallel) scheme is $840\,\text{ns}$ ($1000\,\text{ns}$). (c) Estimated ___ expectation value, ____, measured for the __0__ state versus the duration of the experiment using the pipelined (blue) and the parallel (orange) schemes. We also plot the excited-state probability (right axis) set by the maximum and minimum physical qubit ___. (d) Post-selected fraction of data versus the number of error detection cycles ___ for the pipelined (blue) and parallel (orange) scheme.

### 3.3.5   Pipelined versus parallel stabilizer measurements

A scalable control scheme is fundamental to realize surface codes with large code distance. To this end, we now compare the performance of two schemes suitable for the quantum hardware architecture proposed in Ref. [98]. These schemes are scalable in the sense that their cycle duration remains independent of code distance. The pipelined scheme interleaves the coherent operations and ancilla readout steps associated with stabilizer measurements of type ___ and ___ by performing the coherent operations of ___ (  ) type stabilizers during the readout of ___ (  ) type stabilizers (Fig. 3.4a). The parallel scheme performs all ancilla readouts simultaneously (Fig. 3.4b). The pipelined cycle scheme duration is shorter than the parallel scheme by 16% which can potentially increase the performance of the code. This only occurs if the interleaved readout of ancillas does not result in increased measurement-induced dephasing between them. To compare their performance, we initialize and stabilize $0$ for up to $N = 15$ cycles. We perform refocusing pulses ($\pi$) on the data qubits to correct for coherent errors during the measurement of ancilla qubits. We also separately calibrate the equatorial rotation axis of this gate for each scheme to extract the best performance. At each $N$, we take data back-to-back for the two schemes in order to minimize the effect of parameter drift, repeating each experiment up to $256 \times 10$ times. Figure 3.4c shows the measurement outcome averaged over the post-selected runs. We extract the error-detection rate from the $N$-dependence of the fraction of post-selected data (Fig. 3.4d) using the procedure described in Methods. We observe that the error rate is slightly lower for the pipelined scheme ($\approx 45\%$), most likely due to the shorter duration of the cycle. This superiority is consistent across different input logical states (see Fig. 3.8) with an average ratio $\approx 97\%$.

## 3.4 Discussion

We have demonstrated a suite of logical-level initialization, gate and measurement operations in a distance-2 superconducting surface code undergoing repetitive stabilizer measurements. For each type of logical operation, we have quantified the increased performance of fault-tolerant variants over non-fault-tolerant variants. Table 3.1 summarizes all the results. We can initialize the logical qubit to any point on the logical Bloch sphere, with logical fidelity surpassing Ref. [74]. In addition to characterizing initialized states using full four-qubit tomography, we also demonstrate logical measurements in all logical cardinal bases. Finally, we demonstrate a universal single-qubit set of logical gates by performing logical process tomography, using the concept of a logical-level Pauli transfer matrix. As expected, the fidelity of the fault-tolerant gates is higher than the non-fault-tolerant ones. However, one would expect a sharper difference given the typical error rates of the operations envolved. We believe this could be due to errors introduced by the stabilizer measurements which might be dominant over the errors of the logical gate itself.

With a view towards implementing higher-distance surface codes using our quantum-hardware architecture [98], we have compared the performance of two scalable stabilization schemes: the pipelined and parallel measurement schemes. In this comparison, two main factors compete. On one hand, the shorter cycle time favors pipelining. On the other, the pipelining introduces extra dephasing on ancilla qubits of one type during readout of the other. The performance of both schemes is comparable, but slightly higher for the pipelined scheme. From detailed density-matrix simulations discussed in the Supplementary Material, we further understand that conventional qubit errors such as energy relaxation, dephasing and readout assignment error alone do not fully account for the net error-detection rate observed in the experiment (see Fig. 3.14 and also not for the reduction in Figs. 3.2b,f; see Fig. 3.15). We believe that the dominant error source is instead leakage to higher transmon states incurred during CZ gates. Our data (Fig. 3.13) shows that the error detection scheme successfully post-selects leakage errors in both the ancilla and data qubits. Learning to identify these non-qubit errors and to correct them without post-selection is the subject of ongoing research [77, 128**?**] and an outstanding challenge in the quest for quantum fault-tolerance with higher-distance superconducting surface codes [129], which to this date have yet to be implemented with repeated error correction.

| Logical operation | | Characteristic | Logical fidelity metric | value (%) |
|---|---|---|---|---|
| Init. | 0 | FT | | 99.83 |
| | 1 | FT | | 99.97 |
| | + | Non-FT/FT | | 96.82/99.78 |
| | | Non-FT/FT | | 95.54/99.64 |
| Meas. | | FT | | 99.8 |
| | | FT | | 98.7 |
| | | Non-FT | | 91.4 |
| Gate | L | FT | | 97.9 |
| | L | FT | | 98.1 |
| | L | Non-FT | | 95.6 |
| | | Non-FT | | 97.3 |

Table 3.1: **Summary of logical initialization, measurement, and gate operations and their performance.** Fault-tolerant operations are labelled FT and non-fault tolerant ones Non-FT. Quoted values are those extracted with the accurate method described in the Supplementary Material.

## 3.5   Data availability

The data supporting the plots and claims within this paper is available online at http://github.com/DiCarloLab-Delft/Logical_Qubit_Operations_Data.

## 3.6   Supplemental material

This supplement provides additional information in support of statements and claims made in the main text.

### 3.6.1   Device characteristics

| Qubit | D | D | D | D | A | A | A |
|---|---|---|---|---|---|---|---|
| Qubit transition frequency at sweetspot, $\omega/2\pi$ (GHz) | 6.433 | 6.253 | 4.535 | 4.561 | 5.770 | 5.881 | 5.785 |
| Transmon anharmonicity, $\alpha/2\pi$ (MHz) | -280 | — | -320 | — | -290 | -285 | — |
| Readout frequency, $\omega_r/2\pi$ (GHz) | 7.493 | 7.384 | 6.913 | 6.645 | 7.226 | 7.058 | 7.101 |
| Relaxation time, $T_1$ (μs) | 27 | 44 | 32 | 102 | 38 | 58 | 43 |
| Ramsey dephasing time, $T_2^*$ (μs) | 44 | 55 | 51 | 103 | 55 | 60 | 52 |
| Echo dephasing time, $T_2$ (μs) | 59 | 70 | 55 | 117 | 69 | 79 | 73 |
| Best multiplexed readout fidelity, $F$ (%) | 98.6 | 98.9 | 96.0 | 96.5 | 98.6 | 94.2 | 98.9 |
| Single-qubit gate fidelity, $F_g$ (%) | 99.95 | 99.86 | 99.83 | 99.98 | 99.95 | 99.91 | 99.95 |

Table 3.2: **Summary of frequency, coherence and readout parameters of the seven transmons.** Coherence times are obtained using standard time-domain measurements [58]. Note that temporal fluctuations of several μs are typical for these values. The multiplexed readout fidelity, $F$, is the average assignment fidelity [106] extracted from single-shot readout histograms after mitigating residual excitation using initialization by measurement and post-selection [130, 131].

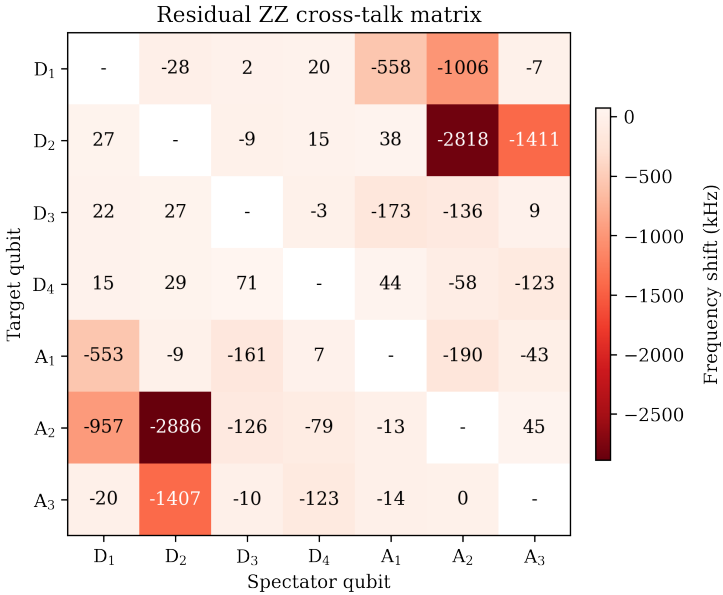

Figure 3.5: **Residual $ZZ$-coupling matrix.** Measured residual $ZZ$ coupling between all transmon pairs at the bias point (their simultaneous flux sweetspot [93]). Each matrix element denotes the frequency shift that the target qubit experiences due to the spectator qubit being in the excited state, $|1\rangle$. The procedure used for this measurement is similar to the one described in Ref. [132].

### 3.6.2 Parity-check performance

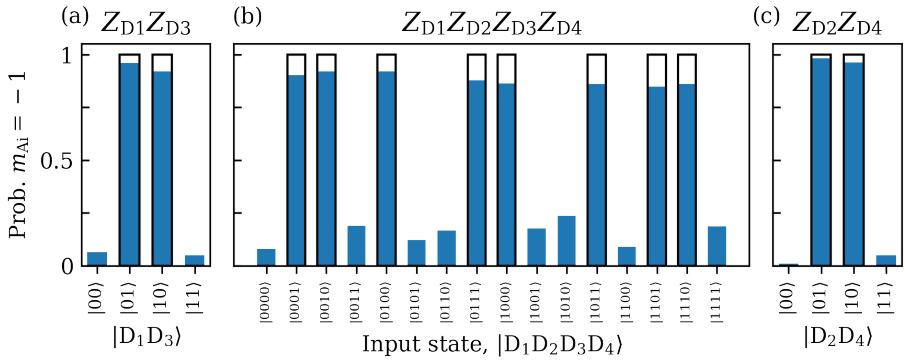

Figure 3.6: **Characterization of the assignment fidelity of -type parity checks.** (a) , (b) * , and (c) parity checks implemented using $A$ , $A$ , and $A$ , respectively. Each parity check is benchmarked by preparing the relevant data qubits in a computational state and then measuring the probability of ancilla outcome $= 1$. Measured (ideal) probabilities are shown as solid blue bars (black wireframe). From the measured probabilities we extract average assignment fidelities $94\,2\%$, $86\,1\%$ and $97\,2\%$, respectively. *This parity check implements the stabilizer measurement with the addition of single-qubit gates on data qubits to perform a change of basis.

### 3.6.3 Process tomography

**3**

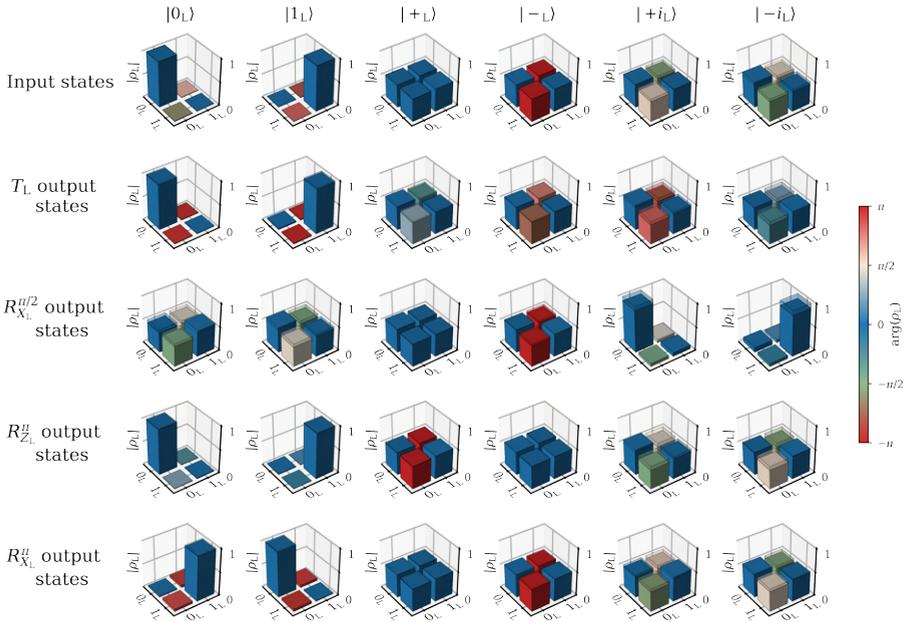

Figure 3.7: **Full set of logical states measured in the logical process tomography procedure.** Measured input and output logical states for each logical gate. Each state is measured using the procedure described in the Methods.

### 3.6.4   Logical state stabilization

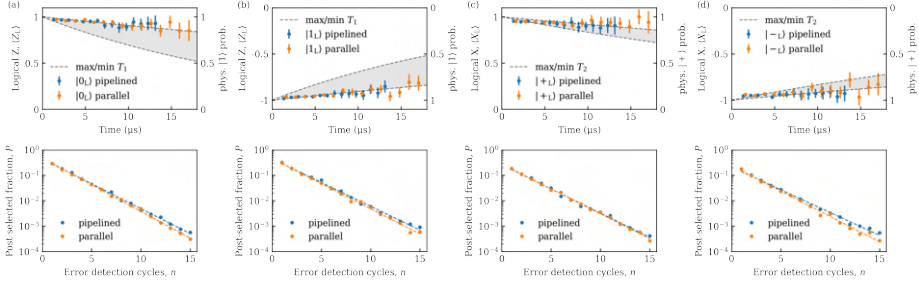

Figure 3.8: **Stabilization of logical cardinal states by repetitive error detection using the pipelined and parallel schemes.** From left to right, the stabilized logical states are $0$ , $1$ , $+$ and . For each logical state, the top panel shows the evolution of the relevant logical operator as a function of number of cycles, , plotted versus wall-clock time. Error bars are estimated based on the statistical uncertainty given by ( ). The shaded area indicates the range of physical qubit    values (a and b) and    values (c and d) plotted on the right-axis. Each bottom panel shows the corresponding post-selected fraction of data, ( ).

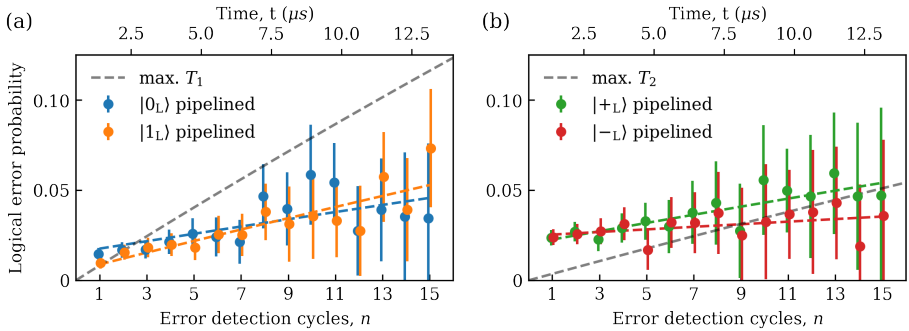

Figure 3.9: **Logical error probability versus number of error detection cycles.** Logical error probability after    cycles of error detection for states $0$ , $1$   (a) and $+$ ,   (b) measured using the pipelined scheme. For comparison, the grey dashed curves in (a) and (b) correspond to the physical error probability of the best    and    respectively. The logical error rate per round of detection, extracted by fitting the data (colored dashed lines), is $0\,43\%$, $0\,67\%$, $0\,49\%$ and $0\,15\%$ respectively.

### 3.6.5 Logical error rate

Here, we study the error rate of the logical qubit and compare it to that of a physical qubit. The probability for a logical error on an eigenstate of      after    cycles is given by

$$= \frac{1 \qquad (\ \ )}{2} \tag{3.21}$$

For eigenstates of      we compare the measured error rate to the error experienced due to      on a physical qubit (Fig. 3.9a). In a physical qubit in the excited state, $1$ , this error is given by

$$= 1 \qquad ^{-} \qquad ^{1} \tag{3.22}$$

For eigenstates of      (Fig. 3.9b) we now consider the error experienced due to      . This error for physical qubits whose state lies on the equator of the Bloch sphere is

$$= \frac{1 \qquad ^{-} \qquad ^{2}}{2} \tag{3.23}$$

We find that the logical error rates for all states, corresponding to the slopes of the colored dashed curves in Fig. 3.9, are lower than the corresponding best physical error rates.

### 3.6.6 Fault tolerance of logical operations

### 3.6.7 Fault tolerance of an operation

We begin by elaborating the definition of a fault-tolerant logical operation. We consider a single fault occurring during the circuit implementing the logical operation, where a fault can refer to any single-qubit Pauli error following a single-qubit gate or an idling step, or any two-qubit Pauli error following a two-qubit gate or a measurement error. Furthermore, a single fault can also refer to any single-qubit error on the input state of the logical operation. Thus we consider the performance of the logical operation either when there is a single error at input *or* a fault in the logical operation. In the context of error detection, the logical operation such as state initialization or gate execution, is fault-tolerant if any such fault either produces a non-trivial syndrome (in case the circuit involves the measurement of the stabilizers) and is thus post-selected out or leads to an outgoing state that is either the desired logical state or any logical state together with a detectable error. This implies that if the logical operation is followed up by a fictitious and ideal measurement of the stabilizers, the detectable error would lead to a non-trivial syndrome and be post-selected out, ensuring that the outgoing state could only be the desired logical state. For a fault-tolerant logical measurement we require that the logical measurement outcome is correct, i.e. if it is applied to a logical state with single error *or* a fault happens during the logical measurement, we either post-select or get the correct outcome.

### 3.6.8  Logical state initialization using stabilizer measurements

We perform fault-tolerant initialization of the logical cardinal states $\overline{0}$, $\overline{1}$, $\overline{+}$ and $\overline{-}$. We focus on the initialization of $\overline{0}$ and $\overline{1}$ and then extend these arguments to $\overline{+}$ and $\overline{-}$.

To prepare $\overline{0}$ ($\overline{1}$) the data-qubit register is prepared in the state $0000$ ($1010$) and a single round of stabilizer measurements is performed. Consider any single-qubit error occurring during the initialization of the data qubits. Any such error will be detected by the following stabilizer measurement and post-selected out.

Then, consider a fault occurring during the stabilizer measurement, implemented by the circuits shown in Fig. 3.4. Any single-qubit error on the data qubits following an idling step either leads to non-trivial syndrome (if it occurs before the two-qubit gate) or constitutes a detectable error (if it occurs after the two-qubit gate). A single-qubit error following any of the single-qubit gates will similarly either produce non-trivial syndrome or lead to an error that is either detectable or an element in $\mathscr{S}$.

A measurement error on ancilla qubits $A_1$ or $A_3$ will always produce a non-trivial syndrome (since the input state is an eigenstate of the measured $Z$-type stabilizers) and will thus be post-selected out. However, a measurement error on $A_2$ can still lead to trivial syndrome (as the input state is not an eigenstate of the $X$-type stabilizer). This will result in the preparation of the desired logical state together with a detectable phase-flip error on any of the qubits.

Any two-qubit Pauli error after each CZ gate involved in the measurement of the $Z$-type stabilizers will lead to the preparation of the desired logical state together with an error that is either in $\mathscr{S}$ (for example in the case when a bit-flip error occurs on the ancilla qubit and phase-flip error on the data-qubit following the first CZ gate of the circuits) or one that is detectable. The same statement holds for any two-qubit error after each of the CZ gates involved in the $X$-type stabilizer check. Here the order of the gates is crucial to ensure that any two-qubit error after the second CZ gate of the circuit is detectable [121].

The fault-tolerant preparation of $\overline{+}$ ($\overline{-}$) involves initializing the data-qubit register in the state $++++$ ($++--$) instead. The arguments for the fault-tolerance of this operation follow closely the ones presented for $\overline{0}$ ($\overline{1}$) with the only difference being that a single measurement error on ancilla qubits $A_1$ or $A_3$ can now lead to a trivial outcome and an outgoing state that involves a detectable (bit-flip) error, while a measurement error on $A_2$ will instead always lead to a non-trivial syndrome.

When initializing arbitrary logical states by preparing the data qubit register in the state given by Eq. 3.15, the procedure is fault-tolerant only when preparing $\overline{0}$ (corresponding to $\theta = 0$ and $\phi = 0$) or $\overline{1}$ (corresponding to $\theta = \pi$ and $\phi = 0$), which are discussed above. When preparing $\overline{+}$ (corresponding to $\theta = \pi/2$ and $\phi = 0$) or $\overline{-}$ (corresponding to $\theta = \pi/2$ and $\phi = \pi$), the input states are $+0+0$ and $+0-0$ respectively. In these cases a single fault (for example a phase-flip error on qubit $D$ on the input state) is not detectable by the stabilizer measurement and instead leads to the initialization of (the opposite states) and $+$ , respectively. The preparation of any other state on the equator of the Bloch sphere is not fault-tolerant either, following the same reasoning: an under- or overrotation of will directly translate to an error at the logical level.

### 3.6.9 Fault-tolerant logical measurements

We now consider the fault-tolerance of the logical measurement, which is performed following the procedure described in the Results (see Fig. 3.2). The only fault to consider in these circuits is a measurement error on one of the data qubits. When measuring or , any such error will result in a non-trivial syndrome (given the assumption that the input state is in the codespace) and the logical measurement outcome is post-selected out. When the fault is instead a single-qubit error on the input state, bringing this state outside of the codespace, the fault-free logical measurement will either detect this error or this error will not have an affect on the logical measurement outcome.

For the non-fault-tolerant measurement of only the value for the stabilizer can be computed and used to detect errors on $D$ and $D$ . Thus a single fault (for example a measurement error on either $D$ or $D$ ) can lead to an incorrect logical measurement outcome, making this operation non-fault-tolerant.

### 3.6.10 Transversal logical gates and non-fault-tolerant gate injection

The logical gates $_L$ and $_L$ (shown in Fig. 3.3d) are clearly fault-tolerant as any single-qubit error following any of the single-qubit gates involved in the circuits is detectable. At the same time the transversal execution of these gates ensures that no single qubit error on the input state can spread to two or more qubits, ensuring that any such fault is detectable. These fault-tolerant properties do not hold when we consider the and $_L$ logical gates implemented by the gate-by-measurement circuits shown in Fig. 3.3b (and Fig. 3.3d). For example a bit-flip error on $A$ following the first single-qubit gate of the circuit will result in a logical error. More generally, any under- or over-rotation in the rotation angle used in preparing the ancilla qubit in translates to a different rotation at the logical level than desired.

### 3.6.11 Quantifying the logical assignment fidelity

We start this section reviewing how the readout fidelity of a physical qubit is standardly quantified from the contrast of a Rabi oscillation when the input states $_\pm$ closest to the eigenstates $_\pm$ of the measured observable have limited fidelity (assumed equal for both).

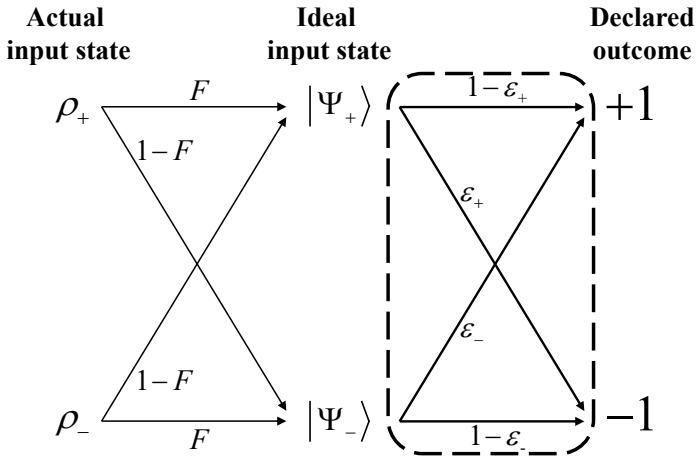

Figure 3.10: **Probability flow diagram for physical qubit readout**. Please see text for the definition of all variables shown. The characterization of physical qubit readout robust to initialization errors determines the probabilities within the dashed box.

Evidently, we want $F$ to quantify the performance of readout only, independent of errors in the input state. To this end, consider the probability flow diagram of Fig. 3.10. We define $F$ as the average probability of proper assignment for perfect input states $|\Psi_\pm\rangle$. Therefore, $F = 1 - (\varepsilon_+ + \varepsilon_-)/2$, where $\varepsilon_\pm$ is the probability of wrongly assigning outcome $\mp 1$ for input state $|\Psi_\pm\rangle$. The positive and negative extremes of the Rabi oscillation are

$$P_+ = F_+ \tilde\varepsilon_+ + \bar F_- \varepsilon_- \tag{3.24}$$

$$P_- = \bar F_- \tilde\varepsilon_- + F_+ \varepsilon_+ \tag{3.25}$$

where $F = 1 - \varepsilon$ and $\tilde\varepsilon_\pm = 1 - \varepsilon_\pm$. Combining these expressions and simplifying terms, it follows that the contrast of the Rabi oscillation (defined as half the peak-to-peak range), is

$$\frac{P_+ - P_-}{2} = (2F - 1)/2 - F + 1 \tag{3.26}$$

Turning over to the logical qubit, we similarly define the logical readout fidelity $F_L$ as the average probability of proper assignment for perfectly prepared logical states $|\Psi_\pm\rangle$ i.e., the logical states that are eigenstates of the measured observable with eigenvalue $\pm 1$. To this end, it is tempting to apply the above equation to the oscillations in Fig. 3.2, simply substituting $F \to F_L$ and $P \to P_L$. However, this approach is not accurate. This is because the probability flow diagram for the logical qubit, shown in Fig. 3.11, is more complex. The quantities we seek to determine are those inside the dashed box, describing logical readout on perfect input logical states: $r_\pm$ is the probability that the experimental logical measurement on $|\Psi_\pm\rangle$ is rejected (R), which occurs whenever the data-qubit outcome string

## Logical assignment butterfly

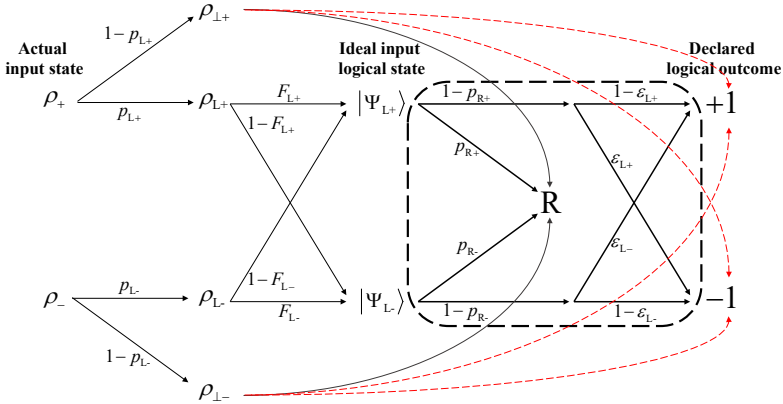

Figure 3.11: **Probability flow diagram for logical readout.** Please see text for the definition of all variables shown. The characterization of logical qubit readout robust to initialization errors determines the probabilities within the dashed box. The probability for states outside the codespace to not be rejected (red dashed curves) is the primary reason why using Eq. 3.26 with the substitutions          and          does not yield an accurate estimate of    .

produces a value of $-1$ on at least one of the stabilizers computable from the string; $\varepsilon_\pm$ is the probability of wrongly assigning logical outcome $-1$ for $\rho_\pm$, conditioned on no rejection. Using these definitions, $\varepsilon = 1 - (\varepsilon_+ + \varepsilon_-)/2$. Outside the dashed box, $\rho_\pm$ is the experimental input state closest to $\rho_\pm$. This imperfect input state has probability $p_\pm$ of being in the codespace and its projection onto the codespace, $\rho_{L\pm}$, has fidelity $F_\pm$ to $\Psi_\pm$. Finally, $\rho_{\perp\pm}$ is the projection of $\rho_\pm$ outside the codespace. We now discuss the more accurate method used to quantify $\varepsilon$ that does not rely on Eq. 3.26. We first consider the transformation of $\rho_\pm$ by the pre-rotations that we perform when measuring in each cardinal logical basis,          . For          there are no measurement pre-rotations, so the states are

$$|0\rangle = \frac{1}{\sqrt{2}}(|0000\rangle + |1111\rangle) \tag{3.27}$$

$$|1\rangle = \frac{1}{\sqrt{2}}(|0101\rangle + |1010\rangle) \tag{3.28}$$

For     , the transformed states are

$$|-\rangle - - - |+\rangle = \frac{1}{2}(|0000\rangle + |0101\rangle + |1010\rangle + |1111\rangle) \tag{3.29}$$

$$|-\rangle - - - = \frac{1}{2}(|0011\rangle + |0110\rangle + |1001\rangle + |1100\rangle) \tag{3.30}$$

Finally, for     , these are

$$|-\rangle |+\rangle = \frac{1}{2}(|0000\rangle - |0111\rangle + |1010\rangle + |1101\rangle) \tag{3.31}$$

$$|-\rangle = \frac{1}{2}(-|0010\rangle - |0101\rangle - |1000\rangle + |1111\rangle) \tag{3.32}$$

The above expressions make clear, for each logical cardinal basis, which data-qubit outcome strings are rejected and which ones are accepted with declared logical outcome $+1$ or $-1$. For completeness, all cases are detailed in Table 3.3.

The key experimental input needed to proceed is the data-qubit assignment probability matrix , shown in Fig. 3.12. Each element of this $16 \times 16$ matrix gives the experimental probability of measuring a string of data outcomes ( ) $\pm 1 \cdots \pm 1$ (varying across rows) when performing simultaneous readout of the data qubits having prepared them in physical computational state , $0 \cdots 1$ (varying across columns). These computational states are prepared by applying the needed parallel combination of pulses on data qubits starting with all qubits (including ancillas) initialized in $0$ . With in

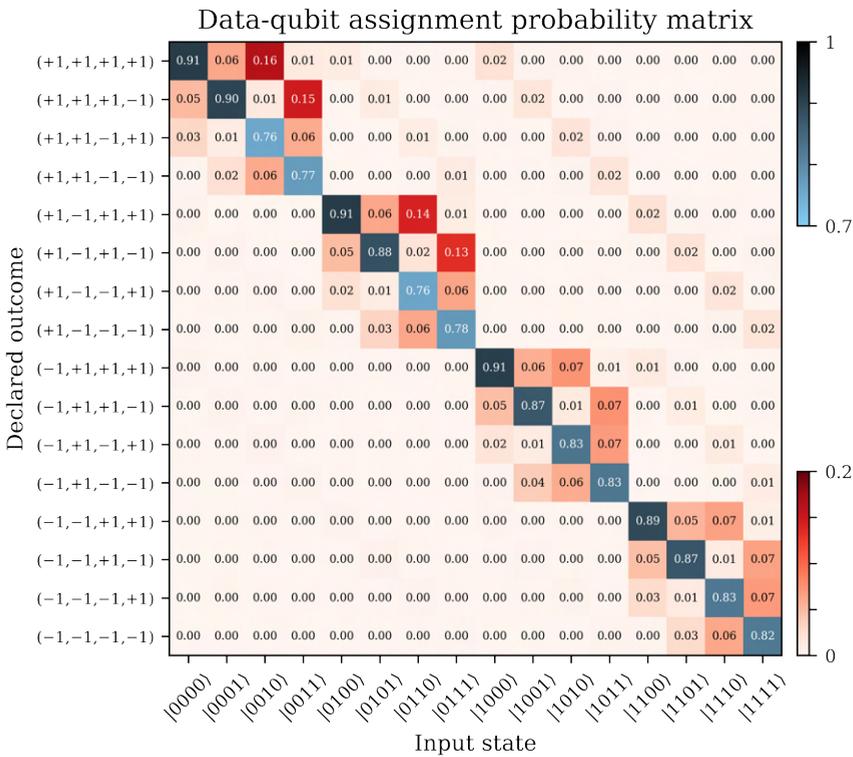

Figure 3.12: **Experimental data-qubit assignment probability matrix.** Each element of gives the experimental probability of measuring outcome string ( ) (varying across rows) when performing simultaneous measurement of the data qubits prepared in , $0 \cdots 1$ (varying across columns).

hand, it is straightforward to compute the probabilities for all strings of data-qubit outcomes for each choice of and $\pm$ . This is given by , where is vector (size 16) whose elements are the probabilities (in the physical data-qubit computational basis) of the corresponding state in Eqs. 3.27-3.32. For example, $= (1\ 2\ 0\ \cdots\ 0\ 1\ 2)$ for and $0$ ,

and $\quad = (1\ 4\ 0\ 0\ 0\ 0\ 1\ 4\ 0\ 0\ 0\ 0\ 1\ 4\ 0\ 0\ 0\ 0\ 1\ 4)$ for $\quad$ and $+\quad$. From and the rejection and logical assignment rules in Table 3.3, it is straightforward to compute all the probabilities within the dashed box of Fig. 3.11. The final results are presented in Table 3.4. The key assumption behind this analysis is that errors induced by single-qubit gates (both during preparation of the physical data-qubit computational states needed for determination of $\quad$ and the measurement pre-rotations when performing logical measurement in $\quad$ and $\quad$) are small compared to the errors induced by data-qubit readout. This assumption is safe given the performance metrics summarized in Table 5.1.

### 3.6.12   Numerical analysis

### 3.6.13   Leakage in experiment

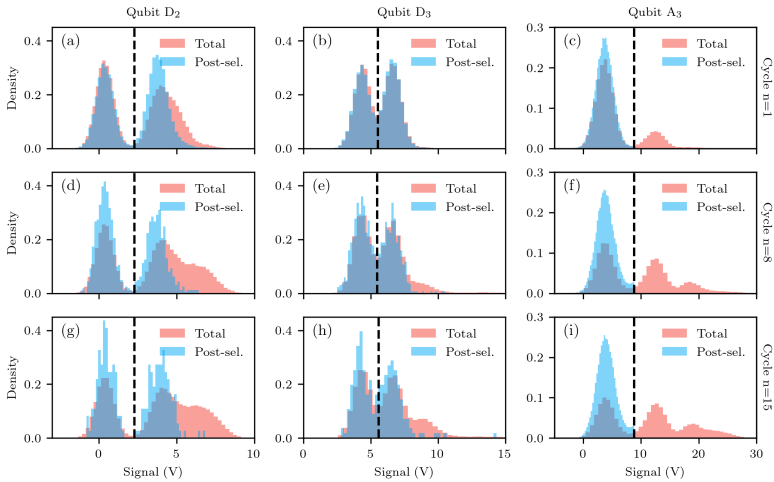

Figure 3.13: **Signature of transmon leakage in experimental data.** Single-shot readout histograms obtained at cycle $\quad$ over all shots (red) and the post-selected shots based on detecting no error in any cycles up to $\quad$ (blue) for $D\quad$ (left), $D\quad$ (middle) and $A\quad$ (right) and at cycle $\quad = 1$ (top row), $\quad = 8$ (middle row) and $\quad = 15$ (bottom row). The dashed black lines indicate the thresholds used to discriminate $0\quad$ from $1\quad$.

We observe a clear signature of leakage accumulation with the increasing number of error-detection cycles in the single-shot readout histograms obtained at the end of each experiment. In Fig. 3.13 we show examples of this accumulation for $D\quad$, $D\quad$ and $A\quad$ at cycles $\quad = 1$, $\quad = 8$ and $\quad = 15$. For dispersive readout, a transmon in state $2\quad$ induces a different frequency shift in the readout resonator compared to state $0\quad$ or $1\quad$. The increased number of data points at $\quad = 8$ and $\quad = 15$ shown in Fig. 3.13, following a Gaussian distribution with a mean and standard deviation different from those observed at $\quad = 1$ is thus a clear manifestation of leakage to the higher-excited states (mostly to $2\quad$). We believe that the dominant source of leakage in our processor are the CZ gates [69, 94]. However, the leakage rate for each gate has not been experimentally characterized, e.g., by performing leakage-modified randomized benchmarking experiments [97, 133]. This is because our CZ tune-up procedure is performed in a parity-check block unit. This maximizes the performance of the parity-check but makes the gate unfit for randomized benchmarking protocols. We can estimate the population $\mathcal{L}$( ) in the leakage subspace $\mathcal{L}$ at cycle from the single-shot readout histograms. We perform a fit of a triple Gaussian model to the histograms from which we extract the voltage that allows for the best discrimination of $2$ from $1$ and $0$. The leaked population $\mathcal{L}$( ) is then given by the fraction of shots declared as $2$ over the total number of shots. Assuming that leakage is only induced by the CZ gates (on the transmon being fluxed to perform the gate) and that each CZ gate has the same leakage rate , we can use the Markovian model presented in Ref. [77] to estimate the value leading to the observed population $\mathcal{L}$( ). This analysis gives a estimate in the approximate range $1$ $4\%$ for most transmons. However, we do not consider these estimates to be accurate due to the low fidelity with which $2$ can be distinguished from $1$ and instead treat as a free parameter in our simulations (see below).

The histograms of the post-selected shots in Fig. 3.13 demonstrate that post-selection rejects runs where leakage on those transmons occurred. Thus, while leakage may considerably impact the error-detection rate in the experiment [77], we do not expect it to significantly affect the fidelity of the logical initialization, and gates.

### 3.6.14  Density-matrix simulations

We perform numerical density-matrix simulations using the *quantumsim* package [134] to study the impact of the expected error sources on the performance of the code. We focus on repetitive error detection using the pipelined scheme and with the logical qubit initialized in $0$ . In Fig. 3.14a, we show the post-selected fraction ( ) as a function of the number of error-detection cycles for a series of models. Model 0 is a no-error model, which we take as the starting point of the comparison. Model 1 adds amplitude and phase damping experienced by the transmon. Model 2 adds the increased dephasing away from the sweetspot arising from flux noise. Model 3 adds residual qubit excitation and readout (SPAM) errors. Finally, Model 4 adds crosstalk due to the residual coupling during the coherent operations of the stabilizer measurement circuits. The details of each model and their input parameters drawn from experiment are detailed below. We find that the dominant contributors to the error-detection rate are SPAM errors and decoherence. However, we also observe that the noise sources included through Model 4 clearly fail to quantitatively capture the decay of the post-selected fraction observed in experiment.

We believe that an important factor behind the observed discrepancy is the presence of leakage, as suggested by the single-shot readout histograms in Fig. 3.13. We consider the leakage per CZ gate as a free parameter and assume the same value for all CZ gates.

**3**

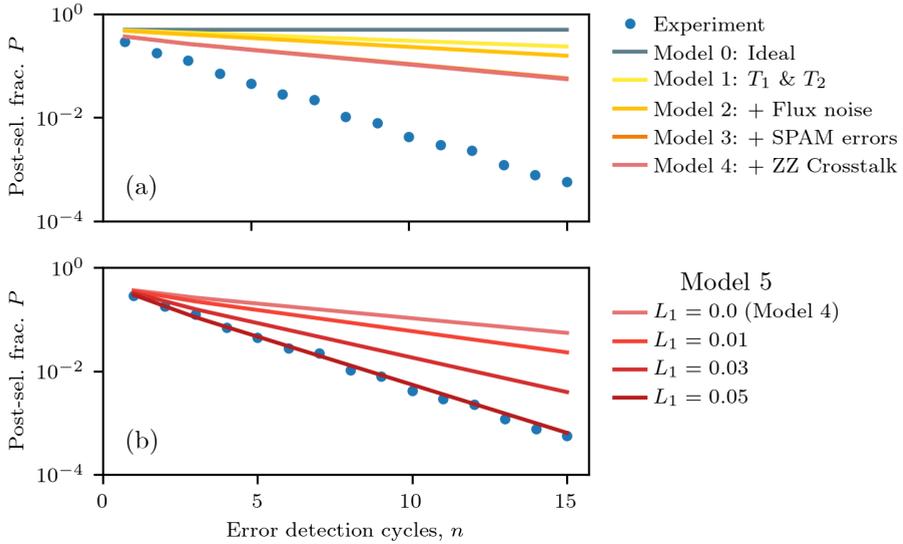

Figure 3.14: **Simulation of error-detection rate.** Post-selected fraction  as a function of the number  of error-detection cycles for  0 . The experimental  (blue dots) is compared to numerical simulation under various models (solid curves). (a) Simulated  obtained by incremental addition of error sources starting from the no-error (Model 0, gray); qubit relaxation and dephasing (Model 1, yellow); extra dephasing due to flux noise away from the sweetspot (Model 2, amber); state preparation and measurement errors (Model 3, orange); and crosstalk due to residual  interactions (Model 4, red). (b) Simulated  for Model 5 adding CZ gate leakage with 4 different values of  , the leakage per CZ gate, assumed equal for all CZ gates.

We simulate the post-selected fraction for a range of  values, shown in Fig. 3.14b. We observe that  $5\%$ produces a good match with experiment, suggesting that leakage significantly impacts the error-detection rate observed. We perform a similar analysis now considering the logical measurement of  experiment depicted in Fig. 3.2f which also finds similar agreement with experimental data (Fig. 3.15). This value of  is significantly higher than achieved in Ref. [69], which used the same device. However, note that in this earlier experiment CZ gates were characterized while keeping all other qubits in  0 . Spectator transmons with residual  coupling to either of the transmons involved in a CZ gate can increase  when not in  0  (which is certainly the case in the present experiment). Note that leakage may also be further induced by the measurement [135], an effect that we do not consider in our simulation. However, the assumption that all CZ gates have the same  , the approximations used in our models, and other error sources that we have not considered here may lead to an overestimation of the true  .

Leakage is an important error source to consider in quantum error correction experiments of larger distance codes, requiring either post-selection based on detection [77] or the use of leakage reduction units [128]. We leave the detailed investigation of the exact leakage rates in our experiment and the mechanisms leading to them to future work.

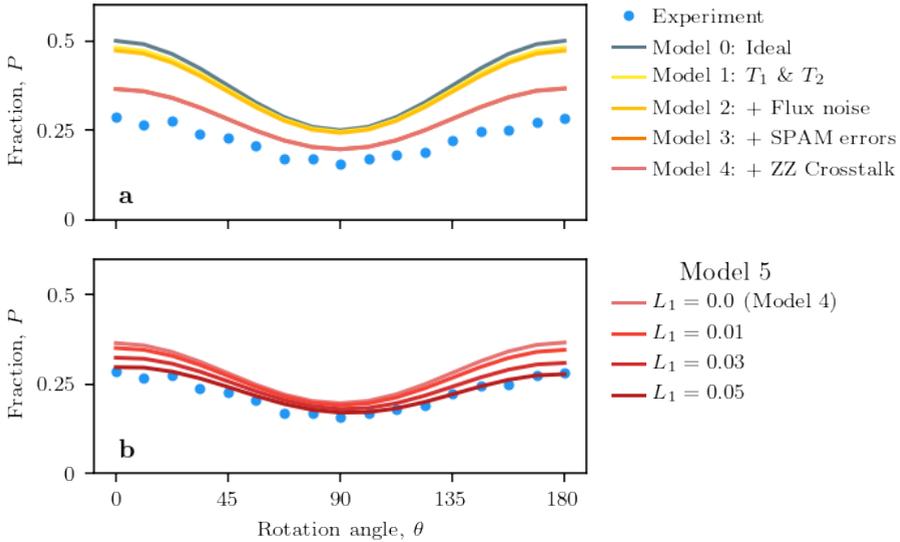

Figure 3.15: **Simulated post-selected fraction.** Post-selected fraction    of Fig. 3.2f for the measurement with the same error models used in Fig. 3.14.

### 3.6.15   Error models

Lastly, we detail the error models used in the numerical simulations in Fig. 3.14.

*Model 1*

We take into account transmons decoherence by including an amplitude-damping channel parameterized by the measured relaxation time      and a phase-damping channel parameterized by the pure-dephasing time at the sweetspot

$$\frac{1}{\;\;\;\;} = \frac{1}{\;\;} \quad \frac{1}{2}$$

where     is the measured echo dephasing time (see Table 5.1). The qutrit Kraus operators defining these channels are detailed in Ref. [77] and we similarly introduce these channels during idling periods and symmetrically around each single-qubit or two-qubit gate (each period lasting half the duration of the gate).

3. SURFACE-7

*Model 2*

We consider the pure-dephasing rate $\Gamma_\phi = 2\sqrt{\ln 2}\, + 1$ away from the sweetspot due to the fast-frequency components of the $1/f$ flux noise, where is the flux sensitivity at a given qubit frequency and is the scaling parameter for the flux-noise spectral density. We use a $\overline{3}$, the average of the extracted values for D, A and A obtained by fitting the measured decrease of as a function of the applied flux bias, following the model described above. This allows us to estimate the dephasing time at the CZ interaction and parking frequencies, which then parameterize the applied amplitude-phase damping channel inserted during those operations [77]. We neglect the slow-frequency components of the flux noise due to the use of sudden Net Zero pulses, which echo out this noise to first order [69, 94].

*Model 3*

We further include state-preparation and measurement errors. We consider residual qubit excitations, where instead of the transmon being initialized in $0$ at the start of the experiment, it is instead excited to $1$ with probability . We extract for each transmon from a double-Gaussian fit to the histogram of the single-shot readout voltages with the transmon nominally initialized in $0$ [131]. We model measurement errors via the POVM operators $= \overline{(\ )}$ for $0\,1\,2$ being the measurement outcome, while $(\ )$ is the probability of measuring the qubit in state when having prepared state . We extract the probability $(Q = ) = \mathrm{Tr}\,^\dagger$ of measuring qubit Q in state from simulation, where is the density matrix, while application of the POVM transforms $^\dagger$ $(Q = )$. In our simulations we condition on the detection of no error and thus we calculate $(Q = 0\,)$ and then apply to the state . We obtain $(0\,)$ for $0\,1$ from the experimental assignment fidelity matrix [102] (where a heralded initialization protocol was used to prepare the qubits in $0$ [130]) and we assume $(0\,2) = 0$, consistent with the observed histograms in Fig. 3.13. At the end of each experiment with error-detection cycles we calculate the probability of obtaining trivial syndromes from the final measurements of the data qubits (see Results). From this and from the probability $(A = 0\,)$ of measuring ancilla A in $0$ at cycle , we calculate the post-selected fraction of experiments defined as $(\ ) = (\ = 0\,)$.

*Model 4*

We consider the crosstalk due to residual interactions between transmons. The CZ gates involved in a parity check are jointly calibrated to minimize phase errors for the whole check as one block (see Fig. 3.6). Instead of modeling this crosstalk as an always-on interaction and taking into account the details of the check calibration, we instead capture the net effect of this noise by including it as single-qubit and two-qubit phase errors in each CZ gate. This assumes that the crosstalk only occurs between transmons that are directly coupled, which the measured frequency shifts observed in Fig. 3.5 validate. We characterize the phases picked up during the CZ gates using $2^{n-1}$ Ramsey experiments for a check involving a total of $n$ transmons (including the ancilla). In each experiment, we perform a Ramsey experiment on one transmon labelled $Q_j$. $Q_j$ is initialized in a maximal superposition using a $\pi/2$ pulse, while the remaining $n-1$ transmons are prepared in each of the $2^{n-1}$ computational states $z$. Following this initialization, the parity check is performed, followed by a rotation of $\pi/2$ (while the other transmons are rotated back to $|0\rangle$) and by a measurement of $Q_j$. By varying the axis of rotation $\varphi$, we extract the phase $\phi_j(z)$ picked up by $Q_j$ with the remaining transmons in state $z$. We perform this procedure for each of the $n$ transmons of the check, resulting in a total of $n2^{n-1}$ measured phases, which are arranged in a column vector $b$. We parameterize each CZ gate used in the parity check by a matrix $\mathrm{diag}(1, \theta_{01}, \theta_{10}, \theta_{11})$. The column vector $\theta_{CZ}$ then contains all of the phases parameterizing each of the $n-1$ CZ gates involved in the parity checks, with $n = 3$ for the $X$ and $Z$ checks and $n = 5$ for the $X$ check. We can express each of the measured phases in the Ramsey experiment as a linear combination of the acquired phases as a result of the CZ interactions between transmons, i.e., $b = M\theta_{CZ}$, where the matrix $M$ encodes the linear dependence. Given the measured $b$, we perform an optimization to find the closest $\theta_{CZ}$ as given by

$$\min_{\theta_{CZ}} \; \lVert M\theta_{CZ} - b \rVert$$

$$\text{subject to} \quad 0 \leq \theta_{CZ} \leq 2\pi$$

The optimal $\theta_{CZ}$ then captures the net effect of the ZZ crosstalk during the parity checks, which we include in the simulation. We do not model phase errors accrued during the ancilla readout, since in our simulation we condition on each ancilla being measured in $|0\rangle$.

*Model 5*

We model leakage due to CZ gates following the model and numerical implementation presented in Ref. [77]. Here, we do not consider the phases picked up when non-leaked transmons interact with leaked ones (the leakage-conditional phases [77]) and we set them to their ideal values. We also neglect higher-order leakage effects, such as excitation to higher-excited states or leakage mobility. Thus, we only consider the exchange of population between $|11\rangle$ and $|02\rangle$ given by $4J_2$, except for the CZ between $A$ and $D$, where the population is instead exchanged with $|20\rangle$ as we use the $|11\rangle - |20\rangle$ avoided crossing for this gate [69].

There remain several relevant error sources beyond those included in our numerical simulation. For example, we do not include dephasing of data or other ancilla qubits induced by ancilla measurement, which we expect to be a relevant error source for comparing the performance of the pipelined and parallel schemes. Also, we only consider the net effect of crosstalk due to residual      interactions during coherent operations of the parity-check circuits, which we include via errors in the single-qubit and two-qubit phases in the CZ gates. Thus, we do not capture the crosstalk present whenever an ancilla is projected to state $1$ by the readout but declared to be in $0$ instead. Furthermore, as      crosstalk does not commute with the amplitude damping included during the execution of the circuit, we are not capturing the increased phase error rate that this leads to.

| Data-qubit outcome | Logical assignment | | | Data-qubit outcome | Logical assignment | | |
|---|---|---|---|---|---|---|---|
| $(+1\ +1\ +1\ +1)$ | $+1$ | $+1$ | $+1$ | $(-1\ +1\ +1\ +1)$ | $R(\mathscr{S})$ | $R(\mathscr{S})$ | $1$ |
| $(+1\ +1\ +1\ -1)$ | $R(\mathscr{S})$ | $R(\mathscr{S})$ | $R(\mathscr{S})$ | $(-1\ +1\ +1\ -1)$ | $R(\mathscr{S}\ \mathscr{S})$ | $1$ | $R(\mathscr{S})$ |
| $(+1\ +1\ -1\ +1)$ | $R(\mathscr{S})$ | $R(\mathscr{S})$ | $1$ | $(-1\ +1\ -1\ +1)$ | $1$ | $+1$ | $+1$ |
| $(+1\ +1\ -1\ -1)$ | $R(\mathscr{S}\ \mathscr{S})$ | $1$ | $R(\mathscr{S})$ | $(-1\ +1\ -1\ -1)$ | $R(\mathscr{S})$ | $R(\mathscr{S})$ | $R(\mathscr{S})$ |
| $(+1\ -1\ +1\ +1)$ | $R(\mathscr{S})$ | $R(\mathscr{S})$ | $R(\mathscr{S})$ | $(-1\ -1\ +1\ +1)$ | $R(\mathscr{S}\ \mathscr{S})$ | $1$ | $R(\mathscr{S})$ |
| $(+1\ -1\ +1\ -1)$ | $1$ | $+1$ | $1$ | $(-1\ -1\ +1\ -1)$ | $R(\mathscr{S})$ | $R(\mathscr{S})$ | $+1$ |
| $(+1\ -1\ -1\ +1)$ | $R(\mathscr{S}\ \mathscr{S})$ | $1$ | $R(\mathscr{S})$ | $(-1\ -1\ -1\ +1)$ | $R(\mathscr{S})$ | $R(\mathscr{S})$ | $R(\mathscr{S})$ |
| $(+1\ -1\ -1\ -1)$ | $R(\mathscr{S})$ | $R(\mathscr{S})$ | $+1$ | $(-1\ -1\ -1\ -1)$ | $+1$ | $+1$ | $1$ |

Table 3.3: **Logical assignments from data-qubit measurement outcomes**. When measuring in the specified logical cardinal basis (as shown in Fig. 3.2a), the final string of data-qubit outcomes is rejected (R) whenever at least one of the computable stabilizers has value -1 (indicated within parentheses). The computable stabilizers are $\mathscr{S}$ and $\mathscr{S}$ when measuring , $\mathscr{S}$ when measuring , and $\mathscr{S}$ when measuring . When the data-qubit outcome string is accepted, the logical assignment (+1 or -1) is given by the appropriate product of data-qubit outcomes: for , for , and for .

| Logical measurement basis | | | |
|---|---|---|---|
| | $0$ | $+$ | $+$ |
| $-$ | $1$ | | |
| | 0.184 | 0.164 | 0.078 |
| $-$ | 0.170 | 0.118 | 0.073 |
| | 0.003 | 0.016 | 0.089 |
| $-$ | 0.002 | 0.010 | 0.083 |
| | 0.998 | 0.987 | 0.914 |

Table 3.4: **Quantified performance of logical measurement.** Final results of the analysis performed to quantify logical measurement in the logical cardinal bases without corruption from initialization errors. See Fig. 3.11 for reference. The extracted logical readout fidelities are those quoted in the main text.

# 4

## AUTOMATIC CALIBRATION AND BENCHMARKING OF A SUPERCONDUCTING DISTANCE-3 SURFACE CODE

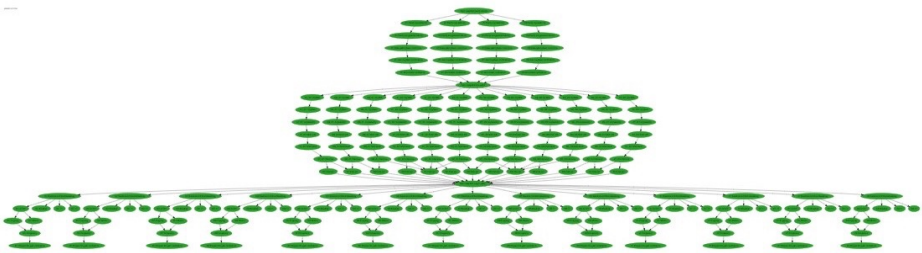

Automating the calibration and benchmarking of superconducting quantum computers is of great interest, especially considering the potential scalability for the number of qubits and the necessity for regular calibration to address continuous drifts in control parameters. This chapter provides details on the calibration strategies and benchmarking of a 17-transmon device using an automatic framework. It offers a hands-off approach for the calibration of single-qubit gates, two-qubit gates, and readout. The chapter also explores common failure modes that necessitate manual control, particularly in cases of parasitic interactions with two-level-system defects. Furthermore, a comparative analysis of individual and simultaneous device performance is presented to illustrate the impact of crosstalk in our quantum device.

## 4.1  Device overview

### 4.1.1  Device layout and performance

Our 17-transmon device [Figure 4.1] is a cQED system, which consists of a two-dimensional (2D) array of 3  3 data qubits and 3  3-1 ancilla qubits, purposely designed for a distance-3 surface code. Qubit transition frequencies are organized into three distinct frequency groups: high-frequency qubits (red) at $6\,7\ \mathrm{GHz}$, mid-frequency qubits (blue/green) at $6\,0\ \mathrm{GHz}$, and low-frequency qubits (pink) at $4\,9\ \mathrm{GHz}$, required for the realization of the pipelined error-correction cycle [98]. The primary advantage of this scheme is its potential scalability beyond 17 qubits.

Each transmon has a microwave drive line (orange) for single-qubit gates and leakage reduction units (see Chapter 5), a flux-control line (yellow) for two-qubit gates, and dedicated pair of resonator modes (purple) distributed over three feedlines (blue) for fast dispersive readout [102, 131]. Nearest-neighbor transmons are statically coupled via resonator buses (sky-blue), whose frequencies are above $20\ \mathrm{GHz}$, mediating the coupling between them [138].

Grounding airbridges (light gray) are fabricated throughout the device to interconnect the ground planes and to suppress unwanted mode propagation. These airbridges are also incorporated at the termination of each resonator, allowing for post-fabrication trimming techniques [55], as further discussed below. Additional resonators (uncolored) are positioned on the left and right feedlines to quantify intrinsic losses of the fabricated superconducting base layer, NbTiN [139].

To coherently manipulate the transmon qubit, it is essential for the qubit transition frequency ( 2 ) to significantly exceed the thermal energy, ensuring          , where  is the reduced Planck constant and     is the Boltzmann constant. Given     2  = 6 GHz, the device should be cooled down to a temperature below $20\ \mathrm{mK}$, easily achieved with a dilution refrigerator. However, to process quantum information, the immediate utilization of a quantum device post-cooldown is not feasible, necessitating a series of basic characterization and calibration experiments. Details about calibration and benchmarking procedures employed are discussed in Section 4.2.

An overview of the achieved performance for a 17-qubit device, nicknamed Uran, is presented in Figure 4.2. This 2D device relies on nearest-neighbor connectivity (black lines) between ancillas (blue and green circles) and data qubits (red and pink circles), as shown in panel (a). As expected, the measured qubit transition frequencies [Figure 4.2. b] clearly exhibit three distinct frequency groups. These values are extracted from qubit spectroscopy techniques, assuming prior characterization of resonator frequencies in the linear regime [140].

Using calibration graphs in Section 4.2, we reach average error rates (shown as dashed lines in panel (c)) of $0\,1\%$, $1\,6\%$, and $1\,2\%$ for single-qubit, two-qubit gates, and single-qubit readout, respectively, across the device. This suggests that two-qubit gate is the dominant error source in this device. It is crucial to note that these are individual benchmarks, and performance might degrade during simultaneous operations due to various crosstalk effects, as discussed in Section 4.2. The individual error rates [Figure 4.2. d-f] are presented for all qubits and pairs.

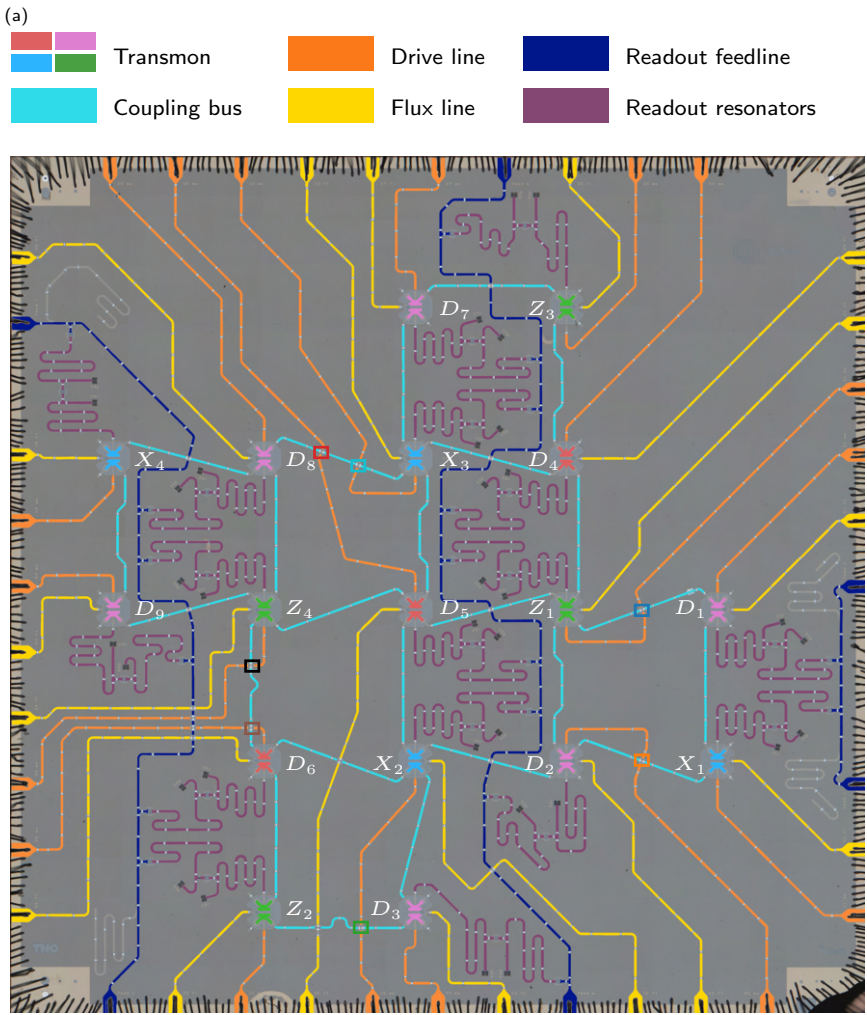

Figure 4.1: **Optical image of the 17-transmon device (Uran).** False colors are added to highlight various essential circuit components of the device (see legend). Detailed description of each component is provided in the main text. Aluminum wirebonds at corners and edges are introduced to interconnect the device to a PCB board. This connects the quantum chip to room-temperature control electronics via semi-rigid cryogenic lines and coaxial cables. Colored squares indicate eight crossovers where a microwave drive line either crosses a qubit-qubit coupler or a feedline. These cause high microwave crosstalk, as discussed in Figure 4.6.

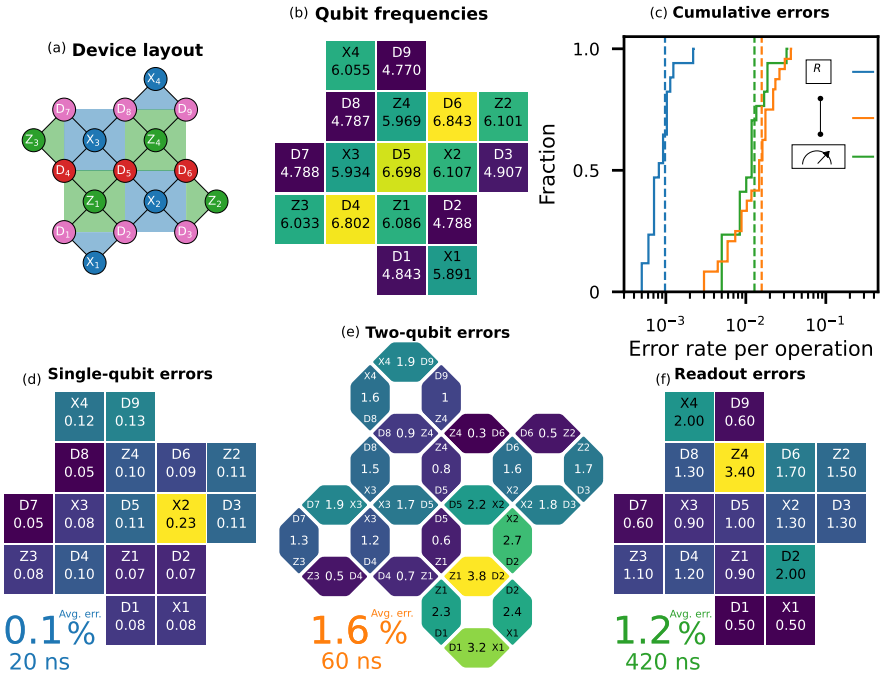

Figure 4.2: **Device layout and overview of the achieved performance.** (a) Device layout showing the connectivity (black lines) between data and ancilla qubits. The color of the plaquettes indicates that this device is purposely designed to implement a distance-three surface code. In addition, the color of each qubit circle corresponds to the intended frequency group. (b) The measured qubit frequencies biased at their simultaneous sweetspot. (c) Cumulative distribution of the individually achieved error rates for single-qubit gates, two-qubit gates and single-qubit readout. Vertical dashed lines indicate the average error rates across the device, highlighting that two-qubit gate is the dominant source of errors. Detailed error rates are presented for the 17 single-qubit gates in (d), 24 two-qubit gates in (e), and 17 single-qubit readout in (f) in $20\,\mathrm{ns}$, $60\,\mathrm{ns}$, $420\,\mathrm{ns}$, respectively. Further details of the calibration process are provided in the calibration section below.

### 4.1.2 Device yield and targeting errors

Superconducting quantum processors are highly engineered systems in which interactions between artificial atoms—transmons—and quantized electromagnetic fields can be tailored.

This proves usefulness for achieving fast and high-fidelity quantum gates and measurements, essential for quantum computation [13, 26, 141]. However, the fabrication of superconducting circuits introduces numerous defects and uncertainties giving rise to critical issues such as poor yield, suboptimal device performance, and imprecise frequency targeting. As the device size scales, these issues escalate, posing significant bottlenecks. For instance, the density of defects tends to grow proportionally with the device size, presenting an immediate risk to the overall performance and reliability of the quantum processor.

In this section, we discuss the challenges arising from fabrication variations and design limitations in the pursuit of fabricating a 'good' 17-qubit device for the implementation of a distance-3 QEC code. We begin by defining "device yield", which refers to the proportion of fabricated components that meet the desired specifications and functionality. For an initial assessment, achieving a $100\%$ yield in our device would require flawless functionality of various components, including 17 transmons, 34 resonators, 24 resonator buses, 17 microwave drive lines, 17 flux control lines, 3 feedlines, 1199 grounding airbridges, and 40 input/output ports. Unfortunately, these criteria cannot be fully verified with room-temperature measurements, and basic characterization of the device at base temperature is essential. This, in turn, entails a time-consuming feedback loop with the fabrication and design teams.

To address this issue, we implemented the following strategies:

1. Automatic detection of defects: our fabrication team invested in developing an image recognition software, known as Pyclq, to autonomously detect defects. Pyclq compares the designed layout (GDS files) with optical images of the fabricated device. Notably, this software can identify minuscule defects that might elude human observation. If a defect is reparable, a localized fabrication process can rectify it; otherwise, the device is deemed unsuitable. Further details and results can be found [142].

2. Two devices in one cooldown: we upgraded our XLD 400 dilution refrigerator's cold finger to allow simultaneous mounting of two Surface-17 devices. The two devices are stacked vertically in order to both fit inside the infrared-radiation and magnetic shields surrounding the cold finger.

3. Automatic and parallel characterization: we developed a framework to perform automatic characterization, calibration and benchmakring of the device, see Section 4.2.

Having a device with a $100\%$ yield allows us to focus on characterizing device performance. As mentioned earlier, imperfections in the fabricated material stack or variations during the fabrication processes can induce fluctuations in device performance. For example, we observe significant variations, up to threefold, in device coherence metrics [Figure 4.3], , , *, across various 17-qubit devices fabricated using similar recipes. The enhanced coherence of device Maximus is positively correlated with high intrinsic quality factor, , of approximately $10$ [Figure 4.3. b]. The measured quality factors are obtained at low input power using the method outlined in [139].

We also conducted measurements of qubit relaxation rates as a function of qubit frequency using static flux biasing (data are not shown here). A linear dependency of $1$      on qubit frequency    suggests that it is primarily limited by dielectric loss, and not via the drive lines     , flux lines       or the Purcell effect [36, 143–145]. These losses are consistent with the loss tangent of thin amorphous oxides at interfaces, where local defects, commonly referred to as two-level systems (TLS), interact with microwave fields. This is an important point of feedback for our fabrication team (as opposed to our design team).

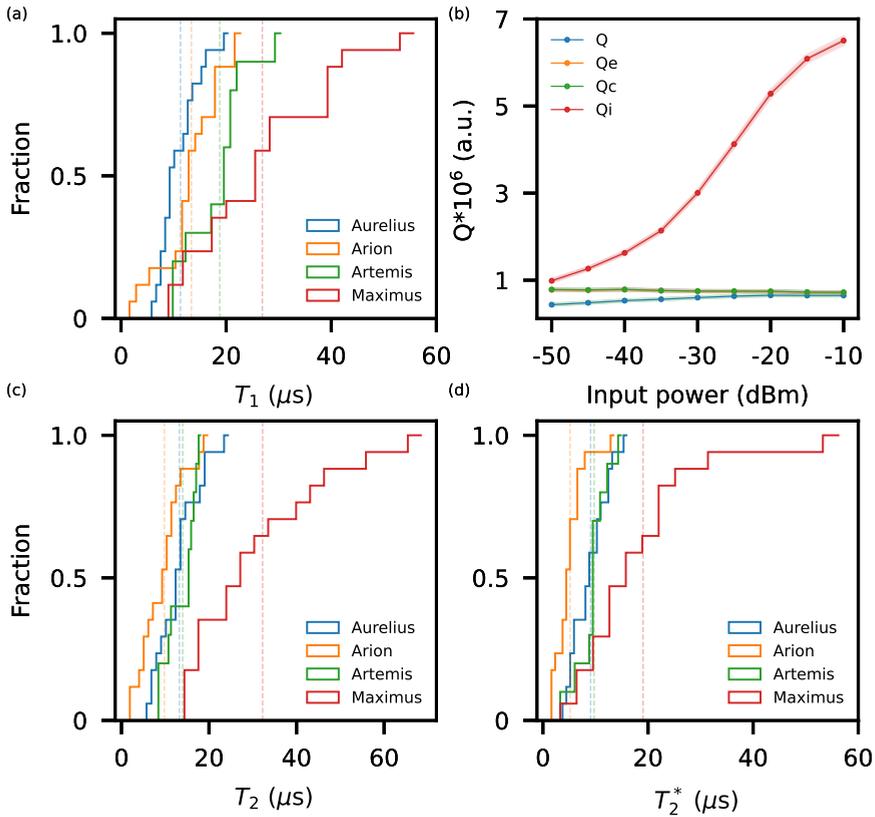

Figure 4.3: **Fluctuations in device coherence of various 17-qubit devices fabricated with similar recipes.** The cumulative distribution of the individually measured device coherence metrics:    in (a),    in (c), and    * in (d) extracted in various cooldowns. Vertical dashed lines indicate the average coherence across the device. The measured quality factors:    ,    ,    , and    of a test resonator in Maximus device as a function of input power. Enhanced device coherence is correlated with high intrinsic quality factor, approximately $10$    at low power. The highlighted shadow around the extracted quality factors display the uncertasinties extracted from the fit [139].

Another manifestation of fabrication uncertainties is the imprecise targeting of frequencies. Inaccurate targeting of resonator and qubit frequencies can give rise to various crosstalk errors, including measurement-induced dephasing, microwave cross-driving, and increased residual-$ZZ$ couplings, as illustrated below. These errors materialize as coherent and correlated errors, posing immediate threats to the performance of this distance-3 surface code [26, 87, 141, 146].

In response to these concerns, we iteratively cycled the basic characterization for 15 different 17-qubit devices. Through this iterative process, we developed a toolbox aimed at refining resonator and qubit frequencies through post-fabrication trimming techniques [55]. This toolbox is crucial for mitigating the impact of imprecise frequency targeting and enhancing the device yield for the tidy requirements of a QEC experiment.

Unwanted frequency overlap between resonators occurs due to fluctuations in the deposited NbTiN thickness, unaccounted capacitive loadings and variations in the CPW phase velocity [55]. The worst-case scenario was found on device Aurelius, where resonator frequencies for qubits X (blue) and Z (green) (both on the central feedline) overlap with each other. Their Purcell-readout resonator pairs are designed to be $100\,\mathrm{MHz}$ apart. However, they overlap significantly [Figure 4.4. b].

Assessing the dephasing impact of this overlap is of crucial importance for surface code, particularly for the pipelined scheme [98]. For this reason, we quantitatively assess the dephasing induced on X by measurement of Z ancilla qubits. This performs a Ramsey (Echo) sequence on X while simultaneously measuring Z [Figure 4.4. c]. By analyzing the amplitude and phase of the resulting oscillation, we extract the coherence element and phase (arg , not shown here) of the Ramsey (Echo) sequence. This experiment is repeated for different measurement amplitudes, and the results are plotted [Figure 4.4. c] as a function of the relative pulse amplitude of the measured ancilla, Z .

The phase shift of the Ramsey (Echo) qubit scales quadratically with the relative measurement amplitude, which is in agreement with dephasing due to AC-Stark shift [147]. We also include different state preparations ($0$ or $1$) of the measured qubit in each case, signifying the impact of residual-$ZZ$ between the two qubits. Overall, we measure considerable dephasing at higher pulse amplitudes than $0.02$ [Figure 4.4. g]. This constitutes a no-go test for Aurelius to implement a pipelined distance-3 surface code.

To tackle this issue, we initially experimented with a simple hack: physically shortening the resonator pair lengths (Purcell and readout resonators) of Z using a wire-bonding technique [Figure 4.4. a]. Using the conversion $1\,\mu\mathrm{m} \approx 2\,\mathrm{MHz}$, we attempted to shorten both resonators by $25\,\mu\mathrm{m}$. Middle-feedline transmission [Figure 4.4. d] after the attempted fix. The Purcell and readout resonators of Z have shifted upwards by $52\,\mathrm{MHz}$, while those for X remain unchanged. We also measured the same coherence times ($\ $ and $\ $) for both Z and X before and after the hack. Therefore, it appears that the hack has been successful. In order to reach a firm conclusion, we assess the measurement-induced dephasing between the two qubits after this fix. Indeed, extracted measurement-induced dephasing [Figure 4.4. i] is clearly reduced, no observable dephasing up to pulse amplitude of $0.04$.

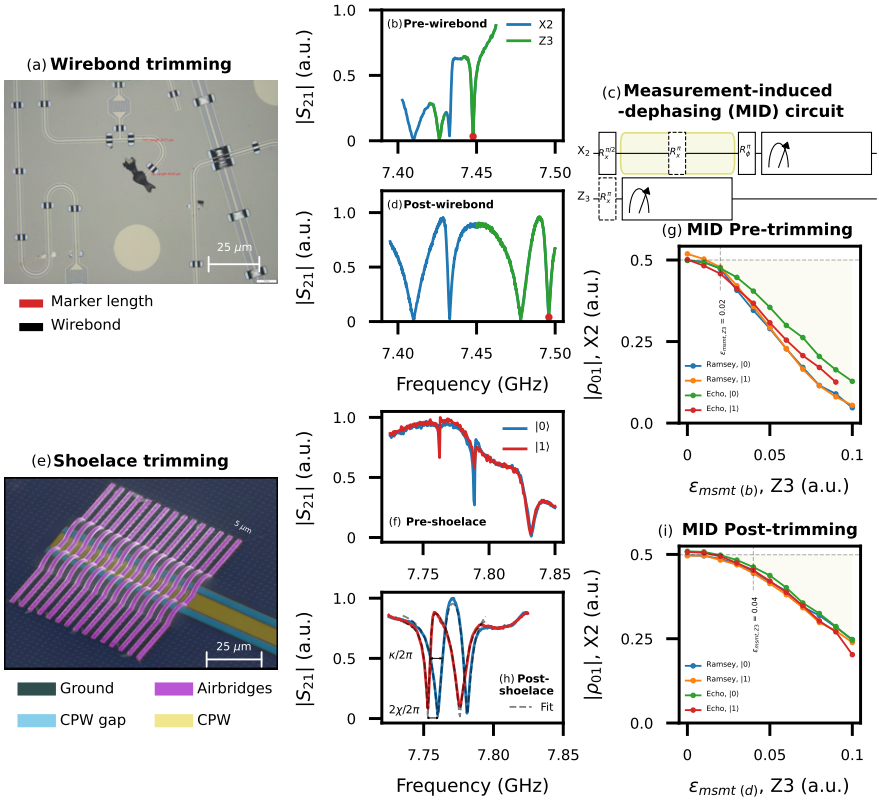

Figure 4.4: **Frequency trimming of readout resonators** using two techniques: wire-bonding (a), and Shoelace (e). The feedline transmission, $S_{21}$, is shown before (b) and after (d) wire-bonding to mitigate the frequency overlap between qubits $X_2$ and $Z_3$ in device Aurelius. Characterization of measurement-induced dephasing in between $X_2$ and $Z_3$ before and after the wire-bonding fix. (c) Ramsey (Echo) experiment to assess the impact of measurement of $Z_3$ on $X_2$. Experimental results obtained for the coherence $|\rho_{01}|$ Ramsey'ed (Echo'ed) qubit as a function of measurement amplitude and measured qubit state ($|0\rangle$, and $|1\rangle$) before trimming in (g), and after trimming in (i). Due to the precision and flexibility of Shoelace method, the hybridization between Purcell and readout resonators can be significantly improved in (h), compared to the reference case (f). This enables fast and high-fidelity readout in the Uran device within a duration of $400$ ns. Details about readout can be found in Section 4.2.4.

The previous fix was manual, quick, and dirty. In order to systematically address this issue, we embedded 10 grounding airbridges [Figure 4.4. e] at the end of the readout resonators, enabling frequency trimming in $100$ MHz range with a resolution of $10$ MHz [55]. Beside avoiding the frequency overlap, it can systematically improve the hybridization between Purcell and readout resonators [Figure 4.4. f-h] for achieving fast and high-fidelity readout.

The imprecise targeting of qubit frequency is a consequence of variations during the fabrication steps of Josephson junctions [148, 149]. To quantify this error, we compare the measured qubit frequencies, at the flux symmetry point, to the ideal target frequencies. The average error across Maximus is approximately $200\,\mathrm{MHz}$ [Figure 4.5. a]. While the measured frequencies distinctly exhibit three well-defined groups, the primary concern lies in the minimal detuning between different frequency groups, especially for qubits that are directly coupled. For example, consider the worst-case frequency collision between qubits D and Z, presenting a low detuning of $335\,\mathrm{MHz}$ at their simultaneous sweetspot. With the knowledge of measured qubit frequency, anharmonicity and transverse coupling in the two excitation manifold, (obtained via qubit spectroscopy techniques [140]), one can analytically calculate the approximated residual- coupling, . This shift is determined by $=$ , using perturbation theory [150]:

$$ = \quad \frac{1}{\rule{2cm}{0.4pt}} + \frac{1}{\rule{2cm}{0.4pt}} \tag{4.1} $$

where, $= 2$ $_6 +$ $_6$, $=$ $_6 +$ $_2$ and $= 2$ $_2 +$ $_2$ for D -Z pair.

This results in an exceptionally high residual- coupling of $6\,977\,\mathrm{MHz}$ at the sweetspot bias, in a reasonable agreement with the measured value [Figure 4.5. b]. The slight overestimation of arises from the approximation used in Equation (4.1). This may lead to high coherent and correlated error between two qubits, which cannot be managed or tolerated by a QEC code.

To address this issue, we employ our homebuilt room-temperature setup capable of automatic probing of junction-pair conductance and millisecond-controlled laser exposure [55, 151, 152]. These tools are integrated into a closed-loop protocol that iteratively measures and thermally anneals the junctions towards a set of predefined target conductance values tuning the Josephson energy of a qubit. The tuning range of our laser annealing setup is up to $7\%$ ($400\,\mathrm{MHz}$), with a precision of $15\,\mathrm{MHz}$. Laser annealing successfully lowers the qubit frequency (red points in Figure 4.5. a) measured in a second cooldown, resulting in significantly reduced residual- couplings [Figure 4.5. c]. Importantly, no significant impact on qubit coherence was observed before and after laser annealing, highlighting the efficacy of this technique in mitigating frequency collisions without compromising coherence.

Recently, we encountered a significant constraint in the lateral layout of our device through the characterization of the 17 17 microwave crosstalk matrix [Figure 4.6] in Uran. Each matrix element, denoted as , is derived from a cross-driving Rabi experiment involving all pairs of drive lines D and qubits Q [133]. Microwave isolation is quantified as the pulse amplitude necessary to implement a $X$ pulse on $Q_j$ by driving through $D_i$, normalized by the amplitude via $D_j$, denoted as $\alpha$:

$$\alpha = 20 \log \frac{A_{ij}}{A_{jj}} \qquad (4.2)$$

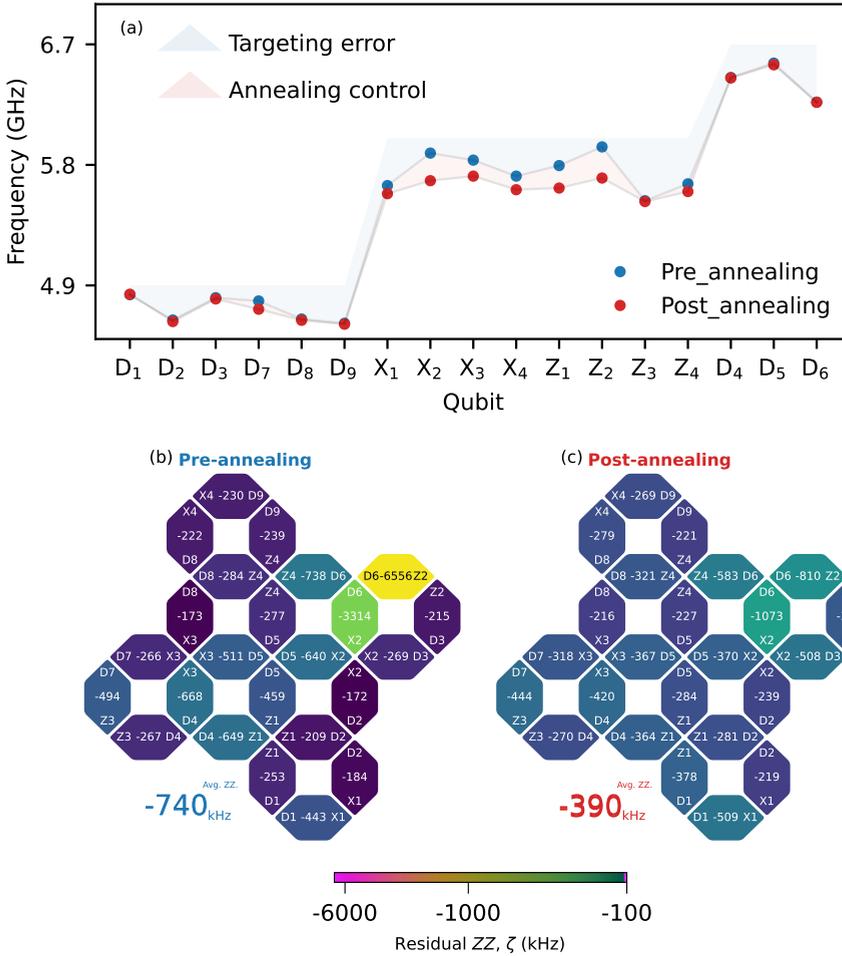

Figure 4.5: **Qubit frequency targeting in Maximus before and after laser annealing.** (a) Measured qubit frequencies at bias point (simultaneous flux sweetspot) before and after laser annealing, in comparison with the ideal target frequencies. These measured frequencies are extracted by qubit spectroscopy technique. Frequency collision between nearest-neighbor qubits results in a shift of qubit frequency, $\zeta$, depending on the state of a spectator qubit. Measured residual-$\zeta$ couplings in Maximus before (b) and after (c) laser annealing. These values are obtained using an Echo-like experiment in [132]. Note the significant improvement for worst-frequency collision pairs: $D_6$-$Z_2$ and $D_6$-$X_2$. In addition, no impact on qubit coherence was observed before and after laser annealing.

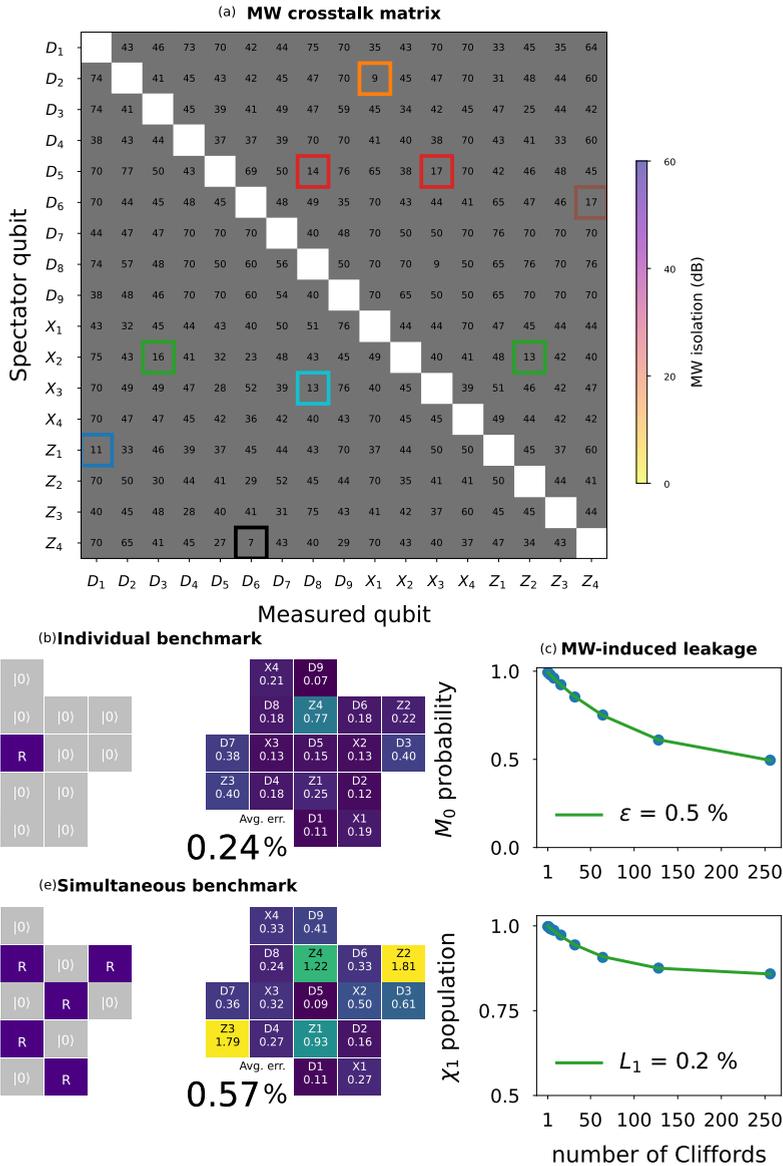

Figure 4.6: **Characterization and impact of microwave crosstalk in Uran.** (a) The measured 17 × 17 microwave crosstalk matrix reveals exceptionally low isolation, correlated with the positions of these crossovers, highlighted as colored squares in Figure 4.1. For more information about the method employed, see the main text (Continued on next page).

In specific instances, certain pairs of drive lines and target qubits show considerably low microwave isolation. Linking these pairs to device geometry [Figure 4.1], allows for the identification of the underlying cause: a microwave signal intended to drive a target qubit X spills

Figure 4.6: (Continued from previous page) This phenomenon is likely attributed to parasitic microwave coupling mediated through the crossover of the drive line of qubit X and the qubit-qubit coupler Z -D . Consequently, this leads to increased average error rates, characterized by simultaneous single-qubit randomized benchmarking (SQRB) between same-frequency groups (e), in contrast to individual benchmarking error rates in Aurelius (c). Microwave crosstalk can also induce leakage when a leakage transition of one qubit overlaps with the transition frequency of another (d). Some of the reported error rates are elevated due to interactions with TLS or various crosstalk effects. We have developed several mitigation strategies to address these errors, successfully improve performance for the Uran device, as demonstrated in Figure 4.2. Further details can be found in Section 4.2.2.

**4**

over to adjacent qubits Z and D . This parasitic coupling is effectively meditated through the crossover of the drive line of qubit X and the qubit-qubit coupler Z -D . Given the low coupling capacitance of qubit X to its own drive line—designed to minimize the Purcell effect—compared to the qubit-qubit coupling, the resulting microwave isolation is exceptionally low. We confirm this low isolation using electromagnetic simulations that model driving a qubit through a drive line crossing a crossover. The simulation predicts an isolation of 7 dB for the worst-case scenario, which is in a good agreement with measured values [Figure 4.6. b].

Poor isolation between qubits can lead to various errors, including coherent rotations around , , and axes of the Bloch sphere. Such errors challenge the effectiveness of stabilizer measurements, which are designed to effectively detect random Pauli errors rather than certain unitary over- or under-rotations. This, in turn, can impact the surface code threshold and the estimation of the logical error rate for different distances of QEC codes [153].

We capture these errors by comparing the individual and simultaneous randomized benchmarking of next-neighbor qubits in Aurelius [Figure 4.6. c and e]. Given that next-neighbor residual- is typically below $200\,\mathrm{kHz}$ in our device [69, 107], we primarily attribute the increased average error rates in simultaneous benchmarking to microwave crosstalk. Moreover, microwave crosstalk can induce leakage ( ) if the transition frequency of a driven qubit closely aligns with the leakage transition of another qubit [26]. This scenario occurred for qubit X , where the leakage transition frequency $_2$ overlaps with the qubit frequency $_1$ of qubit X [Figure 4.6. d]. Possible mitigation strategies of microwave crosstalk include interferometric cancellation [26, 141], dynamical decoupling schemes [154], and potential design modifications to minimize the number of crossovers.

## 4.2 Automatic calibration and benchmakring

### 4.2.1 Graph-based tuneup

Building a scalable quantum processor presents significant challenges, particularly in calibrating and maintaining high performance. The concern is amplified as control parameters of a quantum device consistently drift over time, necessitating frequent recalibration [155]. Moreover, the complexity increases exponentially with the scaling of qubit numbers, making the development of automated calibration routines essential. Such automation is invaluable not only for detecting errors in repetitive algorithms but also for compensating system drift parameters in large-scale qubit arrays [155]. In this work [156], a dependency graph was introduced for automatic calibration. In our group, we extended this concept into a framework known as Graph-Based Tuneup (GBT), initially developed by former PhD student M. A. Rol and master student Timo Abswoude [150, 157, 158]. We have since expanded GBT capabilities to perform automatic calibration and benchmarking, ranging from basic frequency domain characterization to more frequent calibrations of single- and two-qubit operations.

Once a new quantum processor is cooled down to base temperature at $10 - 20 \text{ mK}$, numerous characterization experiments are necessary to provide basic knowledge about device parameters. This includes identifying readout resonator frequencies, qubit transition frequencies, qubit anharmonicity, sweetspot offsets, qubit-qubit couplings and device coherence metrics: , *, and . The first challenge for GBT is to automatically characterize the device with limited knowledge such as room-temperature resistances of the junctions (translating into predication of qubit frequencies) and target parameters from deign.

Additionally, the GBT codebase needs to be modular, fast, easy to track, and sufficiently accurate [158]. The framework must also accommodate common failure modes and be capable of autonomously recalibrating in the event of unsuccessful calibration attempts. GBT manages this by integrating logic, measurement functions, and their dependencies within graph nodes. The progress of the automatic characterization process is visually monitored through a live graph, with colors ranging from green (successful) to red (failed to calibrate).

During the first six months of my PhD, I collaborated with my Bachelor student, A.C. Brandwijk, to enhance the GBT codebase. Together, we autonomously characterized and calibrated Surface-7 devices up to single-qubit operations in approximately $10 - 5 \text{ hrs}$. More detailed results from this phase are documented in this report [159]. Later, my colleague J. F. Marques and I expedited GBT to perform hands-off time-domain measurements, high-level calibration, and regular benchmarking of the fundamental components of the QEC circuit, which will be discussed in next sections.

### 4.2.2   Single-qubit gate calibration and benchmarking

Single-qubit gates are autonomously calibrated and benchmarked with the GBT framework [Figure 4.7. a]. The process begins with the calibration of the $\pi$ pulse amplitude through a Rabi measurement. A Gaussian pulse, with a standard deviation ($\sigma$) of $5\,\mathrm{ns}$, is applied to the dedicated microwave drive line at the transition frequency ($\omega$) between the $0$ and $1$ states. This induces sinusoidal oscillations, exchanging populations between the two states, with the probability of the excited state given by $\sin(\Omega^R t)$, where $\Omega$ is the Rabi strength and $t$ is the gate time ($20\,\mathrm{ns}$). The $\pi$ pulse amplitude is set to maximize the population of the excited state, and other amplitudes, like a ($\pi/2$), are interpolated accordingly. This holds true as far as DAC amplitude is linear [140]. Successful calibration requires a preliminary tuning of the qubit frequency using spectroscopy. Additionally, since microwave signals are synthesized with IQ mixing, calibrating mixer leakage offsets and mixer skewness is crucial for accurately estimating the $\pi$ pulse amplitude.

The next step involves fine-tuning the qubit frequency using a standard Ramsey measurement, as qubit spectroscopy has limited precision due to finite pulse bandwidth and power broadening [140, 160]. After updating the qubit frequency, it is necessary to fine-tune the $\pi$ and $\pi/2$ pulse amplitudes. This is achieved through a standard flipping experiment involving repeated sequences of N-times $\pi$ or $\pi/2$ pulses. The error in the microwave pulse amplitude then accumulates, allowing for precise fine-tuning of all possible rotations on the Bloch sphere. Simultaneously, phase errors and leakage in microwave gates are effectively suppressed with DRAG pulses [64, 65]. If either the flipping or Motzoi nodes fail, the calibration graph will revert to recalibrate Rabi and Ramsey qubit experiments.

For a rapid assessment of the preceding calibration nodes, we employ the standard ALLXY sequence [140]. This sequence simultaneously detects various error syndromes, including qubit frequency detuning, pulse amplitude errors, inaccurate estimation of the DRAG coefficient, and pulse distortions due to reflections. If an ALLXY deviation is below certain threshold, $1\%$, we declare successful calibration and proceed to more accurate benchmark with SQRB [96, 161–163]. This benchmakring technique is a standard practice to benchmark quantum gates due its scalability and insensitivity to SPAM errors [164].

The GBT routine ultimately evaluates various qubit metrics, including qubit coherence times, which are known to drift over time [165], and single-qubit readout, optimized by the readout GBT discussed later. A summary of the performance achieved using this calibration procedure is provided [Figure 4.7. b]. The color of the curves in the figure denotes the level of performance: green indicates low error rates (as seen for qubit D), orange signals a warning level, and red corresponds to higher error rates, suggesting potential issues that require careful manual inspection. This variation in performance can arise due to strong coupling with TLS defects, which manifests differently in single-qubit gates and readout.

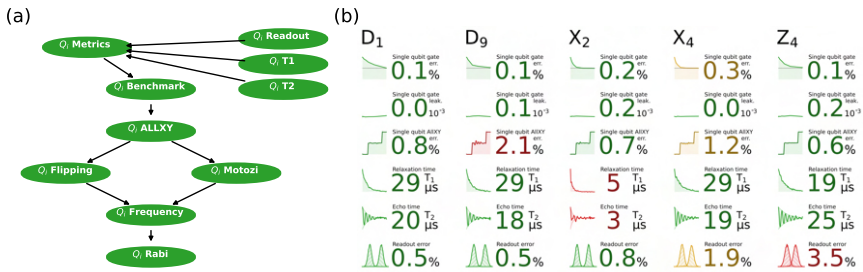

Figure 4.7: **Single-qubit gate calibration and benchmarking.** (a) GBT nodes for automatic calibration and benchmarking of single-qubit gates, assuming prior knowledge of qubit transition frequencies obtained through spectroscopy measurements. (b) A sample of the achieved single-qubit gate performance with this calibration procedure. This routine approximately takes $4\,\mathrm{min}$ per qubit.

The graph for 17 qubits completes in approximately $70\,\mathrm{min}$, averaging $4\,\mathrm{min}$ per qubit. Given the necessity for frequent recalibration due to control parameter drifts, the calibration process becomes tedious and scales poorly with the qubit count. Moreover, control parameters may change when running simultaneous operations or between coupled qubits due to various crosstalk effects. To address these issues, I supervised a master student, Hiresh Y. Jadoe-nathmissier, to parallelize this calibration graph between multiple qubits. This significantly reduced the calibration time for the 17-qubit device to just two steps: one for ancilla qubits and the other for data qubits. Detailed results can be found in this report [166].

Similarly, simultaneous benchmarking provides a more realistic evaluation of gate performance in a multi-qubit processor, as opposed to individual benchmarking. A comparison between individual and simultaneous benchmarking [Figure 4.6. c and e] reveals that error rates can be more than doubled due to parallel operations. Moreover, it can also introduce leakage errors, especially in scenarios with low microwave isolation [Figure 4.6. d]. Future efforts should be dedicated to mitigate the impact of crosstalk effects, especially with 2D lateral layout and static couplings.

Maintaining high performance in single-qubit gates presents challenges due to their strong coupling with TLS defects, which can significantly compromise operational fidelities. The dynamic nature of these defects results in continuous shifts in control parameters, necessitating regular calibration [165, 167]. Various signatures of the TLS impact on single-qubit gates are depicted in Figure 4.8, including coherent oscillations in     measurement, extremely low    . To mitigate these effects, we apply DC biasing to detune the qubit, aiming to minimize the population loss of the excited state as a function of the qubit frequency [Figure 4.17. b and c]. Specifically, by detuning the qubit by     $2 = 100\,\mathrm{MHz}$ from the sweetspot (    $2 = 0$), we can restore good qubit coherence [Figure 4.8. b and d]. Implementing this strategy across the Uran device, we achieve a notable single-qubit error rate of $0.1 - 0.02\%$, as determined by SQRB across the 17 qubits.

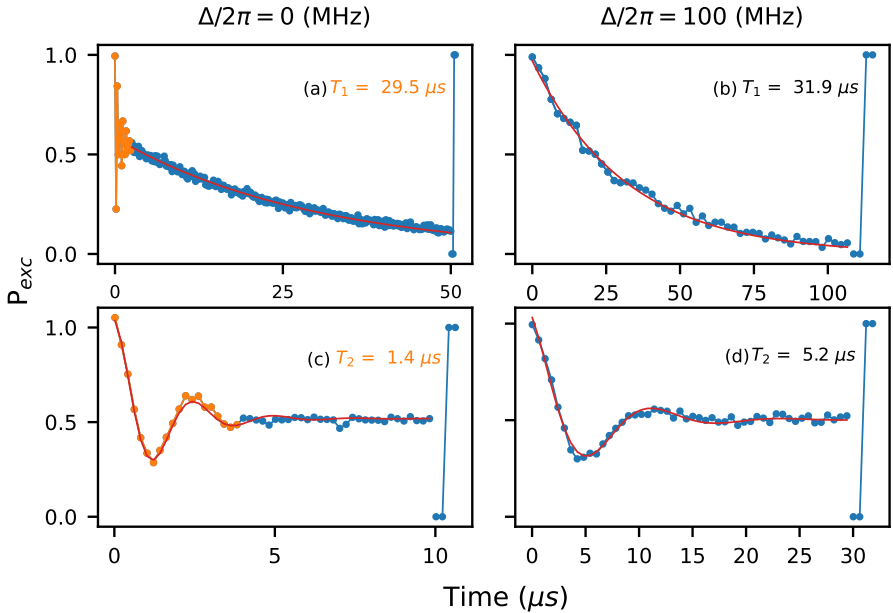

Figure 4.8: **Mitigating of the impact of TLS on single-qubit gates** by statically detuning the qubit frequency away from the interaction zone. $T_1$ and $T_2$ measurements at the sweetspot ($\Delta/2\pi = 0$) in (a) and (c), and at a detuning of $\Delta/2\pi = 100\,\mathrm{MHz}$ in (b) and (d). Orange color indicates the signatures of strong coupling to a TLS mode: coherent oscillation in $T_2$ measurement and extremely low $T_2$ times. Enhanced coherence is evident with qubit detuning.

### 4.2.3 Two-qubit gate calibration and benchmarking

Recent advancements in multi-qubit superconducting quantum processors have emphasized the critical role of achieving high-fidelity two-qubit gates, a key factor in minimizing errors in various quantum algorithms [13, 25, 26, 141, 168]. Achieving and maintaining high performance presents significant challenges. This section discusses these challenges and describes our approach, focusing on the implementation of a dynamical flux-based two-qubit CPhase gate that leverages the transverse coupling $J_2$ between a computational state $11$ and a non-computational state $02$ [66, 67].

Our approach relies on the SNZ CZ gate that operates at the speed limit, $t = \frac{1}{2J_2}$ [69]. Beyond speed, SNZ gate offers a simplified tuneup with clear interrelation between phase and leakage landscapes, and can be easily generalized to realize arbitrary phase gates. These features make SNZ an excellent choice for assembling the flux dance of a distance-3 QEC code [98]. For more details about the SNZ gate, see Chapter 2.

Our primary focus is on optimizing the sequence for implementing the 24 CZ gates required by the $= 3$ QEC cycle. This introduces additional complexities beyond tuning individual CZ gates. For example, when CZ gates run in parallel, they must be temporally aligned to avoid unwanted interaction on the way to or at target avoided crossings. Moreover, simultaneous operation can induce extra errors due to various crosstalk effects such as residual- , microwave cross-driving, and flux crosstalk. To mitigate these intricate errors, we have developed calibration strategies that tune simultaneous CZ gates as block units in both time and space. This approach allows for absorbing some of the crosstalk errors into the tuneup, further detailed in Chapter 6.

A two-qubit GBT procedure (TQG GB in Figure 4.9. a) is designed to autonomously calibrate SNZ CZ gates, assuming prior knowledge of qubit transition frequencies and anharmonicities. The calibration begins with determining polynomial coefficients for flux (voltage) to frequency conversion, known as flux arc [Figure 4.10. a]. This involves a Ramsey-like experiment termed Cryoscope to estimate the qubit detuning upon applying a voltage signal [103]. During the Ramsey evolution, a square pulse generated by a flux AWG is embedded with different amplitudes. This assumes a prior calibration of flux pulse distortion (not shown in TQG GBT) characterized the Cryoscope routine and using real-time digital filters in the AWG. This results in on-chip flux waveforms with rise time on par with that of the AWG $(0\ 5\ \mathrm{ns})$. For more details about this, we follow this reference [103].

The and components of the Bloch vector (blue and orange data points in Figure 4.10. c) are measured by changing the final $2$ rotation. Previously, this flux arc procedure took about $3\ \mathrm{hrs}$ per qubit, as it required measuring long Cryoscope experiments with high precision to accurately estimate qubit detunings. My colleague and I significantly expedited the calibration by avoiding the uploading of redundant waveforms to AWG memory and by measuring fast Cryoscope traces with a low number of points in $20\ \mathrm{ns}$ and low acquisition averages of $2$ . Instead, we achieved high precision by fitting the power spectral density (PSD) of the measured data [Figure 4.10. d], which is more resilient to measurement noise. This new flux arc measurement completes in $5\ \mathrm{min}$ per qubit, using four different flux pulse amplitudes to achieve the required accuracy.

The next step in the TQG GBT procedure is to determine the speed limit from the characteristic Chevron measurement of $2$ -population    $_{ }$) as a function of the amplitude and duration of a unipolar square flux pulse acting on $11$ [69]. In pursuit of automation, the code framework predicts the frequency (DAC voltage) of the $11$ - $02$ avoided crossing based on the qubit frequencies, anharmonicities and the previously obtained voltage-to-frequency conversion. This prediction relies on solving the transmon Hamiltonian and utilizes optimization (minimization) to determine the transmon parameters: and that match the target frequency and anharmonicity. This process provides the detunings for the avoided crossings in the transmon energy spectrum [Figure 4.11. c]. The frequency (DAC voltage) range is defined accordingly to explore the desired avoided crossing. Figure 4.11. a and b show the experimentally mea-sured $11$ - $02$ interaction Chevron, which takes about $3\,\mathrm{min}$ per gate. The data are then fitted to the following function for the extraction of SNZ pulse time and :

$$= \frac{4}{\quad}(1 \quad \cos( \quad + \quad)) \tag{4.3}$$

and

$$= \frac{}{2} \tag{4.4}$$

where, $_{\mathrm{exc}}$ is the population of the $2$ state, is the transverse coupling of $11$ and $02$ , is the frequency of Chevron oscillation, and $_{\mathrm{dist}}$ serves as a phase offset, employed as a free parameter to account for pulse distortion. We note that is likely longer than due to the finite rise time of the pulse and leftover pulse distortion. Overall, this systematic approach enables the automated calibration and analysis of Chevron experiments.

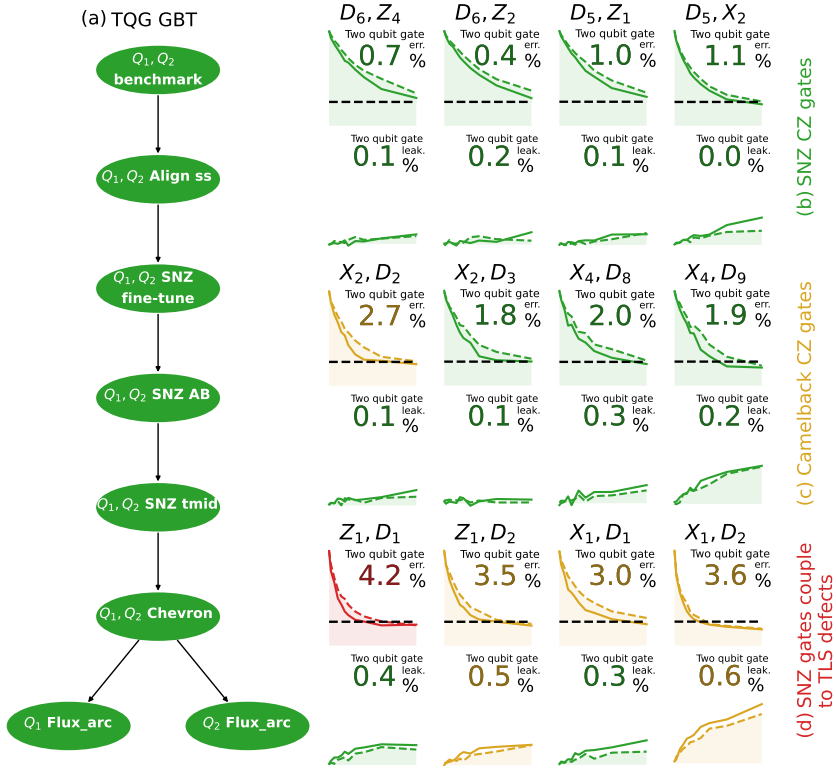

Figure 4.9: **Two-qubit gate calibration and benchmarking.** (a) GBT graph nodes and dependencies are designed for automatic calibration of SNZ CZ gates. Overview of the achieved gate performance obtained by randomized benchmarking protocols for the standard SNZ flux pulse in (b), the camelback-based CZ gate in (c), and SNZ gates that strongly couple to TLS defects on the way to their target avoided crossings in (d). The color of the measured curves encodes the level of performance: green for optimal performance, orange as a warning level, and red indicating low performance. Calibration strategies and additional details about each node are discussed in the main text. The overall calibration time for this procedure, involving 24 CZ gates in a distance-3 surface code, is approximately $15\,\mathrm{hrs}$, averaging around $38\,\mathrm{min}$ per gate. This estimate does not account for potential variations in calibration time, especially when dealing with some problematic gates as a result of overlapping with TLS defect modes (see below).

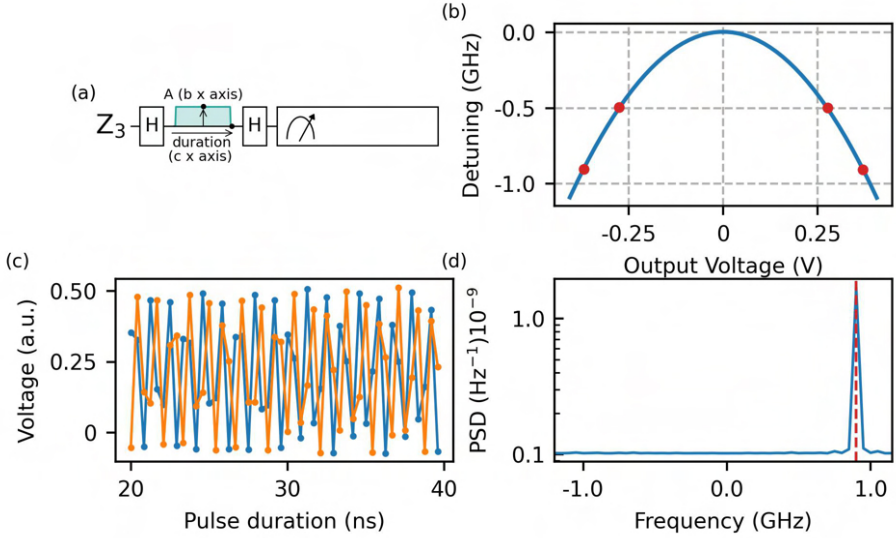

Figure 4.10: **Flux arc calibration for qubit Z in the Uran device.** (a) The quantum circuit involves embedding a fast square pulse into a Ramsey-like experiment before reading out the and components, by changing the final 2 rotation, of the Bloch sphere probing the qubit frequency. (b) The measured qubit frequency detuning as a function of the applied voltage of a unipolar flux pulse generated at the AWG output. The measured points (red dots) are fitted to a second order polynomial equation to extract the corresponding arc (blue curve). (C) The evolution of the measured (blue) and (orange) components of the Bloch sphere at a specific applied voltage of the flux pulse. (d) Each data point is obtained via fitting the power spectral density of the measured time traces in (c), resulting in frequency detunings (red dots in (b)). This approach is more resilient to measurement noise and efficient in calibration runtime, taking approximately $5\,\mathrm{min}$ per qubit.

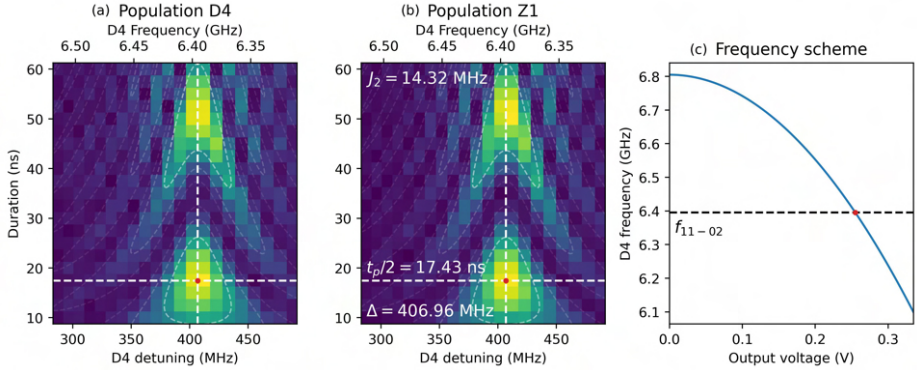

Figure 4.11: $11$ - $02$ **Chevron interaction of gate D4-Z1 in device Uran.** (a) Evolution of the $2$ state population of qubit D4 and (b) the $1$ state population of qubit Z1 as functions of the amplitude and duration of a unipolar square pulse applied to qubit D4. This pulse effectively brings the $02$ and $11$ states into resonance. Key parameters such as $J_2$, $\Delta$, and $t_p$ (white dashed lines) are determined through fitting procedures (refer to the main text). This approach facilitates the use of low measurement averages, coarse step size for control parameters, and efficient measurement runtime. Prior calibration of voltage-to-frequency conversion is necessary in (c), showing the dependency of qubit frequency on the applied voltage through the $11$ - $02$ avoided crossing at $f_{11-02}$. Subsequently, the code uses this interaction frequency to arbitrarily set the main pulse amplitude (A) of the SNZ gate. The perfect symmetry observed in this chevron indicates successful calibration of flux pulse distortions such that .

Now, the groundwork is laid to initiate the calibration of the SNZ gate, which involves two key steps: finding the optimal time between the two square pulses, $t_{mid}$, and determining the main and intermediate flux pulse amplitudes, A and B (AB landscape). Such parameters should yield the optimal CZ gate with a conditional phase, $\phi = \pi (\mathrm{mod}\,2\pi)$, and leakage, $1\%$[69]. To extract the optimal $t_{mid}$, the SNZ flux pulse [Figure 4.12. a] is embedded into the standard conditional oscillation experiment while sweeping A (translated into frequency detuning on the x axis) and $t_{mid}$ on the y axis. The outcomes of the conditional oscillation experiments—the measured and —are plotted in [Figure 4.12. b and c], respectively.

Similar to the Chevron experiment, analytical expressions of gate landscapes are derived to efficiently extract the optimal $t_{mid}$ via fitting. This involves considering two coupled energy levels, $11$ and $02$, with a coupling term $2$ and dynamical detuning $|\rangle - |\rangle$. By solving the time evolution of this simple Hamiltonian for different areas of the SNZ pulse, obtained through the Schrödinger equation and Hamiltonian diagonalization, relevant parametrizations of the gate are identified. Further details can be found in appendix A of this thesis [169]. This significantly simplifies the detection of optimal $t_{mid}$ values for various gate detunings $|\rangle - |\rangle$. This is particularly practical for mid-to-low gates, where $|\rangle - |\rangle$ is comparable to the AWG sampling rate. Moreover, the choice of $t_{mid}$ is now resilient to noise, allowing for the use of low averages (and a low number of points) to extract optimal $_{\text{mid}}$ (blue star in Figure 4.12. c), which completes in $7\,\text{min}$ per gate.

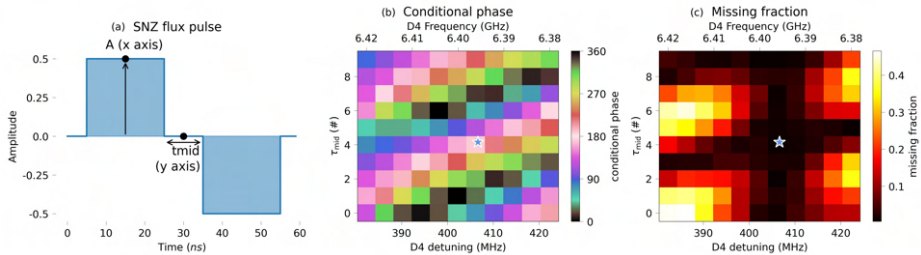

Figure 4.12: **T$_{\text{mid}}$ calibration of the SNZ gate** (a) The SNZ flux control parameters for the $_{\text{mid}}$ landscape include the main pulse amplitude A and the time $_{\text{mid}}$ between the two strong pulses within a $60\,\text{ns}$ gate duration. Conditional oscillation experiments are conducted, and the two-qubit phase in (b), and the missing fraction of the control qubit, in (c) are extracted as functions of flux control parameters, where A is on the x axis and $_{\text{mid}}$ is on the y axis. The optimal $_{\text{mid}}$, indicated by the blue star, is determined through fitting procedures (see main text). This calibration process requires approximately $7\,\text{min}$ per gate.

AB landscape undergoes a similar calibration procedure, but this time the control parameters are A and B, as depicted in [Figure 4.13. a]. As expected, the phase and leakage landscapes [Figure 4.13. b and c] provide useful crosshairs, where the $= 180°$ contour always crosses the minimum leakage, resulting from two main valleys (vertical and diagonal) [69]. The optimal A and B (white point in Figure 4.13. d) are chosen based on a cost function that ensures the best CZ gate, $= \pmod{2}$, and $1\%$. If the time of the flux pulse is much longer or shorter, by several sampling points, than the speed limit, there are significant consequences on the leakage landscapes. The signatures of these two cases are presented in [Figure 2.3]. The possibility for the too-long case to achieve an optimal CZ demonstrates the practical viability of the SNZ pulse. This is particularly useful to satisfy the time constraints of several SNZ gates operating in parallel, where some could intentionally be made longer.

Prioritizing the minimization of leakage over achieving the ideal conditional phase in the cost function can prove advantageous. This is particularly critical due to the coarse tuning of in the previous calibration steps which employ low averages and reduced number of points to expedite the calibration time. Alternatively, precise control of and is achieved by fine-tuning the amplitude B and sweeping the asymmetry between the two main flux pulses of the SNZ gate, as depicted in [Figure 4.14 a. and b], respectively. The final calibration step is important to minimize leakage, compensating for residual pulse distortions or imprecise DC biases. Additionally, a virtual- correction is applied to null the qubit phase errors extracted from conditional oscillation experiment [Figure 4.14. c]. This approach allows precise control of gate parameters, ultimately improving the gate performance.

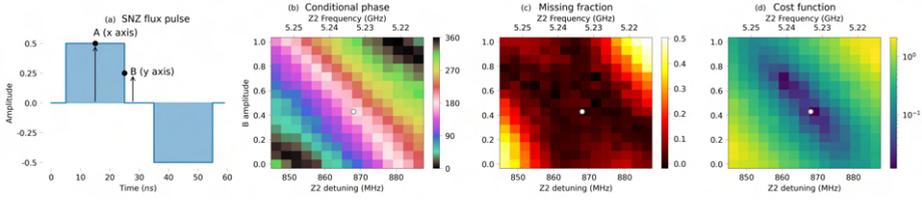

Figure 4.13: **AB calibration of the SNZ gate.** (a) SNZ flux control parameters of the AB landscape are the main pulse amplitude A, and the intermediate amplitude B. Amplitudes A on the x axis (translated to frequency detuning through voltage-to-frequency conversion) and B on the y axis are swept for the calibration of the two-qubit phase, in (b), and the missing fraction, in (c). Optimal control point (white circles) is chosen based on a cost function optimization in (d), achieving the best possible CZ gate: $= 180°$ and $1\%$. The cost function prioritizes optimal points in the middle when the vertical fringe intersects the diagonal one [69]. This approach adds an extra layer of protection and flexibility when fine-tuning the SNZ gate as part of a parity check (see Figure 6.6). With $11$ sweep points and $2$ averages, this measurement requires approximately $40 \text{ min}$ per gate.

Due to the frequency overlap of middle- and low-frequency transmons in our device, parking flux pulses on spectator qubits are necessary whenever adjacent SNZ gates are implemented [74, 98]. The always-on coupling, mediated by resonator buses, introduces coherent phase errors and leakage in the qubits involved in the two-qubit gates [87]. To address this, we conduct a parking optimization experiment [Figure 4.15] (not shown in TQG GBT). This measures various gate parameters such as [Figure 4.15. b], two-qubit phase error ( ) [Figure 4.15. c], single-qubit phase errors [Figure 4.15. d] and [Figure 4.15. e] as a function of the parking frequency on the spectator qubit. is measured when the parked qubit is in states $0$ (blue) and $1$ (red).

Parking optimization experiment reveals two critical features where phase and leakage errors peak. These occur when parked frequency overlaps with one or approximately two anharmonicities away from the higher-frequency qubit [Figure 4.15. c]. This is in qualitative agreement with the observed avoided crossings in the transmon energy spectrum for the three-excitation manifold, $001 - 100$, $011 - 020$, $021 - 030$ [Figure 4.15. a], and the findings of this reference [87]. This work also explains the shift ($127 \text{ MHz}$) between the measured data and calculated avoided crossing $021 - 030$, attributing it to the imprecise assumption that $=$ .

In practice, the codebase sets a parking distance from lower-frequency qubits to at least $700 \text{ MHz}$, ensuring a safe distance from any undesirable peaks. This approach safeguards against the potential impact of coherent phase errors and leakage when a spectator qubit state is excited.

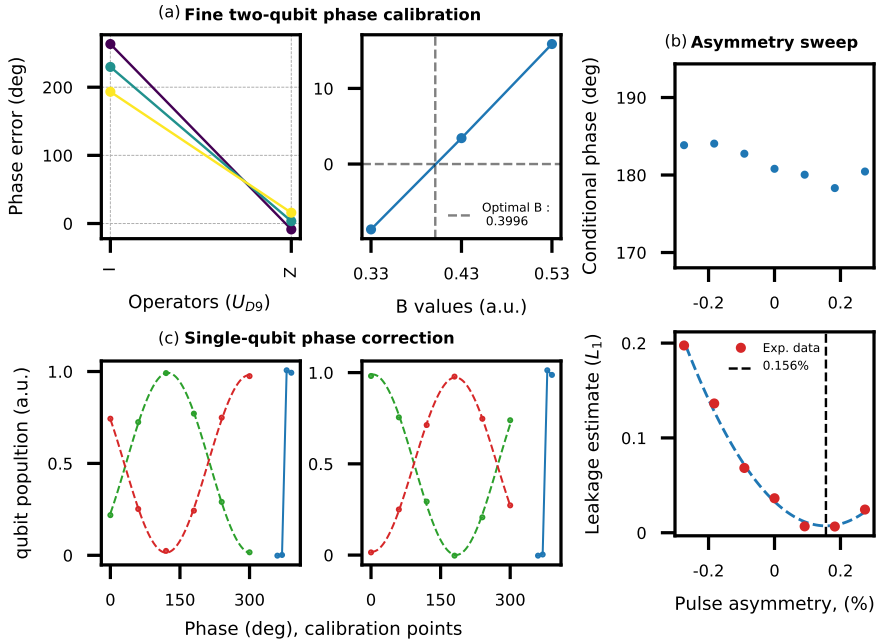

Figure 4.14: **Fine-tuning of two-qubit gate performance.** (a) Precision control of the two-qubit phase achieved through a conditional oscillation experiment by varying the amplitude B of the SNZ gate. The left panel illustrates phase errors when toggling the control qubit D between $0$ and $1$ states. The right panel showcases linear control of the desired conditional phase as a function the amplitude B. (b) Optimization of asymmetry between the two arms of the SNZ flux pulse. This step involves sweeping the asymmetry, expressed in percentage, to minimize (bottom panel) while maintaining almost unaffected (top panel). (c) Elimination of single-qubit phase errors through virtual-Z correction in the microwave AWG. This correction is accomplished by conducting two conditional oscillation experiments, swapping the roles between target and control qubits in the left and right panels, respectively. Calibration points are included at the end to assess the contrast of the obtained data against ideal values. The calibration is completed in approximately $3\,\mathrm{min}$.

Following calibration, the graph advances to the final stage of benchmarking the gates using 2QIRB protocols [96, 97]. The experimentally measured RB curves are presented in [Figure 4.9. b]: reference and CZ-interleaved return probabilities to $00$ (top panel), and population in the computational space (bottom panel). Individual benchmarking of the 24 CZ gates [Figure 4.2. e] achieves an average error of $1\,6\%$ with leakage of $0\,24\%$ in $60\,\mathrm{ns}$ gate time. It should be noted that not all gates are calibrated using the standard SNZ gate scheme; the reason for this is discussed below.

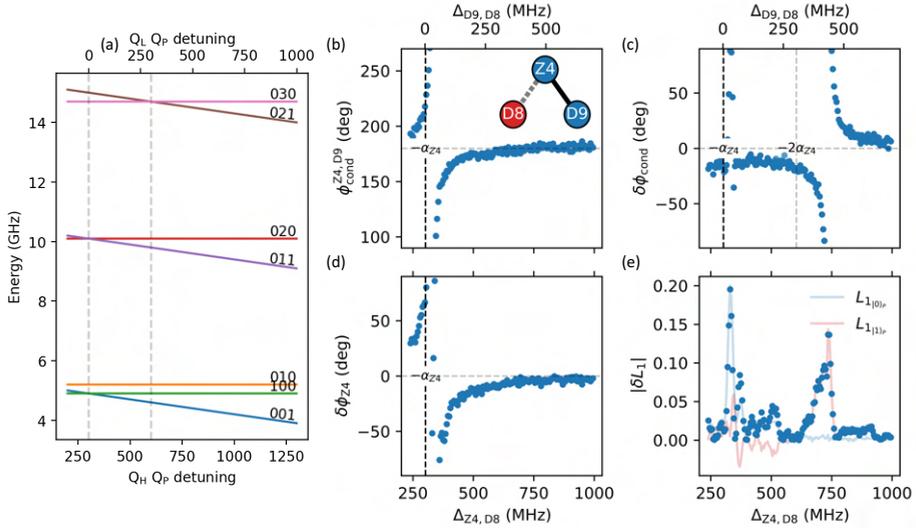

Figure 4.15: **Parking optimization for minimizing spectator effects.** (a) The computed transmon energy spectrum in the three-excitation manifold, denoted as          , reveals avoided crossings when the spectator qubit D toggles between $0$ and $1$. (b) The two-qubit phase,      , and (c) phase error,        , for the gate Z -D  as functions of the parking frequency of qubit D . (d) Single-qubit phase errors of qubit Z , and (e) the difference in      as D  toggles between $0$ and $1$. These parameters are extracted from characteristic conditional oscillation experiments involving bipolar parking pulses parallel to the main SNZ gate Z -D .

Recent QEC experiments highlight the significant challenge in achieving and maintaining high-fidelity two-qubit gates, primarily due to strong interaction with TLS (two-level system) defects [26, 87, 167]. This issue is particularly pronounced with flux-based CZ gates, as they dynamically move transmon qubits in frequency bringing the states $02$ and $11$ into resonance, increasing the likelihood of interacting with TLS defects along the frequency trajectory. Such interaction can lead to a substantial suppression of qubit coherence, severely limiting gate performance. Addressing this issue is complex, especially considering the temporary fluctuations of TLS, emphasizing the necessity of frequent calibration. Here, we discuss our approach for the mitigation of TLS impact on two-qubit gates through the development of novel calibration routines.

There are two scenarios where TLS defects can intersect the trajectory of a two-qubit gate. In the first scenario (TLS overlaps with or on the way to the $11$ - $02$ interaction), we initially utilized the other avoided crossing, $11$ - $20$ (refer to Figure 2.10). However, this approach encountered residual interactions at the $11$ - $02$ and $01$ - $10$ resonances. To mitigate this, we dynamically detune the low-frequency qubit D  using a baseband flux pulse (orange), aligned with the main SNZ flux pulse on the high-frequency qubit Z  (blue), as shown in

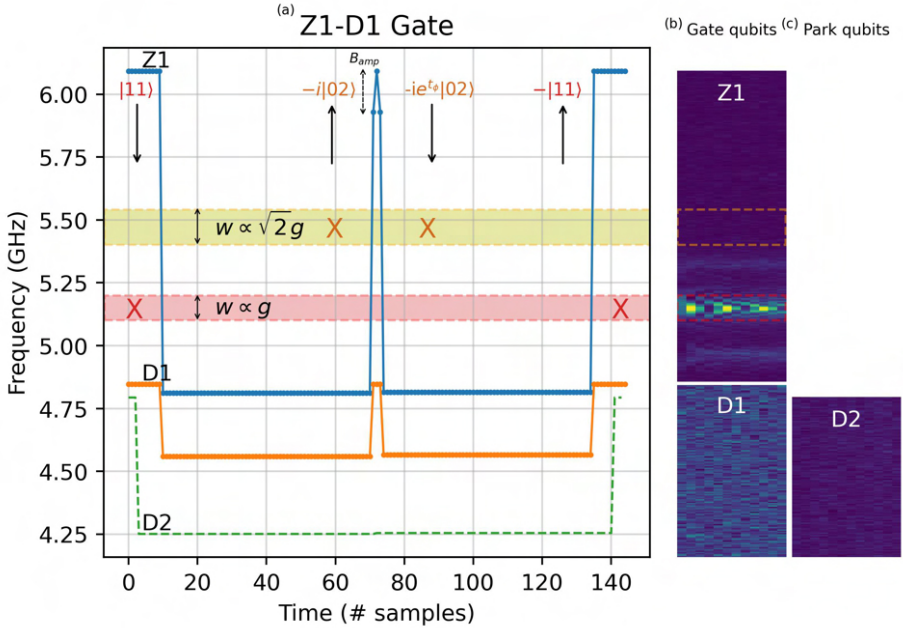

Figure 4.16: **Frequency trajectory of the SNZ gate Z -D .** (a) Frequency trajectory, derived from flux pulse amplitudes and voltage-to-frequency conversion. Initially, the avoided crossing $11$ - $02$ overlaps with a TLS mode (red rectangle width  is proportional to the coupling strength  ), especially relevant when qubit Z  is in the $1$  state at the gate start and end. Of more concern is the halfway point during the gate where leakage to $02$  is maximized, due to stronger coupling to the defect mode (yellow rectangle width  is proportional to the coupling strength  $\overline{2}$  ). Our mitigation technique involves activating the flux pulse on the lower-frequency qubit D , dynamically shifting the target avoided crossing down. This strategy proves effective unless the frequency trajectories of qubit D  (orange pulse) and the parked qubit D  (green pulse) encounter additional defect modes.

Figure 4.9. This technique dynamically shifts the target avoided crossing $11$ - $20$  away from the problematic TLS mode. The detuning frequency for qubit D  chosen based on optimizing gate performance through 2QIRB protocols. Following this strategy, we achieve relatively functional two-qubit gates [Figure 4.9. d].

However, this approach presents several limitations. First, it has higher dephasing rate during the gate, as the low-frequency qubit is no longer at its sweetspot. Second, when the computational state $11$  enters (exits) at the beginning (end) of the SNZ gate, it loses some residual population to the TLS mode [Figure 4.16. a]. Similarly, the non-computational state $02$ , half-away during the SNZ gate, experiences a even stronger interaction with the TLS (yellow rectangle) by a  $\overline{2}$. This approach also fails to consider the possibility of encountering multiple TLS defects, either on the high- or low-frequency qubits, along the trajectory to the intended interaction.

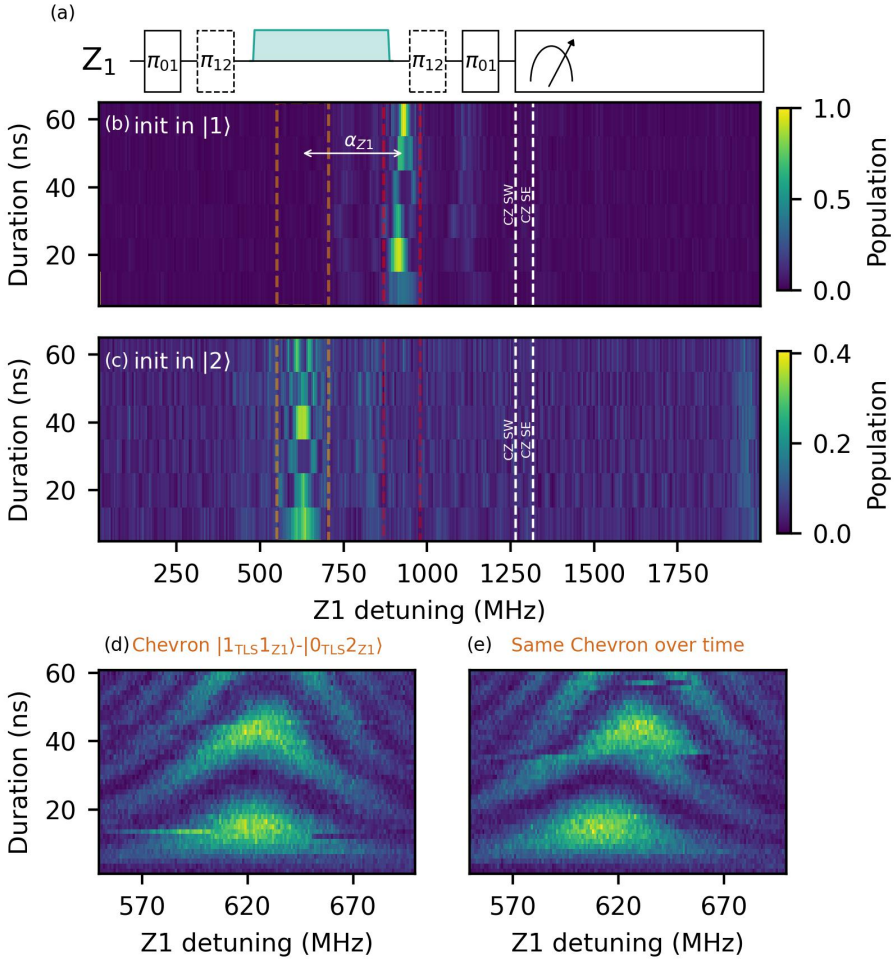

Figure 4.17: **Instability of coherent operations with TLS.** (a) Quantum circuit used to measure thee population loss of states $1$ and $2$ upon applying a unipolar flux pulse. The measured populations when qubit $Z$ is initialized in states $1$ (b), and $2$ (c) as functions of the flux pulse amplitude (translated to frequency detuning on X axis) and pulse duration on Y axis. Two parasitic chevron interactions appear along the frequency trajectory of qubit $Z$ at $620\,\mathrm{MHz}$ and $930\,\mathrm{MHz}$. The difference between these two interactions is exactly one anharmonicity $_1$. (d) and (e) The measurement of the chevron interaction in the two-excitation manifold over time. The instability of this coherent interaction over time is likely due to the flickering nature of the TLS. Key parameters such as $2,$ , and are determined via fitting procedures.

In principle, if a defect mode is strongly coupled to a qubit, coherent exchange might be possible. We explore this possibility in Figure 4.17. Rather than diabatically crossing the TLS on the way to the interaction point, the qubit frequency is specifically tuned to the TLS frequency to maximize coherent interaction. The frequency trajectories of qubit Z , prepared in states $1$ and $2$ , are presented [Figure 4.17. b and c]. Zooming on the interaction chevron of the coupled states $1 \quad 1_1$ and $0 \quad 2_1$ , we observe instability over time [Figure 4.17. d and e]. This instability, attributed to the flickering nature of the TLS, underscores the impracticality of performing coherent operations with defect modes.

In the second and most challenging scenario, a TLS overlaps with the qubit at the sweetspot. For full leakage interference with either SNZ or conventional NZ pulse schemes, symmetry is essential—the qubits involved must be biased to the sweetspot frequency. If this condition is not met, the flux pulse will not average to zero within the gate time, breaking symmetry. An example of an asymmetric flux pulse is shown in Figure 4.18. a, where the high-frequency qubit is biased off-sweetspot (green dot). This configuration fails to fully eliminate the DC component caused by the buildup of long-time scale flux distortion, resulting in poor gate repeatability [94]. Moreover, the right frequency trajectory undergoes a strong interaction as it crosses this large defect mode (red cross) on its way to the interaction point (dark orange). This interaction adversely affects chevron quality, as seen in [Figure 4.18. d], in contrast to the left chevron shown in [Figure 4.18. c]. This underscores the critical importance of symmetry in the NZ gate scheme.

The initial demonstrations of CZ gates utilized unipolar flux pulses, whether employing simple square pulses [67, 83] or fast-adiabatic gate schemes [84, 86]. A significant limitation of these techniques is the accumulation of flux distortion over time, leading to poor gate repeatability. While the NZ scheme [94] addresses this issue, both SNZ and NZ CZ gates face challenges in the presence of TLS at sweetspot, as they fail to maintain the required symmetry. Consequently, there is an ongoing exploration of more robust and flexible gate approaches that account for TLS implications at the symmetry point. Our current focus is on an enhanced version of unipolar gates, which includes the calibration of long-time scale flux distortions. In the following section, we introduce two primary approaches, referred to as DC and camelback flux pulses.

Long-timescale flux distortion arises from various sources including high-pass filtering of a bias tee, skin effect, impedance mismatch, and chip packaging [103, 170]. Notably, the fast characteristic time constant of the bias tee, typically ranging from $5 \quad 50$ s [103, 171], governs the rate at which accumulated charges dissipate, significantly influencing long-time distortion within the coherence time of gate operations. To mitigate these distortions, bias tees are removed from the qubits that cross TLS at or near the sweetspot frequency. Instead, these qubits are biased off-sweetspot, sufficiently distant from strongly coupled modes, utilizing direct offsets provided by the flux AWG.

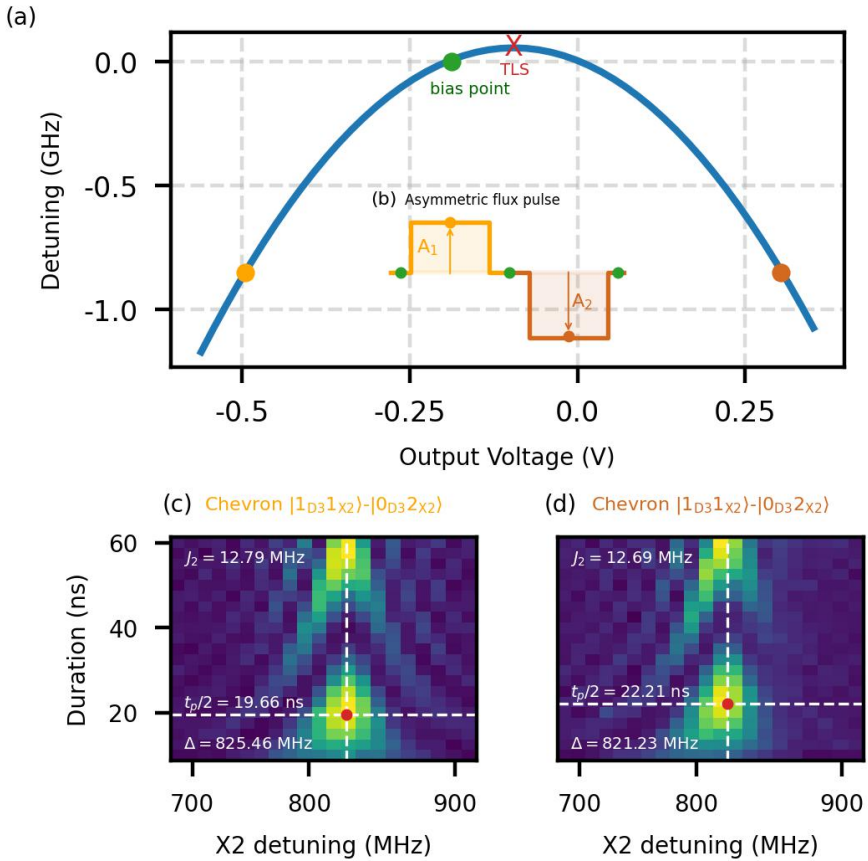

Figure 4.18: **Asymmetric flux gate in presence of a strongly coupled TLS at the sweetspot.** (a) Flux arc of the high-qubit qubit biased off-sweetspot (green point) and (b) Asymmetric flux pulse implementing two interaction points: orange on the left arm and dark-orange on the right arm, crossing the problematic TLS (marked with a red cross). (c) The left and (d) right Chevron interactions of $11$ - $02$ as functions of the unipolar flux pulse amplitudes,         (left) and          (right), and pulse duration. Extracted Chevron parameters are shown on the left and right arms. The poor quality of the right interaction (bad symmetry along the vertical dash line) signifies the impact of crossing the significant defect along the frequency trajectory.

Remaining flux distortion is corrected using a spectroscopy technique [172], as illustrated in [Figure 4.19. a]. This technique involves applying a long square pulse to the dedicated flux line of the qubit while sliding a     pulse with a Gaussian envelop of $10$ ns at various times t, to probe the qubit frequency. Unlike the Cryoscope method, which is strictly limited by qubit coherence [103], this approach can characterize flux distortion up to the millisecond time frame. To compensate for these distortions, we use the 8 built-in exponential filters in ZI HDAWG to iteratively improve the flux step response, as depicted in [Figure 4.19. b]. The iterative optimization results in a constant step response within $0 1\%$ in $30$ s interval [Figure 4.19. c], facilitating the use of unipolar pulse scheme for high-fidelity two-qubit gates.

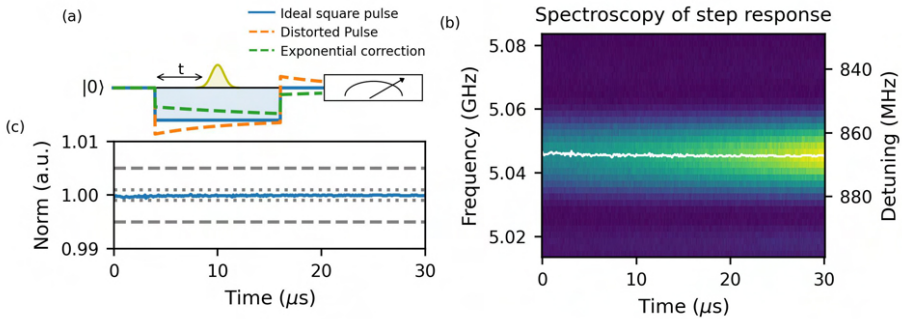

Figure 4.19: **A spectroscopy Cryoscope for the characterization of long-timescale flux pulse distortions.** (a) Implementation of a $30$ s flux pulse, while sliding a pulse with a $10$ ns Gaussian envelope at various times t, to probe the qubit frequency. Due to various sources of distortion (see main text), the intended square pulse (orange curve) may exhibit overshooting. We iteratively improve the step response using the built-in exponential filters in the ZI HDAWG, compensating for overshooting distortions (green curve). (b) The best measured step response (white curve), constant within $0 1\%$ over $30$ s. This response is obtained by Gaussian fitting of the population of the $1$ state, as a function of the sliding time and the qubit frequency detuning centered at $870$ MHz. The color intensity variation corresponds to the qubit coherence decay, particularly noticeable at the beginning when the qubit excitation waits for $30$ s before being measured. (c) (c) A 1D cross-section of the step response, fitted to an exponential decay model (refer to Table S1 in [103]) to extract the characteristic time constant, which varies from $100$ ns to $100$ s.

We have successfully calibrated the truly sudden unipolar gate, nicknamed DC pulse (inset of Figure 4.19. a). Unlike the NZ scheme, this simple pulse implements a Chevron interaction where the ideal two-qubit phase aligns vertically along the minimum leakage zone between the two Chevron blobs. The ideal gate parameters are marked by blue circles in [Figure 4.19. c and d]. Following calibration, gate performance is benchmarked using 2QIRB protocol, achieving a fidelity of $98 5\%$ with $0 1\%$ leakage [Figure 4.19. b and e]. This is likely limited by qubit coherence—since the high-frequency qubit is biased off-sweetspot—and by $-$ flux noise due to direct biasing. Importantly, if the actual flux pulse is unipolar, any associated parking flux pulses should not be bipolar.

Another flexible CZ gate scheme, known as the camelback (inset of Figure 4.21. a), is a hybrid between the DC and SNZ schemes. Unlike the DC pulse, the camelback gate implements two interaction points in the same direction to circumvent the faulty TLS [Figure 4.21. a]. This configuration still preserves full leakage interference, resulting in gate landscapes [Figure 4.21. c and d] akin to those of the SNZ gate, making [Figure 4.9] applicable to this gate as well.

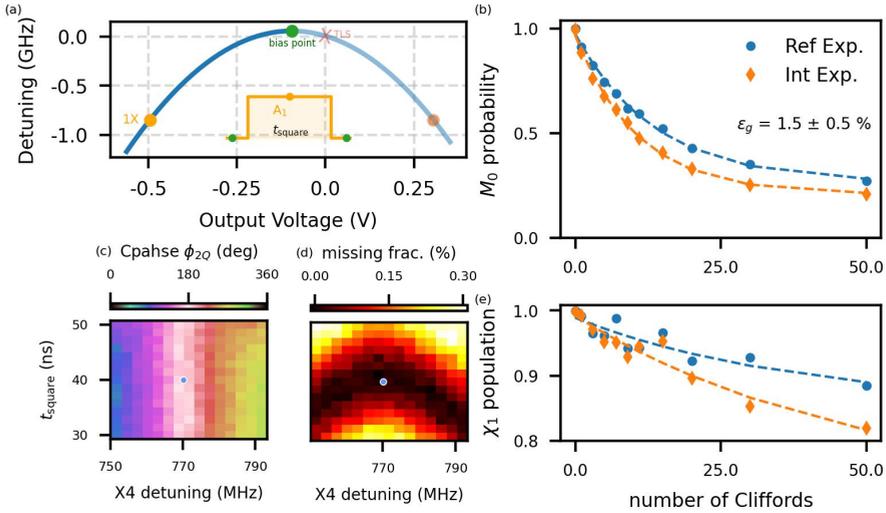

Figure 4.20: **Calibration and benchmarking of the DC CZ gate.** (a) Flux arc and DC flux pulse (inset) implementing single unipolar interaction point (orange point) when higher-frequency qubit is intentionally biased off-sweetspot (green point). Note that the inset figure does not share any axis with the flux arc. Gate landscapes for (c) , and (d) as functions of the DC pulse amplitudes (A, translated to frequency detuning on the X axis) and pulse duration ( on the Y-axis). Optimal gate parameters (blue point) are chosen to achieve $= 180°$ and the minimum possible . (b) Return probability to 00 , and (c) the computational space population are presented for both reference and CZ-interleaved curves. Spectator qubits to this gate remain in the ground state during benchmarking. The extracted (not shown here) due to this gate is $0 19    0 1\%$.

Additionally, the camelback gate facilitates the use of bipolar parking pulses, useful for the flux dance assembly. However, it lacks the built-in echo protection since the high-frequency qubit does not reside at the sweetspot. The benchmarking of this gate shows similar performance to that of the DC pulse, as illustrated in [Figure 4.21. b and e]. Gates calibrated using this scheme are further illustrated in [Figure 4.9. c].

### 4.2.4    Readout calibration and benchmakring

Measurement is a key component of quantum algorithms, particularly in QEC codes that require repeated measurements. Recent QEC experiments highlight the significance of minimizing measurement errors, often dominant tied with two-qubit gate fidelities [26, 141]. Common readout calibration routines involve exploring large parameter space (measurement-based approach) to reduce measurement errors. This process is tedious and time-consuming partic-ularly with scaling the number of qubits. In this section, we introduce our strategy to calibrate optimized readout across a 17-qubit device. To this end, we focus on developing the automated framework with GBT to autonomously calibrate the readout from basic spectroscopy to fine-tuning of readout assignment fidelity and benchmakring measurement QNDness. We also study how TLS defects impact the readout performance. Furthermore, we characterize multiplexed readout performance to evident the significance of readout crosstalk and measurement-induced mixing. Finally, we assess the backaction of ancilla measurements on computational data-qubits, which is of great interest for a QEC experiment.

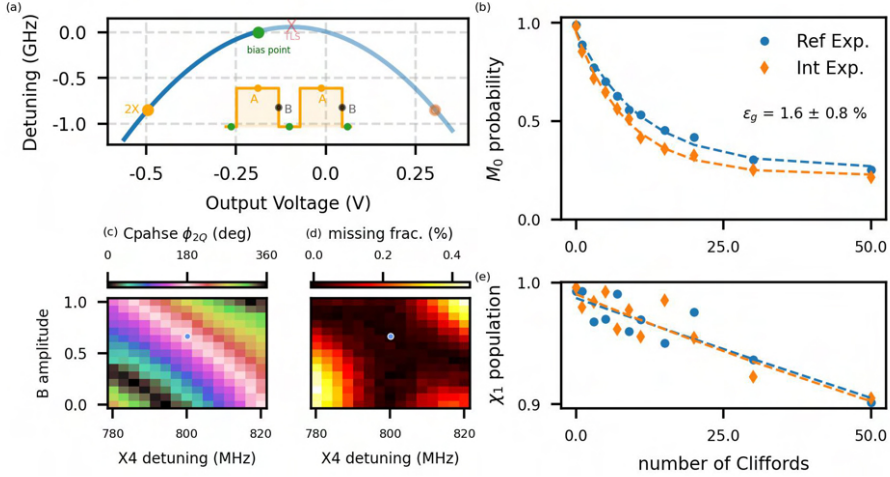

Figure 4.21: **Calibration and benchmarking of the camelback CZ gate.** (a) Flux arc and camelback flux pulse (inset) implementing two interaction points (orange point) in the same direction. The inset figure does not share any axis with the flux arc. Gate landscapes for (c) , and (d)     as functions of the camelback pulse amplitudes (A, translated to frequency detuning on X axis) and B on the Y axis. (b) Return probability     to 00 , and (c) the computational space population    , are shown for both reference and CZ-interleaved curves. Spectator qubits to this gate remain in the ground state during benchmarking. Extracted    (not shown here) is negligible, requiring more averages and number of Cliffords.

In our Surface-17 design, the readout circuitry is distributed over three feedlines, which are coupled through readout resonators to 6, 9, and 2 qubits, respectively. This imbalance distribution is made such as the device layout remains scalable beyond 17 qubits [98]. Each qubit is readout by a dedicated pair of readout and Purcell resonators to eliminate the Purcell effect while enabling fast and high-fidelity readout [102]. Both dedicated pairs are short-circuit end of a $\lambda/4$ CPW resonator. The target detuning between each readout pair of adjacent qubits is at least $100\,\mathrm{MHz}$ to mitigate the impact of the spectral overlap between readout pulses. This significantly reduces measurement-induced dephasing ( ) [147, 173], especially between measured ancillas and data qubits, as shown later in [Figure 4.24].

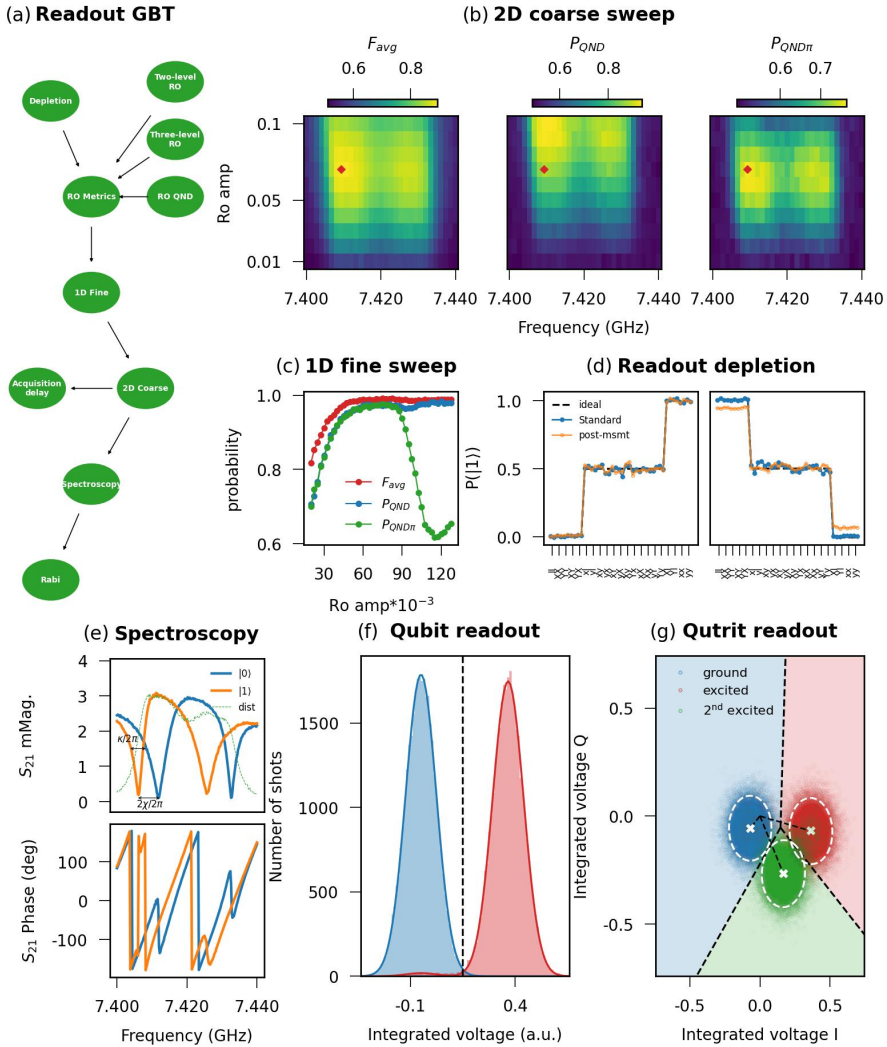

Figure 4.22: **Calibration and benchmarking of readout using GBT.** (a) Readout GBT nodes designed for autonomous calibration and benchmarking of single-qubit readout. (b) 2D coarse optimization of readout pulse frequency and amplitude to maximize the average readout assignment fidelity, $F_{avg}$, and the probabilities of QND: $P_{QND}$ and $P_{QND\pi}$. The last two metrics are derived from [168], where the quantum circuit includes two consecutive measurements ($m_1$ and $m_2$) followed by a $\pi$ pulse and a third measurement ($m_3$). Here, $P_{QND} = P(m_2 = m_1)$ and $P_{QND\pi} = P(m_3 = m_2)$. (c) 1D optimization to fine-tune the readout amplitude. (d) Readout depletion using an ALLXY gate sequence between two measurements. When the qubit is initialized in $|1\rangle$, $P(|1\rangle)$ is reduced due to qubit decay during the first measurement (orange in the most-right panel). (e) Transmission $|S_{21}|$ and phase of readout resonator pairs prepared in $|0\rangle$ and $|1\rangle$ to measure the dispersive shift $2\chi/2\pi$. When the effective linewidth of the readout pairs is equal to $2\chi/2\pi$, optimal readout performance is achieved. (f) The measured two-level, and (g) three-level assignment fidelities extracted from single-shot histograms in $420\,\mathrm{ns}$ in device Uran.

To calibrate readout with GBT, a previously calibrated $\pi$ pulse amplitude, obtained by a Rabi measurement of the qubit, is necessary. Calibration procedure involves three main stages. At first, we calibrate the delay between the readout pulse and the acquisition interval and measure the effective dispersive shift ($2\chi_{ef}$) of the readout resonator using spectroscopy routines. At a fixed readout pulse duration, readout spectroscopy sweeps the frequency of the measurement pulse and acquires the corresponding complex transmission ($S_{21}$) through the feedline, whose magnitude $|S_{21}|$ and phase are shown in [Figure 4.22. e]. This allows to estimate a range where optimal readout frequency is found at the maximum distance between the two complex vectors $S^{|g\rangle}$ and $S^{|e\rangle}$ in IQ plane (green line in Figure 4.22. e). For fast and high-fidelity readout, two conditions must be satisfied: matching resonator pairs (readout and Purcell modes) for high hybridized linewidth ($\kappa$), and $2\chi_{ef} = \kappa$ maximizing the signal-to-noise ratio at fixed photon number ($n_{ph}$) in the readout mode [131, 174]. Shoelace technique can be used to improve the matching between resonator modes in case of pair detuning due to fabrication variations. If not properly adjusted, such mismatches can lead to significantly slow readout speeds, exceeding $1\mu s$ [55].

At the second stage, a 2D optimization of the readout pulse parameters is performed to maximize the average readout assignment fidelity ($F_{ass}$) and the QND probabilities ($P_{QND}$, and $P_{QND}$) [Figure 4.22. b]. More details about this experiment can be found in the caption of this figure. At optimal readout frequency (red point in Figure 4.22. b), readout amplitude is finely tuned through a 1D optimization [Figure 4.22. c]. Optimal readout amplitude is selected where maximum $F_{ass}$ is achieved. Further increasing the readout amplitude leads to a non-QND readout, which can induce transitions to higher leakage states [61, 135, 175].

At the final stage, photon depletion from the resonator is verified within the target readout time of $420\,ns$, using an ALLXY gate sequence between two measurements [Figure 4.22. d] [18]. The ALLXY pattern obtained from these measurements is evaluated against the ideal staircase to determine if the allocated depletion time is sufficient and does not adversely affect subsequent coherent operations on the qubits. After optimal integration weights calibration, various readout metrics are benchmarked, including two-level (qubit readout), three-level (qutrit readout) fidelities [Figure 4.22. f and g] and readout QNDness using measurement butterfly technique [Fig. S5 in [175]]. The average readout assignment fidelity across 17-qubits is $98.8\%$, extracted from the single-shot histograms in device Uran. The readout GBT procedure takes about $5\,hrs$ per qubit. To speed up this process, future effort could benefit from parallelizing this procedure across multiple qubits using multiplexed readout.

One of the common failure modes that the current readout GBT struggles with is the strong coupling to a TLS during readout. This interaction can severely limit the qubit $T_1$ during measurement, leading to a high probability of measurement-induced relaxation. During measurement, the qubit frequency experiences a Stark shift and is dephased at a rate denoted by $\Gamma_{\phi}$ [147, 176], which depends on the photon number $n_{ph}$ in the readout mode. Within this frequency range, the qubit parasitically interacts with a TLS, leading to suboptimal readout performance [Figure 4.23. a]. When this occurs, manual control is taken to detune the qubit

**4**

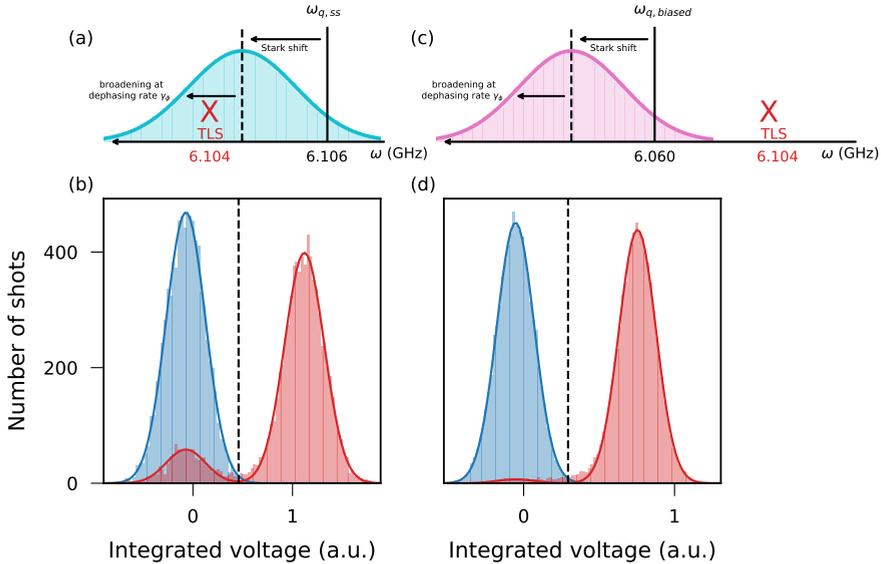

Figure 4.23: **Impact of TLS on single-qubit readout.** (a) Qubit frequency trajectory overlapping with a TLS mode during measurement. The qubit frequency experiences a Stark shift and is dephased at a rate denoted by γ. (b) Suboptimal readout histograms with a significant probability of measurement-induced relaxation. (c) Mitigation involves detuning the qubit frequency away from the TLS interaction. (d) Improved readout histograms by avoiding overlap with the TLS mode.

frequency so that its trajectory during measurement avoids overlapping with the TLS mode. This can indeed recover optimal readout, as shown in [Figure 4.23. b]. Recently, Google AI team [177] proposed a model-based optimization for efficient calibration of readout across a multi-qubit device. Not only is this approach significantly faster and scalable to hundreds of qubits, but it also incorporates various error models, including the relaxation rate of the qubit as a function of frequency, offering a practical solution for enhancing the current readout GBT.

We perform multiplexed readout for ancilla and data qubits in device Uran. By preparing all combinations of input states, assignment probability matrices are constructed in $420\ \mathrm{ns}$ measurement time (shown in Figure 4.24. c and d) [102]. The average multiplexed readout fidelities extracted from single-shot readout histograms, are $98\ 6\%$ for ancilla qubits and $98\ 4\%$ for data qubits, nearly identical to their individual readout fidelities (see Section 4.1.1). Additional off-diagonal features may appear below (above) the main diagonal due to individual qubit decay and measurement-induced relaxation (excitation). Readout crosstalk is further assessed through the cross-fidelity matrix [Figure 4.24. a, and b]. The off-diagonal terms show insignificant crosstalk errors in this device. It is important to note that the quoted fidelities include mitigation of residual excitation effects through post-selection on a pre-measurement, which is below $1\%$ for most qubits in this device.

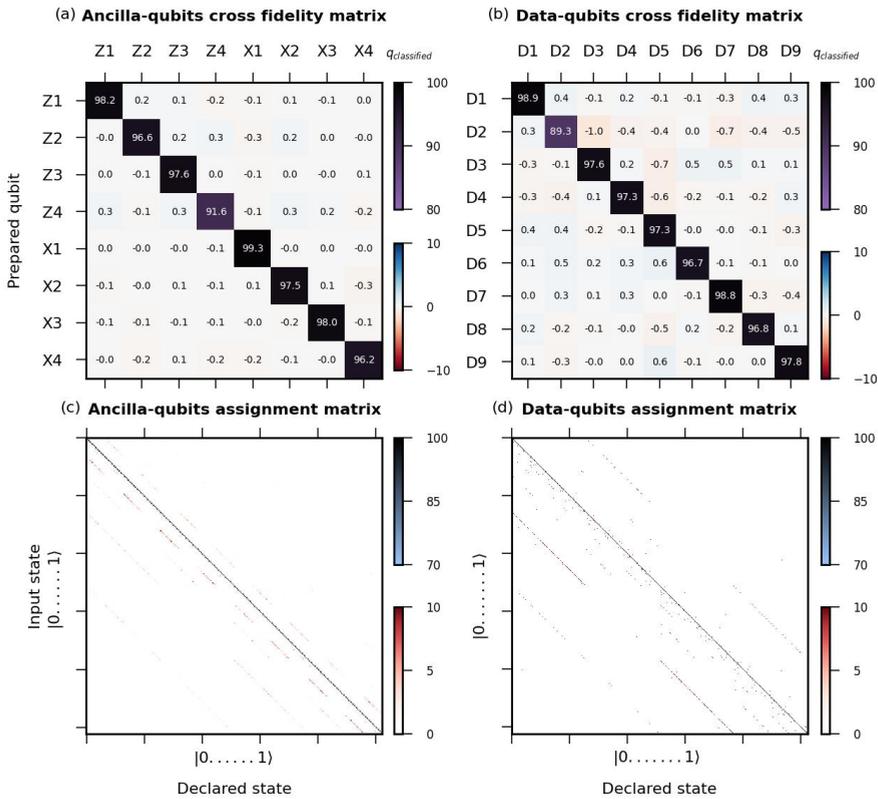

Figure 4.24: **Multiplexed readout of ancilla and data qubits** in (a), and (b), respectively, in device Uran. Experimental cross-fidelity matrices shown in (c), and (d). Readout assignment probabilities between ancilla and data qubits [102]. An end-to-end measurement, performed within $420\,\text{ns}$, includes the pulse excitation time and additional time for full photon depletion from the resonator modes of each pair.

Due to the finite bandwidth of the readout pulse, part of its spectrum may overlap with neighboring readout resonators, a concern particularly for the middle feedline coupled to nine resonator pairs [102]. This issue becomes problematic when readout targeting errors reduce the relative detuning between adjacent resonator modes. In the worst-case scenarios, such overlaps can lead to the complete destruction of the target qubit state when measur-ing the overlapped qubit [Figure 4.4]. To prevent this, post-fabrication techniques such as wire-bonding and Shoelaces are utilized [55], as detailed in Section 4.1.2.

Preserving the state of a computational qubit during the measurement of a stabilizer qubit is critical in QEC codes. For this purpose, the dephasing rate, , is evaluated using the experimental setup shown in [Figure 4.4. c]. The experiment is conducted in four variants: without ancilla measurement, with ancilla measurement, initializing the ancilla qubit in the $0$ state, and initializing it in the $1$ state. These scenarios produce four distinct curves, which are then fitted to extract phase and amplitude parameters. The dephasing rate for each initial state of the "measure" qubit is calculated as:

$$= \log \frac{}{} \qquad \frac{1}{} \qquad (4.5)$$

The average dephasing rate, , of data qubits due to the measurement of ancilla qubits, prepared in $0$ and $1$ , is presented in Figure 4.25. c and d. A Gaussian filtered readout pulse effectively reduces compared to the standard square pulse. Consequently, this pulse shaping is adopted for the ancilla measurement pulses that are repeatedly measured in a QEC code (see Chapter 6).

Another consequence of ancilla measurements in a stabilizer code is the introduction of coherent phase errors on data qubits due to residual- interactions between idling data qubits and actively measured qubits [147]. These errors can be mitigated using various dynamical decoupling techniques [25, 178]. To counteract these coherent errors, single-echo pulses are applied to the data qubits midway through the ancilla measurement. This pulse sequence is employed on data qubits in parallel to ancilla measurements during repetitive QEC cycles, as detailed in Figure 6.1.

### 4.2.5 Summary of GBT calibration and achieve performance

This section summarizes the results and performance metrics achieved through GBT calibration procedures of various operations across the 17-qubit device (Uran). The calibration process was targeted to optimize the performance while maintaining a manageable calibration time. The table below provides an overview of the calibration time for each operation and the performance achieved in terms of best, worst-case, and average scenarios.

The values provided in the table are illustrative and are based on experimental results obtained from the calibration and benchmarking processes detailed in this chapter. The efficiency of the GBT approach is evident in the improvement of system performance, leading to reduced logical error rates over time, as detailed in the Surface-13 experiment [Figure 6.17. d].

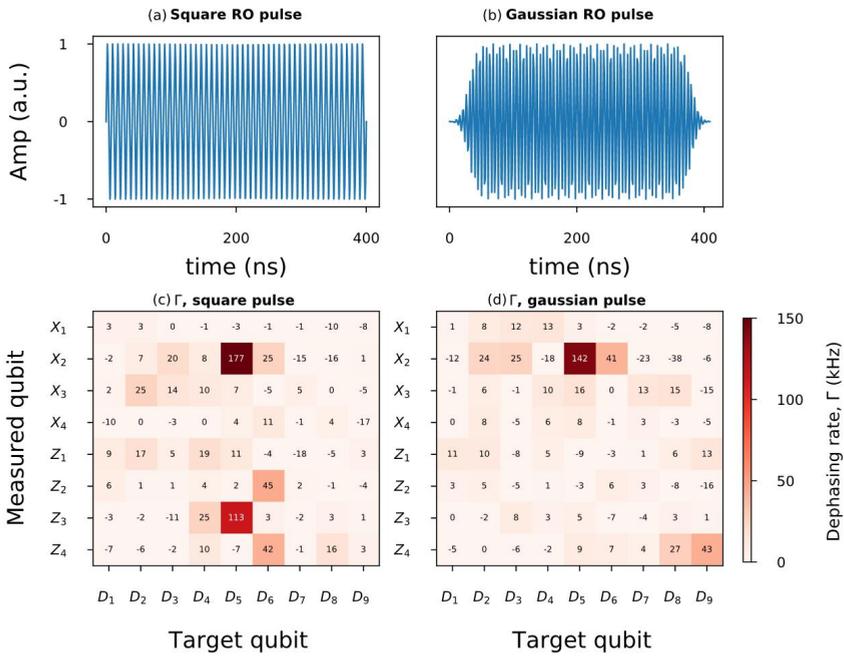

Figure 4.25: **Mitigation of measurement-induced dephasing with Gaussian filtered Pulse** effectively lowers dephasing rates on data qubits during ancilla measurements. (a) Standard readout square pulse, and (b) filtered Gaussian pulse with a sigma of $5$ ns. The dephasing rate, $\Gamma$, of data qubits due to ancilla measurements over $420$ ns for the square (c), and Gaussian pulses (d), respectively.

| Operation | GBT calibration time | Achieved performance (%) | | |
|---|---|---|---|---|
| | | **Best** | **Worst-case** | **Average** |
| Single-qubit gates (20 ns) | 5 mins | 99.9 | 99.7 | 99.9 |
| Two-qubit gates (60 ns) | 38 mins | 99.7 | 96.2 | 98.4 |
| Single-qubit readout (420 ns) | 300 mins | 99.5 | 96.6 | 98.8 |

Table 4.1: GBT Calibration Time and Achieved Performance Metrics in device Uran.

# ALL-MICROWAVE LEAKAGE REDUCTION UNITS FOR QUANTUM ERROR CORRECTION WITH SUPERCONDUCTING TRANSMON QUBITS

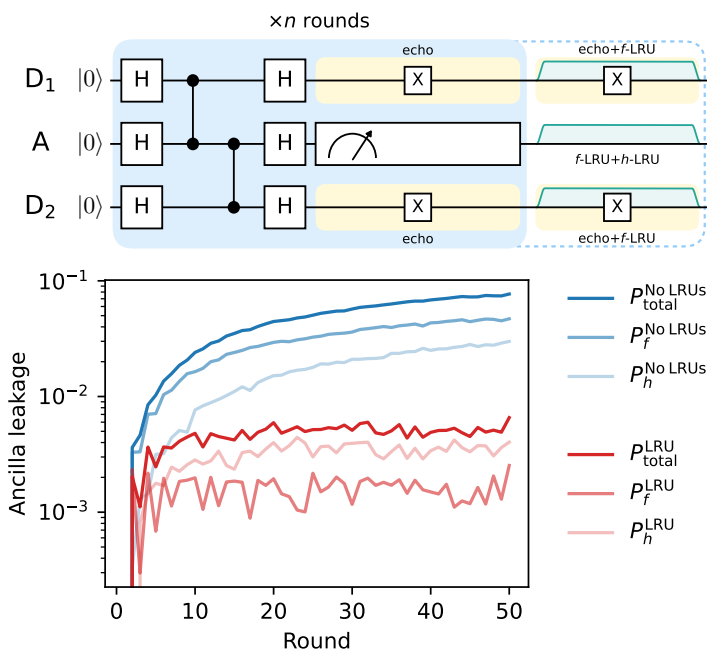

Minimizing leakage from computational states is a challenge when using many-level systems like superconducting quantum circuits as qubits. We realize and extend the quantum-hardware-efficient, all-microwave leakage reduction unit (LRU) for transmons in a circuit QED architecture proposed by Battistel et al. This LRU effectively reduces leakage in the second- and third-excited transmon states with up to $99\%$ efficacy in $220\,\text{ns}$, with minimum impact on the qubit subspace. As a first application in the context of quantum error correction, we show how multiple simultaneous LRUs can reduce the error detection rate and suppress leakage buildup within $1\%$ in data and ancilla qubits over 50 cycles of a weight-2 stabilizer measurement.

## 5.1  Historical context

Superconducting qubits, such as the transmon [60], are many-level systems in which a qubit is represented by the two lowest-energy states     and   . However, leakage to non-computational states is a risk for all quantum operations, including single-qubit gates [64], two-qubit gates [67–69] and measurement [61, 135]. While the typical probability of leakage per operation may pale in comparison to conventional qubit errors induced by control errors and decoherence [25, 69], unmitigated leakage can build up with increasing circuit depth. A prominent example is multi-round quantum error correction (QEC) with stabilizer codes such as the surface code [71]. In the absence of leakage, such codes successfully discretize all qubit errors into Pauli errors through the measurement of stabilizer operators [111, 179], and these Pauli errors can be detected and corrected (or kept track of) using a decoder. However, leakage errors fall outside the qubit subspace and are not immediately correctable [78–80]. The signature of leakage on the stabilizer syndrome is often not straightforward, hampering the ability to detect and correct it [72, 77]. Additionally, the build-up of leakage over QEC rounds accelerates the destruction of the logical information [20, 25]. Therefore, despite having low probability per operation, methods to reduce leakage must be employed when performing experimental QEC with multi-level systems.

Physical implementations of QEC codes [26, 141, 180–183] use qubits for two distinct functions: Data qubits store the logical information and, together, comprise the encoded logical qubits. Ancilla qubits perform indirect measurement of the stabilizer operators. Handling leakage in ancilla qubits is relatively straightforward as they are measured in every QEC cycle. This allows for the use of reset protocols [19, 20] without the loss of logical information. Leakage events can also be directly detected using three- or higher-level readout [141] and reset using feedback [73, 130]. In contrast, handling data-qubit leakage requires a subtle approach as it cannot be reset nor directly measured without loss of information or added circuit complexity [81, 82, 184]. A promising solution is to interleave QEC cycles with operations that induce seepage without disturbing the qubit subspace, known as leakage reduction units (LRUs) [78, 80–82, 128, 146, 185–187]. An ideal LRU returns leakage back to the qubit subspace, converting it into Pauli errors which can be detected and corrected, while leaving qubit states undisturbed. By converting leakage into conventional errors, LRUs enable a moderately high physical noise threshold, below which the logical error rate decreases exponentially with the code distance [80, 81]. A more powerful operation called 'heralded leakage reduction' would both reduce and herald leakage, leading to a so-called erasure error [188, 189]. Unlike Pauli errors, the exact location of erasures is known, making them easier to correct and leading to higher error thresholds [190–193].

In this chapter, we present the realization and extension of the LRU scheme proposed in Ref. [128]. This is a highly practical scheme requiring only microwave pulses and the quantum hardware typically found in contemporary circuit QED quantum processors: a microwave drive and a readout resonator dispersively coupled to the target transmon (in our case, a readout resonator with dedicated Purcell filter). We show its straightforward calibration and the effective removal of the population in the first two leakage states of the transmon (   and   ) with up to   $99\%$ efficacy in $220\,\mathrm{ns}$. Process tomography reveals that the LRU backaction on the qubit subspace is only an AC-Stark shift, which can be easily corrected using a   -axis rotation. As a first application in a QEC setting, we interleave repeated measurements of a weight-2 parity check [72, 73] with simultaneous LRUs on data and ancilla qubits, showing the suppression of leakage and error detection rate buildup.

## 5.2   Experimental results

Our leakage reduction scheme [Fig. 5.1(a)] consists of a transmon with states   ,   and   , driven by an external drive   , coupled to a resonant pair of Purcell and readout resonators [102] with effective dressed states $00$ and $1^{\pm}$ . The LRU scheme transfers leakage population in the second-excited state of the transmon,   , to the ground state,   , via the resonators using a microwave drive. It does so using an effective coupling   mediated by the transmon-resonator coupling,   , and the drive   , which couples states $00$ and $1^{\pm}$ . Driving at the frequency of this transition,

$$
\pm \quad 2 \quad +
\tag{5.1}
$$

transfers population from $00$ to $1^{\pm}$ , which in turn quickly decays to $00$ provided the Rabi rate is small compared to   . Here,   and   are the transmon qubit transition frequency and anharmonicity, respectively, while   is the resonator mode frequency. In this regime, the drive effectively pumps any leakage in   to the computational state   . We perform spectroscopy of this transition by initializing the transmon in   and sweeping the drive around the expected frequency. The results [Fig. 5.1(c)] show two dips in the   -state population corresponding to transitions with the hybrized modes of the matched readout-Purcell resonator pair. The dips are broadened by   $4 \quad 8\,\mathrm{MHz}$, making them easy to find.

To make use of this scheme for a LRU, we calibrate a pulse that can be used as a circuit-level operation. We use the pulse envelope proposed in Ref. [128]:

$$
(\ ) = \begin{cases} \sin & \dfrac{}{\mathrm{r}} & \text{for } 0 \qquad , \\ & & \text{for } \qquad , \\ \sin & \dfrac{\mathrm{p}-}{\mathrm{r}} & \text{for } \qquad , \end{cases}
\tag{5.2}
$$

where   is the amplitude,   is the rise and fall time, and   is the total duration. We conservatively choose   $= 30\,\mathrm{ns}$ to avoid unwanted transitions in the transmon. To measure the fraction of leakage removed,   , we apply the pulse on the transmon prepared in   and measure it [Fig. 5.2(a)], correcting for readout error using the measured 3-level assignment fidelity matrix [Fig. 5.2(c)]. To optimize the pulse parameters, we first measure   while sweeping the pulse frequency and   [Fig. 5.2(d)]. A second sweep of   and   [Fig. 5.2(e)] shows that $R > 99\%$ can be achieved by increasing either parameters. Simulation [128] suggests that $R \approx 80\%$ is already sufficient to suppress most of the impact of current leakage rates, which is comfortably achieved over a large region of parameter space. For QEC, a fast operation is desirable to minimize the impact of decoherence. However, one must not excessively drive the transmon, which can cause extra decoherence (see Fig. 6 in Ref. [128]). Considering the factors above, we opt for $t_\mathrm{p} = 220$ ns and adjust $A$ such that $R \gtrsim 80\%$. Additionally, we benchmark the repeated action of the LRU and verify that its performance is maintained over repeated applications, thus restricting leakage events to approximately a single cycle (Fig. 5.6).

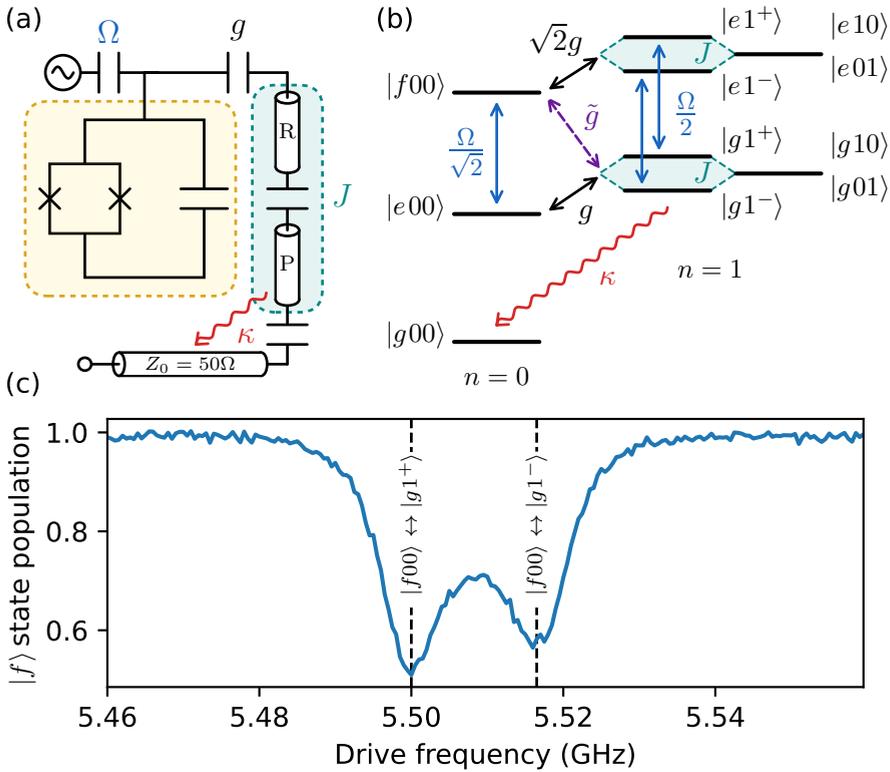

Figure 5.1: **Leakage reduction unit scheme.** (a) Schematic for the driven transmon-resonator system. A transmon ($T$, yellow) with three lowest-energy levels $g$, $e$, and $f$ is coupled to a readout resonator ($R$) with strength $g$. The latter is coupled to a frequency-matched Purcell resonator ($P$) with strength $J$. The Purcell resonator also couples to a $50\,\Omega$ feedline through which its excitations quickly decay at rate $\kappa$. The transmon is driven with a pulse of strength $\Omega$ applied to its microwave drive line. (b) Energy level-spectrum of the system. Levels are denoted as $T\,R\,P$, with numbers indicating photons in $R$ and $P$. As the two resonators are frequency matched, the right-most degenerate states split by $2J$, and $\Omega$ is shared equally among the two hybridized resonator modes $1^-$ and $1^+$. An effective coupling $\tilde{g}$ arises between $f00$ and the two hybridized states $g1^{\pm}$ via $e00$ and $e1^{\pm}$. (c) Spectroscopy of the $f00 \leftrightarrow g1^{\pm}$ transition. Measured transmon population in $f$ versus drive frequency, showing dips corresponding to the two transitions assisted by each of the hybridized resonator modes.

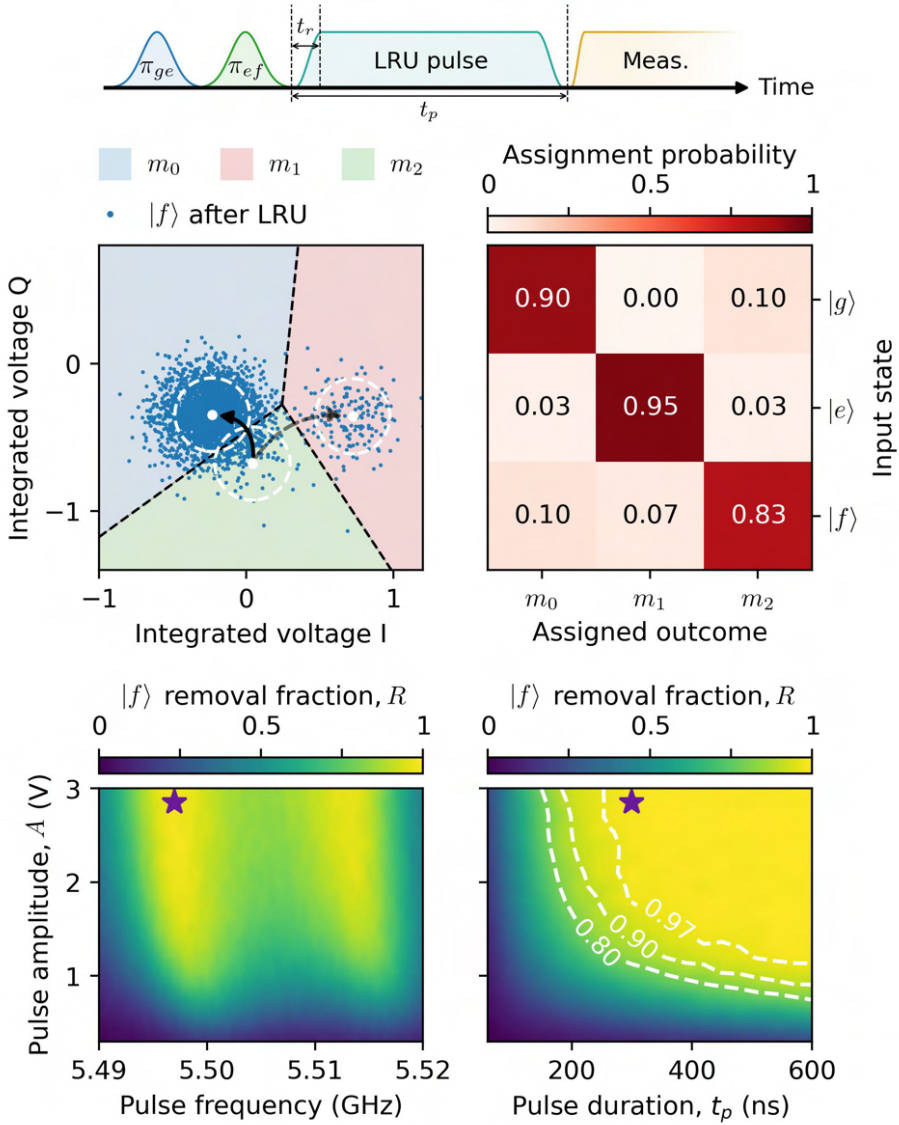

Figure 5.2: **Calibration of the leakage reduction unit pulse.** (a) Pulse sequence used for LRU calibration. (b) Single-shot readout data obtained from the experiment. The blue, red and green areas denote , , and assignment regions, respectively. The mean (white dots) and $3$ standard deviation (white dashed circles) shown are obtained from Gaussian fits to the three input-state distributions. The blue data shows the first $3 \ 10$ (from a total of $2$ ) shots of the experiment described in (a), indicating $99 \ (3)\%$ -state removal fraction. (c) Measured assignment fidelity matrix used for readout correction. (d-e) Extracted -state removal fraction versus pulse parameters. Added contours (white dashed curves) indicate $80$, $90$ and $97\%$ removal fraction. The purple star indicates the pulse parameters used in (b).

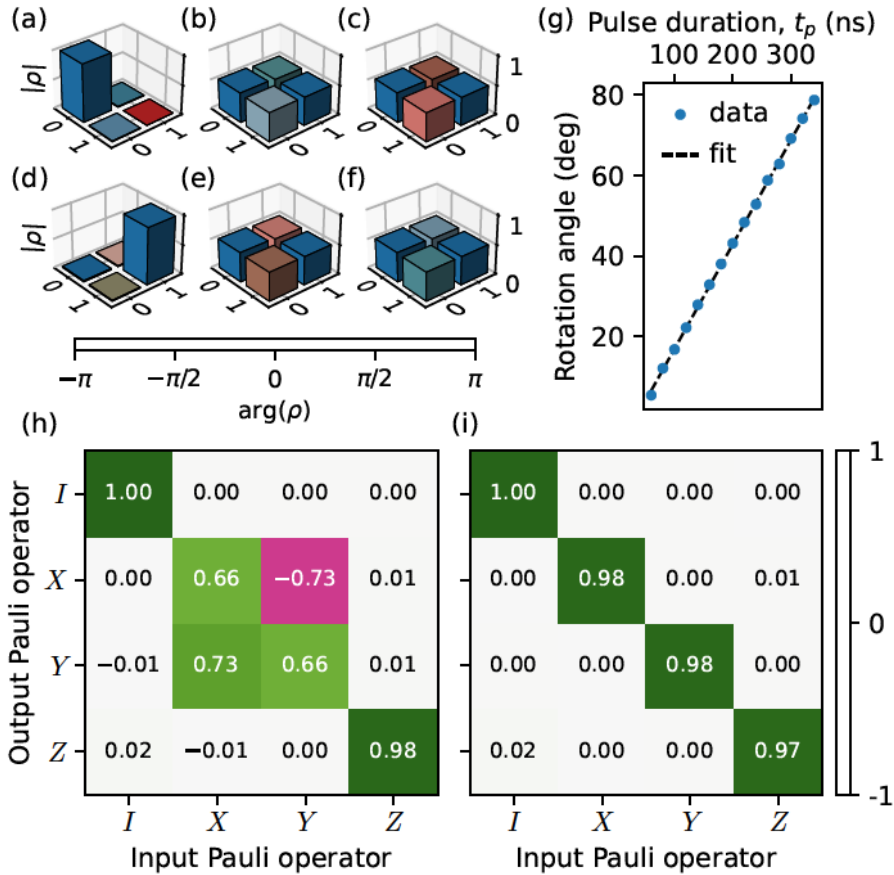

Figure 5.3: **Process tomography of the leakage reduction unit.** (a-f) Measured density matrices after the LRU gate for input states $|0\rangle$, $|+\rangle$, $|+i\rangle$, $|1\rangle$, $|-\rangle$, and $|-i\rangle$, respectively. (g) $Z$-rotation angle induced on the qubit versus the LRU pulse duration. The linear best fit (black dashed line) indicates an AC-Stark shift of $71(9)$ kHz. (h-i) Pauli transfer matrix of the LRU with (i) and without (h) virtual phase correction ($t_\mathrm{p} = 220$ ns and $R = 84.(7)\%$).

With the LRU calibrated, we next benchmark its impact on the qubit subspace using quantum process tomography. The results (Fig. 5.3) show that the qubit incurs a $Z$-axis rotation. We find that the rotation angle increases linearly with $P_e$ [Fig. 5.3(g)], consistent with a $71(9)\,\mathrm{kHz}$ AC-Stark shift induced by the LRU drive. This phase error in the qubit subspace can be avoided using decoupling pulses or corrected with a virtual $Z$ gate. Figures 5.3(h) and 5.3(i) show the Pauli transfer matrix (PTM) for the operation before and after applying a virtual $Z$ correction, respectively. From the measured PTM [Fig. 5.3(i)] and enforcing physicality constraints [120], we obtain an average gate fidelity $F = 98\,(9)\%$. Compared to the measured $99\,(2)\%$ fidelity of idling during the same time ($t = 220\,\mathrm{ns}$), there is evidently no significant error increase.

Finally, we implement the LRU in a QEC scenario by performing repeated stabilizer measurements of a weight-2 $Z$-type parity check [72, 73] using three transmons (Fig. 5.4). We use the transmon in Figs. 1-3, $D_1$, plus an additional transmon ($D_2$) as data-qubits together with an ancilla, $A$. LRUs for $D_2$ and $A$ are tuned using the same procedure as above. A detailed study of the performance of this parity check is shown in the Supplementary Information (Fig. 5.7). Given their frequency configuration [98], $D_1$ and $A$ are most vulnerable to leakage during two-qubit controlled-$Z$ (CZ) gates, as shown by the avoided crossings in Fig. 5.4(a). Additional leakage can occur during other operations: in particular, we observe that leakage into states above $e$ can occur in $A$ due to measurement-induced transitions [135] (see Fig. 5.11). Therefore, a LRU acting on $e$ alone is insufficient for $A$. To address this, we develop an additional LRU for $f$ ($f$-LRU), the third-excited state of $A$ (see Supplementary Information Fig. 5.10). The $f$-LRU can be employed simultaneously with the $e$-LRU without additional cost in time or impact on the $e$ removal fraction, $P_r$. Thus, we simultaneously employ $e$-LRUs for all three qubits and an $f$-LRU for $A$ [Fig. 5.4(a)]. To evaluate the impact of the LRUs, we measure the error detection probability (probability of a flip occurring in the measured stabilizer parity) and leakage population of the three transmons over multiple rounds of stabilizer measurement. Without leakage reduction, the error detection probability rises $\approx 8\%$ in 50 rounds. We attribute this feature to leakage build-up [20, 26, 146]. With the LRUs, the rise stabilizes faster (in $\approx 10$ rounds) to a lower value and is limited to $2\%$, despite the longer cycle duration ($500$ versus $720\,\mathrm{ns}$ without and with the LRU, respectively). Leakage is overall higher without LRUs, in particular for $D_2$ and $A$ [Fig. 5.4(c)], which show a steady-state population of $\approx 10\%$. Using leakage reduction, we lower the leakage steady-state population by up to one order of magnitude to $\approx 1\%$ for all transmons. Additionally, we find that removing leakage on other transmons leads to lower overall leakage, suggesting that leakage is transferred between transmons [77, 146]. This is particularly noticeable in $A$ [Fig. 5.4(c)], where the steady-state leakage is always reduced by adding LRUs on $D_1$ and $D_2$.

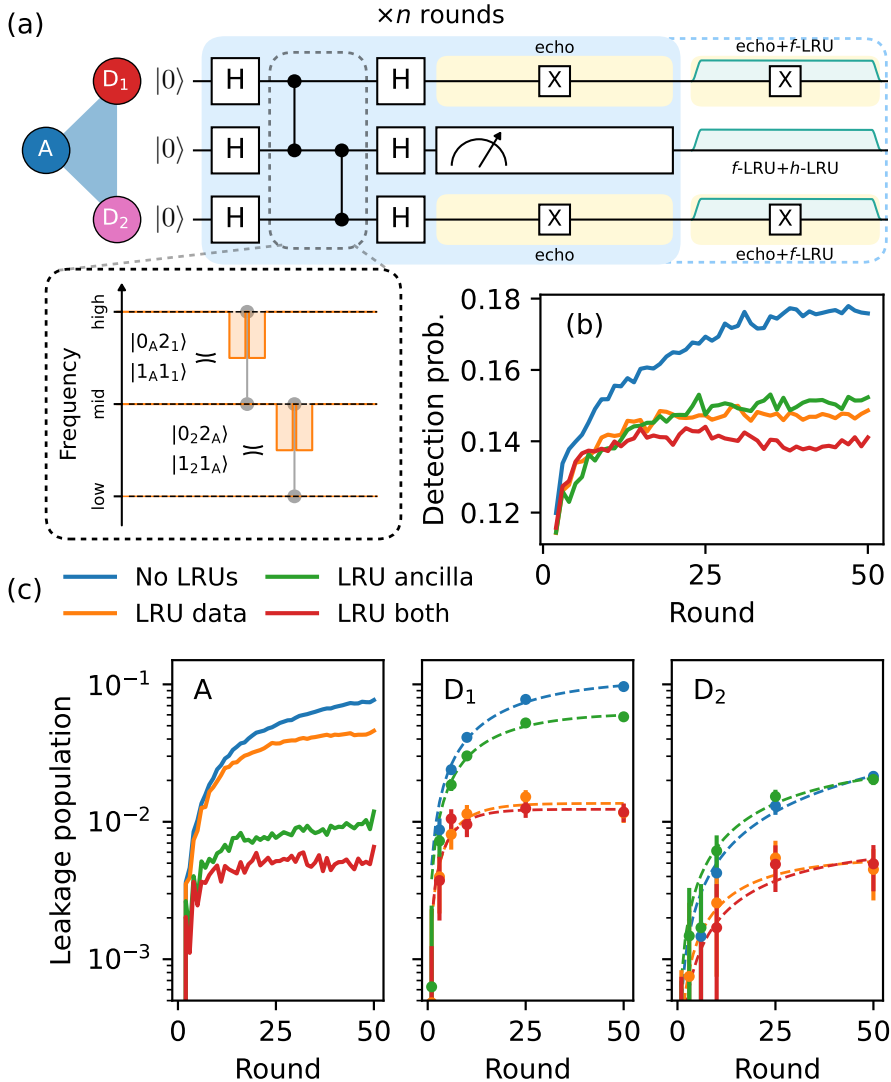

Figure 5.4: **Repeated stabilizer measurement with leakage reduction.** (a) Quantum circuit using ancilla $A$ to measure the -type parity of data qubits $D$ and $D$ . The dashed box shows the frequency arrangement for two-qubit CZ gates. A CZ gate is performed by fluxing the higher-frequency transmon down in frequency to the nearest avoided crossing (orange shaded trajectory). The duration of single-qubit gates, CZ gates, and measurement are $20$, $60$ and $340 \text{ ns}$, respectively, totalling $500 \text{ ns}$ for the parity check (light-blue region). Performing the LRUs extends the circuit by $220 \text{ ns}$ (blue-dashed region). Echo pulses on data qubits mitigate phase errors caused by residual crosstalk and AC-stark shift during the measurement and LRUs (light yellow slots). (b-c) Measured error-detection probability (b) and leakage (c) versus the number of parity-check rounds in four settings. The No LRUs setting (blue) does not apply any LRUs. LRU data (orange) and LRU ancilla (green) settings apply LRUs exclusively on the data qubits and the ancilla, respectively. The LRU both (red) setting applies LRUs on all qubits.

## 5.3 Conclusions

We have demonstrated and extended the all-microwave LRU for superconducting qubits in circuit QED proposed in Ref. [128]. We have shown how these LRUs can be calibrated using a straightforward procedure to deplete leakage in the second- and third-excited states of the transmon. This scheme could potentially work for even higher transmon states using additional drives. We have verified that the LRU operation has minimal impact in the qubit subspace, provided one can correct for the static AC-Stark shift induced by the drive(s).

This scheme does not reset the qubit state and is therefore compatible with both data and ancilla qubits in the QEC context. We have showcased the benefit of the LRU in a building-block QEC experiment where LRUs decrease the steady-state leakage population of data and ancilla qubits by up to one order of magnitude (to $1\%$), and thereby reduce the error detection probability of the stabilizer and reaching a faster steady state. We find that the remaining ancilla leakage is dominated by higher states above (Fig. 5.11) likely caused by the readout [61, 135]. Compared to other LRU approaches [20, 146], we believe this scheme is especially practical as it is all-microwave and very quantum-hardware efficient, requiring only the microwave drive line and dispersively coupled resonator that are already commonly found in the majority of circuit QED quantum processors [26, 141, 181]. Extending this leakage reduction method to larger QEC experiments can be done without further penalty in time as all LRUs can be simultaneously applied. However, we note that when extending the LRU to many qubits, microwave crosstalk should be taken into account in order to avoid driving unwanted transitions. This can be easily avoided by choosing an appropriate resonator-qubit detuning.

## 5.4 Data availability

The data supporting the plots and claims within this paper are available online at http://github.com/DiCarloLab-Delft/Leakage_Reduction_Unit_Data. Further data can be provided upon reasonable request.

## 5.5 Supplemental material

This supplement provides additional information in support of the statements and claims in the main text.

### 5.5.1 Device

The device used (Fig. 5.5) has 17 flux-tunable transmons arranged in a square lattice with nearest-neighbor connectivity (as required for a distance-3 surface code). Transmons are arranged in three frequency groups as prescribed in the pipelined architecture of Ref. [98]. Each transmon has a dedicated microwave drive line used for single-qubit gates and leakage reduction, and a flux line used for two-qubit gates. Nearest-neighbor transmons have fixed coupling mediated by a dispersively coupled bus resonator. Each transmon has a dedicated pair of frequency-matched readout and Purcell resonators coupled to one of three feedlines, used for fast multiplexed readout in the architecture of Ref. [102]. Single-qubit gates are realized using standard DRAG pulses [195]. Two-qubit controlled-  gates are implemented using sudden net-zero flux pulses [69]. Characteristics and performance metrics of the three transmons used in the experiment are shown in Tab. 5.1.

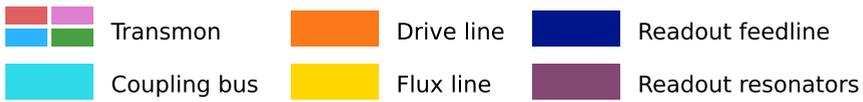

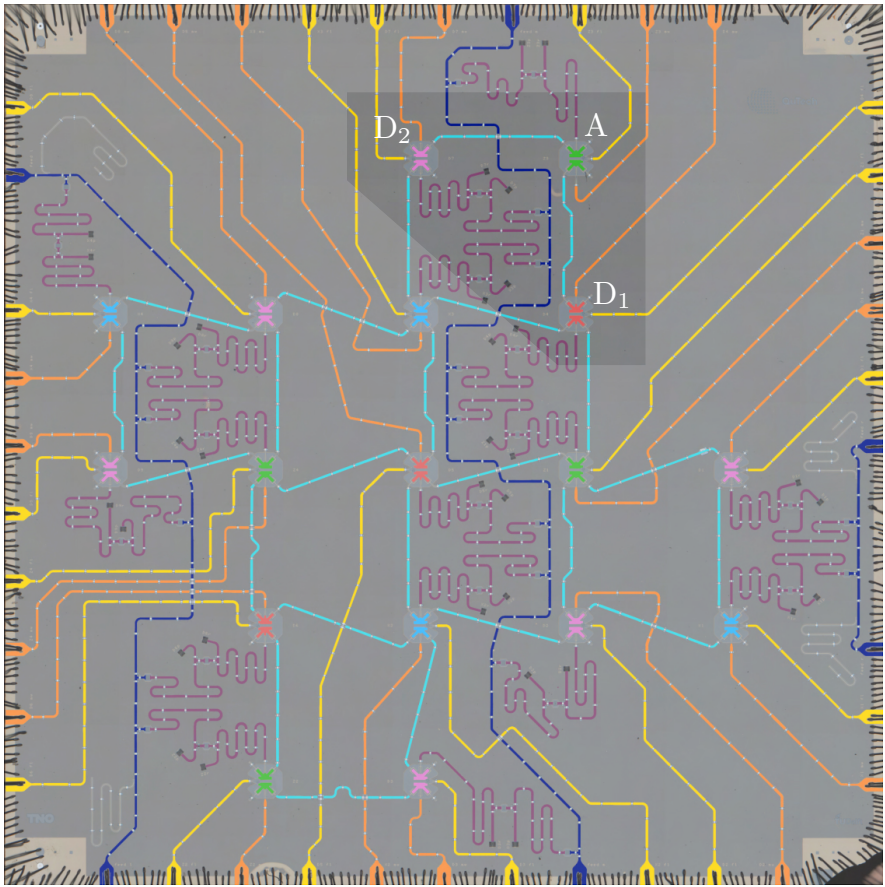

Figure 5.5: **Circuit QED device.** Optical image of the 17-transmon quantum processor, with added falsecolor to highlight different circuit elements. The shaded area indicates the three transmons used in this experiment.

| Transmon | D | A | D |
|---|---|---|---|
| Frequency at sweetspot, $\omega/2\pi$ (GHz) | 6.802 | 6.033 | 4.788 |
| Anharmonicity, $\alpha/2\pi$ (MHz) | -295 | -310 | -321 |
| Resonator frequency, $\omega_r/2\pi$ (GHz) | 7.786 | 7.600 | 7.105 |
| Purcell res. linewidth, $\kappa/2\pi$ (MHz) | 15.(5) | 22.(5) | 12.(6) |
| -LRU drive frequency (GHz) | 5.498 | 4.135 | 2.152 |
| -LRU drive frequency (GHz) | - | 3.496 | - |
| $T_1$ ($\mu$s) | 17 | 26 | 37 |
| $T_2$ ($\mu$s) | 19 | 22 | 27 |
| Single-qubit gate error (%) | 0.1(0) | 0.0(7) | 0.0(5) |
| Two-qubit gate error (%) | | 1.(1) | 1.(9) |
| Two-qubit gate leakage (%) | | 0.3(7) | 0.1(1) |
| -LRU removal fraction, (%) | 84.(7) | 99.(2) | 80.(3) |
| -LRU removal fraction, (%) | - | 96.(1) | - |

| Operation | Duration (ns) |
|---|---|
| Single-qubit gate | 20 |
| Two-qubit gate | 60 |
| Measurement | 340 |
| LRU | 220 |

Table 5.1: **Device metrics.** Frequencies and coherence times are measured using standard spectroscopy and time-domain measurements [58]. Gate errors are evaluated using randomized benchmarking protocols [96, 97, 194].

### 5.5.2   Repeated LRU application

For QEC we require that the LRU performance remains constant over repeated applications. To assess this, we perform repeated rounds of the experiment shown in Fig. 5.6 while idling or using the LRU. In each round apply an $X$-$\pi$ rotation with rotation angle $\theta$ to induce a leakage rate

$$L = \frac{\sin(\theta/2)}{2} \tag{5.3}$$

and choose $\theta$ such that $L = 2\%$. For the purpose of this experiment, we lower the readout amplitude in order to suppress leakage to higher states during measurement [135], The results (Fig. 5.6) show that while idling, leakage in $|f\rangle$ builds up to a steady-state population of about $16\%$. Using the LRU, it remains constant at $L = $ throughout all rounds. This behavior shows that LRU performance is maintained throughout repeated applications and suggests that leakage events are restricted to a single round.

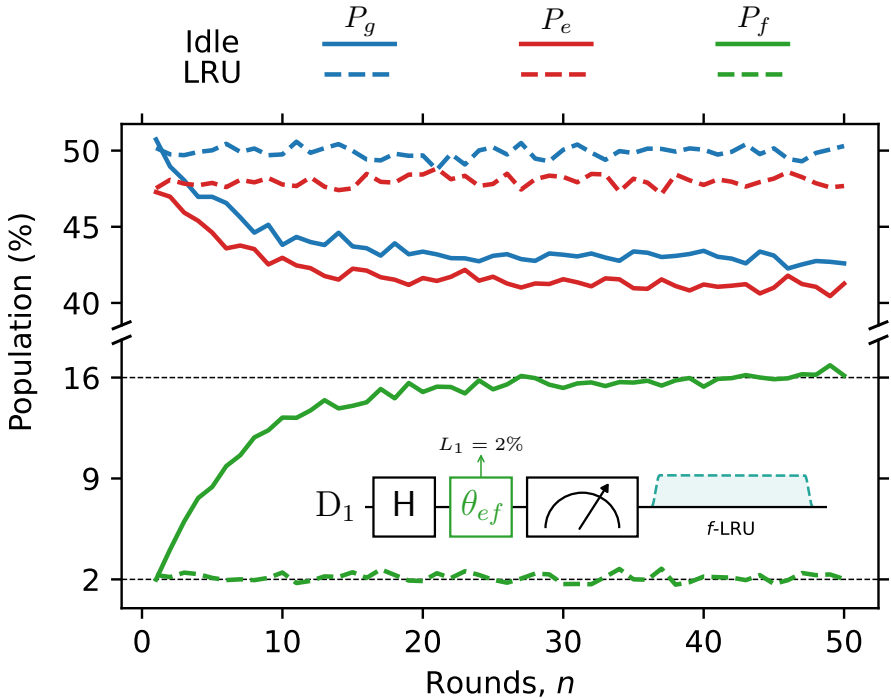

Figure 5.6: **Repeated LRUs on** $D_1$. Computational (top) and leakage (bottom) state population over repeated cycles of the circuit as shown. Transmon $D_1$ is repeatedly put in superposition, controllably leaked with rate and measured. Measurement is followed by either idling (solid) or the LRU (dashed). The horizontal lines denote the steady-state leakage with and without LRUs.

### 5.5.3 Benchmarking the weight-2 parity check

We benchmark the performance of the weight-2 parity check using three experiments assessing different error types. First, we assess the ability to accurately assign the parity of the data-qubit register by measuring the ancilla outcome for all data-qubit input computational states. The results [Fig. 5.7(a)] show an average parity assignment fidelity of 95.6%. Next, we look at errors occurring on the data qubits when projecting them onto a Bell state using a -type parity check [Fig. 5.7(b)]. From data-qubit state tomography conditioned on ancilla outcome, we obtain an average Bell-state fidelity of $97.7\%$ ($96.9\%$ for $= +1$ and $98.5\%$ for $= 1$). For each reported density matrix, we apply readout corrections and enforce physicality constraints via maximum likelihood estimation [120]. Finally, we look at the back-action of two back-to-back parity checks [Fig. 5.7(c)]. Here, we measure the correlation of the two parity outcomes. Ideally, the first parity outcome should be random while the second should be the same as the first. Since our ancilla is not reset after measurement, the proba-bility of both parities being correlated is $P(m_2 = +1) = 90.0\%$ [bar plot in Fig. 5.7(c)]. We can also reconstruct the data qubit state after the experiment. Here, we find that the average Bell-state fidelity drops to $91.0\%$ ($90.6\%$ for $m_1 = +1$ and $91.5\%$ for $m_1 = -1$). This drop in fidelity is likely due to decoherence from idling during the first ancilla measurement.

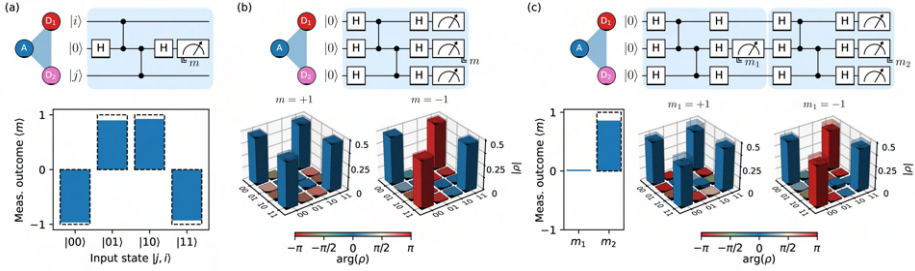

Figure 5.7: **Benchmarking of the weight-2 parity check.** (a) Quantum circuit of the weight-2 $ZZ$-type parity check and bar plot of the average measured ancilla outcome for the different input computational states of the data-qubit register. Dashed bars show ideal average outcome: $m = +1$ ($-1$) for even (odd) data-qubit input parity. (b) Generation of Bell states via stabilizer measurement (top) and corresponding data-qubit state tomography (bottom) conditioned on the ancilla outcome. The obtained fidelity to the ideal Bell states (shaded wireframe) is $96.9\%$ and $98.5\%$ for $m = +1$ and $m = -1$, respectively. (c) Repeated stabilizer experiment. (Bottom left) Average measured ancilla outcome $m_i$ for each round of stabilizer measurement. Ideally, the first outcome should be random and the second always $+1$. The measured probability is $P(m_2 = +1) = 90.0\%$. (Bottom right) Reconstructed data-qubit states conditioned on the first ancilla outcome. The obtained Bell-state fidelities are $90.6\%$ and $91.5\%$ for $m_1 = +1$ and $m_1 = -1$, respectively.

### 5.5.4 Measurement-induced transitions

Previous studies have observed measurement-induced state transitions that can lead to leakage [61, 135]. To evaluate the backaction of ancilla measurement, we model the measurement as a rank 3 tensor $M$ which takes an input state $\rho$, declares an outcome $m$ and outputs a state $\sigma$ with normalization condition,

$$\sum_m \text{Tr}(M_m \rho M_m^\dagger) = 1 \tag{5.4}$$

To find $M$, we perform the experiment in Fig. 5.8(a). For each input state $\rho$, the probability distribution of the measured results $P(m, n)$ follows

$$P(m = \mu, n = \nu) = \tag{5.5}$$

(a)

(b)

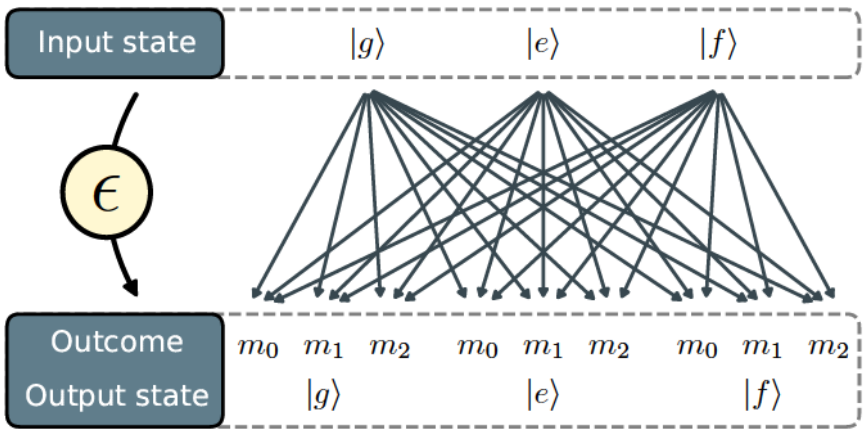

(c)                                        (d)

This system of 27 second-order equations is used to estimate the 27 elements of        through a standard optimization procedure. In this description, the assignment fidelity matrix [Fig. 5.8(b)] is given by

$$=\hspace{10em}(5.6)$$

Furthermore, this model allows us to assess the probability of transitions occurring during the measurement. This is given by the QNDness matrix

$$=\hspace{10em}(5.7)$$

The results [Fig. 5.8(b)] show an average QNDness of $95\ 4\%$ across all states. The average leakage rate $((\hspace{1em}+\hspace{1em})\ 2)$ is $0\ 06\%$, predominantly occurring for input state        .

### 5.5.5   Readout of transmon states

In order to investigate leakage to higher states in the ancilla, we need to discriminate between the first two leakage states,        and        . To do this without compromising the performance of the parity check, we simultaneously require high readout fidelity for the qubit states        and        . We achieve this for the ancilla for the states        through        using a single readout pulse. Figure 5.9(a) shows the integrated readout signal for each of the states along with the decision boundaries used to classify the states. Any leakage to even higher states will likely be assigned to        since the resonator response at the readout frequency is mostly flat for        . The average assignment error for the four states is $7\ (2)\%$ [Fig. 5.9(b)] while the average qubit readout error is $1\ (3)\%$ [Fig. 5.9(c)]. Here, we assume that state preparation errors are small compared to assignment errors.

### 5.5.6   Leakage reduction for higher states

Although the most common leakage mechanisms usually populate the second-excited state of the transmon,        , some operations such as measurement can leak into higher-excited states [135]. We observe the build-up of population in these higher states in the repeated parity-check experiment (Fig. 5.4). Figure 5.11 shows the fraction of total leakage to these higher states for the ancilla. Therefore, leakage reduction for higher states is necessary for ancillas. Similar to the leakage reduction mechanism that drives        [with effective coupling        in Fig. 5.10(a)], one can drive        (with effective coupling        in Fig. 5.10(b)]. This transition can be induced much like the former, with an extra drive at frequency

$$\pm\hspace{2em}2\hspace{2em}+3\hspace{6em}(5.8)$$

$2$   below the   -LRU transition. We then have two LRU mechanisms,   -LRU and   -LRU, increasing seepage from        and        , respectively. We drive both of these transitions simultaneously using two independent drives. Following the same calibration procedure shown

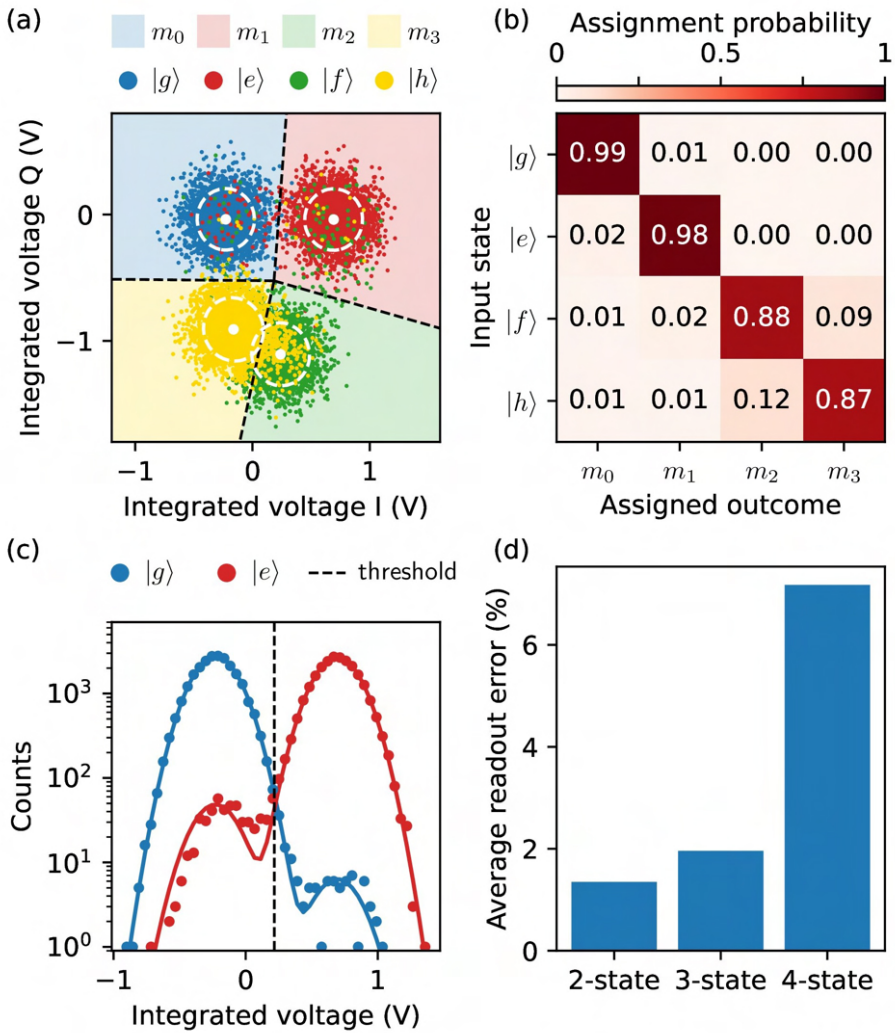

Figure 5.9: **Four state readout.** (a) Single-shot readout data for the four lowest-energy trans-mon states , , and of A. Data are plotted for the first $3 \times 10$ from a total of $2$ shots for each input state. The dashed lines show decision boundaries obtained from fitting a linear discrimination classifier to the data. The mean (white dot) and $3$ standard deviation (white dashed circles) shown are obtained from Gaussian fits to each input state distribution. (b) Assignment probability matrix obtained from classification of each state into a quaternary outcome. (c) Histogram of shots for qubit states taken along the projection maximizing the signal-to-noise ratio. (d) The average assignment errors for 2-, 3- and 4-state readout are $1\ (3), 1\ (9)$ and $7\ (2)\%$, respectively.

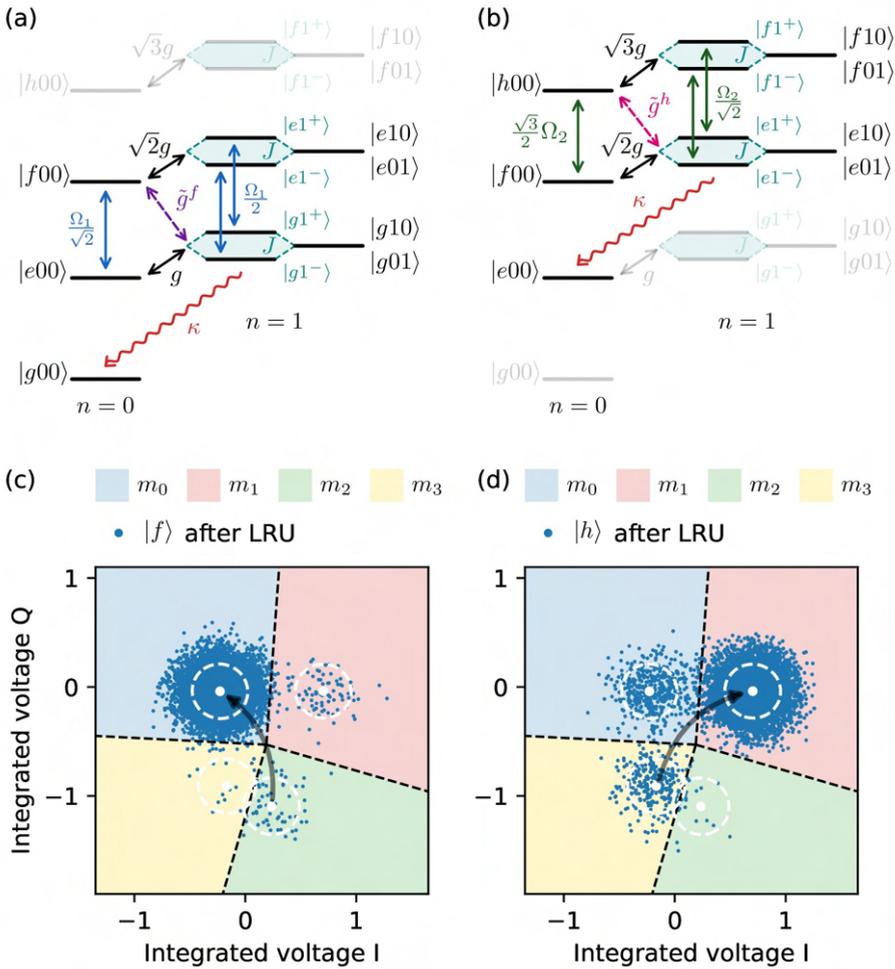

Figure 5.10: **Leakage reduction for     and    .** (a, b) Transmon-resonator system level structure showing the relevant couplings for the    -LRU (a) and    -LRU (b). Each effective coupling,      and      , is mediated by its respective drive       and       and transmon-resonator coupling  . (c, d) Readout data ($2$    shots) of leakage states        (c) and        (d) after applying both LRU pulses simultaneously. The white dots and dashed cicles show the mean and $3$ standard deviation obtained from fitting calibration data for each state.

**5**

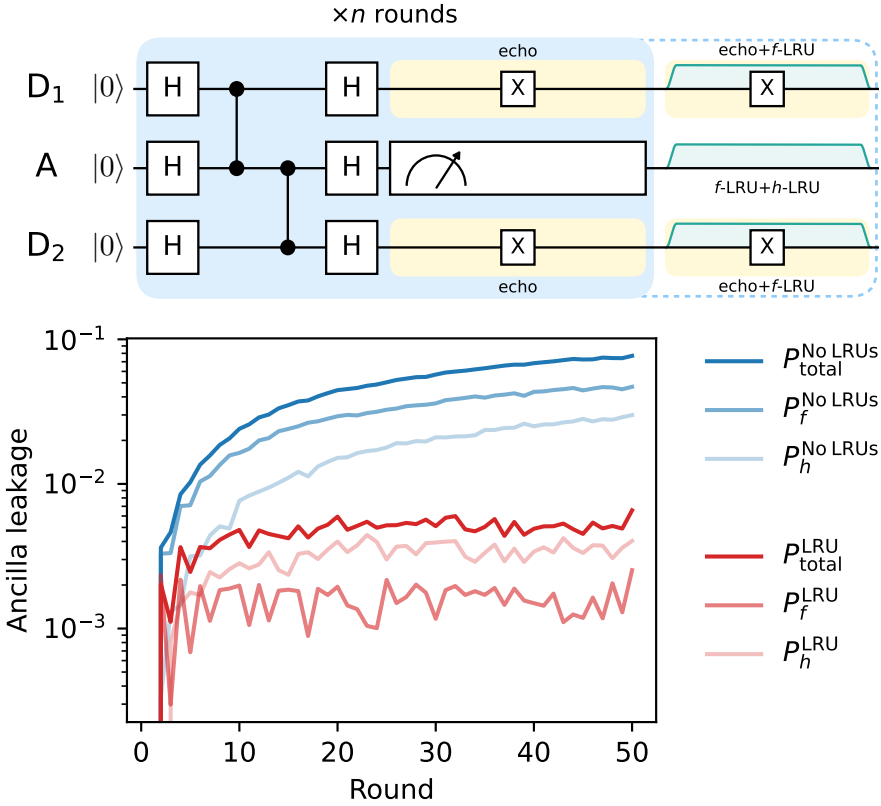

Figure 5.11: **Higher leakage states of the ancilla qubit.** Composition of ancilla leakage during the repeated stabilizer measurement of Figure 5.4.          =          +          denotes the total leakage population without (blue) and with (red) LRUs.

in Fig. 5.2 for the    -LRU, we tune up a pulse for the    -LRU. Figures 5.10(c) and 5.10(d) show readout data for states          and          after performing both LRUs simultaneously. The corresponding removal fraction for each state is          $= 99\,(2)\%$ and          $= 96\,(1)\%$ for    $= 220\,\mathrm{ns}$. Using this scheme, we can effectively reduce leakage in both states (Fig. 5.11). In particular, leakage in          is effectively kept under $0\,2\%$, while that in          sits below $0\,4\%$ (red curves in Fig. 5.11). The former shows a flat curve and therefore corresponds to the          of the cycle (similar to Fig. 5.6). The apparent remaining leakage in          could possibly be due to higher-excited states, which are naively assigned as          by the readout as they cannot be distinguished.

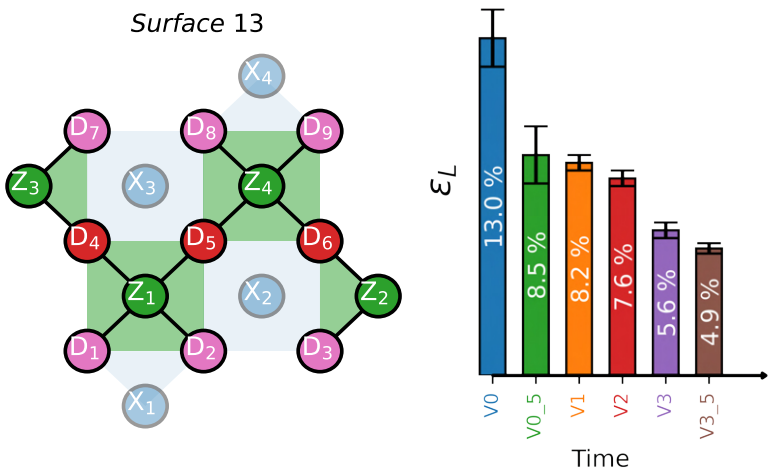

A quantum circuit is sub-optimally constructed if simply compiled from single- and two-qubit operations tuned in isolation. Here, we introduce highly parallel and orthogonal calibration strategies for optimal calibration of the QEC cycle for a distance-3 surface code. This enables the realization of a bit-flip distance-3 surface code with 13 qubits, known as Surface-13. This code serves as a simple testbed for investigating decoding strategies and quantifying the link between logical and physical qubit performance. We also attempt to stabilize a logical state using     and     stabilizers and assess the measured defect rates for multiple rounds.

## 6.1　Historical context

QEC is indispensable for realizing fault tolerance computations, harnessing the full potential of quantum computers. At its core, it involves encoding quantum information redundantly, creating a protected state, known as logical qubit, that can tolerate certain errors without losing the integrity of the quantum computation [4, 11, 22–24]. The basic idea is to distribute the information across multiple physical qubits in such a way that errors affecting individual qubits can be locally detected through stabilizer measurements, and then corrected. The main promise of QEC lies in the exponential suppression of the logical error rate achieved by encoding information in more physical qubits. The logical error rate, , as a function of the code distance is defined as[108, 134]:

$$ = \frac{}{\frac{+1}{2}} \tag{6.1}$$

where is a fitting constant, and is an exponential error suppression factor, proportional to $\frac{\text{th}}{}$, is code threshold, and is a physical error per operation. This holds true if, and only if, the following conditions are met:

1. Fault-tolerance threshold: physical error rates must fall below a threshold , referred to as the fault-tolerance threshold.

2. Locality: errors should be localized, allowing for their detection and correction through stabilizer measurements.

3. Stability: high performance should be maintained consistently over many QEC cycles.

Failure to meet these conditions may result in the total overhead of the QEC scheme surpassing the performance of logical qubits compared to that of the physical one. Therefore, ensuring that these conditions are met is critical for the successful implementation of QEC.

In our research group, we focus on the implementation of QEC with surface code. Surface code is experimentally attractive due to its modest requirements on the quantum hardware-specifically, requiring only nearest-neighbor coupling, and $1\%$ error threshold, significantly higher than those of other codes like Shor and Steane [11, 22, 98]. Despite this modest error threshold, the surface code demands encoding more physical qubits to implement the same code distance. In addition, successful implementation of QEC with surface code is challenging, satisfying the aforementioned conditions and exhibiting extremely low tolerance to yield problems. This underscores the critical importance of ensuring that all physical components operate precisely as intended, as detailed in Chapter 4.

Over the past four years, there has been notable acceleration in the progress of superconducting-based QEC experiments. State-of-the-art experiments have demonstrated QED of both errors using a small 2D surface code, known as Surface-7 ( $= 2$) [25, 74, 107, 168], exponential suppression of one type of errors (either bit- or phase-flips) using 1D repetition codes [25], a correction of both bit- and phase-flip errors in a $d = 3$ surface code [26, 141, 181, 182], and a proof-of-concept suppression of logical errors achieved by scaling the surface code from $d = 3$ to $d = 5$ [26]. These experiments signify the importance of achieving and maintaining high-fidelity quantum operations calibrated as part of a multi-qubit processor, rather than being tuned in isolation. In the next section, I introduce our strategy for the optimal calibration of the QEC cycle as parallel and orthogonal block units.

## 6.2   A distance-3 surface code

A distance-3 surface code (Surface-17) is the smallest code capable of error correction, given that the maximum number of correctable errors is $\frac{d-1}{2}$. This code utilizes nine data-carrying qubits to form one logical qubit, which is stabilized by the eight $X$ and $Z$ stabilizer measurements. Our quantum circuit for this code completes in a total of $960\,\mathrm{ns}$, as illustrated in Figure 6.1. Importantly, it has the ability to detect and correct any arbitrary single-qubit error on both data and ancilla qubits.

The backbone of a QEC code relies on stabilizer measurements, also known as quantum parity checks [11, 24, 72, 73, 98]. They detect bit- or phase-flip errors by repeatedly measuring $X$-type ($Z$-type) parity operators $\prod X_i$ ($\prod Z_i$), where i denotes data qubits on the corners of the green (blue) plaquettes. Standard quantum circuit for measuring $X$- and $Z$-type stabilizers [Figure 6.1] involve a sequence of coherent interactions on an ancilla qubit with its nearest-neighbor data qubits, followed by projective ancilla measurements. The order of two-qubit gates operating in the parity checks is important for two reasons [98, 121]. First, data qubits shared between adjacent plaquettes must complete all their interactions with one ancilla before engaging with the other. Second, the S (N) pattern for $X$-type ($Z$-type) stabilizers imparts resilience to single ancilla-qubit errors, even in a compact surface code like Surface-17.

Our group has previously studied and compared the performance of various stabilizer measurement schemes: fully-parallel (Fig. 3 in [98]), parallel (Fig. 4b in [107]), and pipelined (Fig. 5 in [196]). In parallel schemes, all ancilla measurements are executed simultaneously at the end of each QEC cycle, as depicted in Figure 6.1. In contrast, the pipelined scheme interleaves the coherent operations and the parity measurement of one ancilla type with the other. A stabilizer scheme is scalable if the QEC cycle time is independent of the number of data qubits. In this regard, both pipelined and parallel schemes are scalable, while the fully-parallel scheme necessitates an increasing number of detuning sequences and detuning ranges as the fabric expands. Every stabilizer scheme comes with its own caveats. For instance, the pipelined scheme offers a shorter QEC cycle time. However, the pipelining process introduces additional dephasing on ancilla qubits of one type during the readout of the other. In the case of Surface-17, we aim to choose between the two options: either the parallel or pipelined stabilizer scheme.

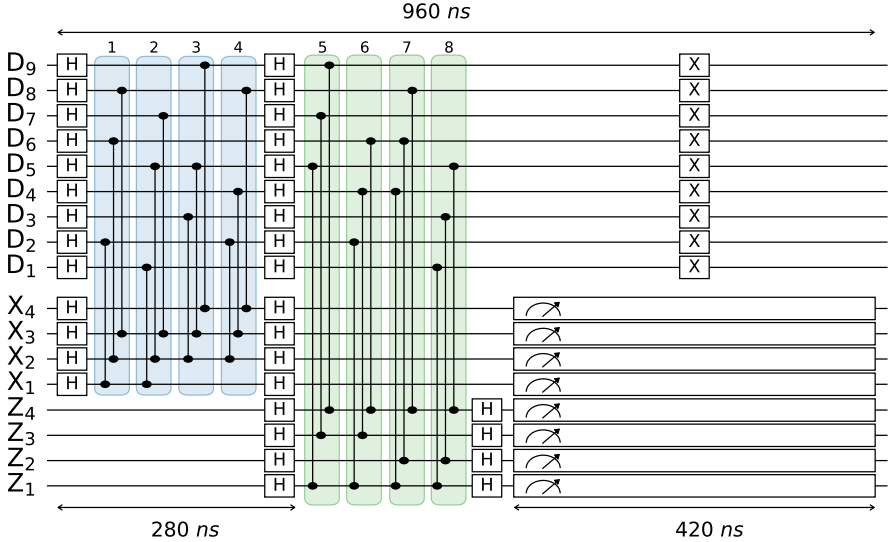

Figure 6.1: **A distance-3 QEC cycle.** The quantum circuit illustrates the parallel scheme for a distance-3 surface code implemented in a 17-transmon device. Initially, the data qubit register is prepared in a product state of all $0$'s. Subsequently, a Logical zero state, $0$, is initialized by executing one round of and optionally stabilizer measurements. The stabilizer measurement can be omitted during this initialization round as it is not strictly necessary. The logical state is then stabilized by iteratively interleaving the coherent part of and stabilizers before conducting projective measurements for all eight ancilla qubits. In parallel, a dynamical decoupling sequence involving single echo pulses is applied to data qubits to alleviate dephasing during measurement. Overall, the total duration of one round is $960\,\mathrm{ns}$, with the coherent segments of single and two-qubit gates taking $280\,\mathrm{ns}$, and the measurement phase completing in $420\,\mathrm{ns}$.

Our proposed stabilizer schemes, both parallel and pipelined, are designed to achieve the following objectives:

1. Optimized coherence: transmons should maximize the use of coherence times at the flux symmetry point, known as the sweetspot. This holds true unless the qubit strongly interacts with a two-level system defect at the sweetspot. In such cases, we bias the qubit off-sweetspot, utilizing a static current offset to minimize any parasitic interactions.

2. Desired avoided crossings: flux-pulsed transmons must avoid crossing any other interaction zones on their way to or from the intended avoided crossings, realizing CZ gates.

3. Fixed flux dance sequences: implementing a fixed number (eight steps for Surface-17) of flux-pulsing sequences, referred to as the flux dance [98], as depicted in Figure 6.2.

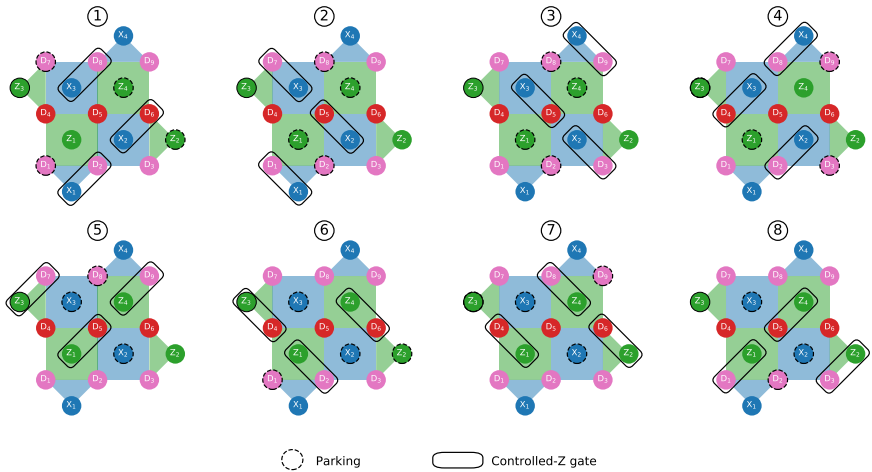

Figure 6.2: **Flux Dance of two-qubit gates in a distance-3 surface code.** This fixed sequence illustrates the order of CZ gates employed to implement the eight and stabilizer measurements in a distance-3 surface code. Additional parking pulses (depicted by dashed circles) are necessary at each step to prevent frequency overlap with the desired avoided crossings, specifically $11 \text{-} 02$, achieved through sudden net-zero gates [69]. The flux parking pulses are sudden and bipolar, synchronized with the three parallel two-qubit gates at each step. The S (N) pattern for -type ( -type) stabilizers imparts resilience to single ancilla-qubit errors, even in this small code.

This pattern is extensible beyond 17 qubits. In each step $(60 \text{ ns})$, three two-qubit gates (solid black frames) are implemented with additional parking pulses (dashed black frames) on spectator qubits to avoid any undesired interactions.

## 6.3  Highly parallel and orthogonal calibration strategies

Successful implementation of QEC hinges on calibrating and maintaining high performance over many QEC cycles. Here, the challenge is to collectively tune the building blocks to their optimal performance in a continuously-running multi-transmon processor. Various forms of crosstalk, such as residual-   , microwave, flux, and measurement crosstalk, introduce significant errors when transmons operate concurrently, are spatially close, or both. Developing calibration routines to mitigate these errors is absolutely pivotal for the successful realization of a QEC experiment. These routines highly impact the overall physical error rates and the incidence of errors in time or/and space dimensions. Ultimately, this contributes to the QEC ability to precisely detect and correct errors, ensuring the stabilization of the logical state.

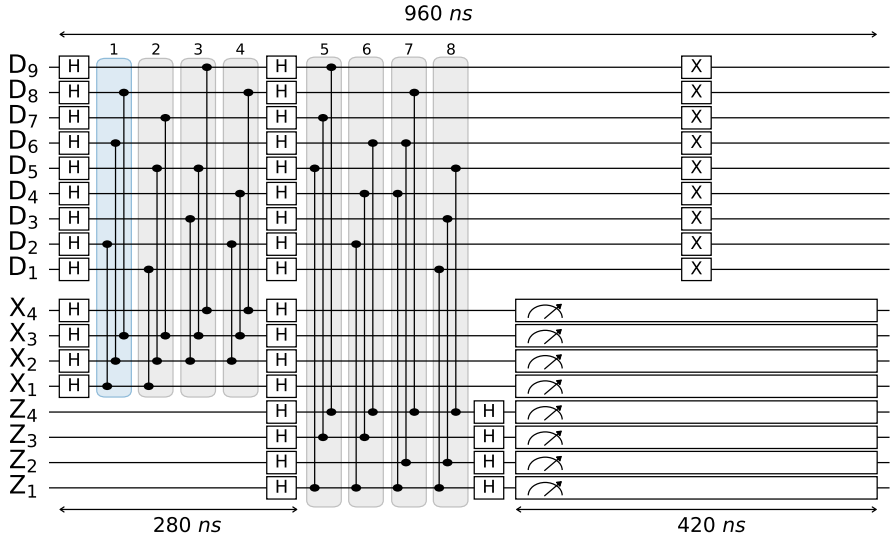

Figure 6.3: **Vertical calibration for a distance-3 surface code** focuses on tuning and align-ing three parallel CZ gates involved in each of the 8 steps, constituting the complete distance-3 flux dance. This improves two-qubit gate performance by mitigating flux crosstalk during si-multaneous operations. This process assumes two preliminary calibrations: the calibration of delay offsets to inter-align all possible gates and measurements, and the calibration of flux pulse distortion using a cryoscope routine [103].

### 6.3.1   Vertical calibration

Vertical calibration (VC) concerns the optimization of parallel two-qubit gates involved the flux dance depicted in Figure 6.2. The sequence involves eight steps, each comprising three parallel CZ gates. Ensuring the proper alignment of the three simultaneous CZ gates (solid frames) and their corresponding parking flux pulses (dashed circles) is essential. This align-ment in time avoids overlapping with other interaction zones during the CZ gates' execution and absorbs errors due to parallelization during the tune-up process.

Before the start of the VC, two prior calibrations are required: fine-timing delays and flux dis-tortion correction. The fine-timing calibration ensures alignment between all possible quantum gates and measurements, while flux pulse distortions are calibrated for all 17 qubits using simultaneous Cryoscope. This routine is a parallelized version of the procedure in [103], which divides the calibration into two stages, one for data qubits and one for ancilla qubits. Now, VC is ready to proceed for calibration of the parallel CZ gates.

Aligning several CZ pairs with different speed limits (e.g., different CZ pairs with different couplings   ) is not straightforward, imposing certain constraints. For instances, for each flux dance step, all flux pulses on the involved qubits, including parking pulses, should be aligned with a common waiting time between the two strong flux pulses [69]. The alignment routine should also ensure that the pair with the longest speed limit does not cross any unwanted interaction zones.

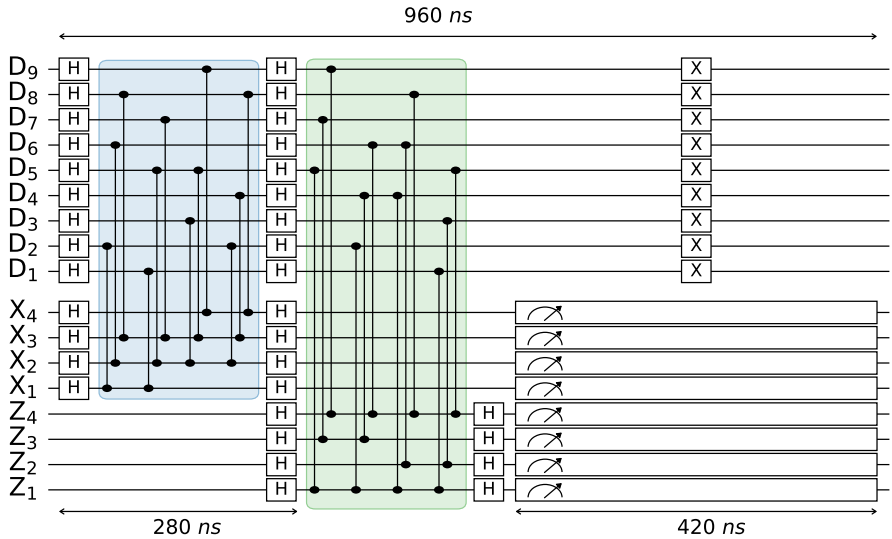

Figure 6.4: **Horizontal calibration for a distance-3 surface code** focuses on calibrating one type of parity check as a parallel block unit. This improves the overall parity check performance by mitigating coherent phase errors resulting from residual     interactions between idling transmons. The method provides orthogonal tuning an linear dependency of the conditional phase of the relevant two-qubit gates on their individual B amplitudes. This facilitates the fine-tuning of CZ gates as part of stabilizer measurements. For additional details and discussion, refer to Section 6.4.1.

In practice, VC is performed in two steps. The first step involves sweeping the timing parameters of the three flux pulses simultaneously and minimizing leakage and conditional phase errors. This is similar to the     landscape in Figure 4.12, but is generalized for the parallel CZ gates. By analyzing the three landscapes, common     and optimal frequency ranges are extracted for the three gates. The second step sweeps the amplitude parameters of each flux pulse for fine control (like AB landscape in Figure 4.13).

### 6.3.2  Horizontal calibration

Horizontal calibration (HC) optimizes each stabilizer measurement as a parallel block unit [Figure 6.4]. This approach enhances the performance of the parity check as it absorbs coherent phase errors due to residual     interaction between idling qubits, as we will see below. We orchestrate all 12 gates required for a specific type of stabilizer measurement—either the blue block for -type or green block for -type ancilla qubits. Whether HC calibration significantly affects fault tolerance is yet to be determined, as it substantially modifies the individual conditional phase for each two-qubit gate from $180°$. This remains an open question that has not been studied in this thesis.

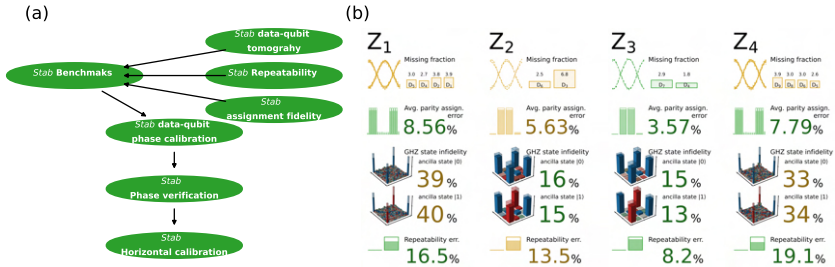

Figure 6.5: **Automatic calibration and benchmarking of stabilizer measurements with GBT.** (a) The calibration graph illustrates nodes and dependencies for autonomously calibrating stabilizer measurements in 17-qubit devices. It initiates with the HC of parity checks as parallel block units. Subsequently, it validates the calibration of desired single- and two-qubit phases in the following two nodes: phase verification and data-qubit phase calibration. The graph concludes by benchmarking the tuned parity checks in three different aspects: parity check assignment fidelity, Bell state generation, and a repeatability experiment, assessing various types of errors. The overview of the achieved performance, exemplified by the -type parity check for Uran device in (b), is then plotted. Similar to the previous graph strategy, the color of the experimentally measured data is threshold-based on predefined error rates in the codebase, providing a visual representation of the achieved performance.

## 6.4   Automatic calibration and benchmakring of stabilizer measurements

### 6.4.1   Dependency graph for stabilizer measurement calibration

In line with graph-based calibrations outlined in Chapter 4, we develop a GBT procedure for the automated calibration and benchmarking of the eight stabilizer measurements in our device, as illustrated in [Figure 6.5. a]. This graph starts calibration with HC for each parity check individually, while performing the full flux dance for each stabilizer type. To understand the optimization process of HC, we parametrize the gate interaction for the weight-2 X   parity check as follows:

$$ =  \quad^-  \quad 1 + \quad^- \quad 1 \quad 1 \quad 1 + \quad^- \quad 2 \quad 1 \quad 2 + \quad^- \quad 3 \quad 1 \quad 1 \quad 2 \qquad (6.2)$$

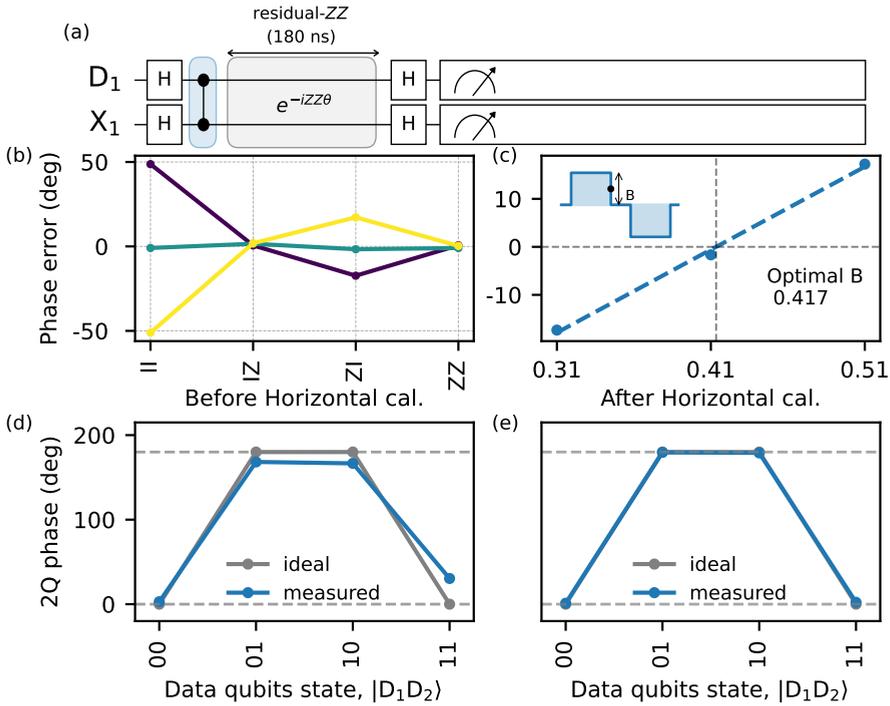

Figure 6.6: **Horizontal calibration for X parity check.** (a) Quantum circuit of the X parity check, performing four steps of $60$ ns each to complete the full -type flux dance. The X - D two-qubit gate is omitted for simplicity. Following the X -D gate [highlighted in shaded blue], there is an idling time of $180$ ns. During this period, an additional coherent phase error ( ) accumulates due to residual- coupling in the X and D qubits. HC mitigates this phase error by treating the parity check as a parallel block unit. (b) Example of the X -D two-qubit phase ( ) orthogonally controlled by sweeping the B amplitude [colormap curves] of its own flux pulse. This ensures that the total two-qubit phase during the four steps of the -type flux dance is $180°$. The terms , , and approach zero, as explained in the main text. (c) The desired two-qubit phase exhibits a linear dependency on the B amplitude, maintaining a minimal leakage zone [moving along the vertical fringe in the SNZ leakage landscape [69]]. This characteristic of the SNZ gate eases the calibration of two-qubit gates within stabilizer measurements, particularly in the presence of always-on couplings. Phases extracted for X parity check compared with the ideal phases of even- and odd-parity computational states in the data qubits before (d) and after HC. The ideal even parity phase is $0 \pmod{2\pi}$, while the ideal odd parity phase is $\pmod{2\pi}$.

In practice, Figure 6.6. b shows that the first single-body-term ( ) in can be corrected with a virtual- gate, as presented in Figure 4.14. The last three-body term is quite small, almost zero provided the parking frequency is correctly chosen [Figure 4.15]. Importantly, each of the two-body terms ( and ) can be orthogonally tuned by sweeping the individual B

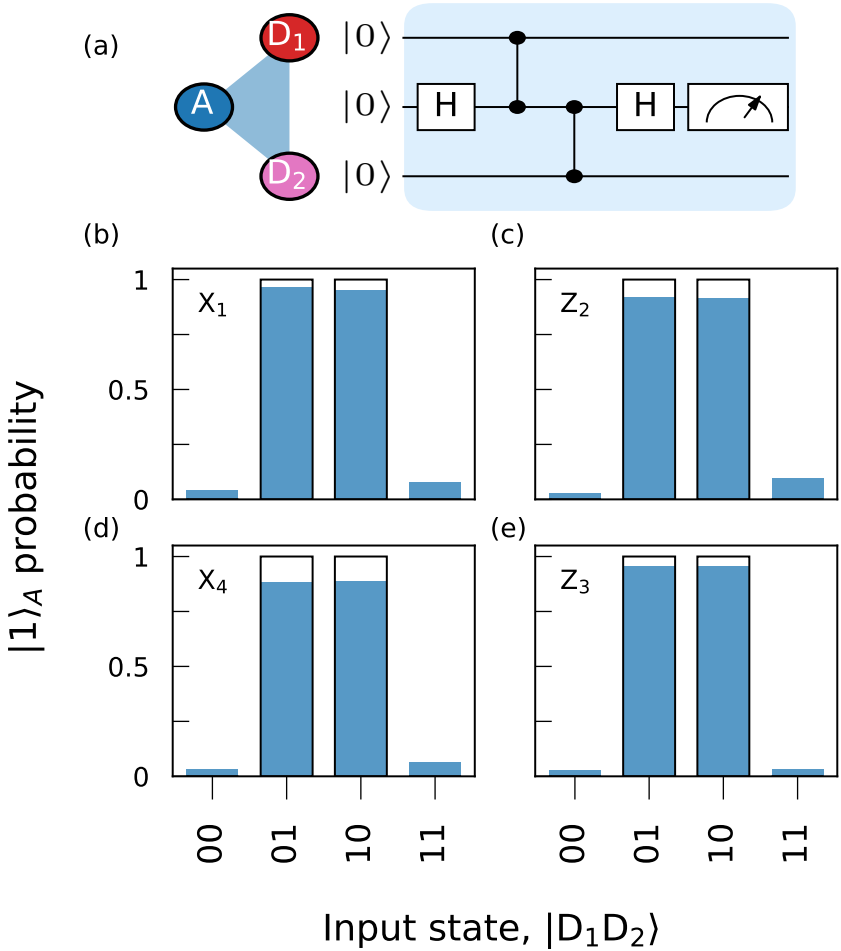

Figure 6.7: **Weight-2 parity check assignment fidelity.** (a) Quantum circuit of the weight-2 $-$type parity check. Characterization of the assignment fidelity of stabilizer measurements in (a) $_1$ $_2$, (b) $_3$ $_6$, (c) $_8$ $_9$, and (d) $_7$ $_4$, implemented using the $X$ , $Z$ , $X$ , and $Z$ qubits in Uran, respectively. Each parity check is benchmarked by preparing the relevant data qubits in a computational state and then measuring the probability of ancilla outcome $_i$. Measured (ideal) probabilities are shown as solid blue bars (black wireframe). From the measured probabilities, we extract average assignment fidelities $95\,1\%$, $92\,8\%$, $91\,7\%$, and $96\,3\%$, respectively.

parameters of the two SNZ flux pulses. The linear dependency of conditional phase on B amplitude [Figure 6.6. c] and still achieving minimum leakage are practical features of the

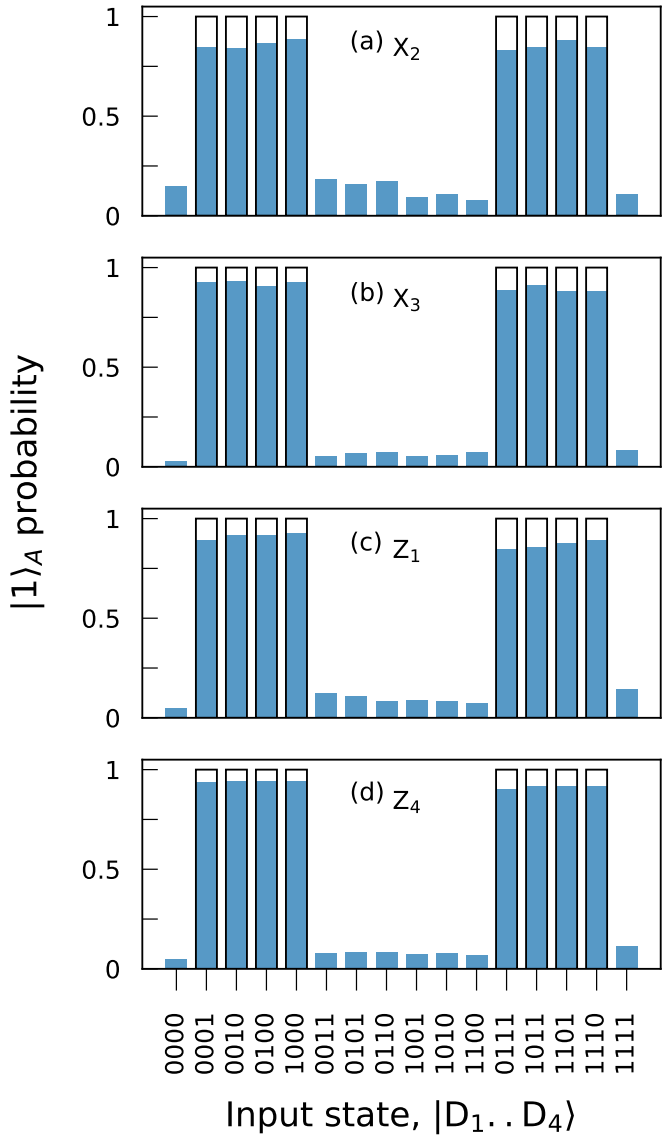

Figure 6.8: **Measured weight-4 parity check assignment fidelities** of X , X , Z , and Z in Uran device. The extracted fidelities are $86\ 2\%$, $92\ 3\%$, $89\ 9\%$, and $92\ 3\%$, respectively, averaged across the 16 prepared data-qubit states. The reported values are corrected for residual excitation effects based on post-selection before measurement.

SNZ gate. Given the other term ( ) remains unchanged, demonstrates orthogonality with very small flux crosstalk effects, 1      2°.

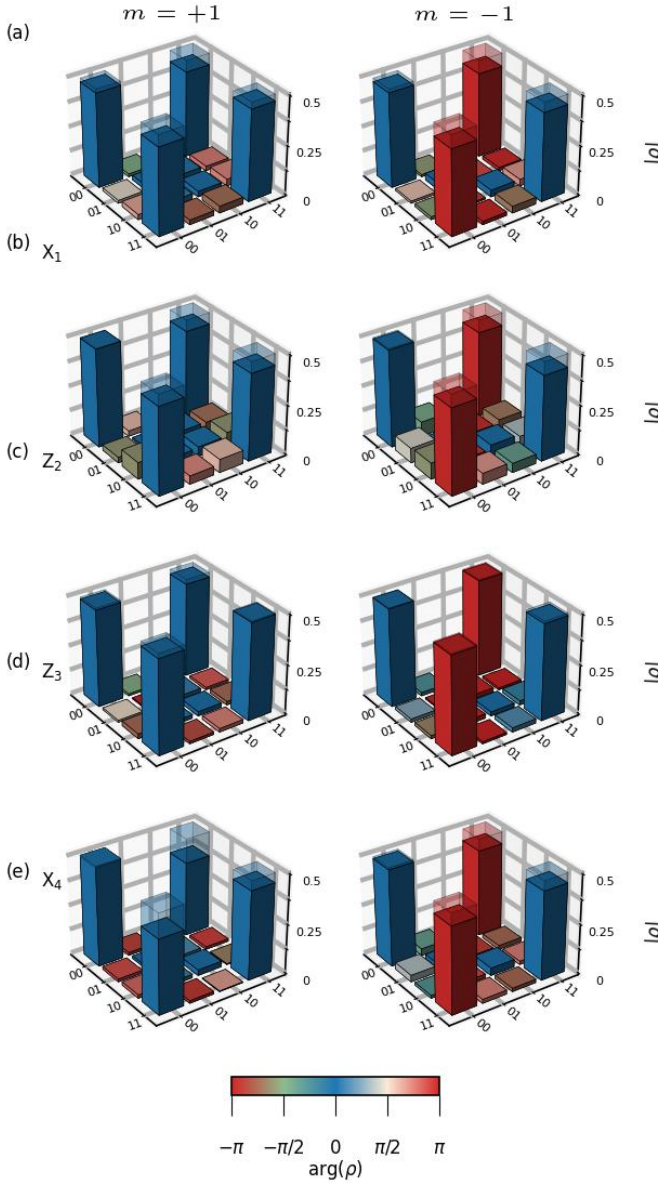

Figure 6.9: **Weight-2 Bell-state generation via stabilizer measurement** assesses the back-action on data-qubit state by generating a Bell-state via weight-2 -type parity check and data-qubit tomography for Uran device in (a). The latter is conditioned on the ancilla outcome: $= +1$, and $= -1$, yielding $\pm = \frac{1}{\sqrt{}}( 00 \quad 11 )$ bell states, each with ideally $50\%$ probability. Compared to the ideal bell state (transparent wireframes), the measured weight-2 Bell-state fidelities are (a) $90\ 7\%$ for X , (b) $90\ 6\%$ for Z , (c) $88\ 3\%$ for X and (d) $97\ 7\%$ for Z stabilizers, averaged across the two ancilla outcomes. The $\_$ bell state usually achieves higher fidelity due to the $1$ relaxation of the ancilla qubit during measurement (false positive).

To verify if X  parity check is calibrated, we compare the measured and ideal even- and odd-parity phases as a function of the input data qubit states, obtained by a standard conditional oscillation experiment in phase verification node. The extracted phases show a significant improvement with HC, demonstrating a nearly perfect weight-2 parity check, as illustrated in Figure 6.6. d and c, respectively. Similar to the calibration process in Figure 4.14, the single-qubit phases on data-qubits are characterized by switching the roles of target and control qubits in the data-qubit phase calibration node [Figure 6.5. a].

### 6.4.2 Benchmarking of stabilizer measurement

Following calibration nodes in GBT, calibrated parity checks undergo benchmarking in three different flavors, each signifying different types of errors. Initially, we evaluate the average parity assignment fidelities for both weight-2 and weight-4 stabilizer measurements in Uran device. This metric assesses the likelihood of correctly assigning the parity operator to the physical computational states of the data-qubit register [Figure 6.7. a]. The results [Figure 6.7 and Figure 6.8] indicate average parity assignment probabilities of $93\%$ and $89\%$ for the four weight-2 and four weight-4 stabilizers, respectively. These values are corrected for residual excitation effects by post-selection on a pre-measurement [107, 175]. Discrepancies in the achieved fidelities are likely due to problematic two-qubit gates, especially those strongly coupled to TLS defects or experiencing suboptimal ancilla readout, as observed in the cases of X  and X  parity checks.

The subsequent benchmark assesses the impact of the stabilizer measurement on the data-qubit state by generating a GHZ state using an    -type parity check and tomography on the data qubits [Figure 6.9. a]. Conditioned on the ancilla qubit outcome    , either $+1$ or    $1$, the reconstructed data-qubit density matrices are displayed in Figure 6.9 and Figure 6.10 for the four weight-2 and four weight-4 stabilizers, respectively. Relative to the ideal Bell state fidelity, we achieve average fidelities of $90\ 6\%$ and $80\ 4\%$, respectively, across the two ancilla outcomes:    $= +1$ and    $=$    $1$. Readout corrections are applied to each of the reported density matrices [120].

The final benchmark evaluates the backaction of two consecutive    -type parity checks, known as repeatability [Figure 6.11. a], while initializing the data-qubit register in the    $0$ state. The first stabilizer measurement projects the state onto one of (    ) each ideally with a $50\%$ probability, as the initialized state of    $0$ is not an eigenstate of the     stabilizer; thus,     is random. The ideal outcome of the second measurement     is $+1$ since ancilla qubit is not being reset. To clarify, if     $= +1$, thus     is also $+1$. Conversely, if     projects onto the stabilizer operator subspace as    $1$,     will undergo two flips, consistently resulting in $+1$. The repeatability of the experiment is therefore directly characterized by the probability of the second ancilla measurement,     $=$  (    $= +1$). On average, we achieve $89\%$ and $80\%$ for the weight-2 and weight-4 stabilizers in [Figure 6.11 and Figure 6.12], respectively. The obtained results account for errors originating from both parity assignment errors and the disturbance on data qubits due to stabilizer measurements. In addition, it is the minimal setup for a repeated QEC code. In fact, it is the ultimate benchmark we calibrate for before executing a higher circuit depth code.

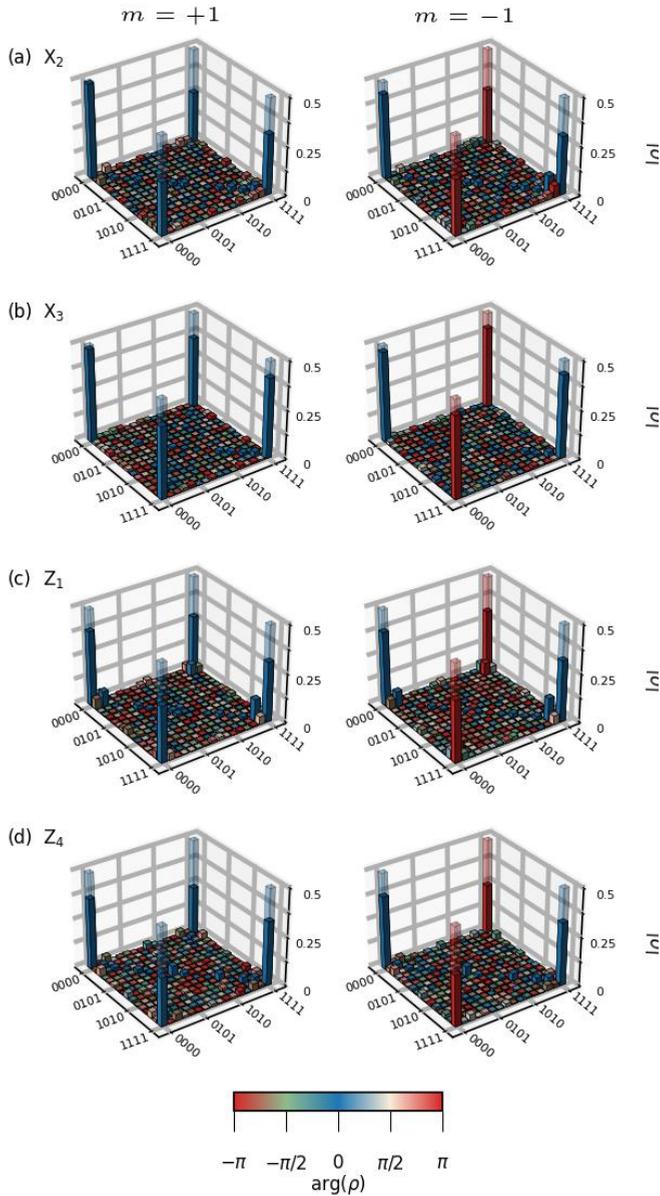

Figure 6.10: **Weight-4 Bell-state generation via stabilizer measurement.** Reconstructed data-qubit density matrices resulting from weight-4 stabilizer measurements of X in (a), X in (a), Z in (a), and Z in (a) for Uran device. The extracted Bell state fidelities are $66\,4\%$, $84\,4\%$, $65\,5\%$, and $73\,6\%$, respectively, averaged across the two ancilla outcomes.

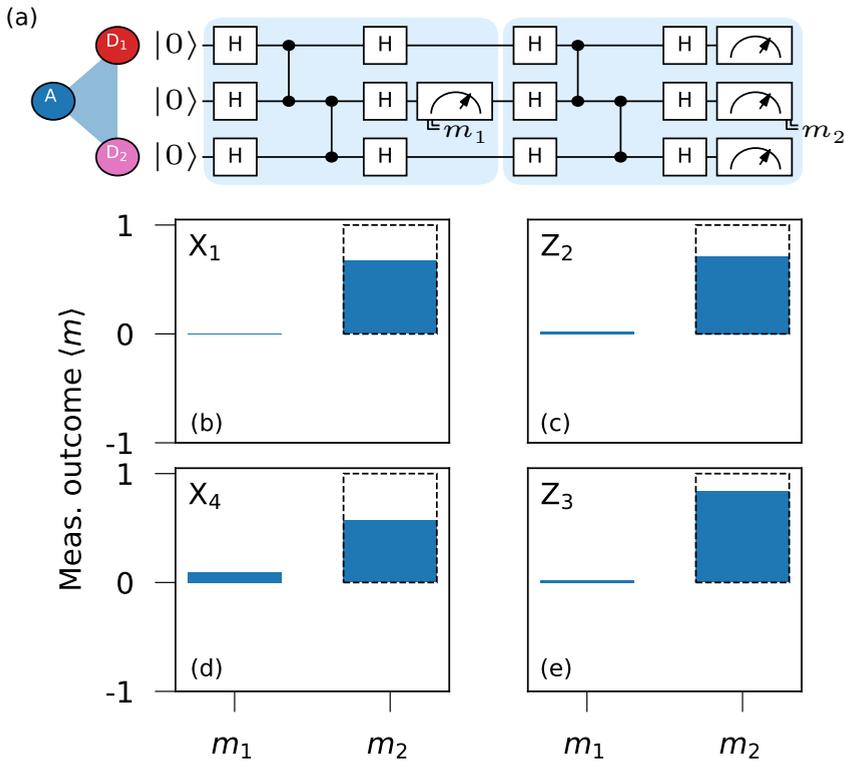

Figure 6.11: **Weight-2 parity check repeatability** quantifies parity assignment errors and the disturbance on data qubits by measuring two back-to-back $X$-type parity checks while initializing the data-qubit register in the $|0\rangle$ state. A dynamical decoupling sequence of gates on data qubits is scheduled halfway during the first ancilla measurement (not shown here). The first measurement outcome $m_1$ is ideally random, while the second outcome $m_2$ is always $+1$ since the ancilla qubit is not being reset. The measured weight-2 parity check repeatability probabilities, defined as $P(m_2 = +1)$, are (a) $83.6\%$ for $X_1$, (b) $85.5\%$ for $Z_2$, (c) $78.5\%$ for $X_4$, and (d) $91.8\%$ for $Z_3$ stabilizers.

To understand the limits of repeatability, we conducted density matrix simulations in collaboration with the Terhal group, using parameters from our experiments. An example of the weight-4 repeatability circuit for Aurelius device is presented in Figure 6.13. Circuit components were incrementally modeled to simulate repeatability. Single- and two-qubit gate processes were parameterized by decoherence, gate fidelity, and leakage rate. The measurement error model was defined by readout fidelity and QNDness probability. Panel .b compares the measured (dashed bars) and simulated (colored bars) repeatability values, revealing a generally reasonable agreement for most stabilizers in Aurelius. However, the repeatability for the $Z_4$ stabilizer is notably underestimated, likely due to measurement-induced transitions during ancilla qubit readout, which were not captured by the previous models.

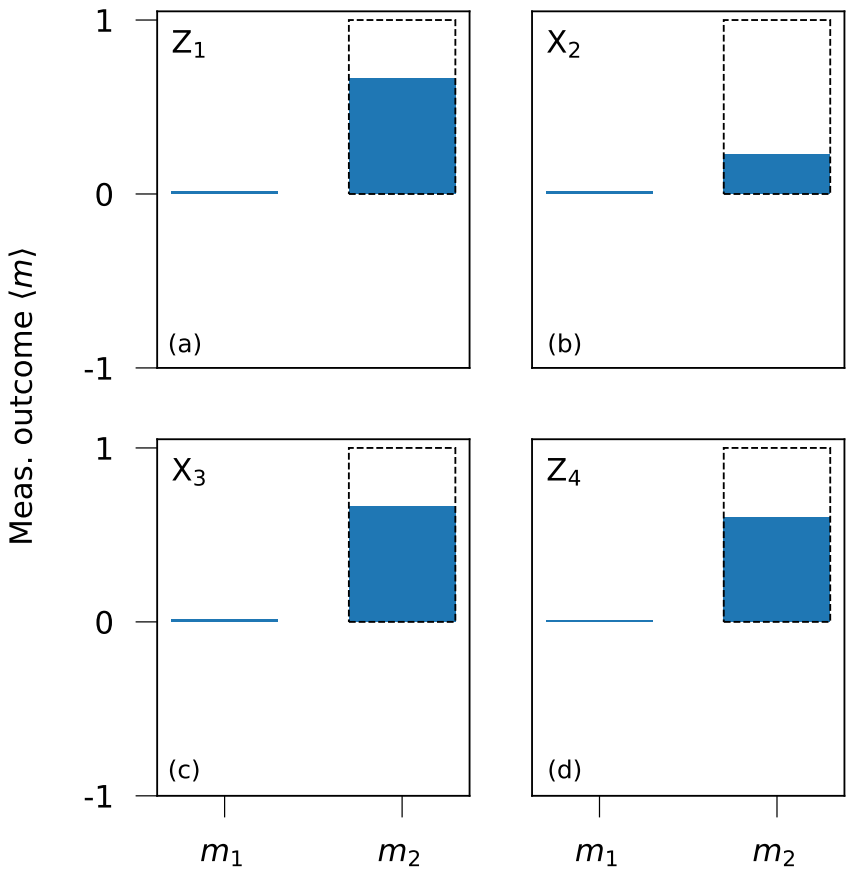

Figure 6.12: **Measured weight-4 parity check repeatability** for (a) $Z_1$, (b) $X_2$, (c) $X_3$, and (d) $Z_4$ stabilizers. The obtained probabilities ($\langle m = +1 \rangle$) are $83.5\%$, $61.4\%$, $81.1\%$, and $80.9\%$, respectively. The extremely low value for the $X_2$ stabilizer is attributed to a strong coupling to a TLS defect mode on qubit $X_2$ at its flux sweetspot. Later on, we developed new two-qubit gates, DC and Camalback flux pulses, to mitigate this issue.

Overall, these findings suggest that decoherence, primarily caused by idling during measurement time of $720\,\mathrm{ns}$, is the leading source of error. To address this issue, recent upgrades to Uran involved enhancing the hybridization of the readout resonators using the shoelace tech-nique (see Section 4.1.2) and fine-tuning the design parameters of the readout, specifically
$2$ and $2$ $2$ [131]. These improvements reduced the measurement time to $420 \, \text{ns}$,
significantly enhancing repeatability achieved in Uran device [Figure 6.11 and Figure 6.12].

**6**

Figure 6.13: **Understanding the limits of repeatability using density matrix simulations.**
(a) Quantum circuit of a weight-4 repeatability for Aurelius device. (b) Comparison of measured
and simulated repeatability values using various error models. Single and two-qubit gate
processes are parametrized by decoherence, gate fidelity and leakage rate. Measurement
error model is defined by readout fidelity and QNDness probability. Decoherence contribution
is likely higher due to reduced     at interaction points during two-qubit gates.

The GBT graph completes in $8 \, \text{min}$ per each stabilizer measurement. This graph also pro-
vides a summary of the achieved performance, exemplified by the     -type stabilizer in Uran
(refer to Figure 6.5. b). Once more, the measured data is color-coded based on predefined
error rates in the code base. Fluctuations in performance may occur due to suboptimal circuit elements. Our attention is directed on identifying and addressing faults in these elements to enhance performance before executing an algorithm.

## 6.5   Small-scale quantum error correction experiments

### 6.5.1   A bit-flip distance-3 surface-code

While our 17-qubit device is purposely designed for the complete implementation of a distance-3 surface code, capable of detecting and correcting single bit- and phase-flip errors through repeated measurements of the     and    stabilizers, strong interaction with TLS defects poses a substantial challenge. This interaction can severely compromise the operational fidelities of single-qubit gates, readout and particularly flux-based two-qubit gates, thereby posing a significant threat to the successful implementation of a QEC code.

In response, we shift our focus to a less demanding code such as a bit-flip distance-3 surface code (Surface-13). This code requires a total of 13 qubits [Figure 6.14. a]: 9 data-qubits and 4  -type ancilla qubits, taking into account potential yield issues with the    -type ancilla qubits. We note that all the strategies developed to mitigate the impact of TLS, as presented in Chapter 4, were recently developed. Importantly, this code serves as a simple testbed for investigating decoding strategies and quantifying the link between logical and physical qubit performance.

The device building blocks are calibrated with dependency graphs (see Chapter 4). Measured qubit frequencies [Figure 6.14. b], which are biased at their simultaneous sweetspot. After benchmarking, We achieve average error rates of $0\,1\%$ for single-qubit gates, $1\,2\%$ for two-qubit gates, and $1\,5\%$ for single-qubit readout [Figure 6.14. c]. The four    -type stabilizer measurements are then calibrated using VC and HC strategies, as previously discussed. The average parity assignment fidelities for the two weight-2 and two weight-4 stabilizer measurements [Figure 6.14. d] are $94\,6\%$ and $91\,1\%$, respectively.

We now investigate the performance of this Surface-13 code using the $3\times3$ data-qubit array to encode the $0$   state. This state is defined as a uniform superposition of the 16 data-qubit computational states as follows:

$$
\begin{aligned}
0 \quad = \quad & 000000000 \; + \; 110000000 \; + \; 000000011 \; + \; 011011000 \\
& + \; 110000011 \; + \; 101011000 \; + \; 000110110 \; + \; 000110101 \\
& + \; 110110110 \; + \; 110110101 \; + \; 101101110 \; + \; 101101101 \\
& + \; 101011011 \; + \; 011101110 \; + \; 011101101 \; + \; 011011011
\end{aligned}
\tag{6.3}
$$

These computational states are eigenstates of the    -type stabilizers and the logical operator,    , of the surface code, with eigenvalue $+1$. Therefore, they can be used to protect a logical qubit from single bit-flip errors, similar to the    $=3$ repetition code [85, 108]. To do this, the data-qubit register is prepared in one of these states (orange box in Figure 6.14. e), and is stabilized against bit-flip errors by performing repeated    -type stabilizer measurements over many QEC rounds. Each round completes in $700$ ns including the coherent step and ancilla measurements. Half-way during ancilla measurements, a X    operator is applied using single echo pulses on the nine data qubits. This alternates between $0$    (in even rounds) and $1$    (in odd rounds) states, symmetrizing the errors and minimizing the dependence on the initialized state. This is useful for the performance of decoders on later stages.

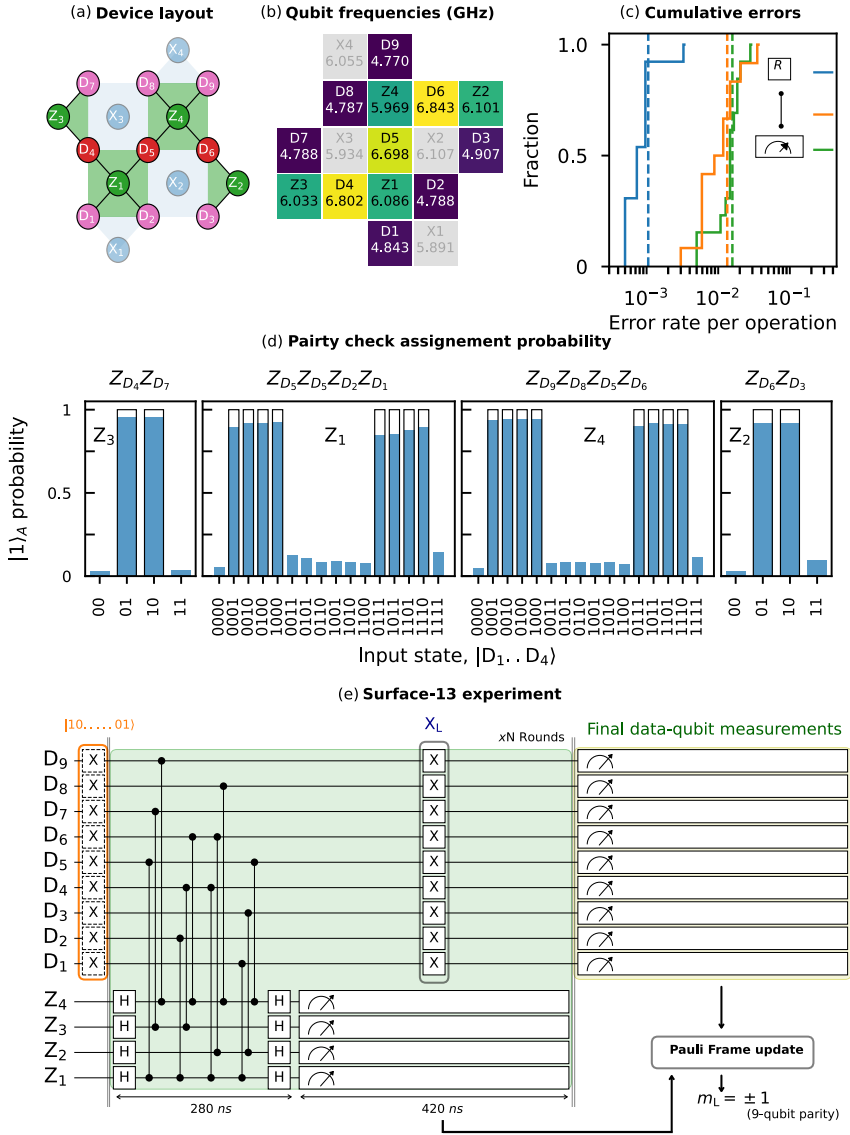

Figure 6.14: **A bit-flip distance-3 surface code.** (a) Device layout with flux-tunable transmons at vertices and fixed coupling resonator buses indicating nearest-neighbor connectivity. Shaded plaquettes and vertices are inactive in this experiment. (b) The qubit frequencies measured at their simultaneous flux sweetspot using spectroscopy techniques. To mitigate spectator effects during two-qubit gates [87], ancilla qubits X and X are statically biased to the lower sweetspot, below $4\,\mathrm{GHz}$. (Continued on next page)

Figure 6.14: (Continued from previous page) (c) Cumulative distribution of individually achieved error rates for single-qubit gates, two-qubit gates, and single-qubit readout. Single- and two-qubit gates are completed in $20$ ns and $60$ ns, respectively, while measurement duration takes $420$ ns (d) Average assignment probabilities of the four stabilizer measurements. The measured probabilities (solid blue bars) are compared with the ideal ones (black wireframe) to obtain average assignment fidelities: $96\,3\%$, $89\,9\%$, $92\,3\%$, and $92\,8\%$, respectively. (e) The quantum circuit used to measure all four    stabilizers for a bit-flip code over many QEC rounds. The initial state (orange) is one of the 16 computational basis states that comprises the logical $0$    state. The correction is tracked using a Pauli frame update based on all intermediate ancilla outcomes and final 9-bit string of the data-qubit measurements to declare a final bit outcome, m $=$ $1$.

After N rounds, a logical    measurement is performed by measuring data qubits in the    basis. The bit outcomes of the data-qubit measurements and all intermediate quantum parity checks are used to perform a Pauli frame update and declare a final bit outcome, m   , as shown in Figure 6.14. e. m    is $+1$ if the 9-qubit    -type parity is even and $1$ otherwise.

For each input state (an example is illustrated in Figure 6.15. a), the ancilla qubit outcome ideally yields an eigenvalue of $+1$. However, due to residual excitation, decoherence and readout errors on both ancilla and data qubits, ancilla qubits have finite probability that they ends up with an eigenvalue of $1$. The information obtained solely from the ancilla measurements is generally insufficient to accurately deduce the errors that have occurred. To isolate individual error events, the error syndrome    $=$    _    is computed, where    represents the vector of binary ancilla qubit readout outcome at round $i$. Unlike previous work where ancilla qubits are reset between rounds [168], in this case, they are not. The defects are defined as    $=$    _    , reflecting changes in the error syndromes (inset of Figure 6.15. b). This quantity reflects the incidence of physical qubit errors accumulated on ancilla qubits throughout the rounds. In addition to the syndromes estimated from the ancilla readout, two additional rounds of syndromes are included when estimating the defects: at the start of the rounds, we append an    _    , under the assumption of an ideally prepared state is a vector of all 0's, and at the last round, calculating a final syndrome from the nine data-qubit measurements.

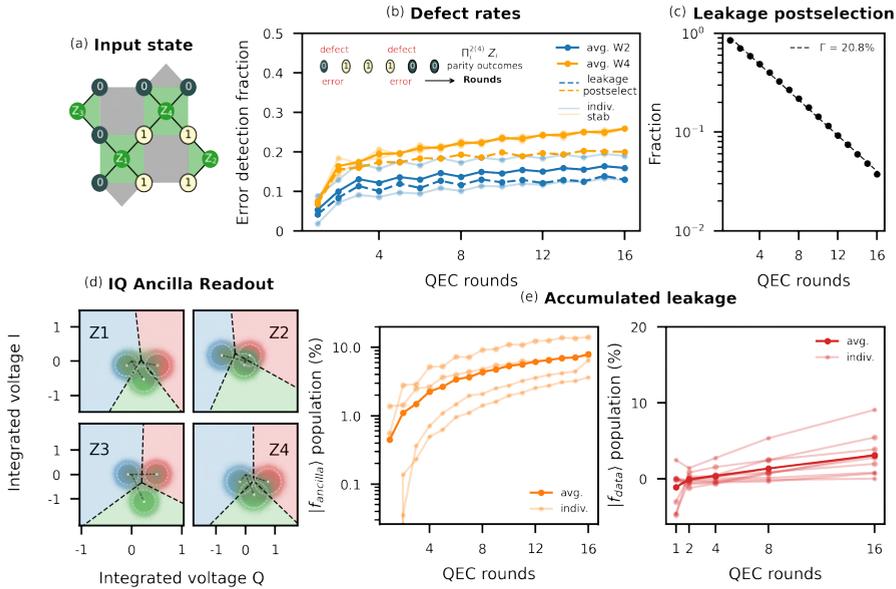

Figure 6.15: **Error-detection fraction and accumulated leakage.** (a) Initialization for one of 16 input states on the data qubits, followed by repeated $Z$-type parity checks to stabilize against bit-flip errors. (b) Error-detection probability, known as defect rate, illustrates the errors detected over 16 rounds for individual (light color) and average (solid color) weight-2 and weight-4 stabilizers. (c) Post-selected fraction of data after discarding runs where leakage was detected in any of the ancilla qubits. (d) Qutrit readout for ancilla qubits showing integrated readout signal data for about $10^4$ shots of each different qutrit state. The white cross and dashed circle denote, respectively, the center and area corresponding to $4\sigma$ obtained from a gaussian-distribution fit to the data. The black dashed lines show the three projection axes used to assign the qutrit state. (e) Accumulated leakage of ancilla (orange) and data qubits (red) over QEC rounds.

The error-detection probabilities for the weight-2 (blue) and weight-4 (orange) stabilizer measurements [Figure 6.15. b] illustrate defect rates over 16 QEC rounds. At the boundary round N = 1, the detection fraction is reduced due to low readout errors of data-qubit initialization. This value is also corrected for residual excitation effects by post-selecting on a pre-measurement. Over rounds, these probabilities incrementally increase, eventually stabilizing at approximately $13\%$ for weight-2 and $23\%$ for weight-4 stabilizers, respectively. This gradual increase over rounds indicates long-term correlated errors, mainly attributed to leakage by various work in literature [25, 26, 141].

As we continue to investigate the observed gradual increase of errors over time, we turn our attention to leakage. Transmons are vulnerable to leakage in most of their operations (particularly for two-qubit gates and measurement). Detecting leakage can be achieved through direct measurements using dispersive readout. In our surface-code implementation, ancilla qubit leakage can be detected by directly measuring them after each stabilizer measurement [141]. To this end, a three-state (or qutrit) readout on ancilla qubits is essential. Using optimized readout pulse and integration parameters for two-state readout, qutrit readout is characterized for the three states: (blue), (red) and (green). The acquired data [Figure 6.15. d] for any two given states are projected along the axis that maximizes signal distinguishability, and a threshold (black dashs) is established to optimize fidelity between each pair of states (see Chapter 4 for more details).

By combining qutrit readout with state assignment using linear regression fitting techniques [197] (shaded areas with linear boundary lines in Figure 6.15. d), we enable the detection of ancilla-qubit leakage in each round of stabilizer measurement. Similarly, accumulated leakage on data qubits can be measured during the final nine data qubit measurements. Measured leakage population on ancilla- and data-qubits are depicted in Figure 6.15. e, assuming all leakage populations go the state. While a substantial buildup of leakage on ancilla qubits is observed, reaching a steady state at approximately $10\%$ on average across the four ancilla qubits (solid curve), the leakage population on data qubits is small, except for one outlier—qubit D . This suggests that the accumulated leakage primarily stems from measurement-induced leakage [61, 135, 146, 175]. It is crucial to note that the extracted leakage populations are corrected for readout errors using the 3 3 qutrit assignment readout matrix for the measured qubits, with diagonal elements $0000$ , $1111$ , $2222$ . This may explain why some leakage rates are negative, potentially indicating a shift in the readout blobs in the IQ plane due to readout crosstalk effects. Ideally, readout correction should be performed using the full multiplexed readout matrix for the qubits measured in parallel.

To assess how leakage impacts error-detection probability, we employ three-level readout data and discard shots where leakage is signaled (dashed curves in Figure 6.15. b). Interestingly, this manifests in two signatures: either a slight offset of the error-detection probability (dashed blue) or flatter the probability curves (dashed orange). The latter suggests that gradual increase in defects over time can be attributed to leakage. These data allow us to extract a leakage detection rate, [Figure 6.15. c], approximately $21\%$. It becomes evident that leakage post-selection is not a scalable approach, as only $3\%$ of data remaining after discarding all leakage runs on both ancilla and data qubits. Additionally, this approach may be compromised by the relatively low qutrit fidelities, leading to more data being discarded than necessary. This excessive data discard is responsible for the observed offset in defect rates. For these reasons, we abandon using post-selection on leakage and instead leverage it as a detection tool for accumulated leakage throughout the rounds. We also use qutrit readout to evaluate the LRU efficacy, see Chapter 5.

Figure 6.16: **A Pairwise correlation matrix** between detection nodes being measured by -type ancilla qubits over 16 QEC rounds. Major ticks correspond to the four stabilizers, while minor ticks indicate the 16 rounds. Two scales are used in this representation: a blue-full scale defined by the maximum correlation coefficient and a red scale, maintaining the same range but saturated at a smaller value to emphasize smaller correlations.

To further understand the link between detected errors and physical error mechanisms, we estimate the pairwise correlations, known as $P$ , between detection nodes across QEC rounds for the four stabilizers [Figure 6.16] by using the following expression [25]:

$$P \quad \frac{}{(1 \quad 2 \quad )(1 \quad 2 \quad )} \tag{6.4}$$

where $= 1$ if there is a defect event and $= 0$ otherwise. The expression assumes correlations are pairwise and total error rates are sufficiently low.

To interpret the matrix [Figure 6.16], we highlight the following features:

1. Light green edges correspond to time-like errors triggering two defects in two consecutive QEC rounds. In this example, the defects, $_1 = _1 = 1$, are due to ancilla-qubit errors on qubit Z .

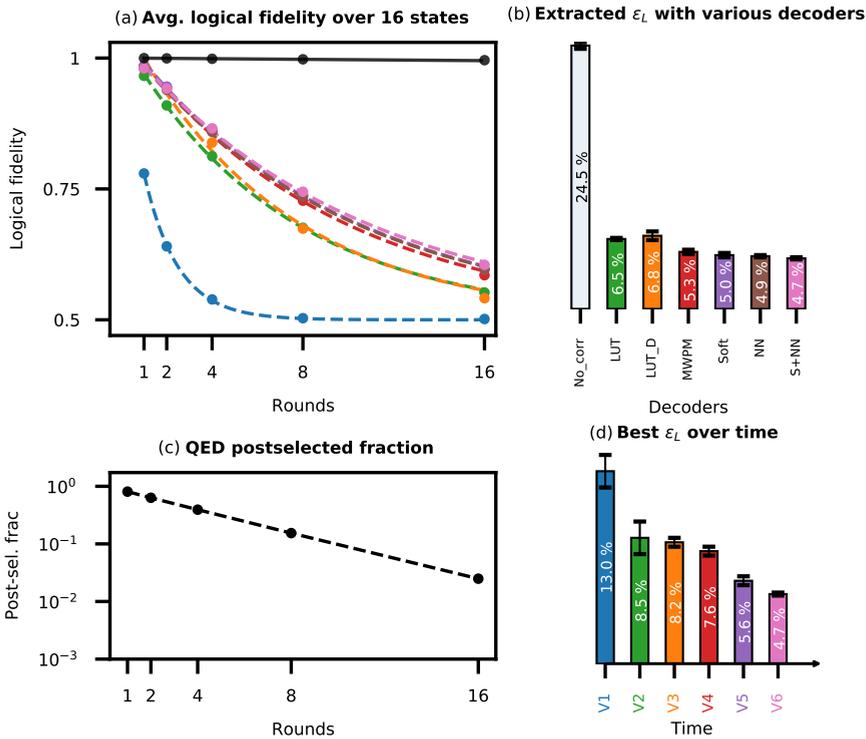

Figure 6.17: **Logical performance using various decoding strategies.** (a) Logical fidelity of the approximated $|0\rangle$ state, obtained by averaging the logical decay of the 16 input states, as a function of QEC rounds. (b) $\varepsilon_L$ for various decoders detailed in the main text. (c) QED post-selection fraction of data as a function of QEC rounds. (d) Best extracted $\varepsilon_L$ over time, highlighting progressive improvements due to optimized calibration strategies.

2. Dark green edges are also time-like errors resulting from measurement errors on ancilla qubits. These lead to two defects in two QEC rounds apart. Here, the defects, $d_4 = d_4 = 1$, are separated by one round since ancilla qubits are not being reset. Otherwise, measurement errors would produce two defects in two consecutive rounds, similar to the previous case [25].

3. Purple edges denote space-like errors, which signal two defects in two different stabilizers. Typically, errors on data qubits produce these correlations, with an exception of boundary data qubits, i.e., low-frequency qubits in this code. The latter leads to single defects signaled by individual stabilizers, known as boundary errors. In the highlighted example, single-qubit errors on qubit $D$ produce two defects on both stabilizers $Z$ and $Z$.

4. Dark gray edges are spacetime-like errors. They show data qubit errors occurring during the CZ gates, which cause detection events offset by one unit in time and space.

5. Dark orange correlations represent the buildup of errors over time. As mentioned earlier, these are attributed to leakage, mainly caused by ancilla measurements and faulty two-qubit gates.

All the aforementioned features represent actual physical error mechanisms, guiding our focus on where to investigate errors. This underscores the effectiveness of this matrix.

We employ various decoding strategies to infer a logical fidelity for the $0$ state [Figure 6.17. a]:

1. No corr (blue) is a logical operator which is obtained by the multiplication of the 9-bit string of the final data-qubit measurements.

2. QED (black) post-selects data where all stabilizer outcomes are $+1$.

3. Look-up table (LUT) decodes ancilla syndromes assuming all errors are on data qubit, i.e., ignoring the presence of errors on the ancilla qubits (space-biased decoder). The LUT decoder (green) is programmed to store most probable correction based on each of the possible combinations of observed defects at a given round. The most probable data qubit correction is estimated from a simple model, assuming that bit-flip errors on each data qubit occur with equal probability.

4. LUTD (orange) is similar to the previous decoder, but it only considers final data-qubit measurements and completely ignores ancilla syndromes. LUTD decoder signifies the role of ancilla syndromes in the decoding process.

5. Minimum weight perfect matching (MWPM) find the most probable errors that result in the observed detection events deduced from a decoding graph. This decoder (red) employs a simulated P matrix derived from a Stim simulation based on simple depolarizating error models, and matches it to the experimentally measured matrix [Figure 6.16].

6. Soft (purple) is a re-weighted MWPM decoder based on soft information of the acquired IQ readout shots. For more details, refer to Chapter 7.

7. NN decoder (brown) uses a recurrent neural network (NN) with experimental training datasets. This approach follows the work in this reference [198].

8. S+NN (pink) combines the two previous decoders to decode experimental datasets for the first time, see Chapter 7.

The averaged logical fidelity of the 16 input states [Figure 6.17. a], approximating $0$, is depicted as a function of QEC rounds using the previous decoding techniques. Each individual data point represents $100\,000$ repetitions. The logical $0$ curve exhibits an exponential decay as bit-flip errors accumulate over the QEC rounds. For each data set, we fit the decay curves to extract , see Chapter 7 for more details.

The extracted are summarized by the bar plot [Figure 6.17. b]. As expected, shows the worst error rate, quickly decaying to the randomization level after 4 rounds, signifying the importance of error correction. The extracted error rates gradually improve from using conventional decoders such as LUT table and MWPM to NN and soft information decoders. S+NN decoder yields the best logical error rate of $4.7\%$, attributed to overall better matching errors causing multiple correlated syndrome defects and exploiting the rich information of the analog readout data to perfectly assign ancilla outcomes. It is crucial to note that, as -type parity checks of the full distance-3 surface code are not measured, the error rate of logical is likely higher. This is likely to underestimate for the $0$ state. However, this does not qualitatively change the nature of the decoding problem.

The logical fidelity curve for the QED approach (excluded in panel b) remains almost flat over 16 rounds, indicating that most errors detected by the four stabilizers are weight-1 errors [74, 107]. The low amount of data left after post-selection (approximately $2\%$ after 16 rounds, as shown in Figure 6.17. c) signifies the impracticality of using this approach. This emphasizes the necessity of an error correction technique.

Experiment is rarely a smooth ride, often requiring multiple attempts to achieve success. They serve as valuable learning experiences, guiding us through a journey of educated trial and error. Throughout this process, we refined our calibration and benchmarking procedures (see Chapter 4), addressing challenges posed by TLS defects. Our decoding strategies also evolved to maximize the benefit of IQ readout data, eventually leading to a gradual reduction in the logical error rate for the $0$ state. The progress is evident in Figure 6.17. d, showing the best extracted for the optimal decoder over six versions. The best decoder of version $6$ (S+NN) demonstrates a notable improvement, roughly threefold, in the extracted .

### 6.5.2   Initial attempt toward the realization of Surface-17

In our initial attempt to implement a Surface-17 code on Aurelius device in June 2022, we integrated essential components for a full distance-3 code. We attempted stabilization of the logical state $0$ by initializing the data qubit register in the $0^{\otimes}$ state and performing repeated and stabilizer measurements over multiple QEC rounds (refer to Figure 6.18. a). By looking at the incidence of physical qubit errors through changes in the stabilizer parity syndrome outcomes, we observed defect probabilities approaching $0.5$ (randomness) after four rounds for most stabilizers, indicating a high rate of physical errors affecting the logical qubit (see Figure 6.18. b).

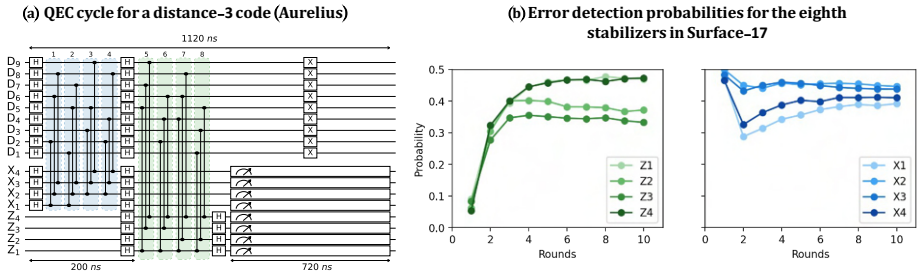

Figure 6.18: **Initial attempt for the realization of a distance-3 QEC code** (a) QEC quantum circuit used to measure the stabilizers of Surface-17. (b) Defect probability as a function of the number of rounds for each stabilizer of the Surface-17 code.

Several factors contributed to these high error rates at the time, although detailed experiments on these aspects are not presented here. First, suboptimal readout with extremely low dispersive shifts with respect to effective readout linewidth led to driving in the non-linear regime. This was done to achieve high fidelity readout, but resulted in non-QND measurements and increased measurement-induced transitions [61, 135, 146, 175]. Second, high residual-couplings (a worst-case scenario of approximately $850$ kHz) induced significant spectator effects, such as high coherent phase errors and elevated leakage rates during two-qubit gates, dependent on the neighboring qubit state.

In our latest Surface-17 design (version 6b), like for Uran device, we addressed these issues by augmenting the dispersive shifts, through the increase the coupling strength between the qubit and readout resonator modes, and lowering       couplings by a factor of four, by reducing the transverse nearest-neighboring couplings by a factor of two. This improved overall device performance, as elaborated in Chapter 4. Upon resolving these issues, our focus turned to prominent source of error in the calibration of sudden net-zero gates. This led us to developing new gate schemes for the mitigation of strong interaction with TLS defects. These improvements significantly enhanced both fidelity and stability in our two-qubit gates.

Although our PhD timeline has concluded, these experiences have provided valuable insights. We now pass on our knowledge to the new PhD generation, who are well-positioned to achieve a successful implementation of the distance-3 Surface code. With improved readout performance, reduced       coupling, enhanced and stable two-qubit gate performance, and leakage reduction units, the prospects for realizing a robust Surface-17 code are brighter.

### 6.5.3   A bit-flip repetition code: scaling from distance-3 to distance-7

Despite suboptimal design choices and the calibration methods applied, we observed exponential suppression of bit-flip errors in a 1D repetition code across different code distances ( $= 3$, $= 5$, and $= 7$) using Aurelius device. We achieved an exponential error sup-pression factor, denoted as     in Equation (6.1), of $1\,15$ for a bit-flip code. To attain this result, our collaborator, Boris M. Varbanov from the Terhal group, employed various decoding strategies to improve the extracted logical error rate for each code distance. Additionally, he performed density matrix simulations to investigate the primary error sources influencing the logical performance. Simulation suggested that the logical error rate is primarily dominated by qubit relaxation times and measurement error rates. Further details about the experiment, the decoding techniques employed, and the performance of the logical qubit are summarized in Chapter 7 of Varbanov's PhD thesis [199].

# REDUCING THE ERROR RATE OF A SUPERCONDUCTING LOGICAL QUBIT USING ANALOG READOUT INFORMATION

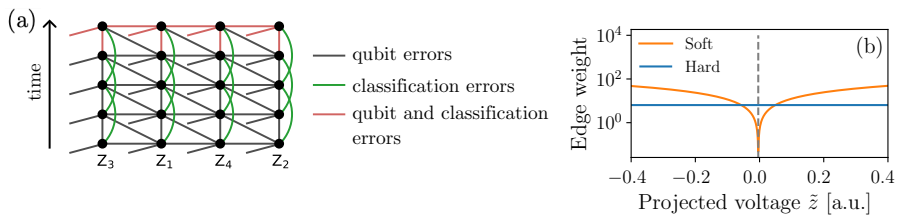

Quantum error correction enables the preservation of logical qubits with a lower logical error rate than the physical error rate, with performance depending on the decoding method. Traditional error decoding approaches, relying on the binarization ('hardening') of readout data, often ignore valuable information embedded in the analog ('soft') readout signal. We present experimental results showcasing the advantages of incorporating soft information into the decoding process of a distance-three ($d = 3$) bit-flip surface code with transmons. To this end, we use the $3 \times 3$ data-qubit array to encode each of the $16$ computational states that make up the logical state $|0\rangle$, and protect them against bit-flip errors by performing repeated $Z$-basis stabilizer measurements. To infer the logical fidelity for the $|0\rangle$ state, we average across the $16$ computational states and employ two decoding strategies: minimum weight perfect matching and a recurrent neural network. Our results show a reduction of up to $6.8\%$ in the extracted logical error rate with the use of soft information. Decoding with soft information is widely applicable, independent of the physical qubit platform, and could reduce the readout duration, further minimizing logical error rates.

## 7.1   Historical context

Small-scale quantum error correction (QEC) experiments have made significant progress over recent years, including fault-tolerant logical qubit initialization and measurement [32, 180], correction of both bit- and phase-flip errors in a distance-three ($= 3$) code [26, 141, 181], magic state distillation beyond break-even fidelity [201], suppression of logical errors by scaling a surface code from $= 3$ to $= 5$ [26], and demonstration of logical gates [30]. The performance of these logical qubit experiments across various qubit platforms is dependent on the fidelity of physical quantum operations, the chosen QEC codes and circuits, and the classical decoders used to process QEC readout data. Common error decoding approaches with access to analog information often rely on digitized (binary) qubit readout data as input to the decoder. The process of converting analog to binary outcomes inevitably leads to a loss of information that reduces decoder performance.

Pattison *et al.* [202] proposed a method for incorporating this analog 'soft' information in the decoding of QEC experiments, suggesting a potential $25\%$ improvement in the threshold compared to decoding with 'hard' (binary) information. The advantage of using soft information has also been demonstrated on simulated data with NN decoders [198, 203]. Soft information decoding has also been realized for a single physical qubit measured via an ancilla in a spin-qubit system [34] and for a superconducting-based QEC experiment with a simple error model assuming uniform qubit quality [182]. The incorporation of soft information with variable qubit fidelity can in theory provide further benefit when decoding experimental data. However, this can be challenging as additional noise sources (e.g., leakage and other non-Markovian effects) add complexity; therefore, the advantage of decoding with soft information is not guaranteed.

In this chapter, we demonstrate the use of soft information in the decoding process of experimental data obtained from a bit-flip $= 3$ code in a 17-qubit device using flux-tunable transmons with fixed coupling. Unlike a typical $= 3$ surface code with both     and     stabilizers, we repeatedly measure    -basis stabilizers and protect against bit-flip errors throughout varying QEC rounds. This approach allows us to avoid problematic two-qubit gates between specific pairs of qubits that occur due to strong interactions of the qubits with two-level system (TLS) defects [43], utilizing 13 out of the 17 qubits in the device [Figure 7.1(a)]. We refer to this experiment as Surface-13. We encode and stabilize each of the $16$ computational basis states that occur in the superposition of the $| 0 \rangle$ state and approximate its performance by averaging across these states. We employ two decoding strategies to extract a logical fidelity: a MWPM decoder and a recurrent NN decoder. For each strategy, we compare the performance across two variants: one with soft information and one without. With soft information, the extracted logical error rates are reduced by $6\text{–}8\%$ and $5\%$ for the MWPM and NN decoders, respectively.

## 7.2 Concept

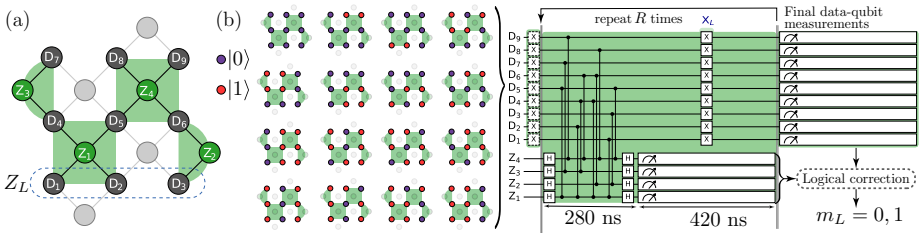

Figure 7.1: **Surface-13 QEC experiment**. (a) Device layout, with vertices indicating flux-tunable transmons, and edges denoting nearest-neighbor coupling via fixed-frequency resonators. Nine data qubits in a $3 \times 3$ array (labeled $D$, dark gray) are subject to 4 $Z$-basis parity checks realized using ancilla qubits (green). Light grey vertices and edges are not used. (b) Quantum circuit for experiment over $R$ QEC rounds, each round taking $700$ ns. During ancilla measurement, a $X$ operation implemented transversally with $\pi$ pulses on all data qubits is applied to average the logical error over the logical subspace. We show the 16 computational states that we average over. We employ various methods in post-processing to decode the measurements and determine the value of the logical observable, $Z_L$ [134].

In our $d = 3$ code [Figure 7.1(a)], the $|0_L\rangle$ state is defined as the uniform superposition of the $16$ data-qubit computational states shown in Figure 7.1(b). These computational states are eigenstates of the $Z$-basis stabilizers and the logical operator, $Z_L$, of the surface code, with eigenvalue $+1$. Therefore, they can be used to protect a logical qubit from single bit-flip errors, similar to the $d = 3$ repetition code [85, 108].

The experimental procedure begins by preparing the data-qubit register in one of the 16 physical computational states, as shown in Figure 7.1(b). Next, repeated $Z$-basis stabilizer measurements are performed over a varying number of QEC rounds, $R$, to correct for bit-flip errors in the computational states of the data qubits. Each round takes $700$ ns, with $20$ ns and $60$ ns for single- and two-qubit gates, respectively, and $420$ ns for readout. The logical state is flipped during the ancilla measurement in each QEC round using the $X_L = X^{\otimes}$ transversal gate to symmetrize the effect of relaxation ($T_1$) errors, minimizing the dependence on the input state. A final measurement of all data qubits is used to determine the observed logical $Z_L$ outcome and compute final stabilizer measurements. The physical error rates of single-qubit gates, two-qubit gates, and readout are $0.1\%$, $1.6\%$, and $1.2\%$, respectively, averaging over the 13 qubits and 12 two-qubit gates used in the experiment. Further details about the device, calibration, and parity-check benchmarking are provided in Sections 7.6.1 and 7.6.2.

We employ two decoding strategies to infer a logical fidelity, $F$, for each input state: a MWPM [204, 205] decoder and a NN decoder [198, 203]. For each strategy, we consider two variants based on using soft or hard information from the recorded IQ readout data, as explained below. The decoder determines whether the outcome $m$ needs to be corrected (flipped) based on the values of combinations of certain measurements (see below), and decoding success is declared if this corrected readout, $\bar{m}$, matches the prepared state. We calculate $p$ for a fixed $N \in \{1, 2, 4, 8, 16\}$ and each input state as the fraction of successfully decoded runs. Finally, $p$ is averaged over the $16$ physical computational states, approximating the logical performance of the $|0\rangle_L$ state.

Qubit readout is performed by probing the state-dependent transmission of a dedicated, dispersively-coupled readout-resonator mode to infer the qubit state, $q$ [102]. The readout pulse for each qubit has a rectangular envelope softened by a Gaussian filter of width $\sigma = 0.5 \text{ ns}$ (see Section 7.6.1 for additional details about readout calibration). After amplification [106], the transmitted signal is down-converted to an intermediate frequency and the in-phase and in-quadrature components integrated over $420 \text{ ns}$ using optimal weight functions [206–208]. The resulting two numbers, $I$ and $Q$, form the IQ signal $v = (I, Q)$ (Figure 7.2(a)) comprising the soft information. The readout pulse envelope and signal integration are performed using Zurich Instruments UHFQA analyzers sampling at $1.8 \text{ GSa/s}$.

To transform an IQ signal $v \in \mathbb{R}^2$ to a binary measurement outcome $m \in \{0, 1\}$, we apply a hardening map which we choose as the maximum-likelihood assignment. The hardened outcome is obtained by choosing $m = 0$ if $P(0|v) > P(1|v)$ and $m = 1$ otherwise, where $P(q|v)$ is the probability that the qubit was in state $q$ just before the measurement given the observed IQ value $v$. Assuming the states $0$ and $1$ are equally likely, one has $P(q|v) \propto P(v|q)$. Here, $P(v|q)$ is the probability density function (PDF) of the soft outcome $v$ conditioned on the qubit being in state $q$ at the start of the measurement. Therefore $m = 1$ if $P(v|1) > P(v|0)$ and $m = 0$ otherwise. We emphasize here that we use 0 and 1 as measurement outcomes for the $0$ and $1$ states, respectively, instead of +1 and $-1$, as we will later allow a measurement outcome of 2 for the $2$ state. To reduce the dimension of the problem, we project the IQ voltages to $u \in \mathbb{R}$ along the axis of symmetry [black dotted line in Figure 7.2(a)]. We obtain the PDFs of the projected data $P(u|q)$ by decomposing $v = (u, u_\perp)$ to a parallel component $u$ and a perpendicular component $u_\perp$, giving $P(v|q) = P(u, u_\perp | q)$. Assuming the IQ responses of the computational basis measurements are symmetric along the axis joining the two centroids [marked by black crosses in Figure 7.2(a)], the projection does not result in information loss. The hardened measurement outcomes are then obtained by comparing $P(u|0)$ and $P(u|1)$. If higher excited states such as $2$ are considered, the symmetry in IQ space breaks, and we consider the full two-dimensional measurement response in classification, see Section 7.6.5 for details.

Our model for the PDF of a measurement response from state $q$ is a mixture of $K$ Gaussians, where $K$ is the number of distinct states in IQ space so that $q \in \{0, 1, \dots, K-1\}$. We find that this heuristic model (see Section 7.6.5) works well in the presence of leakage to the $2$ state and it allows for an easy calculation of $P(v|2)$. To determine the fit parameters of the Gaussian model, we use $1.3 \times 10^4$ calibration shots per state preparation $q \in \{0, 1, 2\}$ for each of 13 transmons. Other, more complex, classification models exist [202, 203] but we omit discussing them here as we found they provided no statistically significant improvement in classification performance for our device (see Section 7.6.5).

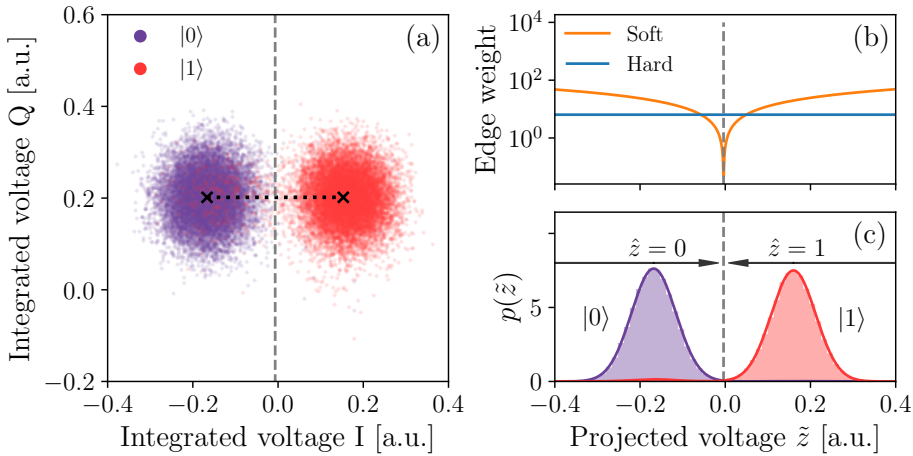

Figure 7.2: (a) The measurement response of the $|0\rangle$ and $|1\rangle$ states in IQ space for data qubit , showing a projection line that connects the means of the two Gaussian peaks (black dotted line). (b) Edge weight as a function of projected voltage for soft and hard measurements; see Equation (7.1). Measurement errors are most likely in the region $\tilde{z} = 0$ where the edge weight is minimized. (c) Histogram and fitted probability density function $p(\tilde{z})$ for state preparations $|0\rangle$, $|1\rangle$.

In QEC experiments, detectors [205] are selected combinations of binary measurement outcomes that have deterministic values in the absence of errors. We refer to a detector whose value has flipped from the expected error-free value as a defect. A decoder takes observed defects in a particular experiment and, using a model of the possible errors and the defects they result in, calculates the logical correction. In Surface-13, assuming circuit-level Pauli noise and with detectors defined as described below, each error results at most two defects. As a result, it is possible to represent potential errors as edges in a graph – the decoding graph – with the nodes on either end representing the defects caused by the error (with a virtual node added for errors that only lead to a single defect). A matching decoder can thus be used, which matches pairs of observed defects along minimum [204, 209, 210] or near-minimum [211] weight paths within the graph, and thereby approximately finds the most probable errors that cause the observed defects. From these, one can deduce whether a logical correction is necessary.

Typically, QEC experiments are described assuming that ancilla qubits are reset following their measurement in every QEC round. However, this is not the case in our experiment. Nevertheless, without resetting qubits, suitable detectors can be chosen as =
_ where $|0\rangle$, $|1\rangle$ is the (hardened) measurement outcome of ancilla in round (see Section 7.6.4 [77] for details). We note that, in our case, the error-free detector values

are always 0. A key difference with the more typical mid-circuit reset case is that ancilla qubit errors (that change the qubit state) and measurement classification errors (where the inferred hardened measurement does not match the true qubit state) have different defect signatures, apart from in the final round (see Appendix 7.6.4). The structure of the decoding graph for the four-round experiment can be seen in Figure 7.3(a).

To construct the decoding graph, one can define the probability of different error mechanisms and use a software tool such as Stim [212]. As we do not have direct knowledge of the noise, this graph may not accurately capture the true device noise. Therefore, we use a pairwise correlation method [25, 168, 213] to construct the graph for the MWPM decoder, whereby the decoding graph edge probabilities are inferred from the frequency of observed defects in the experimental data. This learning-based approach is susceptible to numerical instabilities. We stabilize it using a "noise-floor graph" [25] that has specific minimum values for each edge in the decoding graph, as described in Section 7.6.4. These instabilities can arise due to the finite data and the approximation that there are no error mechanisms resulting in more than two defects, such as leakage. The impact is more pronounced at the boundary of the code lattice where single defects occur [199].

## 7.3 Experimental results

To use soft information with a MWPM decoder, we follow Ref. [202]. The edge corresponding to a measurement classification error is given a weight

$$w = \log \left( \frac{p'(z = 1 | z')}{p'(z | z')} \right) \tag{7.1}$$

where $z'$ is the inferred state after measurement. This is found by taking $z' = 1$ if $p'(z = 1) \geq p'(z = 0)$ and 0 otherwise. The PDFs $p'(z | z')$ are obtained by keeping only the dominant Gaussian in the measurement PDFs – this is to avoid including ancilla qubit errors during measurement in the classification error edge. Therefore, to incorporate soft information, we simply replace the weights calculated for the classification error edges using the pairwise correlation method with the weights from Equation (7.1). This procedure is appropriate in all rounds but the final round of ancilla and data-qubit measurements, where both qubit errors and classification errors have the same defect signature. In these cases, we instead (i) calculate the mean classification error for each measurement by averaging the per-shot errors; (ii) calculate the mean classification error for each edge from those for each measurement; (iii) remove the mean classification error from the edge probability; and (iv) include the per-shot classification error calculated from the soft readout information. Further information is given in Section 7.6.5.

Our second decoder – the NN decoder – can learn the noise model during training without making assumptions about it [198, 203, 214, 215]. NNs have flexible inputs that can include leakage or soft information. Recent work [198, 203] has shown that NNs can achieve similar performance to computationally expensive (tensor network) decoders when evaluated on experimental data for the $= 3$ and $= 5$ surface code. Those networks were trained with simulated data, although Ref. [203] did a fine-tuning of their models with $2 \cdot 10$ experimental samples (while using $2 \cdot 10$ simulated samples for their main training). One may expect that the noise in the training and evaluation data should match to achieve the best performance.

We train a NN decoder on experimental data and study the performance improvement when employing various components of the readout information available from the experiment. We have used two architectures for our NN, corresponding to the network from Varbanov *et al.* [198], and a variant of this model. The variant includes encoding layers for dealing with different types of information [see Figure 7.4(a)]. The inputs for our standard NN decoder are the observed defects. For our soft NN decoder, the inputs are the defect probabilities given the IQ values [see Equation (7.4)] and the leakage flags, one for each measured ancilla qubit. A leakage flag gives information about the qubit being in the computational space, i.e. $= 1$ if $= 2$ and $= 0$ otherwise, where is the hardened value of using the three-state classifier in Section 7.6.5. For the final round when all data qubits are measured, we do not provide the decoder with any soft information to ensure that we do not make the task for the decoder deceptively simple [203]; this directly relates to the drawback of running the decoder only on the logical $0$ state and not on randomly chosen $0$ or $1$ (see the discussion in Section 7.6.3).

Due to the richer information of soft inputs, we can use larger networks than in the (hard-)NN case without encountering overfitting issues during training. The network performance when given different amounts of soft information is included in Section 7.6.3, showing that the larger the amount, the better the logical performance. We follow the same training as Ref. [198], but with some different hyperparameters; and we use the ensembling technique from machine learning to improve the network performance without a time cost, but at a resource cost [216]. For more information on NNs, see Section 7.6.3.

The extracted logical fidelity as a function of for the MWPM and NN decoders is shown in Figures 7.3 and 7.4, respectively. The results are presented for the 16 physical computational states (transparent lines) and for their average (opaque lines), approximating the logical performance of the $0$ state. To find a logical error rate, , the decay of logical fidelity is fitted using the model

$$( ) = \frac{1}{2} \quad 1 + (1 \quad 2 \quad )^{ \quad - \quad 0}$$  (7.2)

where indicates the fitted fidelity to the measured , and is a round offset parameter [134]. For both decoders, including soft readout information enhances the logical performance, resulting in the reduction of by $6 \cdot 8\%$ and $5 \cdot 0\%$ for MWPM and NN, respectively. The improvement in the logical fidelity at the eighth round is statistically significant for all 16 initial states in the case of the MWPM decoder, and some states in the NN decoder case. We note that the error bars are different in the two cases due to the differing ways of splitting the dataset of $9\;10$ samples per round and initial state. With the MWPM decoder, we use half the data to perform the pairwise correlation method to obtain the decoding graph, and half for obtaining the logical error probability. We then swap the data halves and average the logical fidelities, thereby using every shot to obtain the overall logical fidelity. With the NN decoder, $95\%$ of samples are used for training and validation and only $5\%$ of samples are used to obtain estimate logical fidelities, leading to larger error bars.

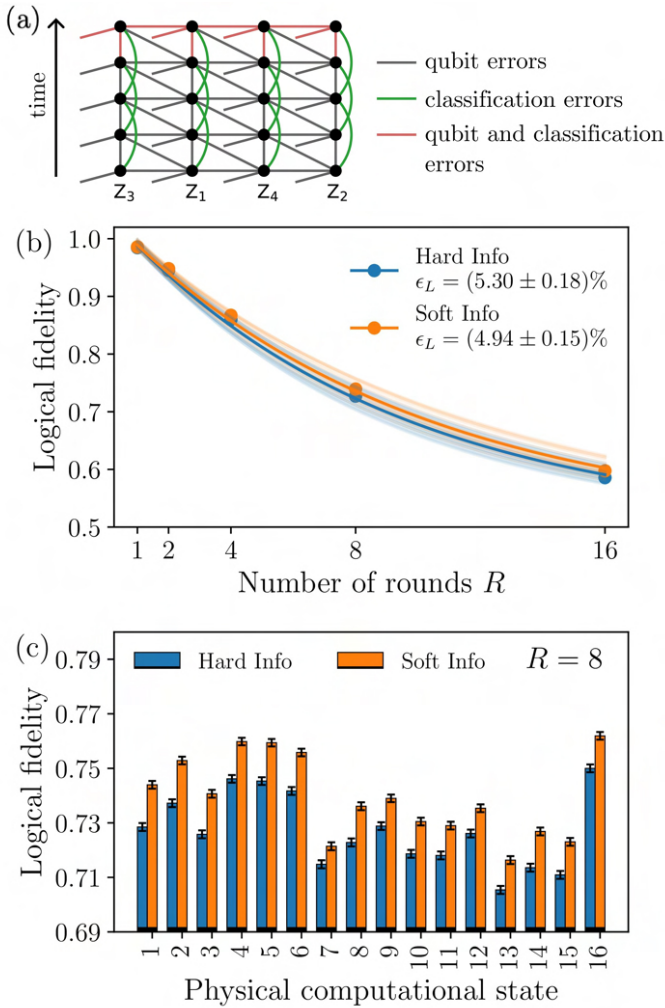

Figure 7.3: (a) Decoding graph showing different types of error mechanisms. The labels indicate the ancilla qubits associated with the detectors in each column. The soft MWPM decoder dynamically updates the weights of edges highlighted in green and red. (b) Logical fidelity of the MWPM decoder as a function of the number of rounds  , shown for each physical computational state that makes up  $0$    (transparent curves) and averaged across all states (opaque curves). (c) Logical fidelity at   $= 8$, shown for each initial computational state.

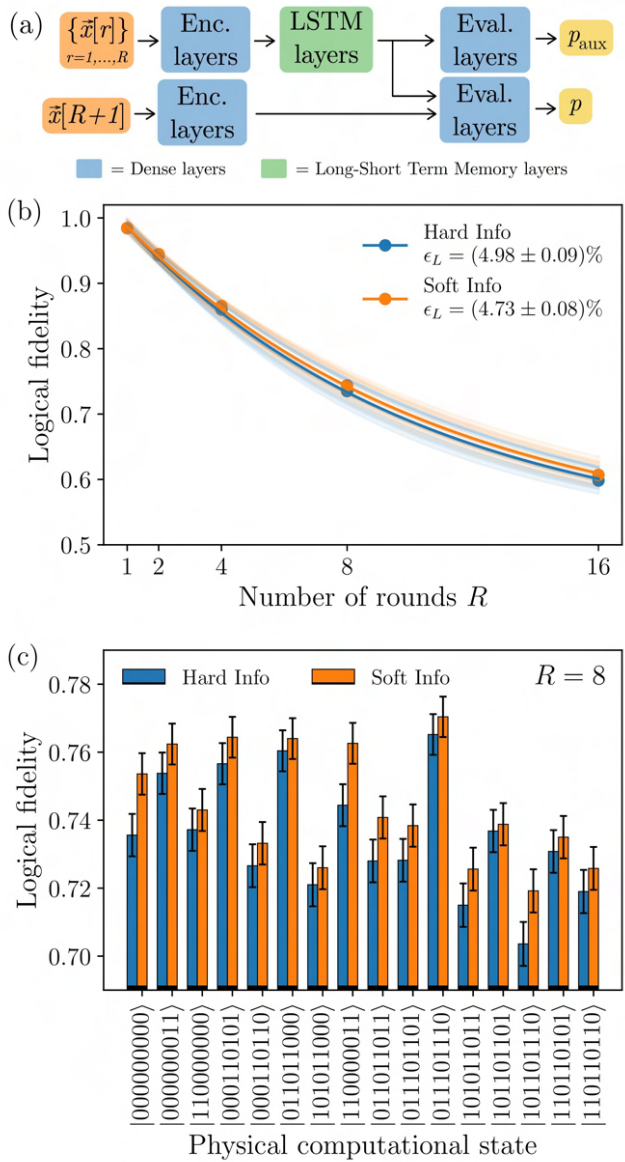

Figure 7.4: (a) Our NN architecture, a variant of Ref. [198]. The input $\vec{x}[r]$ for the soft-NN contains the defect probability and leakage flag data of round $r$, while for the (hard-)NN it only contains the defect data. The vector $\vec{x}[R+1]$ contains the final-round defects. The output $p$ is a probability that estimates if a logical error has happened and $p_{\text{aux}}$ is only used to help in the training (see Section 7.6.3). (b) Logical fidelity of the NN decoder as a function of the number of rounds $R$, shown for each physical computational state that makes up $|0\rangle$ (transparent curves) and averaged across all states (opaque curves). (c) Logical fidelity at $R = 8$, shown for each initial computational state.

## 7.4  Conclusions

We have experimentally demonstrated the benefits of using soft information in the decoding of a $= 3$ bit-flip surface code, utilizing $13$ qubits. With soft information, the NN decoder achieves the best extracted logical error rate of $4\;73\%$. It is crucial to note that, as we do not measure the -basis stabilizers of the typical distance-$3$ surface code, the extracted is likely underestimated (compared with of $5\;4\%$ for the standard $= 3$ surface code by Ref. [141] without leakage post-selection). However, the nature of the decoding problem will be the same and will benefit from the decoding optimizations explored in this paper.

Despite the modest improvement in this work, simulations [202] suggest further advantages of soft information decoding: improvement in the error correction threshold, and increased suppression of logical errors as the code distance increases. We also have not leveraged leakage information on measured qubits with the MWPM decoder. Implementation of a leakage-aware decoder [81] could potentially significantly enhance the logical performance with MWPM, but this has yet to be explored in experiment. With the NN decoder, it is unclear if the defect probabilities and leakage flags are the optimal way to present the information to the network, and this could be the subject of further investigation. Finally, our experiments utilized readout durations that were optimized for measurement fidelity. Calibrating the measurement duration for optimal logical performance [202] instead could potentially lead to higher logical performance.

## 7.5  Data availability

The source code of the NN decoder and the script to replicate the results are available in Ref. [217] and Ref. [218], respectively. The script to analyze the experimental data can be found in Ref. [218].

## 7.6  Supplemental material

This supplement provides additional information supporting statements and claims made in the main text.

### 7.6.1   Device overview

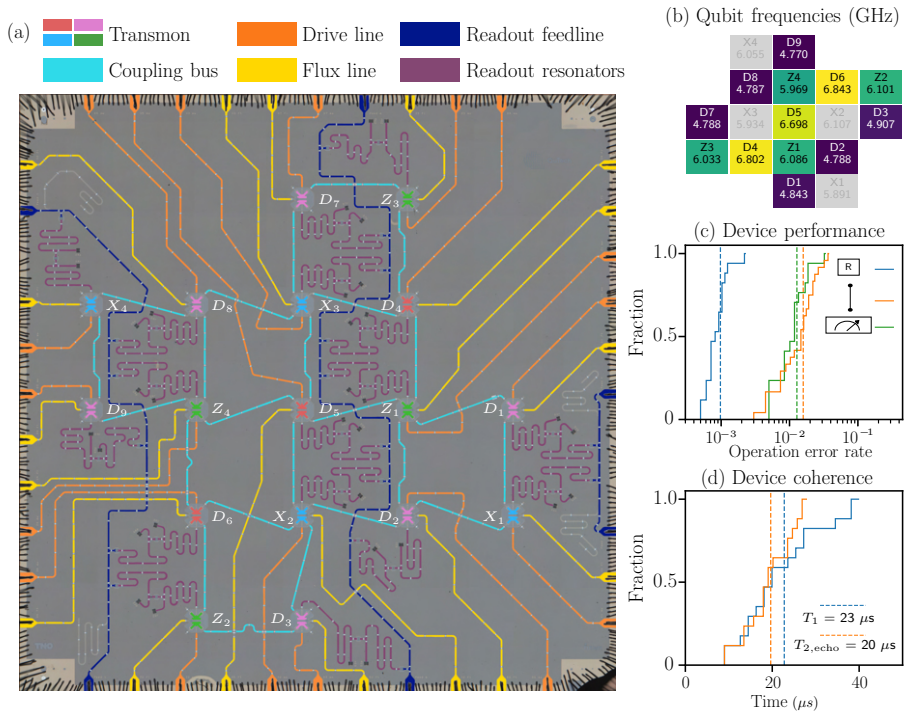

Figure 7.5: **Device characteristics**. (a) Optical image of the 17-transmon device, with added false color to emphasize different circuit elements. The device is connected to a printed circuit board using aluminum wirebonds, visible at the edges of the image. (b) Measured qubit transition frequencies with all transmons biased to their flux sweetspot. $X$-basis ancilla qubits (light gray) are not used in this experiment. (c) Cumulative distribution of error rates for single- and two-qubit gates, obtained by randomized benchmarking protocols with modifications to quantify leakage [97, 194], and average readout assignment fidelities, extracted from single-shot readout histograms [131]. (d) Cumulative distribution of measured qubit relaxation time $T_1$ and echo dephasing time $T_{2,\mathrm{echo}}$. Dashed lines in (c) and (d) indicate the average over the 13 qubits used and the 12 two-qubit gates.

Our 17-transmon device [Figure 7.5(a)] consists of a two-dimensional (2D) array of 9 data qubits and 8 ancilla qubits, designed for the distance-3 rotated surface code. Qubit transition frequencies are organized into three frequency groups: high-frequency qubits (red), mid-frequency qubits (blue/green), and low-frequency qubits (pink), as required for the pipelined QEC cycle proposed in Ref. [98]. Each transmon has a microwave drive line (orange) for single-qubit gates, a flux-control line (yellow) for two-qubit gates, and a dedicated pair of resonator modes (purple) distributed over three feedlines (blue) for fast dispersive readout with Purcell filtering [102, 131]. Nearest-neighbor transmons are coupled via dedicated coupling resonators (sky-blue) [138]. Grounding airbridges (light gray) are fabricated across the device to interconnect the ground planes and to suppress unwanted modes of propagation. These airbridges are also added at the short-circuited end of each readout and Purcell resonator, allowing post-fabrication frequency trimming [55]. After biasing all transmons to their flux sweetspot, the measured qubit frequencies clearly exhibit three distinct frequency groups, as depicted in Figure 7.5(b). These values are obtained from standard qubit spectroscopy. The average relaxation ( ) and dephasing ( ) times of the 13 qubits used in the experiment are $23$ and $20$ s, respectively [Figure 7.5d].

To counteract drift in optimal control parameters, we automate re-calibration using dependency graphs [155]. The method, nicknamed graph-based tuneup (GBT) [158], is based on Ref. [156]. Single-qubit gates are autonomously calibrated with DRAG-type pulses to avoid phase error and to suppress leakage [64, 65], and benchmarked using single-qubit randomized benchmarking protocols [163]. The average error of the calibrated single-qubit gates [Figure 7.5(c)] across $13$ qubits reaches $0$ $1\%$ with a leakage rate of $10^-$ . All single-qubit gates have $20$ ns duration.

Two-qubit controlled-  (CZ) gates are realized using sudden Net-Zero flux pulses [69]. The QEC cycle in this experiment requires 12 CZ gates executed in 4 steps, each step performing 3 CZ gates in parallel. This introduces new constraints compared to tuning an individual CZ gate. For instance, parallel CZ gates must be temporally aligned to avoid overlapping with unwanted interaction zones on the way to, from, or at the intended avoided crossings. Moreover, simultaneous operations in time (vertical) and space (horizontal) may induce extra errors due to various crosstalk effects, such as residual      coupling, microwave cross-driving, and flux crosstalk. To address these non-trivial errors, we introduce two main calibration strategies into a GBT procedure: vertical and horizontal calibrations (VC and HC). These tune simultaneous CZ gates in time and space as block units [136]. This approach absorbs some of the flux and residual-     crosstalk errors. After calibration, GBT benchmarks the calibrated gates with two-qubit interleaved randomized benchmarking protocols with leakage modification [97, 194]. The individual benchmarking of the $12$ CZ gate reveals an average error of $1$ $6\%$ with a $0$ $24\%$ leakage. All CZ gates have $60$ ns duration.

Readout calibration is performed in three main steps, realized manually and not with a GBT procedure. In the first, readout spectroscopy is performed at fixed pulse duration ($200$ ns-$300$ ns) to identify the optimal frequency maximizing the distance between the two complex transmission vectors   $|$ $\rangle$ and $|$ $\rangle$ in the IQ plane. The second step involves a 2D optimization over pulse frequency and amplitude. The goal is to determine readout pulse parameters that minimize a weighted combination of readout assignment error ( ), and measurement quantum non-demolition (QND) probabilities (      and      ). These probabilities are obtained using the method of Ref. [168]. The final step verifies if photons are fully depleted from the resonator within the target total readout time, $420$ ns, using an ALLXY gate sequence between two measurements [18]. By comparing the ALLXY pattern obtained to the ideal stair-case, we can determine if the time dedicated for photon depletion is sufficient to not affect follow-up gate operations.

After calibrating optimal readout integration weights [18], we proceed to benchmark various readout metrics such as        and standard readout QND (      ) using the measurement butterfly technique [175]. The average       [Figure 7.5(c)] is $1\,2\%$, extracted from the single-shot histograms. We also perform simultaneous multiplexed readout of all 13 qubits, constructing assignment probability and cross-fidelity matrices [102]. The average multiplexed readout error rate is $1\,6\%$, indicating that readout crosstalk is small. Moreover, the average        for the four    -basis ancillas is $95\,3\%$ considering a three-level transmon [175]. This also yields an average leakage rate due to ancilla measurement of $0\,14\%$, predominantly from $1$ .

### 7.6.2  Benchmarking of parity checks

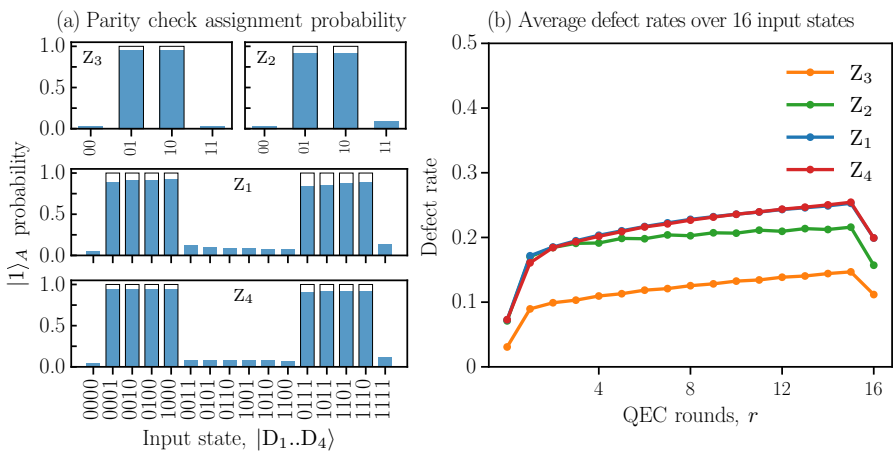

Figure 7.6: **Parity check benchmarking**. (a) Benchmarking of the assignment fidelity for four stabilizer measurements:      $_4$    $_7$'    $_6$    $_3$'    $_4$    $_5$    $_2$    $_1$, and     $_9$    $_9$    $_5$    $_6$. (b) The average defect rate as a function of QEC rounds for each of the four    -basis stabilizers across the 16 input states.

With the individual building blocks calibrated, we proceed to calibrate the four    -basis stabilizer measurements as parallel block units using VC and HC strategies, as discussed in Section 7.6.1. The average probabilities of correctly assigning the parity operator        are measured as a function of the input computational states of the data-qubit register. The measured probabilities (solid blue bars) [Figure 7.6(a)] are compared with the ideal ones (black wireframe) to obtain average parity assignment fidelities, $96\,3\%$, $92\,8\%$, $89\,9\%$, and $92\,3\%$ for $Z$ , $Z$ , $Z$ , $Z$ , respectively. These results are obtained after mitigating residual excitation effects by post-selection on a pre-measurement [175].

The defect rates (see the main text for the definition of a defect), reflect the incidence of physical qubit errors (bit-flip and readout errors) detected throughout the rounds $r$, where $r \in \{1, 2, \dots\}$. For each of the four $Z$-basis stabilizers, the defect rate [Figure 7.6(b)] is presented over 16 QEC rounds and averaged across the 16 physical computational input states. The sharp increase in the defect rate between rounds $r = 1$ and $r = 2$ is due to the low initialization error rates and the detection of errors occurring during the ancilla-qubit measurements in the first round. At the boundary round $r = 16$, the defects are obtained using the final data-qubit measurements, which, given the low readout error rates, lead to the observed decrease in the defect rate. Over rounds, defect rates gradually build up until leveling at approximately $15\%$, $22\%$, $25\%$, and $25\%$ for $Z_1$, $Z_2$, $Z_3$, $Z_4$, respectively. The build-up is attributed to leakage [25, 26, 141, 175].

### 7.6.3 Details of the neural network decoder

*NN inputs, outputs and decoding success*

The inputs provided to the NN decoder consist of the defects, the defect probabilities, and the leakage flags [see Figure 7.4(a)]. The only element not described in the main text is the defect probabilities. These are obtained following Ref. [198] and using the two-state readout classifier from the main text and Section 7.6.5. First, we express the probability of the measured qubit 'having been in the state' $m$ ($m \in \{0, 1\}$) given the IQ value $c$ as

$$P(m \mid c) = \frac{P(c \mid m)\, P(m)}{P(c \mid 0)\, P(0) + P(c \mid 1)\, P(1)} \tag{7.3}$$

with $P(m)$ the probability that the qubit was in state $m$. Recall that the detectors in the bulk are defined as $d_r = m_r \oplus m_{r-1}$, and that they have error-free values of 0. We estimate the probability that the defect $d_r$ has been triggered given $c_r$ and $c_{r-1}$ (named *defect probability*) by

$$P(d_r = 1 \mid c_r, c_{r-1}) = P(m_r = 0 \mid c_r)\, P(m_{r-1} = 1 \mid c_{r-1}) \\ + P(m_r = 1 \mid c_r)\, P(m_{r-1} = 0 \mid c_{r-1}) \tag{7.4}$$

where $m_r$ refers to the state of the ancilla qubit $q$ just before the measurement at round $r$. Note that we have used $P(m_r)$ instead of $P(d_r)$ as the latter is always $0$ or $1$ because $d_r$ is completely determined by $c_r$. Using $P(m_r)$ allows us to infer the reliability of the defects, e.g. $P(m_r = 0 \mid c_r) \approx P(m_r = 1 \mid c_r)$ leads to $P(d_r = 1 \mid c_r, c_{r-1}) \approx 1/2$ (in between 0 and 1, thus unsure). Remember that we do not use the defect probabilities of the final round as explained in Section 7.6.3 and that we assume $P(0) = P(1) = 1/2$ when calculating $P(d_r = 1 \mid c_r, c_{r-1})$.

For completeness, we study the performance of the NNs given four combinations of inputs: (a) defects, (b) defects and leakage flags, (c) defect probabilities, and (d) defect probabilities and leakage flags. The results in Figure 7.7 show that the networks can process the richer inputs to improve their performance. The reason for not giving the network the soft measurement outcomes as input directly (but the defect probabilities instead) is that we have found that the NN decoder does not perform well on the soft measurements [198]; effectively, the NN has to additionally learn the defects, which is possible with larger NNs as in Ref. [203].

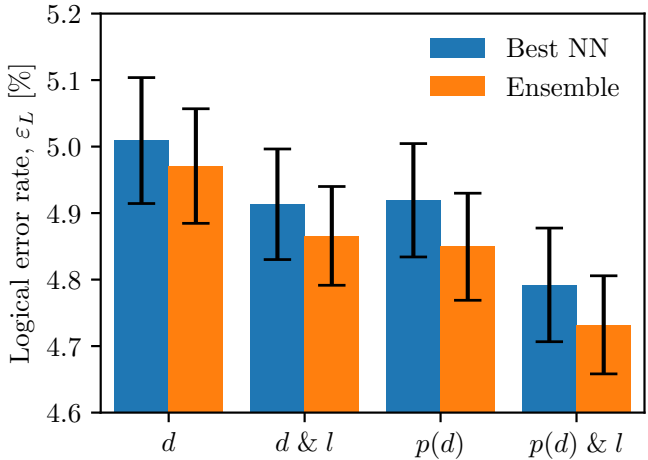

Figure 7.7: Logical error rates of the NN decoders when given different inputs: the label $d$ corresponds to defects, $l$ to leakage flags, and the label $p(d)$ corresponds to defect probabilities given the soft information. "Best NN" (blue) and "Ensemble" (orange) correspond to the architectures with the lowest logical error rate for a single NN and for an ensemble of 5 NNs, respectively, as described in Section 7.6.3.

The NN gives two outputs, $p_L$ and $p_L^{\prime}$, which correspond to the estimated probabilities that a logical flip has occurred during the given sample. The output $p_L$ has been calculated using all the information given to the NN, while $p_L^{\prime}$ does not use the final round data; see Figure 7.4(a). We train the NN based on its accuracy in both $p_L$ and $p_L^{\prime}$ because the latter helps the NN to not focus only on the final round but decode based on the full QEC data [198, 219]. Note that the outputs correspond to physical probabilities (i.e., $p_L, p_L^{\prime} \in [0, 1]$) thanks to using sigmoid activation functions in the last layer of the NN architecture.

To determine whether the neural network decoded the QEC data correctly or not, the output $p_L$ is used as follows. If $p_L \geq 1/2$, we set the logical flip bit to $f = 1$; if $p_L < 1/2$, we set the logical flip bit to $f = 0$. On the basis of the final data qubit measurements, we compute the (uncorrected) logical $m$ as a bit $m \in \{0, 1\}$. We take the logical input state $n$ (in our experiments $n = 0$ always as we prepare $|0\rangle_L$ in all cases) and mark the run as successful when $m \oplus f \oplus n = 0$, and unsuccessful when $m \oplus f \oplus n = 1$.

*Learning the final logical measurement*

In machine learning, one needs to be careful about the information given to the network. For example, the neural network could predict the logical bit flip correction   without using the defect information gathered over multiple rounds. In particular, given that in our experiment we only start with state $0$, the neural network could just output the value of   such that   $= 0$. It can do this by learning       , and hence the network should get no explicit information about       . While it is true that we do not directly provide the data-qubit measurement outcomes to the NN decoder, there still might be some partial information provided by the defect probabilities or the leakage flags, e.g. when a qubit in state $2$ is more likely declared as a $1$ than a $0$; see Figure 7.10 for qubit  . Note that this issue does not occur if one randomly trains and validates with either $0$ or $1$  since then the optimal   $=$       and   is a random bit unknown to the network.

We indeed observe an abnormally high performance for a single network when using soft information in the final round, with   $4\,2\%$ when using defect probabilities and   $4\,1\%$ if we include the leakage flags too. Such an increase suggests that the NN can partially infer   and thus 'knows' how to set  . This phenomenon does not occur when giving only the leakage flags of data qubit measurements, which instead leads to   $4\,9\%$. Nevertheless, due to the reasons explained above, we decided to not include the leakage flags in the final round to be sure we do not provide any information about   to the NN.

We note that Ref. [203] also opted to just give the defects in the final round because of the possibility that the NN decoder can infer the measured logical outcome from the final-round soft information.

*Ensembling*

Ensembling is a machine-learning technique to improve the network performance without costing more time, as networks can be trained and evaluated in parallel [216]. It consists of averaging the outputs,   , of a set of NNs, to obtain a single more accurate prediction,   . One can think that the improvement is due to "averaging out" the errors in the models [216]. Reference [203] trained 20 networks with different random seeds and averaged their outputs with the geometric mean. In this work, the output 'average',   , is given by

$$\log\left(\frac{1}{\phantom{xxx}}\right) = \frac{1}{5}\sum\log\left(\frac{1}{\phantom{xxxx}}\right) \tag{7.5}$$

with   the predictions of 5 individual NNs of the logical flip probability; see Section 7.6.3. This expression follows the approach from the repeated qubit readout with soft information [220], which is optimal if the values are independently sampled from the same distribution. Once   is determined, we threshold it to set the flip bit   as described in Section 7.6.3.

*Network sizes, training hyperparameters and dataset*

Due to the different amounts of information in each input, the NNs in Figures 7.4 and 7.7 have different sizes to maximize their performance without encountering overfitting issues. In Figure 7.8, the size of the network is increased given a set of inputs until overfitting degraded the performance or there was no further improvement. The specific sizes and hyperparameters of the NNs shown in the figures are summarized in Table 7.1. These hyperparameters are the same as in Ref. [198] but with the following changes: (1) reducing the batch size to avoid overfitting, as the experimental dataset is smaller, and (2) decreasing the learning rate for the large NNs. For comparison, the NN in Ref. [203] for a $d = 3$ surface code uses 5.4 million free parameters, a learning rate of $3\,5 \cdot 10^{-}$, and a batch size of 256. The number of free parameters in a NN is related to its capacity to learn and generalize from data, the learning rate is related to the step size at which NN parameters are optimized, and the batch size is the number of training samples used in a single iteration of gradient descent.

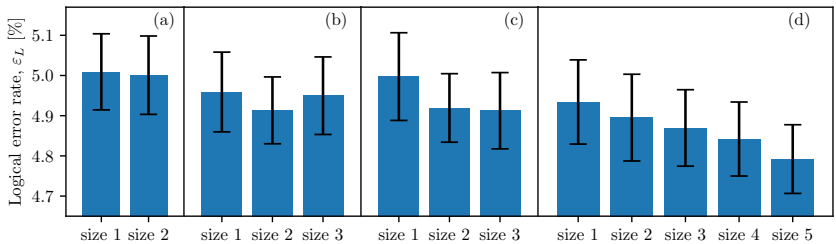

Figure 7.8: Logical performance of the NNs given the four input combinations and different network sizes. The NN input $[\ ]$ in panel (a) are the defects, in (b) the defects and leakage flags, in (c) the defect probabilities, and in (d) the defect probabilities and the leakage flags. The NN sizes are summarized in Table 7.1.

| Label | Enc | Enc | LSTM | LSTM | Eval | Eval | # free parameters | batch size | learning rate | dropout rate |
|-------|-----|-----|------|------|------|------|-------------------|-----------|---------------|--------------|
| size 1 | | | 2 | 90 | 2 | 90 | $\sim 115$k | 64 | $5 \cdot 10^{-4}$ | 20% |
| size 2 | 2 | 32 | 2 | 100 | 2 | 100 | $\sim 160$k | 64 | $2 \cdot 10^{-4}$ | 20% |
| size 3 | 2 | 64 | 2 | 120 | 2 | 120 | $\sim 250$k | 64 | $2 \cdot 10^{-4}$ | 22% |
| size 4 | 2 | 90 | 3 | 100 | 2 | 100 | $\sim 285$k | 64 | $2 \cdot 10^{-4}$ | 22% |
| size 5 | 2 | 100 | 3 | 100 | 2 | 100 | $\sim 290$k | 64 | $2 \cdot 10^{-4}$ | 20% |

Table 7.1: NN sizes and the training hyperparameters used in this work. refers to the number of layers in block Enc LSTM Eval and the dimension of these layers. If $_{\text{Enc}}$ is not specified, the network does not have Encoding layers. The blocks are shown in Figure 7.4(a). The number of free parameters depends on the given input combination, but the changes are of the order of 5k.

The splitting of the experimental dataset into the three sets of *training*, *validation*, and *testing* was done as follows. For each initial state and each number of rounds, (1) randomly pick $5\,10$ samples from the given data and store them in the *testing* dataset, (2) randomly

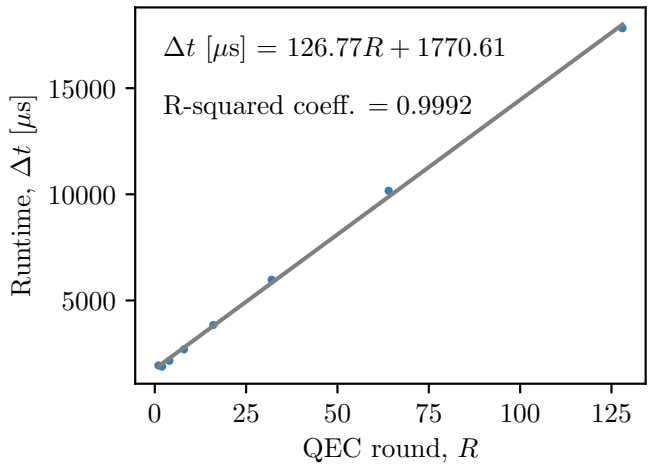

$$\Delta t \, [\mu s] = 126.77R + 1770.61$$

R-squared coeff. $= 0.9992$

Figure 7.9: Evaluation runtime for the size-5 network with batch size = 1. The line corresponds to a linear regression where the R-squared coefficient shows that the fit is appropriate. The inputs for the NN in this calculation are created at random (not based on experimental data) because we are not interested in the logical performance and the number of operations the NN needs to perform only depends on the number of rounds. In particular, the number of operations in the LSTM and encoding layers grows linearly with , but for the evaluation layers it is constant. Therefore, we can associate the   intercept as the time required for the evaluation layers and the slope as the time spent on the encoding and LSTM layers. Each point is the average of $5 \quad 10$   samples.

select 90% of the remaining samples and store them in the *training* dataset, and (3) store the rest in the *validation* dataset. The reason for this choice is to ensure that the datasets are not accidentally biased towards an initial state or number of rounds. After the splitting, we have a training dataset consisting of $6 \ 9 \quad 10$   samples, a validation dataset with $7 \ 6 \quad 10$  , and a testing dataset with $4 \quad 10$  . Note that the NN has been trained and tested on the same number of rounds because the experiment only goes until $= 16$. However, in longer memory experiments, the NN should be trained only up to a "low" number of rounds to avoid long trainings.

The training for each single NN was carried out on an NVIDIA Tesla V100S GPU and lasted around 10 hours for the smallest size and 23 hours for the largest one when using the training dataset consisting of $6 \ 9 \quad 10$   samples with $7 \ 6 \quad 10$   samples for validation. The evaluation of the network performance was done on an Intel Core(TM) i7-8650U CPU @ 1.90GHz  4. We estimate that it takes   $127$  s per QEC cycle and $1 \ 77$ ms for the final round; see Figure 7.9. For comparison, the NN in Ref. [203] for a  $= 3$ surface code takes   $20$  s to decode a QEC round, but it was evaluated on a Tensor Processing Unit (TPU).

### 7.6.4 The decoding graph

In this Appendix, we describe in further detail the effect of not using mid-circuit resets on the decoding graph. Here, we consider specifically Surface-13, but similar ideas apply to the full surface code and other stabilizer codes.

As discussed in the main text, in order to detect errors in stabilizer codes, it is typical to define detectors, combinations of binary measurement outcomes that have deterministic values in the absence of errors. We refer to detectors whose value has flipped from the expected error-free value as defects. A decoder takes observed defects in a particular experiment and, using a model of the possible errors and the defects they result in, predicts how the logical state has been affected by the errors. It is desirable for errors to result in a maximum of two defects, as this enables a matching decoder to be used, which can efficiently find the most probable [204, 209, 210] errors that cause the observed defects.

In Surface-13, the stabilizers are, in the bulk, weight-four $\quad$ operators. On the boundaries, the stabilizers are of lower weight. In general, each data qubit is involved in two Z-type stabilizers, or one on the boundary. The experiment proceeds by using ancilla qubits to measure the stabilizers for some number of rounds, $\quad$, with $\quad$ the $\quad$th stabilizer outcome in the $\quad$th round. An $\quad$ error on a data qubit in round $\quad$ will change the value of the outcomes of the stabilizers involving that data qubit from round $\quad$ onwards. We define the detectors $\quad$ to be the difference between stabilizer measurements in adjacent rounds, so that

$$\quad = \quad \; - \; \quad \tag{7.6}$$

We begin Surface-13 with initial data qubit states that are eigenstates of the Z-type stabilizers with eigenvalues +1; therefore, we set $\quad = 0$. At the end of the experiment, the data qubits are measured in the Z-basis; these measurements can be used to construct outcomes for the stabilizers.

With the definition in Equation (7.6), an $\quad$ data qubit error before a QEC round results in a maximum of two defects, as desired. We note that, in general, there may be multiple errors that result in the same defect signature. The probabilities of these errors are typically combined to give a single edge weight.

We now consider the effect of ancilla qubit and measurement classification errors. An ancilla qubit error in round $\quad$ on the qubit used to measure the $\quad$th stabilizer will change the stabilizer measurement outcome in the $\quad$th round, thus resulting in defects $\quad$ and $\quad$. The effect of classification errors depends on how the stabilizer outcomes are obtained. If the ancilla qubits are reset after measurement, $\quad = \quad$, where $\quad$ is the hardened measurement outcome on the $\quad$th ancilla at round $\quad$, and therefore a classification error in round $\quad$ results in the same defects as an ancilla qubit error in round $\quad$. However, if the ancilla qubits are not reset after measurement, as is the case in this paper,      =           _ . Therefore, with respect to measurements, the detectors are defined as

$$= \qquad \_ \tag{7.7}$$

for $2$           . In this case, a classification error in measurement      causes defects and         . In both cases, a data qubit measurement classification error looks like a data qubit Pauli error between rounds     and    $+ 1$. We also define      =      .

In this error model, we assume that two independent events are possible during measurement – an ancilla qubit error and a classification error. In practice, these are not independent events as both are affected by      processes. However, should an error occur that causes the qubit to decay from the $1$ to the $0$ state during measurement, it can either be viewed as an ancilla qubit error before measurement (if the inferred hardened measurement outcome is 0) or an ancilla qubit error after measurement (if the inferred hardened measurement outcome is 1). Therefore, these coupled events can be viewed as a single ancilla qubit error. We note, however, that such qubit errors are not symmetric, and thus using a single edge weight is an approximation.

We note that we have not discussed mid-round qubit errors that result in so-called hook errors, as we have focused on explaining the differences between the decoding graphs with and without mid-circuit reset. The hook errors in the two cases will be identical.

*Noise-floor graph*

The lower-bound error parameters used in the construction of the experimentally-derived decoding graph are given in Table 7.2. The operation times are taken to be the same as the real device. We note, however, that the other parameters are not the same as those stated in the previous section Section 7.6.1. This is because the parameters here set a lower bound on the error probabilities and should only be used when the pairwise correlation method gives unfeasibly low values. Whilst extensive exploration of the parameters was not undertaken, the ones stated here were found to give good performance and several other options did not result in significant changes to the results.

Each probability,    , is the probability of a depolarising error after the specified operation. These are defined so that, for single-qubit gates, resets and measurements, the probability of applying the Pauli error      is given by

$$= \overline{3} \tag{7.8}$$

for                  . For two-qubit gates, the probability of applying the Pauli      is given by

$$= \overline{15} \tag{7.9}$$

where $\qquad$, excluding $\quad = \quad = \quad$. Idle noise is incorporated by Pauli twirling the amplitude damping and dephasing channel to give, for an idling duration $\tau$, [221]

$$( ) = \quad ( ) = \frac{1}{4} \quad 1 \quad ^{-} \quad ^1 \tag{7.10}$$

$$( ) = \frac{1}{2} \quad 1 \quad ^{-} \quad ^2 \quad \frac{1}{4} \quad 1 \quad ^{-} \quad ^1 \tag{7.11}$$

We note that, as $\quad = \quad$ in our model, $( ) = ( ) = ( )$. The noise-floor graph is derived from the circuit containing the appropriate parameters using Stim [212].

| Parameter | Value |
|---|---|
| single-qubit gate error probability | $0\,5\quad 10^{-}$ |
| two-qubit gate error probability | $5\quad 10^{-}$ |
| reset error probability | $0\,0$ |
| measurement qubit error probability | $1\quad 10^{-}$ |
| measurement classification error probability | $1\quad 10^{-}$ |
|  | $30\quad$ s |
|  | $30\quad$ s |
| single-qubit gate time | $20$ ns |
| two-qubit gate time | $60$ ns |
| measurement time | $420$ ns |

Table 7.2: Lower-bound noise parameters used in the experimental graph derivation. These parameters are used to fix the minimum values of each edge.

### 7.6.5 Soft information processing

The processing of the soft information [i.e., measurement edge weights in Equation (7.1), the defect probabilities in Equation (7.4), and leakage flags], all use the probability density functions $( \quad )$ which can be found from experimental calibration. These PDFs are obtained by fitting a readout model to the readout calibration data, consisting of a set of IQ values for each prepared state $\quad$. The performance of the soft decoders is limited by the accuracy of the readout models used, thus in this section we describe the models employed in this paper and their underlying assumptions about the qubit readout response.

*Probability density function fits*

Each of the 13 transmons in the device has a characteristic measurement response in IQ space, requiring a unique PDF to be fitted for each. In this section, we detail two models of PDFs used for this purpose, comparing their logical performance when decoding soft information using the MWPM decoder. The first model, described in Section 7.6.5 is an implementation of the amplitude damping noise model presented in [202]. The second model, described in Section 7.6.5, is a heuristic model representing a Gaussian mixture that generalizes for an arbitrary number of distinct states including leakage to $2$. We find that the heuristic model performs equally well to the physics-derived model and enables an easy leakage probability calculation which is used as neural network input. We exclusively use the Gaussian mixture model in the results presented in the main text. The amplitude damping noise model as detailed below is given for future reference.

AMPLITUDE DAMPING NOISE MODEL    Here we consider an amplitude damping model that more accurately captures the     processes that affect the IQ response over the course of measurement [202, 203]. However, we found it led to no statistically significant improvement in classification performance for our device. Furthermore, as the amplitude damping noise model does not easily generalize to leaked states, and is sensitive to state preparation errors in the calibration stage, we opt for the Gaussian mixture model. Nevertheless, we present it here for completeness.

We first project the 2D shots for each qubit along an axis of symmetry     $=$          where     for     $0$ $1$  is the mean in IQ space of the experimental distributions for state     . In this transformation, vectors         $\mathbb{R}$  are mapped to the one-dimensional vector         $\mathbb{R}$. The projection does not result in significant information loss as long as the IQ responses for $0$ and $1$ share the same fluctuation strength in all directions, modeled as 2D Gaussian distributions with standard deviation     $=$          . In the presence of leakage to $2$ , the IQ response is no longer symmetric around a single axis [146] and classification must be performed with the full 2D dataset rather than a projection. We fit the ground state probability density function     $(\quad 0)$ in projected coordinates using a Gaussian distribution of form

$$( \; \qquad 0) = \frac{\overline{\quad}}{\quad^2} \exp \quad \frac{}{\quad^2} (\qquad )  \tag{7.12}$$

where     is the mean of the Gaussian in projected coordinates with a standard deviation     $= 1$ $\overline{\quad}$. A qubit where the readout time is short will see a wider distribution in     $(\quad 0)$ due to the increased effect of fluctuations in the integrated IQ signal. The model response of the $1$ state is given [202] by

$$( \; \qquad 1) = \frac{\overline{\quad}}{\quad^2} \exp \quad \frac{}{\quad^2} (\qquad ) \qquad \frac{}{\quad^4} \exp \quad \frac{}{\quad^8} + \frac{}{\quad^2} (\qquad )$$
$$\mathrm{erf} \quad \frac{\overline{\frac{1}{\quad^8}} + (\qquad )}{} \qquad \mathrm{erf} \quad \frac{\overline{\frac{1}{\quad^8}} + (\qquad )}{}  \tag{7.13}$$

where     is the mean of a Gaussian in projected coordinates in the absence of amplitude damping and     controls the strength of the amplitude damping process via     $= \frac{}{\quad_1}$ given a measurement time     and amplitude damping time     . For small     we have a     $(\quad 1)$ distribution that is a Gaussian centred at     with standard deviation     $= 1$     $\overline{\quad}$, while increasing     leads the distribution to decay and move closer to     .

The primary limitation of the amplitude damping model is that it does not easily generalize to further excited states. The mean measurement response of a $2$ state is not guaranteed to be on the same axis in IQ space as $0$ and $1$, hence a projection to one dimension does not allow one to accurately distinguish the $2$ state. While a generalization of the amplitude damping noise model to the $2$ state is possible, it requires a larger number of fitting parameters than the 1D projected model. A secondary problem is that, in the presence of state preparation errors, the calibration response of a $0$ state may include some shots where the qubit is instead in the $1$ state and vice versa. This can be accounted for by fitting a linear combination of the $(\quad 0)$ and $(\quad 1)$ states instead of only using one of the distributions. Furthermore, with the amplitude damping model, it becomes more difficult to separate the pure classification error from qubit errors.

TWO-STATE GAUSSIAN MIXTURE MODEL   For discrimination between $0$ and $1$, we can use projected coordinates as defined in Section 7.6.5, where the PDFs $(\quad)$ for $0\ 1$ have the form

$$(\quad) = (1\quad)\ (\ ;\quad) + \ (\ ;\quad) \tag{7.14}$$

Here, $(\ ;\quad)$ is a 1D Gaussian distribution with mean $\quad$ and standard deviation $\quad$, and $[0\ 1]$ is an amplitude parameter that determines which normal distribution is dominant in the mixture. For $\ = 0\quad = 1$ the model represents a readout response with a single dominant component (i.e. no state preparation errors), while $\quad 0\quad\quad 1$ represents a measurement response where, due to state preparation errors, there are two distinct components to the measurement response.

The Gaussian mixture model allows us to discard state preparation errors from the $(\quad 0)$ by fitting the parameters $\quad$ and $\quad$ for $\quad$ from Equation (7.14) and then setting $\ = 0$. This assumption holds on the condition that no $0\quad\quad 1$ processes are present over the course of the measurement time – if significant amplitude damping occurs over the course of the measurement, the PDF found using the Gaussian method is inaccurate. When comparing experimental results for logical fidelity with and without setting $\ = 0$ for the ground state distribution $(\quad 0)$ we find no statistically significant difference. We assume the absence of a fidelity improvement is due to the rarity of $0$ state preparation errors which are mitigated by heralded initialization.

As mentioned in the main text, whilst we include both Gaussians in the PDF in order to classify the measurement, we use only the main peak in calculating the soft edge weights. This removes the component of qubit error that occurs during measurement from the edge associated with measurement classification error.

We find minimal difference between the logical error rates for the Gaussian mixture and the amplitude-damping PDFs. Using the MWPM decoder, the logical error rate for the amplitude damping model as calculated via Equation (7.2) is found to be $(5\ 30\quad 0\ 18)\%$ when decoding hard information and $(4\,94\quad 0\,16)\%$ for soft information. The result is nearly identical for the Gaussian mixture, with $(5\,30\quad 0\,18)\%$ and $(4\,94\quad 0\,15)\%$ for hard and soft information, respectively. This observation gives us confidence that the Gaussian PDF is no worse as an approximation of the device physics for the $13$ qubits used.

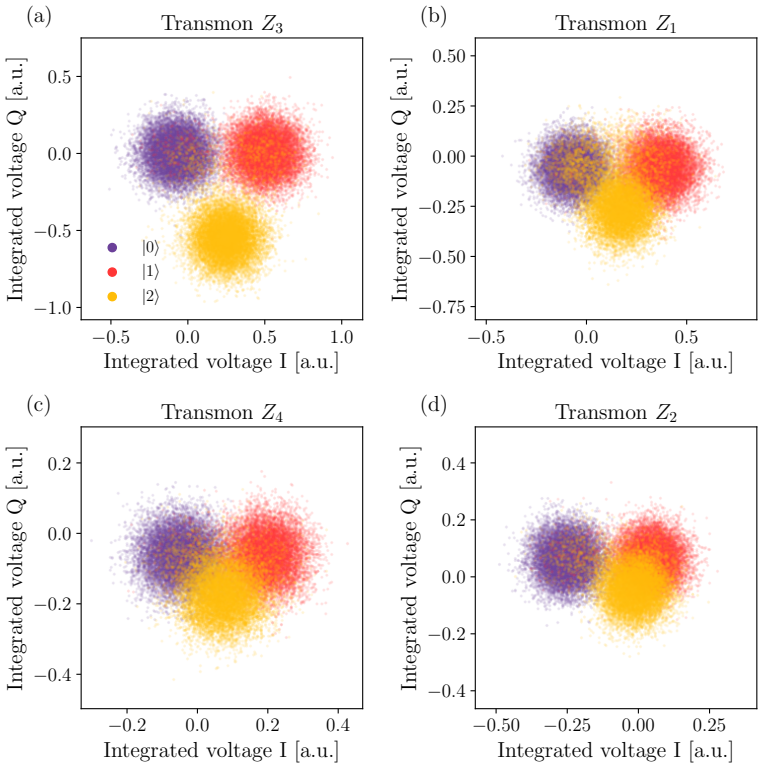

Figure 7.10: Calibration shots for experimental IQ voltages, shown for each ancilla prepared in $0$, $1$ and $2$. The cluster for $2$ is off-axis when compared to the clusters for $0$ and $1$, and its separation from the two other clusters varies from transmon to transmon. State $0$ has the cleanest response, as it does not decay to any other state, while $1$ and $2$ decay partially to lower energy states over the course of measurement.

THREE-STATE CLASSIFIER   The 1D projected model is unable to characterize leakage to $2$, which has its own characteristic response in the two-dimensional IQ space, shown in Figure 7.10. To model this three-state regime and discriminate leakage, we fit a mixture of 2D Gaussians to normalized histograms of the calibration data, giving PDFs $(\quad)$ for $\mathbb{R}$, $0\,1\,2$ as follows:

$$(\quad) = (\;\quad) + (\;\quad) + (\;\quad) \tag{7.15}$$

where $(\;\quad)$ is the PDF of a 2D Gaussian distribution with mean and covariance matrix, and parameters, and are to be fitted for each state.

In Figure 7.10, we observe that the $2$ state has a measurement response that is off the axis, forming a distinct constellation in the IQ space below the other two measurement responses $0$ and $1$. The ground state response is centred around , while the $1$ response is distributed between a dominant peak around and a small number of data points closer to , indicating $1 \to 0$ decay. The response of the $2$ state can be seen to decay to both $0$ and $1$ states, most notably for transmon where the effect can be clearly seen.

Given this simple model, the three-state classifier used to set the leakage flags as input for the NN decoder works as follows. Maximum likelihood classification means that given , we should pick $= 0$ $1$ $2$ which maximizes $(\ )$ in Equation (7.3). The denominator in Equation (7.3) can be dropped as it solely depends on . For the numerator, we need to know $(\ )$ which we assume to be independent of (which is not completely warranted as $= 2$ is much less likely), and hence $(\ )$ $(\ )$ with $(\ )$ in Equation (7.15). These arguments identically apply to the two-state classifier discussed in the main text.

*Combining soft information with the pairwise correlation method in the final round*

As discussed in the main text, we obtain the decoding graph edge weights from experimental data. These weights will include some averaged probability of a classification error that we wish to remove and replace with a soft-information-based weight on a per-shot basis. In the bulk of the experiment, this is straightforward – the weight of the edge corresponding to a classification error can simply be replaced with that calculated using the soft information, following Equation (7.1), as classification and qubit errors have different defect signatures.

However, in the final round, both ancilla and data qubit errors result in the same defects as classification errors. Therefore, we expect the total (averaged across shots) edge probability to be given by

$$= (1\ \ ) + (1\ \ ) \qquad\qquad (7.16)$$

where is the averaged probability of any classification error and is the probability of qubit errors. The validity of this assumption depends on the degree to which we have made correct assumptions about the possible error channels, including that there is no correlated noise. In our case, is obtained by the pairwise correlation method, but it could be obtained by other means. We note that, in general, errors on multiple qubits may result in the same defects and thus contribute to the same edge probability. In our Surface-13 experiment, this only occurs for certain final-round data qubit measurements – the pairs D1 and D2, and D8 and D9.

We wish to retain the edge contributions due to the qubit errors, , and replace only the averaged classification contribution, , with the per-shot value. We thus have several steps to calculate the soft-information-based weight for final-round edges:

1. For each measurement , calculate the mean classification error, , by averaging the per-shot errors.

2. For each edge, calculate the total edge classification error, , from the individual measurement classification errors, , through

$$= \frac{1}{2}\left[1 - \prod(1 - 2)\right] \tag{7.17}$$

where the product is over all measurements whose classification errors result in the edge defects.

3. For each edge, remove the mean classification error from the edge probability by rearranging Equation (7.16) to find

$$= \frac{}{1 - 2} \tag{7.18}$$

4. For each edge, include the per-shot classification error calculated from the soft readout information by combining it appropriately with .

We now explain the above steps in more detail.

STEP (I): CALCULATING THE MEAN CLASSIFICATION ERROR FOR EACH MEASUREMENT   For each experiment shot , we have a soft measurement outcome for each measurement , given by , and associated inferred state after measurement, . As discussed in the main text, in our case, this is found by taking $= 1$ if $( \quad 1) \quad ( \quad 0)$ and $0$ otherwise. The PDFs $( \quad )$ are obtained by keeping only the dominant Gaussian in the measurement classification PDFs. We can calculate the estimated averaged probability of a classification error in measurement , , via

$$= \frac{1}{} \quad = \frac{1}{} \quad \frac{(1 \quad )\ '( \quad 1 \quad )}{( \quad )\ ( \quad ) + (1 \quad )\ ( \quad 1 \quad )} \tag{7.19}$$

where $( \quad )$ is the overall probability, across all shots, that the qubit is in state ; for example, if $= 0$, then $( \quad ) = ( \quad 0)$. In this work, we take $( 0) = ( 1) = 1 \quad 2$. Therefore, we have

$$= \frac{1}{} \quad \frac{( \quad 1 \quad )}{( \quad ) + ( \quad 1 \quad )} \tag{7.20}$$

STEP (II): CALCULATING THE MEAN CLASSIFICATION ERROR FOR EACH EDGE   Using the estimated values for  obtained with Equation (7.20), we use Equation (7.17) to obtain the mean classification error, , for each edge. In the case where an edge is only due to a single measurement classification error, we simply have $= $, for the relevant measurement . The only other case of relevance for Surface-13 is where two classification errors contribute to an edge; in this case, $= {}^1(1 \quad {}^2) + (1 \quad {}^1) \ {}^2$ for the relevant measurements  and .

STEP (III): REMOVING THE MEAN CLASSIFICATION ERROR FROM EACH EDGE PROBABILITY    We now calculate   for each edge using Equation (7.18).

STEP (IV): INCLUDING THE PER-SHOT CLASSIFICATION ERROR    We now wish to combine the per-shot soft information with   in order to obtain the full edge weight. In doing so, we mostly follow Ref. [202], with the difference that we merge edges that result in the same defects into a single edge. In everything that follows, we are considering a single experiment shot and thus, to reduce notational clutter, drop the index   from above.

We begin by considering the full decoding problem we wish to solve. We have a set of possible errors $\mathbb{E}$ and consider a single error,   , to consist of all events that contribute to the same edge. This includes both classification errors and other errors. We further have a set of (labelled) soft measurement outcomes, $\mathbb{Z}$, which is the union of all sets $\mathbb{Z}$ , where $\mathbb{Z}$ is the set of measurements whose incorrect classification leads to the same defect combination as   . We wish to find the combination of errors $\mathbb{D}$ that explains the observed defects and maximizes

$$(\mathbb{D} \mid \mathbb{Z}) = \frac{(\mathbb{D} \quad \mathbb{Z})}{(\mathbb{Z})} \qquad (\mathbb{D} \quad \mathbb{Z}) \tag{7.21}$$

where $\mathbb{D} \quad \mathbb{Z}$ is the event that the combination of errors $\mathbb{D}$ occur and the soft measurements outcomes $\mathbb{Z}$ are obtained. We can ignore the denominator   $(\mathbb{Z})$ as it is a constant rescaling of all probabilities   $(\mathbb{D} \quad \mathbb{Z})$ and thus does not need to be considered in order to find the most likely error.

Assuming independence of events, we split   $(\mathbb{D} \quad \mathbb{Z})$ into individual terms for each edge so that

$$(\mathbb{D} \quad \mathbb{Z}) = \prod_{\in \mathbb{D}} ( \quad \mathbb{Z} ) \prod_{\in \mathbb{D}} ( \quad \mathbb{Z} ) \tag{7.22}$$

Rearranging, we find

$$(\mathbb{D} \quad \mathbb{Z}) = \prod_{\in \mathbb{E}} ( \quad \mathbb{Z} ) \prod_{\in \mathbb{D}} \frac{( \quad \mathbb{Z} )}{( \quad \mathbb{Z} )} \prod_{\in \mathbb{D}} \frac{( \quad \mathbb{Z} )}{( \quad \mathbb{Z} )} \tag{7.23}$$

where, again, we can drop the term that is common to all error combinations. Maximizing   $(\mathbb{D} \quad \mathbb{Z})$ is equivalent to minimizing

$$-\log[ (\mathbb{D} \quad \mathbb{Z})] = \sum_{\in \mathbb{D}} \log \left[ \frac{( \quad \mathbb{Z} )}{( \quad \mathbb{Z} )} \right]_{\in \mathbb{D}} \tag{7.24}$$

where we have defined

$$= \log \left[ \frac{( \quad \mathbb{Z} )}{( \quad \mathbb{Z} )} \right] \tag{7.25}$$

Let us now consider a particular error,   , and its   associated soft measurements   for   $= 1$       . We recall that, in the Surface-13 case,   , which is the number of classification errors that contribute to edge   , is a maximum of two, and we use this below. In order to calculate   , we split   into two:   , which consists of classification errors only, and   which consists of all other errors. There are now two ways in which can occur: (i) occurs and does not occur (i.e. there are an even number of classification errors); (ii) does not occur and does occur (i.e. there are an odd number of classification errors). Therefore,

$$( \quad \mathbb{Z} ) = ( \ ) ( \quad \mathbb{Z} ) + ( \ ) ( \quad \mathbb{Z} ) = ( \quad \mathbb{Z} ) + (1 \quad ) ( \quad \mathbb{Z} )$$

(7.26)

$$( \quad \mathbb{Z} ) = ( \ ) ( \quad \mathbb{Z} ) + ( \ ) ( \quad \mathbb{Z} ) = ( \quad \mathbb{Z} ) + (1 \quad ) ( \quad \mathbb{Z} )$$

(7.27)

where we have defined $( \ ) = $ . The probability of obtaining the observed soft measurement outcomes and having an *odd* number of classification errors is

$$( \quad \mathbb{Z} ) = \begin{cases} ( \quad 1 \quad ) & = 1 \\ ( \quad 1 \quad ) ( \quad ) + ( \quad ) ( \quad 1 \quad ) & = 2 \end{cases}$$

(7.28)

and the probability of obtaining the observed soft measurement outcomes and having an *even* number of classification errors is

$$( \quad \mathbb{Z} ) = \begin{cases} ( \quad ) & = 1 \\ ( \quad ) ( \quad ) + ( \quad 1 \quad ) ( \quad 1 \quad ) & = 2 \end{cases}$$

(7.29)

These expressions can easily be extended to larger values of , but we omit the general expressions here for brevity. From these, we calculate the edge weight using Equation (7.25) and find

$$= \begin{cases} \log \dfrac{1}{1} & = 1 \\ \log \dfrac{1 \quad 2 \qquad 1 \quad 2}{1 \quad 2 \quad 1 \quad 2} & = 2 \end{cases}$$

(7.30)

where

$$= \frac{}{1}$$

(7.31)

$$= \frac{( \quad 1 \quad )}{( \quad )}$$

(7.32)

### 7.6.6　Calculation of logical error rate

To extract the logical error rate from experimental data, we calculate the logical fidelity ( ) for each round of the experiment and fit the data to a decay curve of the form given in Equation (7.2). The error in the logical fidelity is given by $_L = (1 \quad )$ , with the number of samples for the given [203], which we propagate through the fitting process to get estimates of uncertainty in  and the offset  .

### 7.6.7   Additional logical error rate figures

We show additional plots of the logical fidelity of the soft and hard MWPM decoders for each round of the experiment in Figure 7.11. To illustrate the improvement that soft information gives to logical fidelity, Figure 7.12 shows the absolute difference in logical fidelity  ( ) between the soft and hard MWPM decoders for each round of the experiment. The average performance is shown in solid lines, and the fidelity for each individual state preparation  0 is shown in the transparent lines.

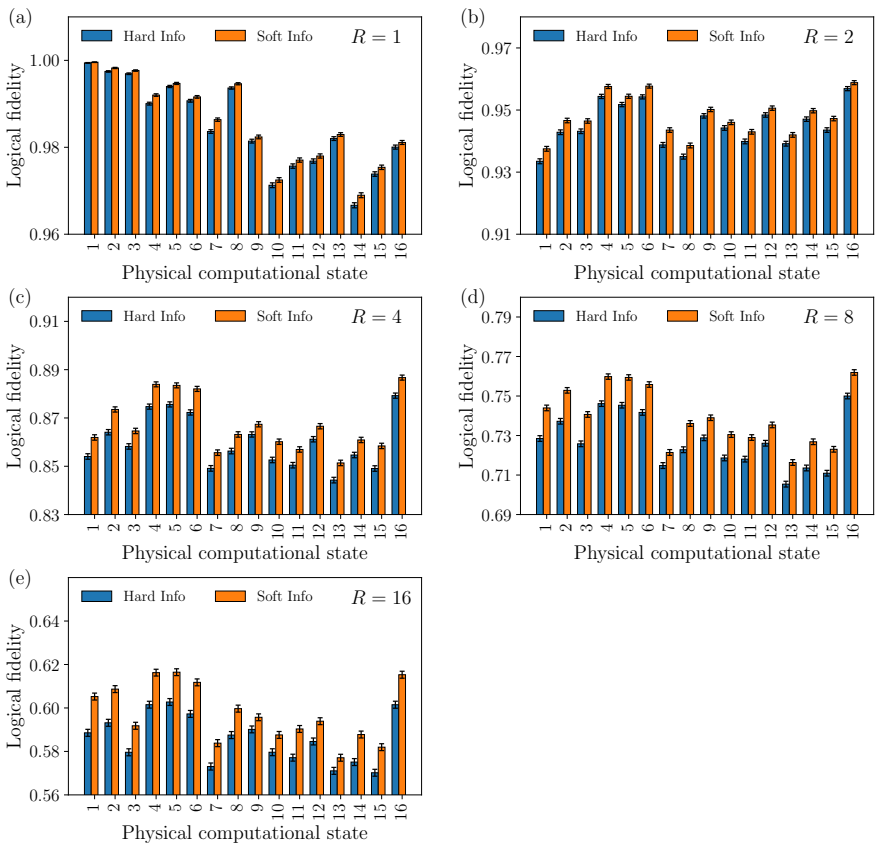

Figure 7.11: Logical fidelity using soft versus hard MWPM decoder for each round of the experiment.

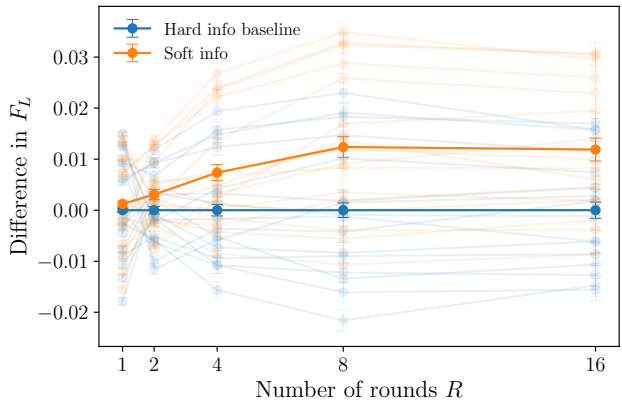

Figure 7.12: Absolute difference in logical fidelity     as a function of rounds, shown for each individual logical state preparation $|0\rangle$ (transparent) and on average across 16 states (opaque) for soft versus hard MWPM.

# CONCLUSION & OUTLOOK

<div style="text-align: right">8</div>

In this final chapter, I provide a summary of the findings from each chapter presented in this thesis. Driven by the obtained results and discussions, I reflect on the key challenges and issues that must be addressed to pave the way for the development of a practical quantum computer in the future. It is important to note that these reflections represent my personal opinions derived from the work presented in this thesis.

**Conclusions**

The work presented in the thesis focuses on advancing the implementation of the surface code for quantum computation based on superconducting transmon qubits. To this end, this section summarizes the main results and conclusions of each chapter.

- Chapter 1 introduced quantum computation and essential components to build a small-scale quantum processor based on surface-code architecture and superconducting transmon qubits.

- Chapter 2: The two-qubit gate is often the dominant source of errors in many quantum algorithms. This chapter introduced a diabatic two-qubit scheme called SNZ gate, designed for implementing flux-based CPhase gates in four pairs of nearest neighbors within a subset of a 7-qubit device. Beyond speed, the key advantage of SNZ was tuneup simplicity, owing to the regular structure of conditional phase and leakage as a function of the flux pulse parameters. SNZ could also be generalized for arbitrary CPhase gates, proving beneficial for QEC in presence of high residual $\quad$ couplings (see Chapter 4), and NISQ applications. The highest performance achieved from one 2QIRB characterization had $99\,93\quad0\,24\%$ fidelity and $0\,10\quad0\,02\%$ leakage.

- Chapter 3: Leveraging the capabilities of a 7-qubit device, referred to as Surface-7, in this chapter, we realized a distance-2 QED code using post-selection techniques. Furthermore, we showed a complete suite of logical operations, including initialization, measurement, and single-qubit gates, with higher performance observed for fault-tolerant variants over non-fault-tolerant ones. We also demonstrated process tomography of logical gates, using the notion of a logical Pauli transfer matrix. At the time of its introduction, this integration of high-fidelity logical operations with a scalable scheme for repeated stabilization reached a milestone on the road to QEC with higher-distance superconducting surface codes.

- Chapter 4: Automating the calibration and benchmarking of superconducting quantum computers is crucial for scalability and addressing continuous parameter drifts. This chapter detailed calibration strategies and benchmarking for a 17-transmon device, utilizing automatic framework known as GBT. This approach streamlined the calibration of single-qubit gates, two-qubit gates, readout, and stabilizer measurements. Common failure modes were discussed, particularly regarding strong interaction with two-level-system defects. Additionally, the chapter presented a comparative analysis of individual and simultaneous device performance to highlight the impact of crosstalk.

- Chapter 5: Leakage hampers the ability of QEC to detect and correct errors. In this chapter, we implemented and extended a simple all-microwave scheme to minimize leakage known as LRU for transmons in a circuit QED architecture, initially proposed by Battistel *et al.*, [PRX Quantum 2021]. This LRU effectively reduced leakage in the second- and third-excited transmon states with up to $99\%$ efficacy in $220\,\mathrm{ns}$, with minimum impact on the qubit subspace. As a result, this LRU scheme could be simultaneously applied to both data and ancilla qubits. As a showcase in the context of a QEC setup, we showed how multiple simultaneous LRUs could reduce the error detection rate and suppress leakage buildup within $1\%$ in data and ancilla qubits over 50 cycles of a weight-2 stabilizer measurement.

- Chapter 6: A quantum circuit is sub-optimally constructed if simply compiled from one- and two-qubit gates and single-qubit readout tuned in isolation. This chapter introduced highly parallel and orthogonal calibration strategies known as VC and HC, for the optimal tuneup of the QEC cycle in a distance-3 surface code. This approach facilitated the implementation of a bit-flip distance-3 surface code with 13 qubits, referred to as Surface-13. This code served as a testbed for investigating decoding strategies and quantifying the link between logical and physical qubit performance. Additionally, we attempted to stabilize the logical state $0$ , using   and   stabilizers and assess the measured defect rates for multiple rounds.

- Chapter 7: Traditional error decoding approaches, relying on the binarization ('harden- ing') of readout data, often ignore valuable information embedded in the analog ('soft') readout signal. In this chapter, We used soft information into the decoding process of a distance-3 bit-flip surface code with transmon qubits. To this end, we used the 3 3 data-qubit array to encode each of the 16 computational states that comprise the logical state $0$ , and protect them against bit-flip errors by performing repeated  -basis sta- bilizer measurements. To infer the logical fidelity for the $0$  state, we averaged across the 16 computational states and employed two decoding strategies: minimum weight perfect matching and a recurrent neural network. Our results showed a reduction of up to $6 8\%$ in the extracted logical error rate with the use of soft information.

**Outlook**

The key challenges discussed in this section are not ordered by their relative importance but rather follow the structured approach outlined in Section 1.3 for building a full-stack super- conducting quantum computer. This progression moves from design choices and fabrication to the calibration of high-fidelity operations, all of which are essential for realizing a scalable quantum computer and achieving fault-tolerant quantum computing.

## 8.1   From microsecond to millisecond   1

Using density matrix simulation, our group has previously explored ways to optimize the logical performance of Surface-17 and extended to Surface-49 [134]. Given the current physical per- formance discussed in Chapter 4, both Surface-17 and Surface-49 fall short of reaching the computational break-even point, making fault-tolerant quantum computing (FTQC) unattain- able. This holds true even without accounting for more challenging errors such as leakage and residual   coupling. To achieve FTQC, Surface-49 necessitates improvements in readout speed and increased qubit relaxation times. This section is dedicated to understanding and improving qubit coherence in our surface-code architecture.

Dielectric losses, primarily attributed to TLS defects, have consistently been identified by the superconducting community as the dominant relaxation channels [36, 143–145, 222]. The qubit decay rate, denoted as $\Gamma$, is defined as a function of the intrinsic losses (or loss tangents $\tan\delta$) of materials in the bulk and across interfaces [144]:

$$\Gamma = \Gamma_i + \Gamma_{other} = \omega_q \, p \tan\delta + \Gamma_{other} \tag{8.1}$$

where $\omega_q = 2\pi f_q$ is the qubit transition frequency, $p$ is participation ratio, and $\Gamma_{other}$ represents all other decay channels including radiative decay [36, 144], cosmic ray events [25, 223–225], and more.

If the qubit coherence is primarily limited by the loss tangent of the bulk materials, which is typically in the range of $10^{-6} : 10^{-8}$ at cryogenic temperature [86], this should allow for several milliseconds of $T_1$ [145]. In contrast, amorphous native oxides at interfaces exhibit much larger loss tangent $10^{-3}$ [86]. This strongly suggests that qubit coherence is predominantly limited by microwave losses due to uncontrolled defects at surfaces and interfaces. Qubit relaxation is also influenced by various other factors such as participation losses [144], transmon geometry [226], capacitor design [86], quasiparticle tunneling [227], coupling to spurious modes [228], flux coupling, and radiative decay [36]. However, these factors are beyond the focus of this section.

As a future area of investigation, I propose focusing on dielectric losses at interfaces. The critical interfaces in our design include substrate-metal (SM), substrate-air (SA), metal-air (MA), and metal-metal (MM) interfaces. While our group has previously studied and improved intrinsic losses by SA and SM interfaces [139], I suggest focusing on the optimization of the MA interface in this proposal.

Many groups have been studying the natively-grown oxides on top of various superconductors: niobium [144, 229] (Nb), tantalum (Ta) [145, 154], and aluminum (Al) [145]. They have also investigated the impact of the superconducting film morphology and various deposition techniques [230]. Notably, recent reports have reached sub-milliseconds $T_1$: $0.114$ ms by Rigetti [144], $0.3$ ms by the Houck group [145], and $0.5$ ms by Wang *et al.* [154]. Regardless of the fabricated superconducting film materials, the key question may lie in the grown complex oxide on top of the metal-to-air interface. This probably explains the improvement for tantalum film, known to have less grown oxides than others superconductors [145, 154].

In our group, we attempted to improve qubit coherence by cleaning the chip surface from its oxides just before cooldown in the fridge [231]. However, this was an uncontrolled processor-like shot in the dark. I believe more can be done in this direction. For example, I propose the following:

1. Conduct a controlled study using characterization tools such as X-ray photoelectron spectroscopy (XPS), transmission electron microscopy (TEM), and energy dispersive spectroscopy (EDS) measurements to analyze the complex oxides on top of the NbTiN surface in our chips. This analysis is crucial for extracting information about the oxide components, their thickness, and their oxidization rate. Such information is highly needed to design a controlled recipe for removing the target layers. Otherwise, there is a risk of increasing surface roughness, which could lead to worse performance after this treatment [229].

2. To simplify the experimental overhead, utilizing a single lithographic layer, specifically the patterned NbTiN film in our Surface-17 layout, is sufficient. This approach aims to quantify losses by only measuring the intrinsic quality factors of the test resonators, following the procedure outlined in [139].

3. I believe that the NbTiN contact profile, rather than just the surface, likely influences coherence times. The Rigetti group has demonstrated an enhancement of   by more than $50\%$ through careful engineering of the contact profile of the Nb film [144]. I anticipate that similar improvements can be achieved for NbTiN films.

4. Another potential avenue is exploring the passivation of the metal surface with thin layers of noble metals such as gold (Au) or silver (Ag) or with self-assembled monolayers (SAMs). The latter is extensively used in organic chemistry and thin-film electronics to control the surface chemistry on the metal contact [232]. By growing well-controlled and single-crystalline monolayers, the goal is to passivate dangling bonds and prevent the formation of native oxides. The choice of monolayer material should be carefully considered to have a lower loss tangent than that of complex amorphous oxides.

In the investigation of the qubit coherence limits in our device, I think it is crucial to re-evaluate the imposed Purcell limit due to the microwave drive line. This is particularly important as we have progressively increased the microwave coupling capacitance from $0\,1$ fF to $0\,25$ fF over the course of my PhD, ensuring fast single-qubit gates in $20$ ns. However, using a simple expression for a grounded transmon:

$$\frac{1}{\quad\quad} \tag{8.2}$$

where   is the total capacitance seen by the transmon,   is the characteristic impedance,   is the qubit transition frequency, and   is the microwave coupling capacitance. Through finite element simulation (COMSOL), we extract an approximate value of $65$ fF for  . With   $= 50\quad$,   $= 0\,25$ fF, and   $= 6$ GHz, this suggests   $16\quad$s. While this is likely an oversimplification, considering our transmon is not grounded (floating), it highlights a concerning limit imposed on   .

This is likely not the dominant factor affecting qubit coherence in our device, as we observed a linear dependency of device coherence on qubit frequencies, indicating that dielectric loss is the main suspect. However, Purcell decay due to coupling to a drive line can still set an upper bound on device coherence. Therefore, conducting a controlled experiment where the microwave coupling capacitance is systematically varied across the same device would be valuable. The sensitivity of this experiment depends on     ; higher coherence reflects a more realistic impact of microwave coupling capacitance.

## 8.2   Two-level systems defects

Strong interaction with TLS defects currently stands as a major bottleneck for the successful implementation of large-scale QEC experiments using transmon-based quantum processors [26, 141, 167, 200]. As the number of qubits scales, the probability of encountering these defect modes highly increases, posing a significant challenge for the future of superconducting-based quantum computers.

The presence of TLS defects is commonly explained within the framework of the standard tunneling model (STM) in amorphous materials at low temperatures [43]. These defects do not necessarily result from impurities in the material but emerge due to disordered structures of the barrier oxides in the Josephson junctions. Typically characterized by low energy, these defects add an additional degree of freedom and can dominate at low temperatures, while they become saturated at high temperatures. Moreover, TLS defects commonly switch between their two eigenstates and exhibit dynamics due to decoherence, coupling to the environment, or thermal activation. This dynamic behavior causes fluctuations in both time and frequency [233, 234].

Current techniques to mitigate the impact of TLS defects are quite short handed. These include waiting for the TLS to go out of the interaction, detuning the qubit away in frequency from the TLS (similar to developed techniques in Chapter 4), or playing the thermal roulette through a thermal cycle and hope the device will cool down in a different configuration. Many research groups have characterized TLS dynamics using qubit spectroscopy techniques [141, 233–235] (similar to Fig 4. 17 in Chapter 4). They have also studied the response of TLS to applied strain [222], DC electric fields [222, 234, 236, 237], temperature variations [139], and background ionizing radiation [225]. Several studies have localized strong defects in the tunnel barrier or at the edge of widely used Al/AlO$_x$/Al Josephson junctions [234, 238]. However, the existing techniques, at best, can induce jumps and scramble the TLS locations in frequency without providing guarantees against their return, the emergence of new TLS defects, or individual control (rather than applying external control to the entire chip).

As a future study, I think it will be interesting to integrate a laser annealing setup in the fridge in which the qubits can be annealed in-situ. The idea is to use the laser beam to manipulate the atomic arrangement in the capacitor pads of the transmon and across the junction [151, 239–241]. The aim is to explore whether this process could induce local rearrangement or elimination of TLS defects, providing an avenue to study their behavior at low temperatures. However, it is not yet quite clear if laser annealing will effectively help the re-arrangement of atoms across the junction, where strong TLS defects concentrate. It is also fair to say that the impact of laser annealing technique is not thoroughly understood at the microscopic level. Additionally, engineering challenges also abound, particularly in applying a laser beam in a low-temperature setting without excessive heating and precisely aligning the laser to target the qubit area.

Alternatively, placing a bottom electrode below the qubit chip can be used to induce a local electric field, potentially tuning coherent defects away from qubit transition frequency as suggested by Lisenfeld *et al.* [237]. To minimize crosstalk, the qubits should be spaced at least $100 \, \mu m$ apart, which is a viable spacing with current design. However, this approach introduces additional sources of decoherence due to capacitive coupling with the gate electrode, and would require complex fabrication techniques for implementation.

## 8.3 The Path to a 100-qubit device and beyond

The path to achieving a 100-qubit device and beyond requires the implementation of 3D integration and packaging technologies. This need arises from the limitations of lateral and planar layouts, including issues such as high crosstalk couplings, interconnect crowding, spurious packaging modes, and limitations in chip size [242–244]. Introducing a third dimension offers advantages in signal routing and interconnects, enabling clever integration of multi-layers, e.g., readout and control lines on separated chips than the qubits.

However, 3D integration introduces additional fabrication steps, that can impact device coherence by increasing surface interfaces, and correspondingly higher defect density. Moreover, the precise alignment between multilayers becomes crucial to avoid significant targeting errors, ensure optimal readout, and minimize crosstalk. These pose significant engineering challenges, and only a few groups worldwide have achieved this milestone, including IBM Quantum [48], Google Quantum AI [26], MIT Lincoln Labs [242, 243], Rigetti [245], and recently the ETH group [244].

As a first step of 3D integration, our fabrication team attempted to fabricate planar surfaces for 7- and 17- qubit devices with vertical input/output (I/O) features employing 3D interconnects, i.e, through-silicon vias (TSV) in flip-chip architectures [98]. We experimented with TSV fabrication both before and after the deposition of the superconducting base layer, NbTiN, to evaluate the impact on device yield and qubit coherence. In presence of big 3D objects like TSVs, which are typically a few millimeters in size, the team conducted numerous optimizations in the receipt to improve device yield and coherence. We managed to achieve an average device coherence between $10$ and $20 \, \mu s$; however, variations in photoresist thickness resulted in significant targeting and yield issues. Eventually, the complexities and extensive time requirements of fabricating multi-qubit devices with TSVs led us to revert to simpler lateral layouts. While this decision was pragmatic and suitable for a few-qubit device, it is not conducive to the future goal of achieving 100-qubit device and beyond. A reassessment of this vertical I/O approach is now imperative.

As a future study, i think it is an interesting avenue to investigate the less common microstrip resonators structures, leveraging the benefits of 3D integration. This approach offers several advantages, including simplified interconnects, an airbridge-free process, the ability to clean the surface just before cooldown, and reduced sensitivity to surface defects. Moreover, the microstrip structure allows for tunable coupling by adjusting the position within the waveguide. As a showcase, Zoepfl *et al.* [246] demonstrates a high intrinsic quality factor, exceeding 1 million, in the low-power regime.

## 8.4   Automatic calibration

Chapter 4 outlined our approach to autonomously calibrate the building blocks of the QEC cycle in isolation using an automatic framework. However, extending this to a large-scale, continuously-running quantum computer poses significant challenges beyond the capabilities of the current routine. Several factors contribute to this limitation. First, the performance of individually-tuned quantum operations can degrade due to various crosstalk effects during simultaneous implementation. Second, the continuous drift of control parameters and the need for regular re-calibration make the calibration process tedious and time-consuming. Third, as the device scales up, interactions between qubits and defect modes become inevitable, limiting gate performance. Last, the challenge of microwave hygiene, particularly in presence of increasing spurious images for individual qubit control is also significant. Additional challenges include frequency crowding problems, dependencies of various errors on qubit frequency trajectories, especially during two-qubit gates, the order of the calibration process, inter- and state dependencies between neighboring qubits.

The complexity of the calibration process and the aforementioned challenges highlight the necessity for a parallel optimizer. Recently, the Google Quantum AI team demonstrated their optimized calibration strategies for measurement [177] and two-qubit gates [167] using a home-built optimizer. Instead of exploring a large parameter space through a slow change and measurement loop, they rely on a model-based approach to guide their calibration process. This substantially reduces calibration time, considers the complexity of simultaneous operations, and is scalable beyond hundreds of qubits.

I advocate for a similar approach to be highly essential for practical quantum computers. Fortunately, we have developed various realistic models using density-matrix simulations to predict and understand dominant errors in our device [18, 69, 72, 77, 94, 103, 107, 134, 175, 247]. I propose integrating these models into a GBT framework with input parameters from experiments. These can namely include qubit transition frequencies, anharmonicities, readout frequencies, TLS densities and more. Implementing a modular software requires substantial software development expertise, a task that exceeds the capabilities of a research lab. Instead, I suggest leveraging the benefits of the existing quantum ecosystem in Delft and collaborating with OQS and Qblox to build such an optimizer.

## 8.5   Measurement tuneup and measurement-induced transitions

Measurement is an essential building block of any quantum algorithm, especially those involving repeated measurements like QEC codes. State-of-the-art QEC experiments highlight the significance of minimizing measurement errors, often dominant tied with two-qubit gate fidelities [26, 141]. Currently, readout calibration routines often involve exploring large parameter space, measurement-based approach, to reduce measurement errors. This process becomes tedious and time-consuming, especially when scaling up the number of qubits. Additionally, it frequently identifies readout control parameters in the non-linear regime which may cause severe measurement-induced transitions to higher leakage states and not necessarily to the second-state [61, 135, 146, 175]. Constraining readout amplitudes below a certain threshold, known as the critical photon number, is an oversimplified approximation and an inadequate strategy for readout calibration in a multi-qubit quantum processor [61]. Other challenges in calibrating readout include poor hybridization between readout resonator modes, measurement crosstalk, limitations of classical hardware, and more. Due to all previous reasons, readout calibration remains a non-trivial problem.

As a future study, I propose employing a model-based approach for readout calibration in a regime where the resonator frequency is above the qubit frequency. This model should consider various error models, including qubit relaxation, average photon number, photon depletion time, off-resonance transitions of the coupled transmon/resonator system, two-level system defects, and more. Some of these models rely on acquiring experimental data through small-scale experiments which then incrementally incorporated to guide readout calibration. This approach has proven useful for the Google Quantum AI team, where they calibrate 17-qubit and 72-qubit devices using a similar model-based approach but in a slightly different regime, where the resonator frequency is below the qubit frequency [61, 177]. Another interesting avenue is to analyze the evolution of IQ shots for the acquired voltages over time and use them as input to a global optimizer like a neural network to search for a global minimum. While this approach doesn't rely on physical error models, it may get stuck in local minima or take a long time to achieve optimized readout.

## 8.6   Leakage

Leakage poses a significant challenge for QEC codes as they rely on detecting binary outcomes (either zero or one) for error detection and correction. Furthermore, leakage can result in long-time correlations that persist over many QEC rounds, leading to a substantial accumu-lation of errors as circuit depth increases [26, 141, 146, 175]. In Chapter 5, we implemented and extended a simple all-microwave leakage reduction units (LRUs) for transmons in a circuit QED architecture, originally proposed by Battistel *et al.* [128]. In a more advanced QEC setup (Chapter 6), LRUs effectively removed leakage in a 13-qubit system, stabilizing the logical zero state at lower error rates.

Simultaneous reduction of leakage across multiple qubits can be challenging in presence of high microwave crosstalk. The latter can cause LRUs to introduce leakage if the LRU drive aligns with a leakage transition. In addition, optimized control parameters of the LRU pulse may change when operating simultaneously, in case the induced AC Stark shift is larger than the linewidth of the LRU transition. In this case, a simultaneous optimization is needed.

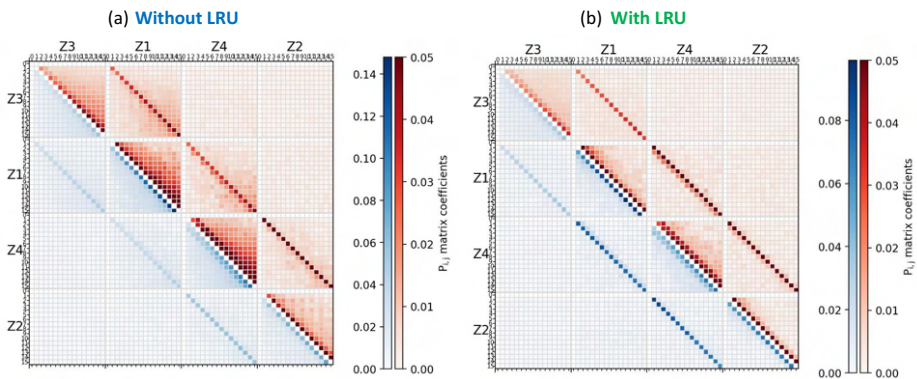

Figure 8.1: **Correlation matrix without and with LRU** Stabilization of the logical zero state without (a) and with (b) LRU applied on high frequency qubits, D , D , D . The application of LRU extends the QEC cycle by $240\,\mathrm{ns}$, with the P  matrix in (b) showing reduced long-term correlations when LRU is applied.

From a hardware perspective, LRUs are quantum-hardware efficient as they do not require additional quantum hardware changes. However, They are certainly not classical-hardware efficient. The microwave drives for single-qubit gates and LRUs may have a significant frequency difference, often beyond the bandwidth of typical AWG. As a result, these can not be synthesized with same hardware components—such as mixers, AWG channels, and microwave sources—thus increasing the complexity of the hardware setup. Achieving high microwave hygiene becomes crucial, given the risk of spurious images overlapping with qubit trajectories or interfering with unwanted transitions. The on-chip microwave isolation also imposes constraints on the selection of LRU drive frequencies.

Developing an automatic optimizer for LRU calibration on multiple qubits is beneficial, taking into account several factors such as unwanted transitions for each qubit, qubit-qubit avoided crossings, coupled qubit/resonator interactions, worst-case microwave crosstalk, and spurious mixer images. The optimizer may need to perform iterative calibration to fine-tune LRU performance in parallel. Additionally, incorporating microwave filters with appropriate cutoff frequencies is crucial to highly attenuate the unwanted frequency range.

To investigate the impact of LRU on the long-lived correlations, we measure the P matrix, with and without LRU when initializing the logical zero state and perform 15 QEC rounds (Figure 8.1). As expected, long-term correlations are effectively reduced by applying the LRU, especially for weight-4 stabilizers Z and Z, as shown in panel b. Additionally, the magnitude of correlations decreased from $14\%$ to approximately $9\%$. Next, the new PhD generation is in excellent position to perform a distance-3 surface code and investigate the impact of applying LRU on the logical qubit performance. As LRU converts leakage events into Pauli errors (can then be corrected), one would expect logical error rate to improve as applying LRUs on all qubits. However, simultaneous LRU may induce bad impact as earlier mentioned.

## 8.7 Crosstalk

### 8.7.1 Mitigation of microwave drive crosstalk

In Chapter 4, our investigation into microwave crosstalk in a 17-qubit device revealed specific pairs where crosstalk levels reached as high as 7 dB. By linking such pairs to the device geometry, I identified crossovers involving microwave drive lines bridging over resonator buses to be the main reason. Poor isolation can lead to multiple types of error [26, 153]. When a pulse is applied to a qubit through its dedicated drive line, other qubits in the same frequency group experience both transversal and longitudinal errors. Qubits in the other two frequency groups experience mainly a longitudinal error (from the AC Stark shift). When qubits are simultaneously driven (which happens repeatedly when executing the surface-code QEC cycle), the errors are generally both transversal and longitudinal regardless of the frequency relation between the qubits. Microwave crosstalk can also induce leakage if the qubit transition frequency of a driven qubit is close to the leakage transition of another qubit [26, 141].

In collaboration with our design team, microwave simulations have confirmed that the transmission from a qubit microwave drive line is only 7.7 dB greater than the transmission from a line that bridges a connected resonator bus. This poses a significant limitation in our planar device design. Using simulation, we explored the effects of various factors on microwave crosstalk, including the dimensions (width and height) of crossover airbridges, the use of double airbridges, and their placement at voltage nodes. Our results showed only minor impact on calculated microwave crosstalk, with a marginal reduction observed when increasing the airbridge height from $2$ m to $10$ m. The challenge lies in the fact that while the coupling between two bridged lines is minimal (approximately -60 dB), so too is the coupling of the microwave drive line to the qubit (about -52 dB), especially when compared to the coupling of the qubit to the bus over which some microwave drive lines are routed.

Previous work has proposed calibration techniques to mitigate high microwave crosstalk through interferometric cancellation [26, 141]. However, this method faces scalability challenges, as it necessitates n physical connections and 2n control knob tuning parameters, where n represents the number of qubits.

As future prospective, I suggest the following research studies:

1. Decrease the coupling of the crossover airbridges by increasing their corresponding heights or lengths. Simulation predicts a few dB reduction in microwave crosstalk.

2. Increase the target coupling of the microwave drive line to its own qubit. However, this approach may impose a Purcell limit on qubit relaxation times, as discussed in Equation Equation (8.2).

3. Explore the use of TSVs to connect on-chip ground planes, thereby eliminating the need for crossover airbridges. It is important to note that fabricating TSV introduces additional fabrication steps that may impact qubit relaxation times and device yield.

4. Place the crossover airbridge at a voltage node may not fully resolve the issue, as suggested by simulation. While this could cancel the first mode of the transmitted field to the crossed resonator bus, the zero-mode may still couple the drive line to the crossed resonator. To address this, one can explore cancellation of the zero-mode by filtering it out using on-chip filter or by adjusting the dimensions of the co-planar waveguide structure asymmetrically to break the symmetry and thus suppressing the zero mode.

### 8.7.2 Mitigation of residual      coupling

Residual-      crosstalk between qubits involved in a two-qubit gate and their spectators can introduce coherent phase errors and significant leakage errors [248]. This manifests in dramatically increased detection error rates when implementing simultaneous stabilizer measurements in a distance-3 code (see Chapter 6). We attribute these higher error rates to the elevated      couplings experienced when running X- and Z-stabilizers in parallel.

In our latest Surface-17 designs, we have taken steps to mitigate this issue by reducing      couplings by a factor of four. This reduction is achieved by halving the transverse nearest-neighboring couplings, resulting in a slowdown of two-qubit gate times by a factor of two. Despite these efforts, current      couplings remain relatively high, leading to correlated errors that violate locality and compromise the ability of QEC scheme in locally detecting and correcting errors. Furthermore, residual      couplings during two-qubit gates are significantly higher at interaction points compared to bias points, often by few times to an order of magnitude. These couplings can not be effectively refocused. The presence of always-on static couplings further complicates parking frequency calibration, making it challenging to choose optimal parking frequencies for spectator qubits during two-qubit gate operations.

Several strategies have been proposed in literature to mitigate     crosstalk including employing hybrid physical qubit platforms with opposite anharmonicities [92, 249, 250], the use of engineered multi-path coupling with fixed [251] and tunable elements [252], introducing tunable coupler schemes to dynamically turn the interaction on and off [13, 89, 90, 253], and the use of simultaneous AC-Stark shift on coupled qubits [254]. All previous strategies come at the price of increasing hardware or software complexities. Tunable coupler schemes have proven to work with tens of qubits [13, 26]. Although it increases hardware complexity and scales up poorly, it highly suppresses residual     errors. I think it is the best strategy so far and compatible with larger qubit devices.

**8**

# ACKNOWLEDGEMENTS

My PhD has been an incredible journey, filled with learning, challenges, and unforgettable moments. But above all, it was the people I met along the way who made it so special. I am deeply grateful to everyone who supported, challenged, and laughed with me throughout this experience.

First, I would like to thank my promoter, **Leo**. Thank you for teaching me quantum physics and circuit-QED. Coming from a different background, my PhD was anything but conventional, and I truly appreciate your patience (and occasional pressure!) in pushing me toward a deeper understanding. I have always admired your Socratic approach—asking about things that seemed trivial to me, only to reveal they had been the focus of entire research efforts years ago! Your dedication, commitment, and motivation to work hard every day have been truly inspiring. We had a challenging PhD project, but I always valued the moments when we were open, honest, and truly listening to each other. While the project may not have been as rewarding as we initially hoped, I have no regrets—I am really glad I pursued this PhD with you in this field. Thanks also go to my co-promoter, **Barbara**. I am grateful for the many insightful discussions that bridged theory and experiment. Experimentation is tricky—many factors can influence the outcome—but you and your group were always available, offering guidance and valuable insights. I truly appreciate how you encouraged us to run controlled experiments, which helped in understanding and fitting experimental data.

A PhD journey is never complete without good companions, and I was lucky to have amazing ones. **Jorge** & **Andreia**, you were more than just colleagues—you were PhD buddies and great friends. Jorge, thanks for introducing me to the Matplotlib gangster mindset and helping me make my plots look good! Your presentations were always top-notch, and I admired your ability to think through problems and simplify life. It was really a joy working with you, my PhD buddy. Andreia, thank you for all the great times outside the lab. Hosting us at your place, bringing us together—it was always so much fun, and I will always cherish those memories. You know Jawahir and I miss you both and wish you all the best! **Miguel** & **Leonor** – Miguel, the most organized person I have ever met! Winning an argument against you is nearly impossible, as you are always well-prepared and knowledgeable. Your dedication and organizational skills (labeling everything!) are truly impressive. I wish you all the best in your PhD! Leonor, thanks for all the great times we spent together and for hosting us at home—it was always a blast. I truly enjoyed our gatherings, the great BBQ, and food.

**Thijs** – the guy who did everything in his PhD! I always loved discussing physics with you. Your deep understanding of fundamental physics and our scientific discussions were truly valuable. I really enjoyed our conversation on how to get things better in fab! **Boris** – The theory gangster! You are incredibly resourceful and can talk about research for hours—and I love that about you! Our discussions were always insightful, and whenever I struggled to grasp something, I knew I could count on you to help me figure it out. I have no doubt you will have a fantastic postdoc experience and a successful career ahead. **Victor** – My first project buddy! Working on the SNZ project together was intense but rewarding. I will never forget our dedication—taking shifts to operate the device 24/7 and writing the entire SNZ manuscript in a week! Your hard work and determination are truly admirable.

**Ruggero** & **Sara** – Ruggero, the Italian gangster! The King of Sicily! Even though we only overlapped for a short time, it feels like we have known each other forever. You are one of the most fun people to be around, from teaching me Italian hand gestures to discussing everything from food to how to get a good Chevron and do intense cardio! Now that you are in your third year—enjoy the power, you can do whatever you want! Sara, though our time together was brief, it was always a joy. Thanks for the fun moments we shared! **Santi** – The Spanish dude! We have known each other since your MSc, so it feels like you have been around for my entire PhD. I still remember how you took ownership of your own fridge back then! Your MSc project was tough, but you handled it like a champ, and I always enjoyed our scientific discussions. I also loved our open conversations about everything—it felt like a good venting session sometimes! Your leadership skills are on another level, and I see how much your lab mates value that about you. I am sure you have a bright future ahead!

I was also fortunate to share this journey with fantastic lab members, who made both work and life in Delft more enjoyable. **Sean** & **Zalyna**, thank you for having so much fun together inside and outside the lab that made it an amazing experience. **Ramiro**, I really appreciate your help during a difficult time of my PhD, introducing me to PycQED and helping me to run quantum measurements. For everyone I had the pleasure of working with during my time in the DiCarlo lab and beyond, **Tim**, **Nand**, **Ale**, **Marc**, **Nadia**, **Matvey**, **Martijn**, **Christos**, **Niels**, **Adriaan**, **Chris**, **Wouter**, **Bart**, **Rebecca**, **Berend**, **Tumi**, **Pim**, **Joost**, **Yuejie**, **Marios**, **Viswanath**, **Kishore**, **Matt**, **Marc**, **Francesco**, **Ophelia**, and **Earl**—it was truly a pleasure working with you all. Thank you for the amazing time we had together inside and outside the lab!

To the talented students I had the privilege of supervising during their BSc and MSc degrees—**Anouk**, **Hiresh**, **Olexyi**, and recently **Bart**—thank you for all the time we spent together solving cutting-edge research problems and delving into the intricacies of our devices. Anouk, being my first student, we both navigated the complexities of quantum physics and circuit-QEDs together. I truly appreciate our efforts in automating single-qubit gate calibration using the GBT framework—a tool I still rely on today. Hiresh, it was a pleasure working with you on the Surface-17 characterization. Although it was during COVID times, you showed up every day in the lab and demonstrated dedication to work. I also enjoyed our discussions on gym tricks to avoid injuries and build muscle. Olexyi, I had so much fun working with you on surface-17 calibration and particularly developing simultaneous calibration routines for tuning parity checks as parallel block units. Your competence, analytical skills, and programming expertise were invaluable.

I would like to express my sincere gratitude to the QuTech administration team: **Jenny**, **Marja**, and **Romina**. I deeply appreciate your support, ranging from handling visa paperwork, extensions, and other tedious administrative tasks. A special thanks to the electronics and cryogenic technical support team: **Raymond**, **Matt**, and **Olaf**. Your expertise in reducing mixed nonlinearities, as well as in the operation and upgrades of the BlueFors fridges, has been instrumental. Thank you for your dedication and assistance throughout.

Outside the lab, many people made my PhD journey enjoyable and memorable. First, I want to thank my roommate and great friend, **Hamed**. I probably would not have made it without you. You introduced me to the wonderful Delft family, who have now become dear friends. Whenever I was feeling down, you were there for me. I truly appreciated your company and our conversations about everything—from the potential of quantum computing to personal matters. You also introduced me to some great restaurants across the Netherlands. We spent so much time together outside the lab, especially during the challenging COVID times. I am really grateful for every moment and hope our paths cross again in the future. To my great friend, **Saad**, we have known each other for over ten years now, and I could write full paragraphs about our memories together. I want to say that I had an incredibly enjoyable time with you and your family. Thanks for being there for the lengthy calls and advising me about the right decisions, paperwork, and life. When we first met, your cooking skills were horrible—I appreciated the effort but not so much the food! Now, you are much more experienced, and I really enjoy your cooking.

To my Delft family—**Fathi**, **Abdulqader**, **Khobayb**, **Muath**, **Mustafa**, **Marouen**, **Abdullah**, **Motasem**, **Kadim**, **Abid**, and **Yasser**—you have been more than friends; you have been brothers. From our epic PlayStation battles to cooking together, every moment was a blast. Our trips and outings were unforgettable, and knowing you were always there meant so much for me. Thank you for all the laughter, support, and memories. To my lifelong friends—**Belal**, **Omar**, **Yehia**, **Ramy**, and **Abdullah**—thank you for always being there for me. Even though our paths have diverged, our connection remains strong. Every time we meet, we relive the amazing and most enjoyable times we have shared. Each of you has shaped my personality in unique ways. You stood by me during the most challenging times of my life, never giving up on me, and I will remember this forever. As one quote beautifully puts it, "True friendship resists time, distance, and silence."

Transitioning to life in Europe was made infinitely smoother by the great support of my friends, **Mazen**, **Hazem**, and **Ayman**. You ensured I always felt at home, sharing experiences, successes, and challenges. Especially during times when families back home would gather, we created our own family here. I deeply appreciate every moment we have shared. During one of the darkest periods of my life, I was fortunate to have the unwavering support of **Mai**, **Shimaa**, **Prof. Allam**, **Amal**, **Asmaa**, **Sakr**, **Mustafa**, and **Prof. Dessouki**. Your encouragement and assistance were instrumental in helping me navigate through those challenging times. From patiently explaining complex lecture material to helping me grasp intricate concepts, your dedication and support made a world of difference. Without your support, my journey would have been completely different. I am deeply grateful and will forever cherish your kindness.

I would also like to express my sincere gratitude to my friends at imec: **Takashi**, **Tuni**, and **Cedric**. Your guidance in conducting thorough research, enhancing my presentation and writing skills, and fostering a critical approach to our work has been invaluable. Thank you for sharing your expertise and supporting my growth during our time together. I would like to also thank my Delft friends, **Anta**, **Mohsen**, **Maria**, **Gustavo**, **Monique**, and little **Cisilia**, for the wonderful times we have shared. From our outings and football matches to having dinner together, each moment has been filled with joy and pleasure. Thank you for making my time in Delft truly memorable.

To my beloved family back home, **Mohamed**, **Heba**, **Eman**, **Tasneem**, **Abdullah**, **Mariam**, **Samah**, **Mahmoud**, **Abdelrahman**, **Ahmed**, **Ehsan**, and **Samia**, I cannot thank you enough for your unconditional love, constant support, and the countless unforgettable memories we have shared. I am truly blessed to have each of you in my life, and I am deeply grateful for the bond we share. The moments we have shared and the love that binds us will forever be cherished in my heart. Without you, life would lack meaning and purpose. You have shaped me into the person I am today, and for that, I will always be thankful. Your love and guidance have been my strength, and I will carry these memories with me always. To the rest of my extended family, thank you for being a part of my life. I am incredibly lucky to have you all.

To my incredible wife, **Jawahir**, and my precious little one, **Zayn**, I feel truly fortunate to have you both in my life. Your unwavering love, constant support, and boundless patience have been the foundation of this success. Through every late night and every weekend when I had to focus on work, telling you that I needed to stay late or push through to finish something, you both were always there, understanding and encouraging. If my thesis had co-authors, partners, or contributors, without a doubt, you would be the ones by my side. Your sacrifices, your encouragement, and your belief in me have made all the difference. Words like "thank you" seem too small to convey the depth of my gratitude, but know that they come from the bottom of my heart. The life we share together is a constant reminder of how lucky I am to have you both with me. Thank you for being my rock, my inspiration, and my everything.

Hany Ali

12–07–1992     Born in Ismailia, Egypt.

## Education

1998–2003     Primary School, Ismailia

2003–2006     Middle School, Ismailia

2006–2009     High School, Ismailia

2009–2015     Bachelor of Science in Electrical Engineering

Suez Canal University, Ismailia, Egypt.

*GPA:* A, First Class Honors

*Thesis title:* Smart solar irrigation systems

*Thesis description:* Design and fabrication of electronic circuits for a solar-powered irrigation system.

*Supervisors:* Prof. Mohamed Dessouki

2016–2018     Master of Science in Micro and Nano Electronics

ESIEE Paris, Université Paris-Est, Paris, France and imec Leuven, Belgium.

*GPA:* A, First Class Honors

*Thesis title:* Integrated high-performance organic thin film transistors (OTFTs)

*Thesis description:* Fabrication, characterization and optimization of novel organic semiconductor molecule into a circuit technology based on a bottom gate bottom contact thin film transistor topology.

*Supervisors:* Dr. Cedric Rolin and Dr. Tung Huei Ke

2019–2024     PhD in Applied Physics

Delft University of Technology, Delft, the Netherlands.

*Thesis title:* Surface-code superconducting quantum processors: from calibration to logical performance

*Thesis description:* Characterization, calibration and benchmakring of surface-code superconducting quantum processors

*Promotors:* Prof. dr. L. DiCarlo and Prof. dr. B. M. Terhal

## Professional Experience

| | |
|---|---|
| 2015–2016 | Technical Support Engineer at Elnady Company, local agent for COMSOL Multiphysics in the MENA area. |
| | Provided modeling, simulation, and technical support for COMSOL Multiphysics users. |
| 2016–2016 | Research Assistant at the American University in Cairo. |
| | Simulated and fabricated vertically aligned black silicon nanowires. |
| 2017–2018 | RD Integration Intern at imec Leuven, Belgium. |
| | Fabricated, characterized, and optimized organic thin film transistors (OTFTs). |
| 2024–now | Senior Quantum Measurement Engineer at QuantWare, Delft. |
| | Measuring and analyzing quantum devices to improve performance and scalability of quantum computers. |

## Awards

| | |
|---|---|
| 2013–2014 | Third position in "Make In Egypt" national competition for the "Smart Solar Irrigation System" project, the Egyptian Engineering Day 13th edition. |
| 2016–2018 | Postgraduate fellowship by the Al Alfi Foundation with a total of 26,000 euros for two years. |